\documentclass{aa}  

\usepackage{graphicx}
\usepackage[format=plain,labelsep=period,labelfont=bf,justification=default,singlelinecheck = false,font=small]{caption}
\usepackage{subfig}
\usepackage{txfonts}
\usepackage{float}
\begin{document}

   \title{Differences in physical properties of coronal bright points and their ALMA counterparts within and outside coronal holes}

   \author{F. Matkovi\'c\inst{1}, R. Braj\v sa\inst{1}, M. Temmer\inst{2}, S. G. Heinemann\inst{3}, H.-G. Ludwig\inst{4}, S. H. Saar\inst{5}, C. L. Selhorst\inst{6}, I. Skoki\'c\inst{1}, \and D. Sudar\inst{1}}
        \authorrunning{Matkovi\'c et al.}
        \titlerunning{Differences in physical properties of CBPs within and outside CHs}

   \institute{Hvar Observatory, Faculty of Geodesy, University of Zagreb, Ka\v ci\'ceva 26, 10000 Zagreb, Croatia
         \and
             Institute of Physics, University of Graz, Universit\"atsplatz 5, 8010 Graz, Austria
        \and
                Max-Planck-Institut f\"ur Sonnensystemforschung, Justus-von-Liebig-Weg 3, 37077 G\"ottingen, Germany
        \and
                Landessternwarte, Zentrum f\"ur Astronomie der Universit\"at Heidelberg, K\"onigstuhl 12, 69117 Heidelberg, Germany
        \and
                Harvard-Smithsonian Center for Astrophysics, 60 Garden Street, Cambridge, MA 02138, USA
        \and
                NAT - N\'ucleo de Astrof\'isica, Universidade Cruzeiro do Sul/Universidade Cidade de S\~ao Paulo, S\~ao Paulo, SP, Brazil
             }

   \date{Received date, accepted date}
      %\date{}

% \abstract{}{}{}{}{} 
% 5 {} token are mandatory
 
  \abstract
   {}%context
  % aims heading
   {This study investigates and compares the physical properties, such as intensity and area, of coronal bright points (CBPs) inside and outside of coronal holes (CHs) using the Atacama Large Millimeter/submillimeter Array (ALMA) and Solar Dynamics Observatory (SDO) observations.}
  % methods heading
   {The CBPs were analysed using the single-dish ALMA Band 6 observations, combined with extreme-ultraviolet (EUV) 193 \AA\space filtergrams obtained by the Atmospheric Imaging Assembly (AIA) and magnetograms obtained by the Helioseismic and Magnetic Imager (HMI), both on board SDO. The CH boundaries were extracted from the SDO/AIA images using the Collection of Analysis Tools for Coronal Holes (CATCH) and CBPs were identified in the SDO/AIA, SDO/HMI, and ALMA data. Measurements of brightness and areas in both ALMA and SDO/AIA images were conducted for CBPs within CH boundaries and quiet Sun regions outside CHs. Two equal size CBP samples, one inside and one outside CHs, were randomly chosen and a statistical analysis was conducted. The statistical analysis was repeated 200 times using a bootstrap technique to eliminate the results based on pure coincidence.}
  % results heading
   {The boundaries of five selected CHs were extracted using CATCH and their physical properties were obtained. Statistical analysis of the measured physical CBP properties using two different methods resulted in a lower average intensity in the SDO/AIA data, or brightness temperature in the ALMA data, for CBPs within the boundaries of all five CHs. Depending on the CBP sample size, the difference in intensity for the SDO/AIA data, and brightness temperature for the ALMA data, between the CBPs inside and outside CHs ranged from between 2$\sigma$ and 4.5$\sigma$, showing a statistically significant difference between those two CBP groups. We also obtained CBP areas, where CBPs within the CH boundaries showed lower values for the measured areas, with the observed difference between the CBPs inside and outside CHs between 1$\sigma$ and 2$\sigma$ for the SDO/AIA data, and up to 3.5$\sigma$ for the ALMA data, indicating that CBP areas are also significantly different for the two CBP groups. We also found that, in comparison to the SDO/AIA data, the measured CBP properties in the ALMA data show a small brightness temperature difference and a higher area difference between the CBPs within and outside of CHs, possibly because of the modest spatial resolution of the ALMA images.}
  % conclusions heading
   {Given the measured properties of the CBPs, we conclude that the CBPs inside CHs tend to be less bright on average, but also smaller in comparison to those outside of CHs. This conclusion might point to the specific physical conditions and properties of the local CH region around a CBP limiting the maximum achievable intensity (temperature) and size of a CBP. The need for the interferometric ALMA data is also emphasised to get more precise physical CBP property measurements at chromospheric heights.}

   \keywords{Sun: corona -- Sun: chromosphere -- Sun: radio radiation -- Sun: UV radiation}

   \maketitle
%
%-------------------------------------------------------------------

\section{Introduction}
Coronal bright points (CBPs) are one of the most frequent activity phenomena in the solar atmosphere. They consist of low-corona small-scale plasma loops that connect two magnetic flux concentrations of opposite polarities in the photosphere (\citeauthor{Madjarska} \citeyear{Madjarska}). \cite{Reale} gives insight into the nature of coronal loops as magnetic flux tubes with hot and dense confined plasma, where the CBPs occupy the low end of their size spectrum. The formation and evolution of CBPs is associated with newly emerging magnetic flux  (\citeauthor{Mou_2016} \citeyear{Mou_2016}; \citeauthor{Chen} \citeyear{Chen}; \citeauthor{Nobrega} \citeyear{Nobrega}), called ephemeral regions (\citeauthor{Harvey_1975} \citeyear{Harvey_1975}), or with the chance encounters of the converging magnetic flux (\citeauthor{Harvey_1984} \citeyear{Harvey_1984}, \citeyear{Harvey_1985}; \citeauthor{Nobrega} \citeyear{Nobrega}).

Coronal bright points can be found in quiet Sun regions, within coronal holes (CHs), and in the vicinity of active regions (\citeauthor{Madjarska} \citeyear{Madjarska}). Due to their large spread over the whole solar disk, CBPs were used for various purposes, one of them being as tracers of the solar rotation (\citeauthor{Brajsa2002} \citeyear{Brajsa2002}, \citeyear{Brajsa_2004}, \citeyear{Brajsa_2008}, \citeyear{Brajsa_2015}; \citeauthor{Skokic_2016} \citeyear{Skokic_2016}, \citeyear{Skokic_2019}; \citeauthor{Sudar_2015} \citeyear{Sudar_2015}, \citeyear{Sudar_2016}; \citeauthor{Wohl} \citeyear{Wohl}). In this study we focus only on CBPs and their physical properties in the quiet Sun regions and within CHs. The quiet Sun is regarded as the region with a diffuse emission devoid of sunspots and active regions (\citeauthor{Bellot} \citeyear{Bellot}; \citeauthor{Del_Zanna} \citeyear{Del_Zanna}). At coronal heights, the temperature of the quiet Sun is found to be about 1 MK, or even more (\citeauthor{Del_Zanna} \citeyear{Del_Zanna}), and the average electron density is found to be higher than $4\times$ 10$^8$ cm$^{-3}$ (\citeauthor{Dere} \citeyear{Dere}). The quiet Sun regions have a mixed-polarity magnetic field and are spattered with small bipolar regions that could give rise to CBPs ((\citeauthor{Del_Zanna} \citeyear{Del_Zanna}). CHs are regions of the Sun that appear dark in extreme-ultraviolet (EUV) and X-rays due to the cooler and less dense plasma than in the surrounding regions (\citeauthor{Cranmer} \citeyear{Cranmer}). They are known for the abundance of an open magnetic field covering only a small fraction of up to 10\% from the entire CH area (\citeauthor{Hofmeister_2017} \citeyear{Hofmeister_2017}, \citeyear{Hofmeister_2019}; \citeauthor{Heinemann_2018} \citeyear{Heinemann_2018}). The expansion of open magnetic structures within CHs is found to be much stronger compared to the quiet Sun (\citeauthor{Tian} \citeyear{Tian}). With the abundance of an open magnetic field, CHs still have regions of a mixed-polarity magnetic field that enable CBP formation (\citeauthor{Wiegelmann_2005} \citeyear{Wiegelmann_2005}).

Coronal bright points are also known to have an enhanced emission in the EUV and X-ray spectrum. They were first discovered in X-rays in 1969 in a series of rocket flights and were named X-ray bright points (XBPs) due to their point-like X-ray feature (\citeauthor{Vaiana} \citeyear{Vaiana}). In X-ray observations analysed by \cite{Golub}, the size (diameter) of compact X-ray CBP features ranged from between 20" and 30", with a bright core of 5"$-$10" in diameter.

The first EUV CBP observations were analysed by \cite{Habbal_1981} using the Harvard EUV experiment on the Skylab Apollo Telescope Mount (\citeauthor{Golub1} \citeyear{Golub1}), showing that CBPs are composed of magnetic loops rooted in the chromosphere. They also suggested that the heating of the CBP plasma occurs at coronal heights and is carried to the chromosphere by thermal conduction. \cite{Habbal} used spectroheliograms from the previous experiment to compare the morphological structure and emission variations of CBPs inside and outside CHs in the quiet Sun region. These authors found that short-time variations in spectral lines were not always co-spatial, suggesting that CBPs are composed of loops of various sizes and temperatures (\citeauthor{Madjarska} \citeyear{Madjarska}). Additionally, quiet Sun and CH CBPs are found to range from 10" to 40" in diameter in both the CH and the quiet Sun region and no difference between those CBPs related to the properties of the observed region was found. Another work done by \cite{Habbal_1991} using similar data showed that depending on the associated magnetic field strength, some CBPs could be composed of loops that cannot reach coronal temperatures. Both previous works reported that the CBP formation and existence is independent of the overlying background coronal magnetic structure, but not the CBP evolution.

More CBP size measurements were conducted in recent years, and based on the full lifetime evolution of CBPs done by \cite{Mou} using the SDO Atmospheric Imaging Assembly (AIA) EUV data, CBPs that formed from magnetic flux emergence were initially 5" in diameter and they reached a maximum size of up to $\sim$60". The value of 60" as a maximum CBP size is used as an upper size limit when considering CBPs in most of the studies, but there are rare exceptions going even up to $\sim$100" in diameter (\citeauthor{Madjarska1} \citeyear{Madjarska1}). Based on the obtained results, values between 5" and 60" could be considered as typical CBP sizes, as is considered in the present paper. In the measurements of the CBP areas, based on 41 CBPs analysed in the quiet Sun region with a size of $780$"$\times780$" in the SOHO EIT 195\AA\space passband, \cite{Zhang} found an average CBP area of about 196 arcsec$^2$. On the other hand, \cite{Alipour} analysed statistical properties of CBPs in the SDO/AIA 193 \AA\space data using the maximum CBP size threshold of 56" and found a higher average CBP area of about 225 arcsec$^2$, but this result should be taken with caution because the authors detected some small bright CBP features as a new CBP, which could give an overall smaller area.

Radio observations of CBPs using the Very Large Array (VLA) at 6 cm (4.8 GHz) in 1977 showed small-scale compact sources between 9" and 25" in diameter with a peak brightness temperature of $6-8\times 10^4$ K with respect to the background of $2\times 10^5$ K (\citeauthor{Marsh} \citeyear{Marsh}). At least half of those sources were associated with bipolar magnetic CBP features. More VLA observations at 6 cm revealed that CBP radio emission shows rapid temporal and spatial variations similar to those in X-rays and EUV (\citeauthor{Fu} \citeyear{Fu}). In the following years, 6 cm and 20 cm observations confirmed that the observed CBP emission comes from electron-ion free-free thermal bremsstrahlung (\citeauthor{Kundu} \citeyear{Kundu}). Based on the comparison of CBPs observed in soft X-rays by the Yohkoh Soft X-ray Telescope and at 20 cm by the VLA done by \cite{Nitta}, half of the 33 observed radio sources were associated with XBPs, and the rest were just overlaying magnetically unipolar regions. More radio observations done by \cite{Gopalswamy}, and later by \cite{Oliveira}, used the Nobeyama Radioheliograph (NoRH) at 17 GHz to study enhanced microwave brightenings inside polar CHs, where some of them were associated with the presence of intense unipolar magnetic fields. Both previous VLA and NoRH observations show that CBPs are not the only radio source in the chromosphere.

The construction of the Atacama Large Millimeter/submillimeter Array (ALMA) enabled detailed observations of the solar chromosphere needed to better understand this layer and all of the features there (\citeauthor{Wedemeyer} \citeyear{Wedemeyer}). ALMA provides observations using both single$-$dish (\citeauthor{White} \citeyear{White}) and interferometric (\citeauthor{Shimojo} \citeyear{Shimojo}) observing modes, with a wavelength range between 0.3 mm and 9 mm (\citeauthor{Wedemeyer} \citeyear{Wedemeyer}). So far, mostly $\sim$1 mm (Band 6) and $\sim$3 mm (Band 3) wavelengths have been used in solar observations in both single$-$dish and interferometric modes, and some of the Band 6 data are presented in this study. New observations made by ALMA since late 2015 show great promise in CBP observations at millimetre and submillimetre wavelengths, enabling us to study the CBPs at chromospheric heights.

Earlier ALMA CBP observations done by \cite{Shimojob}, and later by \cite{Rodger}, did not study CBPs directly, but instead used the Band 3 channel to study a solar plasmoid ejection from a CBP. Using the ALMA Band 6 data, \cite{Brajsa_2018} recently reported the first analysis of solar structures in the 1.21 mm full$-$disk solar ALMA images. They compared a full$-$disk solar ALMA image, taken on 18 December 2015, with simultaneous images in the optical (H$\alpha$), infrared (He I 10830 \AA\space), and EUV (AIA 1700 \AA, 304 \AA, 211 \AA, 193 \AA,\space and 171 \AA) spectrum as well as with the SDO Helioseismic Magnetic Imager (HMI) magnetogram. CBPs visible in the observed data showed a very good match with the ALMA bright features, where 82\% of all CBPs from the EUV image corresponded to the ALMA 1.21 mm bright points. A continuation of this work was done by \citet{Brajsa} with an emphasis on CBPs in the ALMA Band 6 data. In the quiet Sun, four CBPs were identified, with other small-scale ALMA bright features most likely being associated with magnetic network elements and plages. It was also found that enhanced emission in the ALMA data is almost always associated with a strong line-of-sight magnetic field. In the active region, using the ALMA Band 3 interferometric data and by comparing them with other wavelength images, out of the 14 randomly selected small-scale ALMA bright features, five CBP candidates were found.

In the present paper, which is a continuation of the work done by \cite{Brajsa}, we present an analysis of mean brightness and area of CBPs within five different CHs and outside of them in the quiet Sun. We first describe data and methods used for CH extraction and CBP identification, measurement, and statistical analysis (Sect. \ref{DM}). Next, we present the results of CBP measurements and a statistical analysis of the measured CBP properties (Sect. \ref{result}), then we discuss and compare the important results (Sect. \ref{Discussion}) and finally finish with plans for future work (Sect. \ref{conclude}).

\section{Data and methodology}
\label{DM}
\subsection{ALMA single-dish data}
From several hundred full-disk solar images of the brightness temperature (measured in kelvins (K)) taken by ALMA between 23 March 2017 and 13 April 2019, a total of five images were chosen that contain different CHs near the central region of the solar disk. The five chosen full-disk solar maps for 16 April 2017, 22 April 2017, 17 April 2018, 3 May 2018, and 25 December 2018 were obtained by scanning the solar disk with a 12 m single-dish total power antenna at Band 6 frequencies 230 GHz ($\lambda=1.3$ mm), 248 GHz ($\lambda=1.21$ mm), 232 GHz ($\lambda=1.29$ mm), 248 GHz ($\lambda=1.21$ mm), and 230 GHz ($\lambda=1.3$ mm) respectively in a double circle pattern (\citeauthor{Philips} \citeyear{Philips}, \citeauthor{White} \citeyear{White}). We restricted ourselves to only Band 6 because of a better spatial resolution in comparison to other currently available bands used for full-disk solar observations. Single-dish beam sizes for the obtained images are 28.3", 26.7", 28.2", 26.7", and 28.4" respectively with a pixel scale of 3".

Before we could analyse the selected ALMA Band 6 images, they had to be corrected for the limb brightening effect. For this purpose we used a second-order polynomial fit for the centre-to-limb brightness function following the procedure given in \cite{Sudar}. The limb brightening correction procedure was done using $limb.py$ Python script made by \cite{Sudar} for the limb brightening correction of the ALMA data.

\subsection{SDO observations}
From the available SDO data, we took the SDO/AIA 193 \AA\space intensity filtergrams (\citeauthor{Lemen} \citeyear{Lemen}) taken at the corresponding observational times of the chosen ALMA images. The spatial resolution for all the selected SDO/AIA EUV images was 0.6" per pixel. Since the limb brightening effect in the SDO/AIA 193 \AA\space data is very low at distances smaller than 0.7 solar radii from the solar disk centre (\citeauthor{Verbeeck} \citeyear{Verbeeck}), there is no need for the limb brightening correction to be made for the purpose of the CBP measurements if we only search for CBPs in the central regions of the solar disk, as is done in the present paper. The same SDO/AIA EUV images were used to extract the CH boundaries using the intensity threshold based software called Collection of Analysis Tools for Coronal Holes (CATCH) following the procedure described in \cite{Heinemann}.

The next set of SDO data was the SDO/HMI data taken at the same times with the spatial resolution of 0.5" per pixel (\citeauthor{Scherrer} \citeyear{Scherrer}). The SDO/HMI data shows magnetograms of the line-of-sight (LOS) magnetic field with a time cadence of 45 s for the whole solar disk. For the purpose of our analysis, the values of the magnetogram intensities were saturated at the values of $\pm$120 G to highlight the magnetic flux sources in the solar photosphere.

\subsection{CBP identification}
\label{CBP_identification}
\begin{figure*}[!htp]
\captionsetup[subfloat]{farskip=1pt,captionskip=1pt}
\centering
\subfloat{\includegraphics[width=0.75\textwidth]{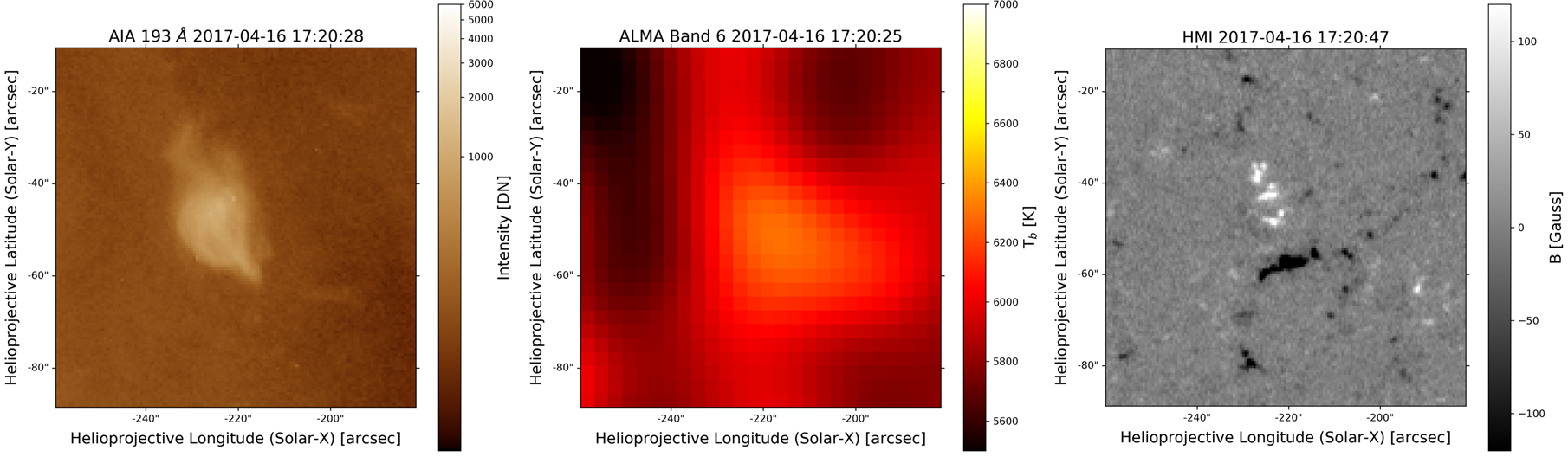}}\\\vspace{0.5em}
\subfloat{\includegraphics[width=0.75\textwidth]{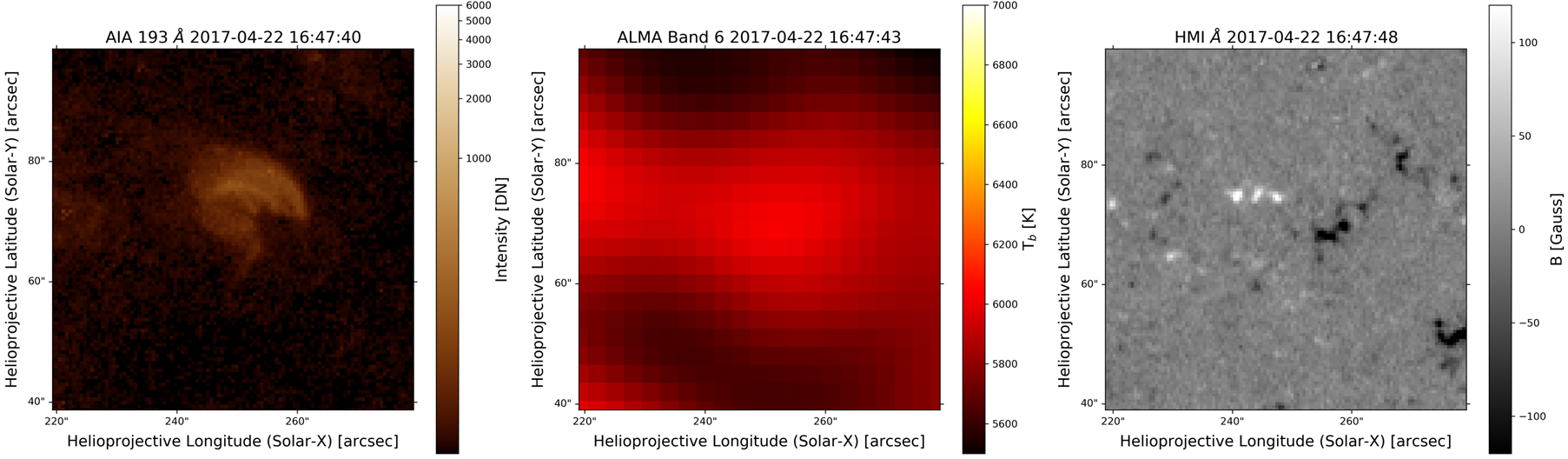}}
\caption{Example of CBP outside CH (upper row) and within CH (bottom row). The upper panel shows the SDO/AIA 193 \AA\space (left) intensity filtergram and the ALMA Band 6 (middle) brightness temperature ($T_\mathrm{b}$) image, with the SDO/HMI magnetogram (right). The intensity values for the corresponding images are clipped between 20 and 6 000 DN, 5 500 and 7 000 K, and -120 and 120 Gauss, respectively.}
\label{CBP_examples}
\end{figure*}

We based our CBP identification on the SDO/AIA 193 \AA\space filtergrams and the SDO/HMI magnetograms. On the SDO/AIA maps (left panels of Fig. \ref{CBP_examples}), CBPs can be recognized as small-scale bright loop-like structures linking two opposite polarity magnetic flux concentrations visible in magnetograms (right panels of Fig. \ref{CBP_examples}). For visualization of the SDO images we used the SunPy\footnote{\url{https://sunpy.org/}} software package in Python. Our goal was to search for bright loops connecting different polarity fluxes, which gave us a strong indication that the observed feature is a CBP. Because CBPs have different morphologies and intensities, in order to determine with confidence that the observed feature is truly a CBP, we used JHelioviewer software (\citeauthor{Muller} \citeyear{Muller}). With JHelioviewer we were able to visualise the evolution of shape and intensity of the features of interest through time a couple of hours before and after the time of the investigation and confirm if the observed feature was a CBP or not. The JHelioviewer was also used to check if at the time of investigation there were any microflaring or minieruption events happening. We require the CBPs to be in a quiet state, hence if these transient phenomena were found for CBPs, those CBPs (in our case about $5-10$\% of them) would be discarded from the analysis.

After the CBP identification was carried out for SDO, we analysed the respective ALMA Band 6 images (middle panels of Fig. \ref{CBP_examples}) to identify the same CBPs there. The CBPs visible in the obtained full-disk ALMA images can be identified as bright ellipsoidal features, mainly because of a poor spatial resolution of the available images. The identification was first carried out within the CH boundaries for both ALMA and SDO/AIA data, and based on the number of CBPs found inside the CH, a much larger sample of CBPs was selected in the quiet Sun region outside the CH far from the limb of the solar disk.

\subsection{CBP edge detection and background subtraction}
\label{edge_back}
Once the CBPs are identified, in order to measure their physical properties, we extract their boundaries. Here we describe a method of edge extraction with background subtraction written in Python, which we developed for the purpose of our analysis.

Before edge extraction, we first take a square-shaped cutout image, with a width and height of about twice the apparent CBP size or more and a CBP approximately in the centre. Then we smooth the cutout image using a Gaussian filter with a standard deviation of $\sigma=1$ pixel to eliminate possible high value noise stemming from a single pixel or very few pixels only. This step is not needed for the ALMA images due to their large beam size.

\begin{figure*}[!htp]
\captionsetup[subfloat]{farskip=1pt,captionskip=1pt}
\centering
\subfloat{\includegraphics[width=0.65\textwidth]{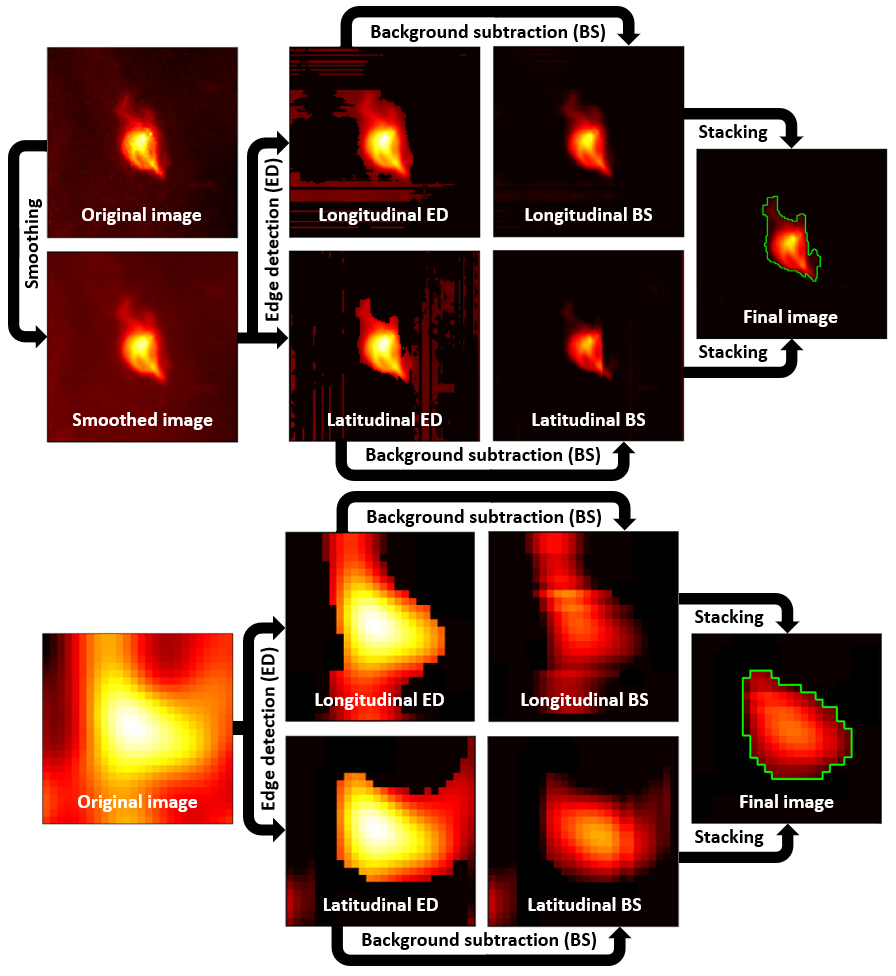}}
\caption{Scheme of extraction procedure for CBP example outside CH (upper row of Fig \ref{CBP_examples}). The upper panel shows the procedure for the SDO/AIA 193 \AA\space image of the CBP, consisting of smoothing, longitudinal and latitudinal edge detection (ED) and background subtraction (BS), and finally stacking step. The bottom panel shows the same procedure for the ALMA Band 6 image of the same CBP, but with no smoothing required. The green contour outlines the edge of a CBP of interest. The same procedure for a CBP inside of CH, given in the bottom row of Fig. \ref{CBP_examples}, is given in Fig. \ref{ED_ALMA_SDO2}. All cutout images have the same spatial scale as the corresponding cutout images in Fig. \ref{CBP_examples}, but are centred on the brightest CBP feature.}
\label{ED_ALMA_SDO1}
\end{figure*}

The second step is the edge detection or extraction. Here we divide the edge detection into longitudinal (row) and latitudinal (column) direction. This is because our method does not know a priori what a CBP is. If the edge detection is done only in longitudinal (latitudinal) direction, there is a possibility for a leftover background emission above and below (left and right) of a CBP. Therefore, by doing the edge detection along two separate directions individually, one direction will eliminate the background emission in a region where the other direction cannot. This direction separation is crucial for the final step in our method, which is described below. For longitudinal (latitudinal) direction of the edge detection, the intensity light curve for each row (column) of pixels in an image is extracted and intensity gradient at the position of each pixel in the corresponding row (column) is calculated. A pixel where the intensity starts to rise rapidly is then taken to be a possible CBP edge, and we call this pixel a lift-off pixel. In this way, in the case of a CBP, we obtain two lift-off pixels on opposite sides of a CBP for the corresponding row (column), indicating the CBP edge. As a result of this step, we obtain two edge detection images, one in longitudinal and the other in latitudinal direction (Fig. \ref{ED_ALMA_SDO1}), where the pixels away from the edges were deleted (black pixels).

It is important to note that part of the CBP emission we observe comes from the background emission, mostly emission of the underlying plasma. Therefore, we need to subtract this background from the foreground CBP emission in order to obtain just the emission coming only from the CBP. In order to estimate the background emission, we first select the two opposite lift-off pixels in each image row (column) that were obtained in the edge detection and we take a $3\times3$ pixel$^2$ area centred on the lift-off pixels in the smoothed image. The mean intensity inside of these small areas were then calculated and the two values for each row (column) were then averaged to obtain the background for each corresponding row (column). These values were then subtracted from the pixel values in the corresponding rows (columns) of the longitudinal (latitudinal) edge detection images. The pixels with values under the measured background values were afterwards deleted. As a result, we get two images with longitudinally and latitudinally detected edges and subtracted background.

In order to obtain a final CBP image with estimated edges and no background emission present, we take the two images with longitudinally and latitudinally detected edges and subtracted background and we stack them together. The stacking is done in a way that the intensity values of the common pixels (i.e. pixels at the same position, both with a non-vanishing value) of the two images are averaged, and the rest is deleted. This procedure eliminates most if not all of the background emission around a CBP that was left from the previous procedures, especially close to a CBP, giving us a much clearer CBP edge.

The complete above procedure is visually presented in Fig. \ref{ED_ALMA_SDO1} and \ref{ED_ALMA_SDO2}, and was carried out for all CBPs for both ALMA and SDO data. Only the emission within green contours outlining a CBP boundary, for example in Fig. \ref{ED_ALMA_SDO1}, is considered in the further analysis. The important thing to note here is that for the ALMA Band 6 data the Rayleigh–Jeans law is valid, which results in the brightness temperature to be proportional to intensity (\citeauthor{Kuhar} \citeyear{Kuhar}). Therefore, throughout our work, especially in the CBP extraction procedure, and later measurements, we consider the brightness temperature to be only a measure of the intensity. It may or may not correspond to the plasma temperature. We should also note that due to a large beam size of the ALMA single-dish images, the intensity light curves do not show a clear intensity gradient lift-off, resulting in a very blurred edge. The procedure described above will therefore only give a possible location of the CBP edges, which might result in a larger CBP size than it probably is in reality.

\subsection{Obtaining CBP physical properties}
In this work, we have chosen to study mean brightness and projected area of CBPs. The measurements of these two properties were carried out on CBPs (e.g. outlined CBP in Fig. \ref{ED_ALMA_SDO1}) in the final cutout images obtained using the extraction procedure described in Subsection \ref{edge_back}. From these images, the mean brightness was measured by calculating the average of the SDO/AIA 193 \AA\space intensity values, and the ALMA Band 6 brightness temperature values, of all of the pixels within the edges of the extracted CBP. The CBP area is measured by counting how many pixels are within the edges of the extracted CBP and this number is then multiplied by the single pixel area to obtain a total CBP area. This again is done for all CBPs in and out of CHs for both ALMA Band 6 and SDO/AIA 193 \AA\space data.

\subsection{Statistical difference of the measured physical properties}
To see if there is a statistically significant difference in the measured properties between CBPs within and outside CHs, we use two different methods. The first method uses the expression (e.g. \citeauthor{Brajsa_1999} \citeyear{Brajsa_1999}):
\begin{equation}
\Delta\omega=\omega_1-\omega_2>N(M(\omega_1)+M(\omega_2)),
\label{Nsigma}
\end{equation}
where $\omega_{1,2}$ represent mean values of the observed property for two different samples, $M(\omega_{1,2})$ are the corresponding standard errors and $N=$ 1, 2, 3, etc. A difference of the measured means $\omega_1$ and $\omega_2$ is statistically significant on the $N\sigma$ level if the above criterion is fulfilled for the largest natural number $N$ possible.

The second method uses the unequal variances $t$-test (\citeauthor{William} \citeyear{William}). Here we also try to determine if the means of two data sets are significantly different from each other and by how much. In $t$-test statistics, this is characterized by two quantities, a $t$-value, corresponding to a distance between the two means in terms of standard deviations, and a $p$-value, corresponding to a probability of obtaining test results at least as extreme as the observed results, assuming that the null hypothesis (in our case, it states that the two mean values are equal) is true. We should note here that, based on our CBP sample ordering, the negative (positive) $t$-value means that the mean value of the measured property for the CBPs outside (within) the CH boundaries is higher. For the purpose of our work, if $p$-value < 0.05, then our two data sets have statistically significant difference between their means. Both $t$-value and $p$-value are obtained using the function $ttest\_ind$\footnote{\url{https://docs.scipy.org/doc/scipy/reference/generated/scipy.stats.ttest\_ind.html}} under SciPy\footnote{\url{https://scipy.org/}} software package in Python.

Both methods were conducted on 200 randomly generated samples of both the detected CBPs inside and outside of CHs. With that we obtain 200 sample pairs of two individual equal size CBP samples, which we use for comparison between CBPs inside and outside of CHs. This procedure is called a bootstrap technique (\citeauthor{Efron} \citeyear{Efron}), and it was used for all the measured CBP properties. On top of that, we repeated the whole procedure on variety of CBP sample sizes to see how the result changes when different numbers of CBPs are included.
\begin{figure*}[!htp]
\captionsetup[subfloat]{farskip=1pt,captionskip=1pt}
\centering
\subfloat{\includegraphics[width=0.82\textwidth]{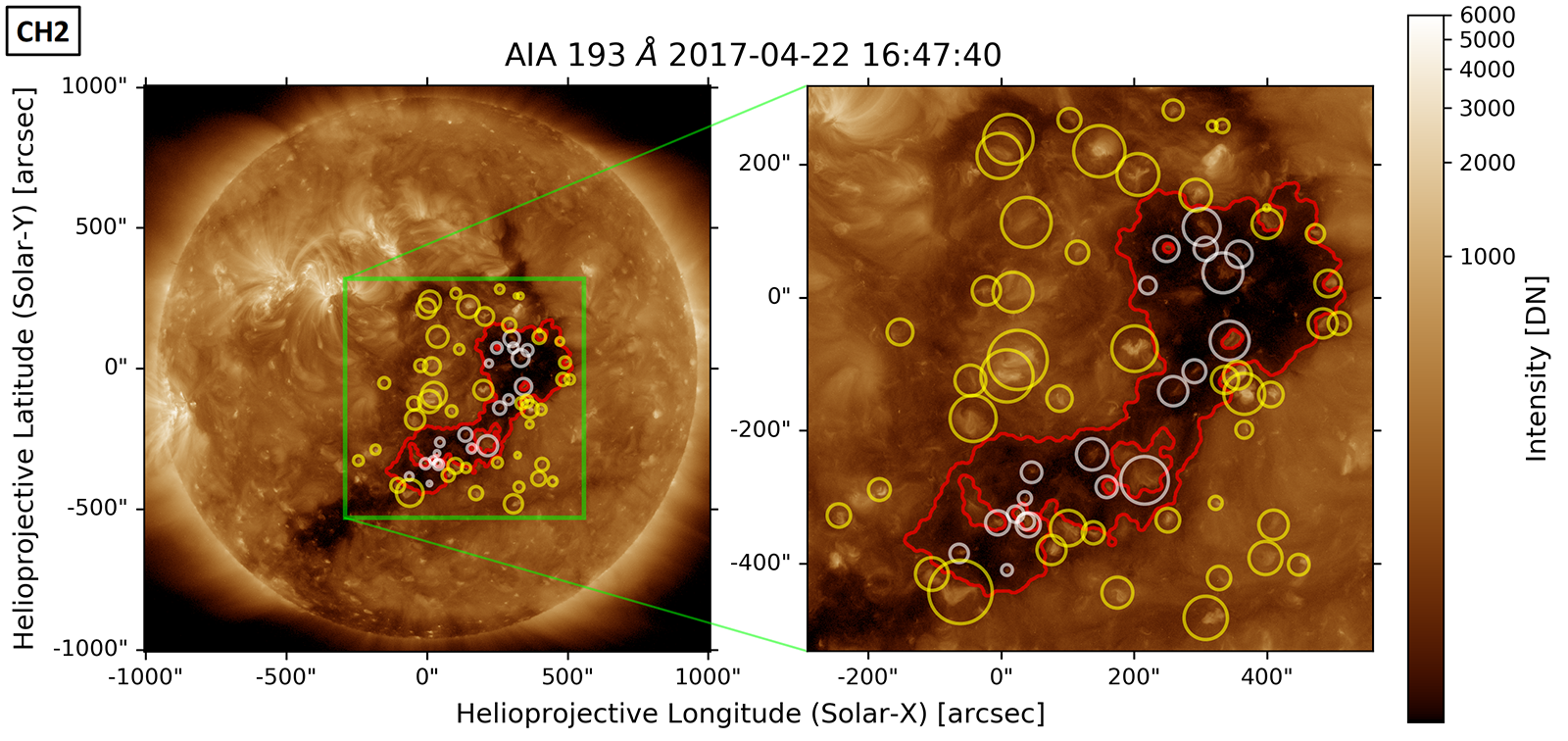}}\\\vspace{0.5em}
\subfloat{\includegraphics[width=0.82\textwidth]{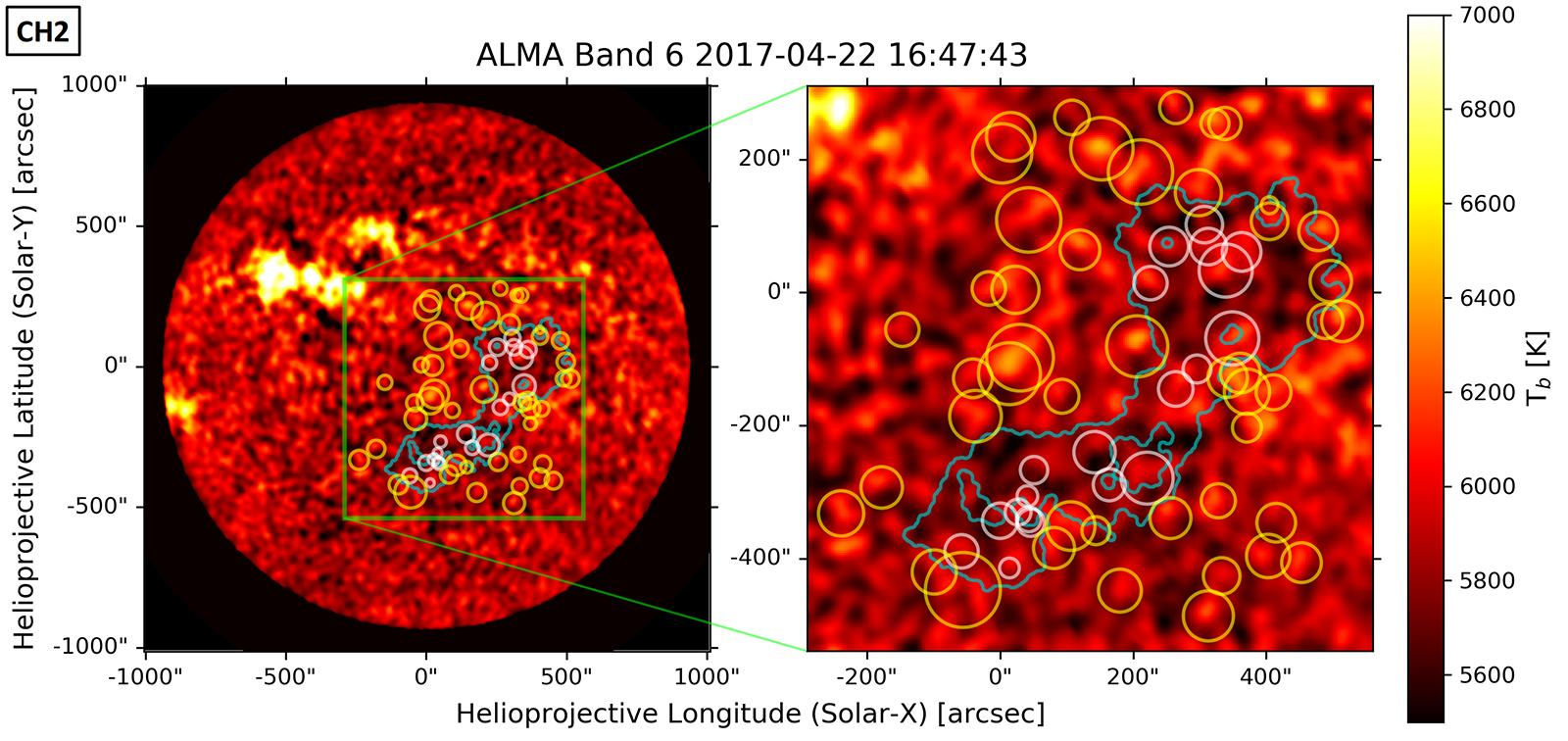}}\\\vspace{0.5em}
\subfloat{\includegraphics[width=0.82\textwidth]{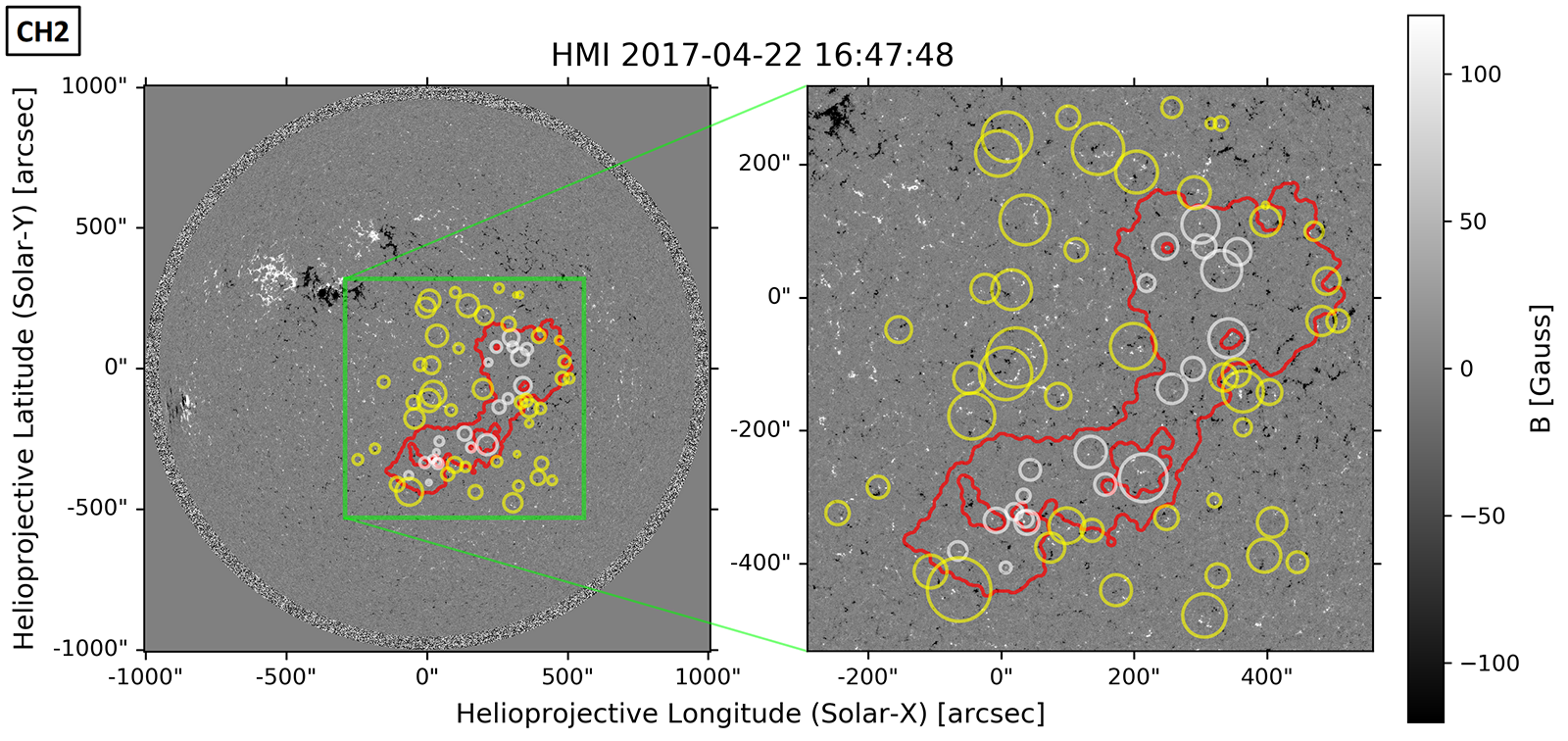}}
\caption{SDO/AIA 193 \AA\space intensity filtergram (top), ALMA Band 6 brightness temperature ($T_\mathrm{b}$) image (middle) and SDO/HMI magnetogram (bottom) of full solar disk (left) with zoomed region around CH2 (right), both containing extracted boundary of CH2 (red - AIA, HMI, cyan - ALMA), obtained with CATCH using SDO/AIA 193 \AA\space image. The selected CBPs are marked with circles (white - within CH2, yellow - outside CH2), where the radius of each circle equals half of the maximum length of a CBP multiplied by a factor of 1.7. The SDO/AIA intensity is clipped between 20 and 6 000 DN, ALMA brightness temperature between 5 500 and 7 000 K, and SDO/HMI magnetic field between $-$120 and 120 Gauss. Similar figures for CH1, CH3, CH4 and CH5 are given in Appendix \ref{intensitymaps}.}
\label{CH_ALMA_SDO1}
\end{figure*}

\section{Results}
\label{result}
\subsection{CH extraction and CBP detection}
\begin{table*}[!htp]
\caption[]{General properties of five chosen CHs denoted as CH1, CH2, CH3, CH4, and CH5.}  
\centering   
\scalebox{0.85}{ 
\begin{tabular}{c c c c c c l l l l l}     
\hline         
 &CH1&CH2&CH3&CH4&CH5\\
\hline
        Date of observation [y-m-d]&2017-04-16&2017-04-22&2018-04-17&2018-05-03&2018-12-25\\
        Time of observation [h:m:s]&17:20:28&16:47:40&16:59:16&16:20:52&12:57:52\\
        Area [10$^{10}$ km$^2$]& 2.74 $\pm$ 0.23&7.65 $\pm$ 0.49&8.16 $\pm$ 1.24&9.24 $\pm$ 0.65&5.93 $\pm$ 0.59\\ 
        Mean Intensity [DN]& 20.13 $\pm$ 0.90&17.83 $\pm$ 0.75&18.29 $\pm$ 1.13&19.60 $\pm$ 0.87&20.48 $\pm$ 0.99\\ 
        Signed mean magnetic field [G]& 2.46 $\pm$ 0.03&-1.90 $\pm$ 0.04&-1.12 $\pm$ 0.14&-2.01 $\pm$ 0.01&1.31 $\pm$ 0.07\\ 
        No. of confirmed CBPs inside CH& 7 & 20 & 18 & 32 & 18\\ 
        No. density of CBPs inside CH [10$^{-10}$ km$^{-2}$]& 2.55 $\pm$ 0.21& 2.61 $\pm$ 0.17& 2.21 $\pm$ 0.34& 3.46 $\pm$ 0.24& 3.04 $\pm$ 0.30\\ 
        No. of selected CBPs outside CH& 36 & 46 & 43 & 52 & 53\\
\hline\\ 
\end{tabular}
}
\label{CH_property} 
\end{table*}
Five chosen CHs outlined with a boundary obtained with CATCH are shown in Fig. \ref{CH_ALMA_SDO1}, \ref{CH_ALMA_SDO2}, \ref{CH_ALMA_SDO3}, \ref{CH_ALMA_SDO4}, and \ref{CH_ALMA_SDO5}. In all of the figures, the analysed CBPs are marked with small circles. Additionally, if a CBP was excluded from a CH by CATCH due to intensity threshold, but was still inside the most outer CH boundary, in our analysis we considered it to be inside the CH.

From CATCH we obtained the area, the mean AIA 193 \AA\space intensity, and the signed mean magnetic field strength within the obtained boundaries of the chosen CHs. Uncertainties for the CH properties were estimated from calculating these properties for the chosen and for slightly lower and higher intensity thresholds around the chosen threshold, while minimising the area uncertainty (\citeauthor{Heinemann} \citeyear{Heinemann}). The obtained properties for all five CHs, including a number of detected CBPs inside and a number of additionally selected CBPs outside of CHs, are given in Table \ref{CH_property}.

Based on the obtained CH area, the smallest of the five CHs is CH1 with 7 confirmed CBPs inside, and the largest one is CH4 with 32 confirmed CBPs inside. Since larger CHs contain larger numbers of CBPs, as is expected, here we find that CH3 is an exception from this behaviour, because it is the second largest CH, but it contains the lowest number of CBPs out of the five CHs. If we do a linear fit between the number density of CBPs in CHs ($N_{\mathrm{in}}/A_{\mathrm{CH}})$ and the CH area ($A_{\mathrm{CH}}$), and we calculate the Spearman's correlation coefficient $r_\mathrm{S}$, by ignoring CH3 we get the following:
\begin{equation}
\frac{N_{\mathrm{in}}}{A_{\mathrm{CH}}/10^{10} \mathrm{km}^2}\sim 4.57\times\frac{A_{\mathrm{CH}}}{10^{10} \mathrm{km}^2};\;\; r_\mathrm{S}=0.69
.\end{equation}
This result shows a modest tendency of the number density of CBPs in CHs to increase with CH area, as is expected. When adding CH3 into consideration, the Spearman's correlation coefficient immediately drops to $r_\mathrm{S}=0.29$, showing a rather weak correlation between the number density of CBPs and CH area. By analysing the corresponding CH area errors, we see that four CHs have the relative area uncertainty less than 10\%, with CH2 having the lowest relative area uncertainty. We find high area uncertainty only for CH3 of about 15\%. This large uncertainty for CH3 is possibly due to its peculiar shape, with a southern patchy structure, that could have also affected the number of CBPs inside of it, thus resulting in the previous weak correlation.

Going forward in the results, we mostly focus our discussion on the results obtained for CH2 as a good example for the measured CBP properties out of the five CHs, but we still compare the results between all five CHs. Results of the statistical analysis for the measured CBP properties for CH1, CH3, CH4, and CH5 can be found in Appendix \ref{CH1}, \ref{CH3}, \ref{CH4}, and \ref{CH5}.

\subsection{Mean CBP brightness}
\subsubsection{ALMA data}
\label{mean_int_alma_data}
Based on the selected CBPs, the maximum values of the mean brightness temperatures for the CBPs within and outside all five CHs in the ALMA Band 6 images are presented in Table \ref{mean_int_table}.
\begin{table}[h!] 
\caption{Mean values of mean measured ALMA Band 6 brightness temperatures ($\langle\overline{I}\rangle_{\mathrm{ALMA}}$) and mean values of mean measured SDO/AIA 193 \AA\space intensities ($\langle\overline{I}\rangle_{\mathrm{ALMA}}$) of CBPs within and outside five chosen CHs, with corresponding standard errors.}   
\label{mean_int_table}
\centering
\scalebox{0.85}{       
\begin{tabular}{c c c l l}
\hline   
&$\langle\overline{I}\rangle_{\mathrm{ALMA}}$ [K]&$\langle\overline{I}\rangle_{\mathrm{SDO}}$ [DN]\\
&within CH / outside CH&within CH / outside CH\\
\hline
        CH1&78 $\pm$ 10 / 124 $\pm$ 8&28 $\pm$ 5 / 104 $\pm$ 10\\
        CH2&97 $\pm$ 7 / 157 $\pm$ 7&29 $\pm$ 3 / 91$ \pm$ 8\\
        CH3&88 $\pm$ 6 / 149 $\pm$ 8&29 $\pm$ 5 / 109 $\pm$ 12\\
        CH4&111 $\pm$ 7 / 141 $\pm$ 5&51 $\pm$ 10 / 99 $\pm$ 9\\
        CH5&107 $\pm$ 9 / 140 $\pm$ 6&32 $\pm$ 5 / 72 $\pm$ 5\\
\hline 
\end{tabular}
}
\end{table}
\begin{figure*}[h!]
\captionsetup[subfloat]{farskip=1pt,captionskip=1pt}
\centering
\subfloat{\includegraphics[width=0.36\textwidth]{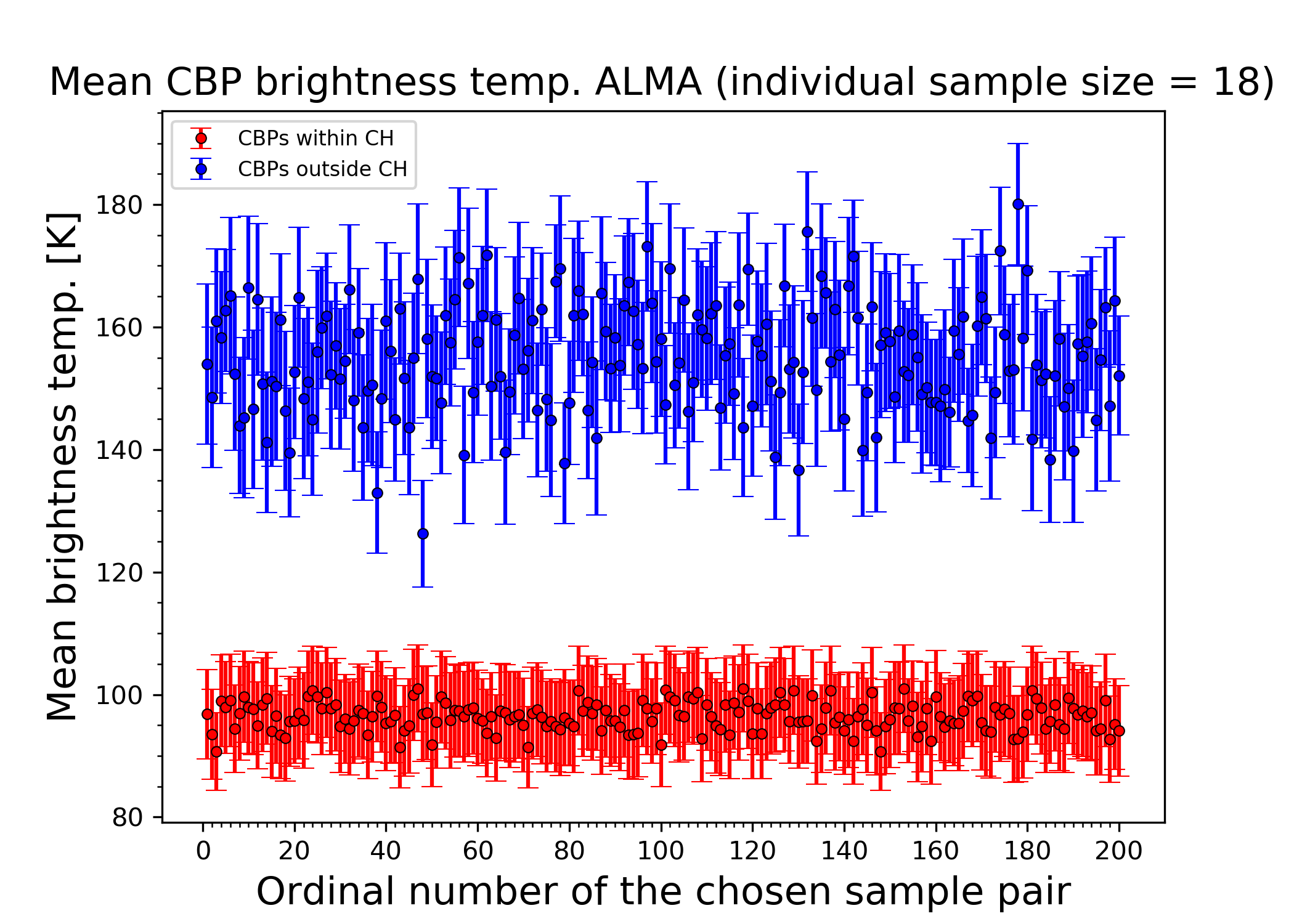}}
\subfloat{\includegraphics[width=0.36\textwidth]{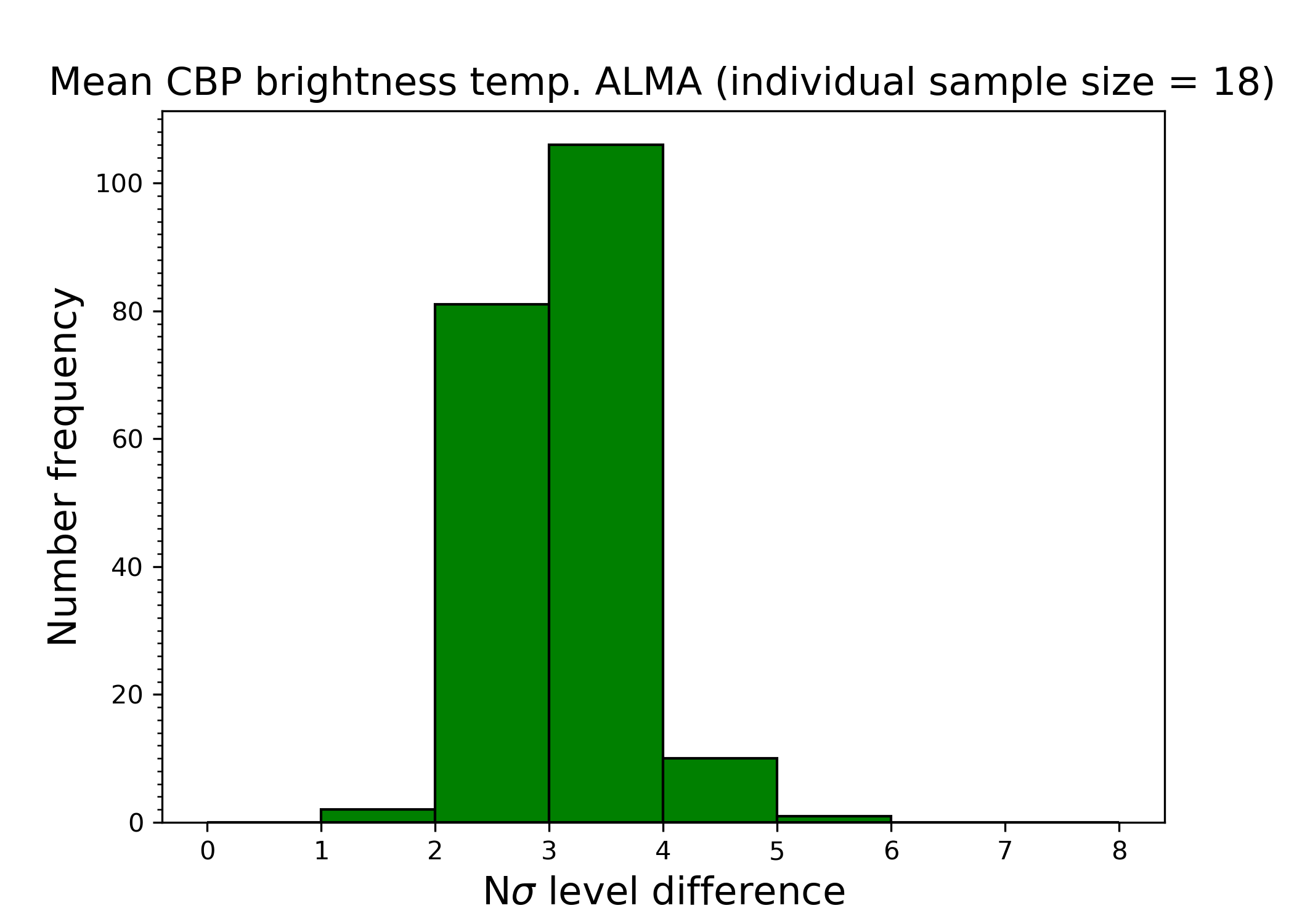}}\\
\subfloat{\includegraphics[width=0.36\textwidth]{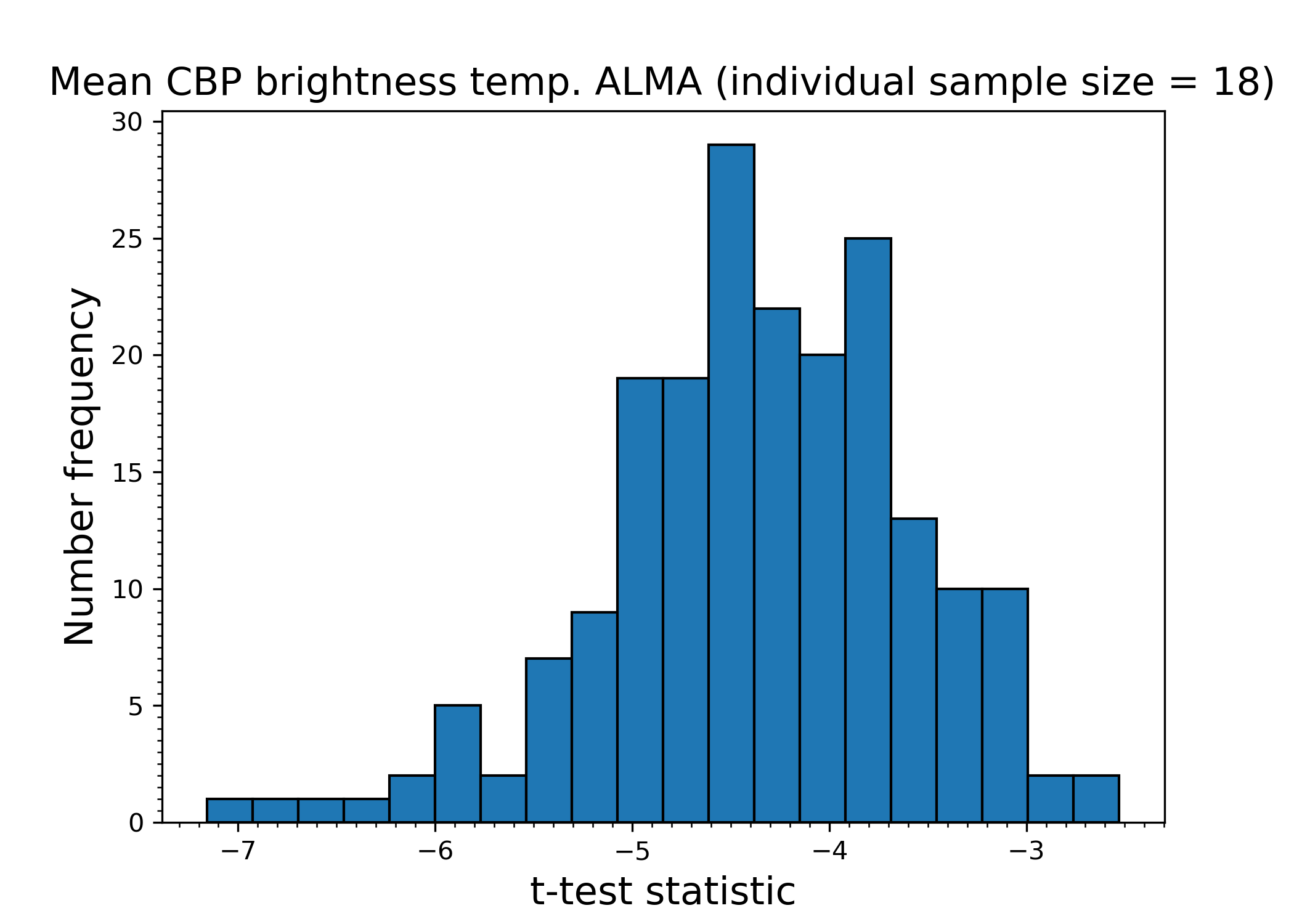}}
\subfloat{\includegraphics[width=0.36\textwidth]{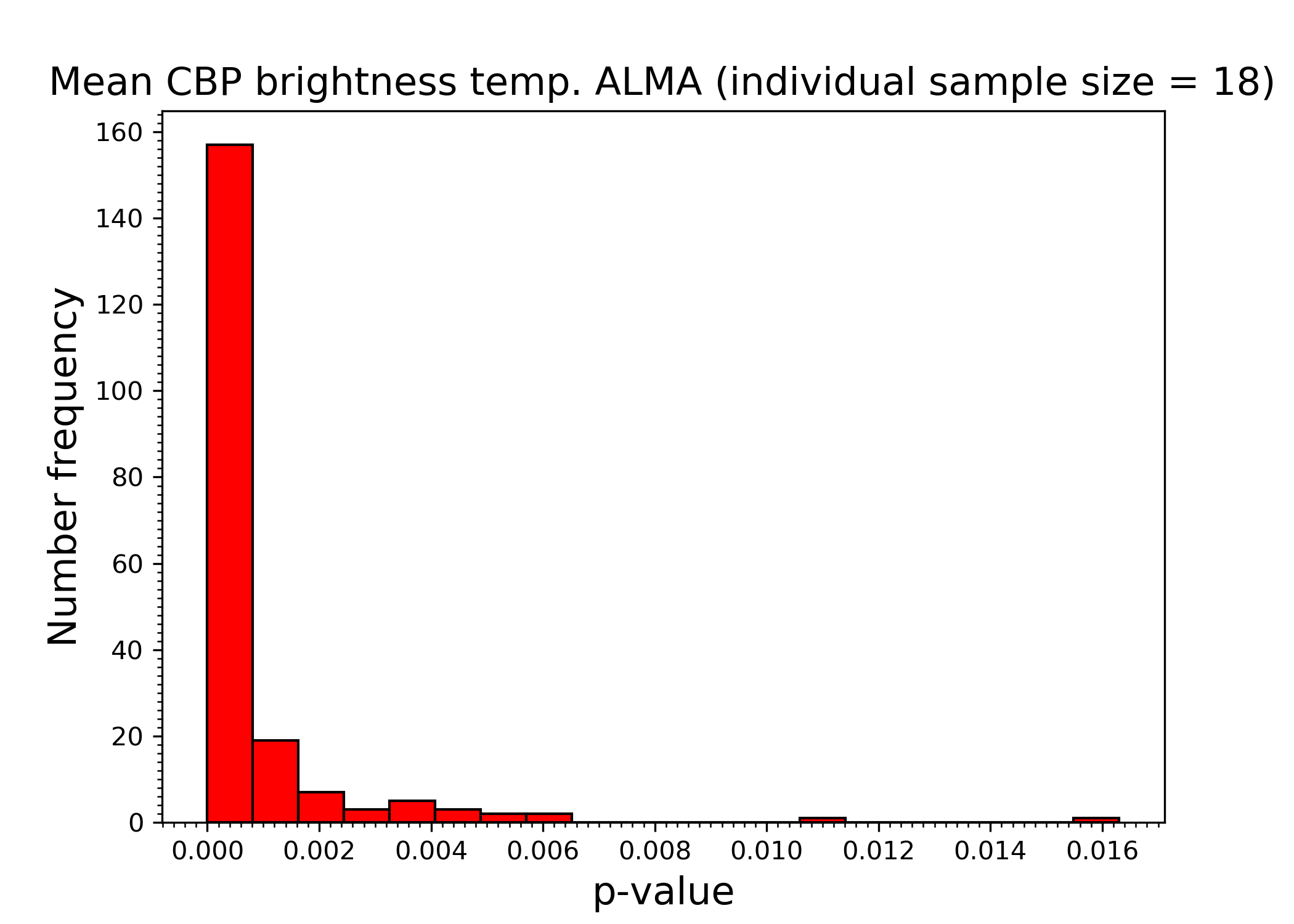}}
\caption{Statistical analysis of mean ALMA Band 6 CBP brightness temperature. Top row: Left panel shows the mean values of the mean CBP intensities in the ALMA Band 6 image with corresponding standard errors of 200 randomly chosen equal size CBP sample pairs, with one sample containing CBPs within (red) and the other outside (blue) the CH2. The right panel shows histogram of the largest N for which the relation (\ref{Nsigma}) holds true. Bottom row: Left panel shows histogram of the $t$-test statistic values ($t$-values) and the right panel shows the histogram of the $p$-values obtained for the mean values of CBP mean intensities in the ALMA Band 6 image. Individual CBP sample contains 18 randomly chosen CBPs out of the many selected CBPs either within or outside the CH2.}
\label{mean_int_fig}
\end{figure*}
First column of Table \ref{mean_int_table} shows that CBPs outside the chosen CHs can reach higher values of the mean brightness temperature than the CBPs inside CHs. In all five cases, the average mean brightness temperature (Table \ref{mean_int_table}) for the CBPs outside the CHs is higher than for those outside, and it reaches values above 120 K. On the other hand, for the CBPs inside the CHs, the average brightness temperature is below the previous value, and in the first three cases, even below 100 K.

The top left panel of Fig. \ref{mean_int_fig} shows very clearly the separation between CBPs inside and outside of CH2, with the ones inside having a lower mean brightness temperature, with a smaller dispersion around a general mean value (Table \ref{mean_int_table}). By increasing the sample size for bootstrapping, we find a more pronounced separation. Moreover, larger overlaps of the mean brightness temperatures were found for two coronal holes, CH1 and CH5, when having smaller sample sizes.

The histogram in the top right panel of Fig. \ref{mean_int_fig}, which was obtained using the expression \ref{Nsigma}, shows that most of the CBP sample pairs have their brightness temperatures differing between 2$\sigma$ and 4$\sigma$. Even larger difference is seen in the bottom row of Fig. \ref{mean_int_fig}, where in the left panel we see a large number of sample pairs having the mean brightness temperature difference between 3.5$\sigma$ and 5$\sigma$, with a tendency towards 4$\sigma$. The $p$-values in the right panel show that almost all of the sample pairs have a $p$-value under 0.05, clearly showing a statistically significant mean brightness temperature difference between CBPs inside and outside the CH2. Similar results were found for the remainder of the CHs as well (Fig. \ref{mean_int_fig_ch1}, \ref{mean_int_fig_ch3}, \ref{mean_int_fig_ch4}, and \ref{mean_int_fig_ch5}), with a higher or smaller difference visible between CBP inside and outside of CH depending on the CH of interest, but with difference values around 2$\sigma$ or more. By changing the individual CBP sample size, we find more significant differences for larger sample sizes.

\subsubsection{SDO/AIA data}
\label{mean_int_sdo_data}
Moving onto the SDO/AIA data, second column of Table \ref{mean_int_table} presents the results for the overall mean value of the mean SDO/AIA 193 \AA\space CBP intensities obtained for all of the selected CBPs. In the second column of Table \ref{mean_int_table}, we see similar CBP brightness behaviour for CH2 as it was the case for the ALMA Band 6 data (first column of Table \ref{mean_int_table}). The mean SDO/AIA intensity for the CBPs within the CH2 is lower in value than we found for the CBPs in the quiet Sun outside the CH2. Moreover, we found that all of the average mean intensity values are surpassing 70 DN for CBPs in the quiet Sun outside the CH2 for all five CHs, while for the CBPs inside the CHs the value is under this value for all the CHs, or more specifically under 51 DN. 

\begin{figure*}[h!]
\captionsetup[subfloat]{farskip=1pt,captionskip=1pt}
\centering
\subfloat{\includegraphics[width=0.36\textwidth]{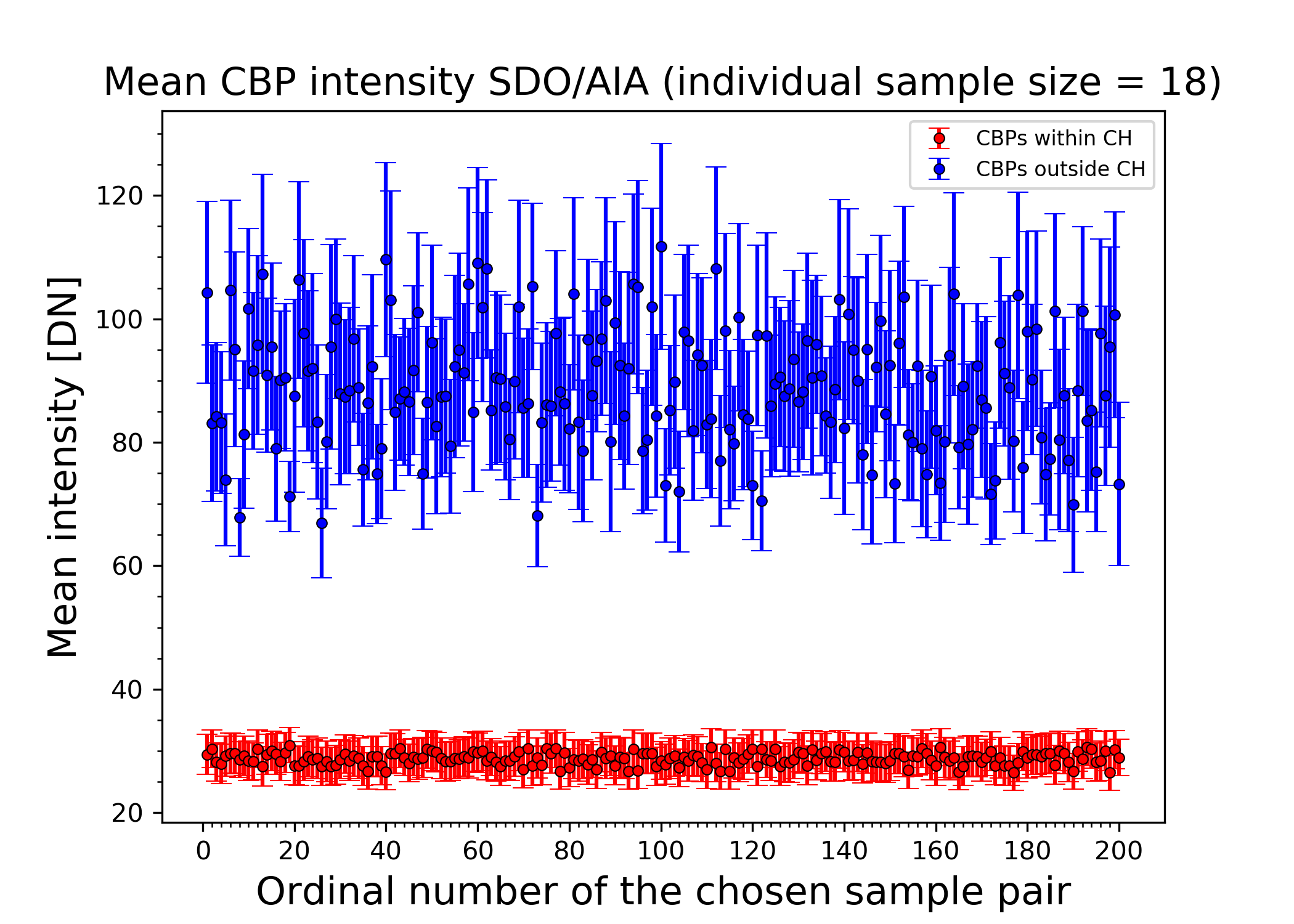}}
\subfloat{\includegraphics[width=0.36\textwidth]{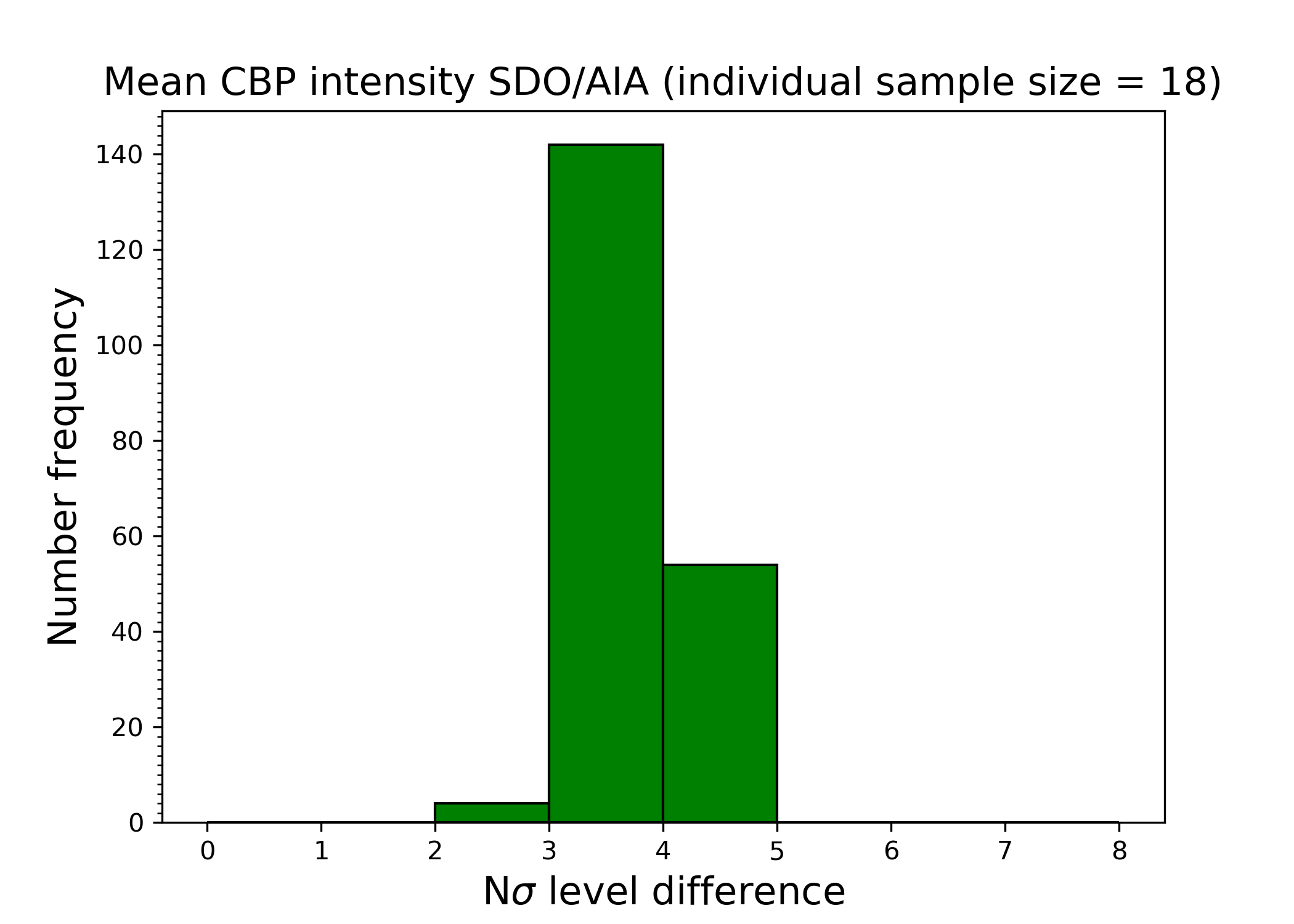}}\\
\subfloat{\includegraphics[width=0.36\textwidth]{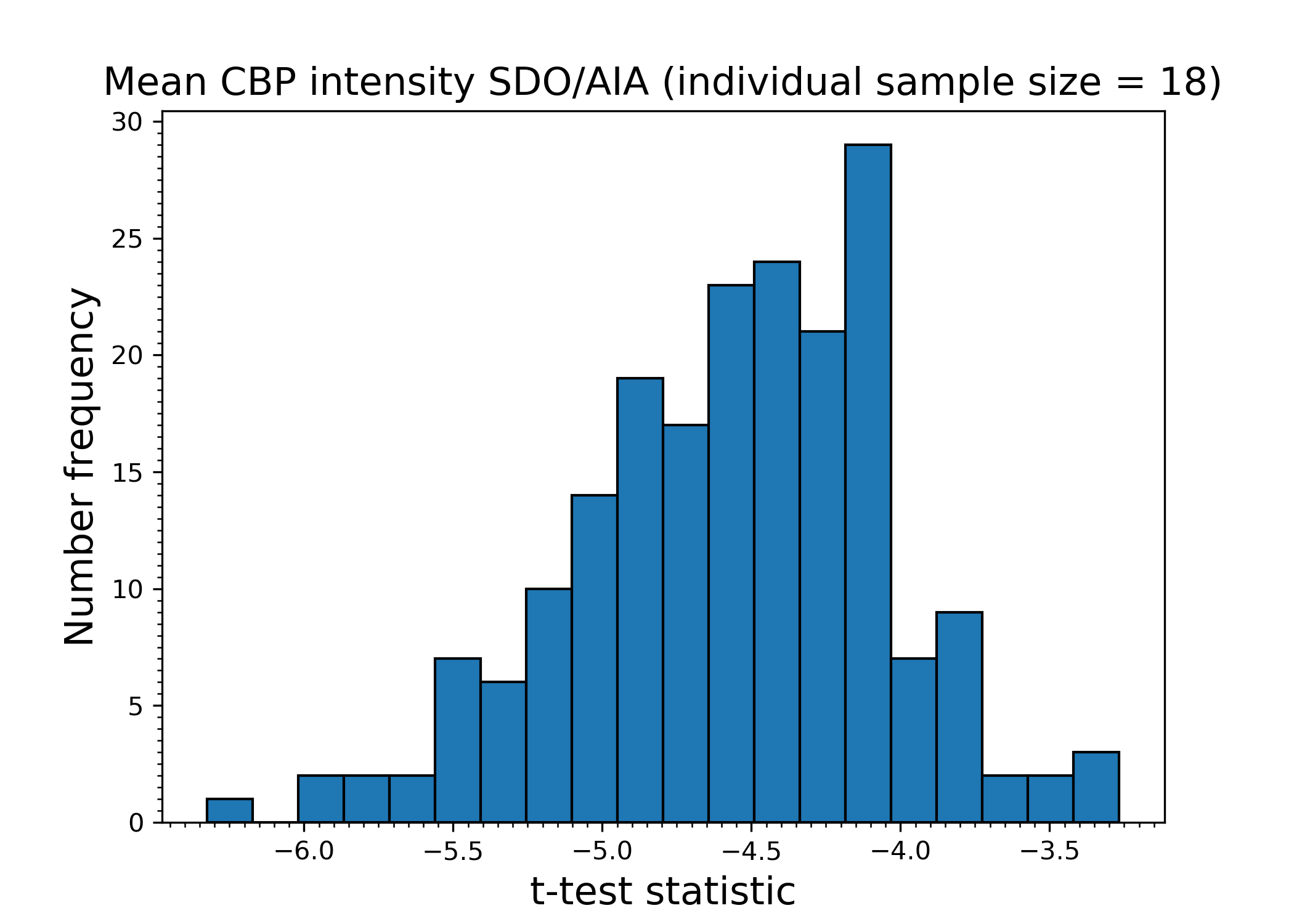}}
\subfloat{\includegraphics[width=0.36\textwidth]{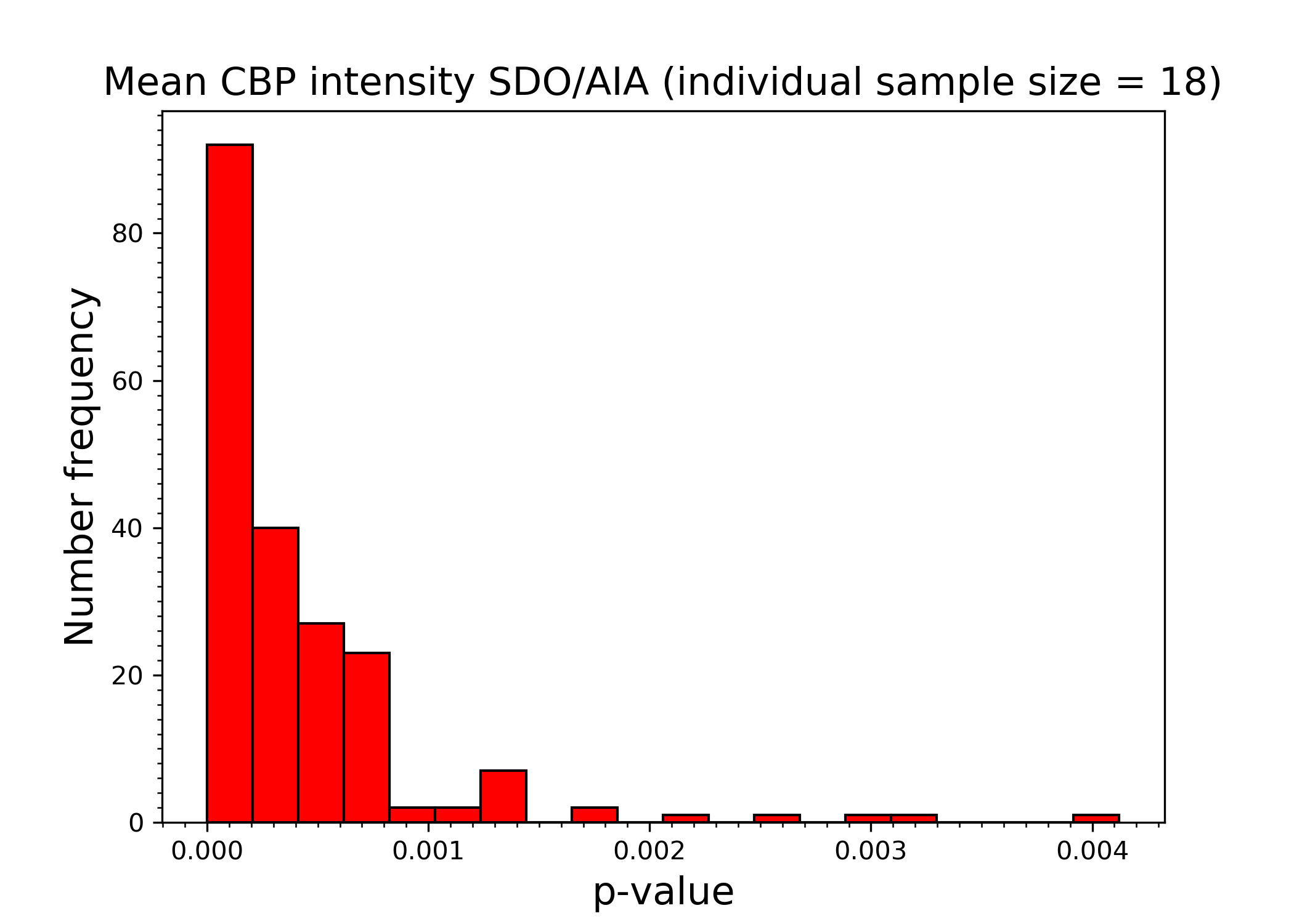}}
\caption{Same as Fig. \ref{mean_int_fig}, but for mean SDO/AIA 193 \AA\space CBP intensity.}
\label{mean_int_fig1}
\end{figure*}

The top left panel of Fig. \ref{mean_int_fig1} shows a separation between the CBPs within and outside of CH2 to be slightly more prominent than in the case of ALMA (Fig. \ref{mean_int_fig}). The dispersion of the mean SDO/AIA intensity for CBPs within CH2 is very narrow, around 10 DN in width, while for the CBPs outside of the CH2, it is much higher. Also, when taking different sample sizes, we found the same effect on the results as it was the case for the ALMA data. Just by looking at the top left panel of Fig. \ref{mean_int_fig1}, we see a very significant difference in the mean SDO/AIA intensity between the CBPs inside and outside the CH, not only for the current case of CH2, but also for the rest of the selected CHs (Fig. \ref{mean_int_fig_ch1_1}, \ref{mean_int_fig_ch3_1}, \ref{mean_int_fig_ch4_1}, and \ref{mean_int_fig_ch5_1}).

Next, in the top right panel of Fig. \ref{mean_int_fig1}, we see that the histogram shows very similar shape as the one for the mean ALMA brightness temperature (Fig. \ref{mean_int_fig}), but with a shift of about 1$\sigma$ to higher values. This time, the largest number of CBP sample pairs have the individual sample mean SDO/AIA intensities differing by about 3$\sigma$ to 5$\sigma$, which suggests this result has a negligible possibility for being coincidental. Bottom row of the Fig. \ref{mean_int_fig1} shows even higher difference between the two groups of CBPs, where $t$-values show that the largest number of CBP sample pairs are grouped between 4$\sigma$ and 5$\sigma$ difference, with a maximum 4$\sigma$ and 4.5$\sigma$. Moreover, the bottom right panel shows that all 200 randomly chosen sample pairs have a $p$-value well under 0.05 threshold. This is a clear indication that the difference in the SDO/AIA intensity between CBPs inside and outside of the CH2 is statistically significant. This was also the case for the rest of the selected CHs (Fig. \ref{mean_int_fig_ch1_1}, \ref{mean_int_fig_ch3_1}, \ref{mean_int_fig_ch4_1}, and \ref{mean_int_fig_ch5_1}), even for small individual sample sizes like the one seen for CH1 (Fig. \ref{mean_int_fig_ch1_1}).

\subsubsection{Correlation between mean CBP brightness and mean CH properties}
\label{correlation}
Based on the mean of the mean CBP brightness values within and outside of CHs in Table \ref{mean_int_table}, we derived linear relations and correlations between those CBP brightness values and mean CH properties in Table \ref{CH_property}. Firstly, we derived the relation between relative brightness ratios of the CBPs within and outside of CHs ($\langle\overline{I}_{\mathrm{in}}\rangle/\langle\overline{I}_{\mathrm{out}}\rangle-1$) for the ALMA Band 6 and SDO/AIA 193 \AA\space data of a form:
\begin{equation}
\frac{\langle\overline{I}_{\mathrm{SDO, in}}\rangle}{\langle\overline{I}_{\mathrm{SDO, out}}\rangle}-1\sim0.79\times\Bigg(\frac{\langle\overline{I}_{\mathrm{ALMA, in}}\rangle}{\langle\overline{I}_{\mathrm{ALMA, out}}\rangle}-1\Bigg);\;\; r_\mathrm{S}=0.97
.\end{equation}
The result shows a very strong correlation that could point to the plasma heating being transferred between two regions at different heights of a CBP loop.

Moving onto CH area, we find a linear relation between the relative ratio of CBP brightness values inside and outside of CHs and CH area ($A_{\mathrm{CH}}$). The obtained linear relations for the ALMA Band 6 and SDO/AIA 193 \AA\space data have a form:
\begin{equation}
\frac{\langle\overline{I}_{\mathrm{ALMA,in}}\rangle}{\langle\overline{I}_{\mathrm{ALMA,out}}\rangle} -1\sim 6.67\times\Bigg(\frac{A_{\mathrm{CH}}}{10^{10}\mathrm{km}^2}\Bigg);\;\; r_\mathrm{S}=0.24
\end{equation}
\begin{equation}
\frac{\langle\overline{I}_{\mathrm{SDO,in}}\rangle}{\langle\overline{I}_{\mathrm{SDO,out}}\rangle}-1\sim10.24\times\Bigg(\frac{A_{\mathrm{CH}}}{10^{10} \mathrm{km}^2}\Bigg);\;\; r_\mathrm{S}=0.45
.\end{equation}
The $r_\mathrm{S}$ coefficient for the ALMA data shows a weak correlation between the two properties, but for the SDO/AIA data we see a higher correlation. If we exclude CH3, for the ALMA data, we get $r_\mathrm{S}=0.51$, and for the SDO/AIA data, $r_\mathrm{S}=0.72$, where we now have a much stronger correlation, showing that the relative brightness ratio rises with CH area, thus indicating to the possibility that larger CHs allow brighter CBPs to form within them.

Moreover, by comparing the CBP brightness values and the absolute values of the signed CH magnetic field ($|B_{\mathrm{sign}}|$), we find:
\begin{equation}
\frac{\langle\overline{I}_{\mathrm{ALMA,in}}\rangle}{\langle\overline{I}_{\mathrm{ALMA,out}}\rangle}-1\sim0.05\times\Bigg(\frac{|B_{\mathrm{sign}}|}{\mathrm{Gauss}}\Bigg);\;\; r_\mathrm{S}=0.01
\end{equation}
\begin{equation}
\frac{\langle\overline{I}_{\mathrm{SDO,in}}\rangle}{\langle\overline{I}_{\mathrm{SDO,out}}\rangle}-1\sim -0.17\times\Bigg(\frac{|B_{\mathrm{sign}}|}{\mathrm{Gauss}}\Bigg);\;\; r_\mathrm{S}=-0.03
.\end{equation}
Due to $r_\mathrm{S}$ being close to zero, there is no visible correlation between the CBP brightness ratio and $|B_{\mathrm{sign}}|$. However, by excluding CH3, we get $r_\mathrm{S}=-0.54$ for the ALMA data and $r_\mathrm{S}=-0.53$ for the SDO/AIA data, where we now see a modest anticorrelation between the two properties. This could mean that the higher the $|B_{\mathrm{sign}}|$ is, the lower is the brightness of CBPs within CHs. Therefore, the CH magnetic field has an opposite effect on the CBP properties than the CH area has.

\subsection{Projected CBP area}
\subsubsection{ALMA data}
\label{area_alma_data}
Finally, we come to the CBP area measurements, where the overall means of the projected CBP areas for ALMA Band 6 data are presented in Table \ref{area_table}.
\begin{table}[h!]
\caption{Mean values of measured projected ALMA Band 6 CBP areas ($\langle A\rangle_{\mathrm{ALMA}}$) and mean values of measured projected SDO/AIA 193 \AA\space CBP areas ($\langle A\rangle_{\mathrm{SDO}}$) within and outside five chosen CHs, with corresponding standard errors.}     
\centering 
\scalebox{0.85}{     
\begin{tabular}{c c c l l}
\hline   
&$\langle A\rangle_{\mathrm{ALMA}}$ [arcsec$^2$]&$\langle A\rangle_{\mathrm{SDO}}$ [arcsec$^2$]\\
&within CH / outside CH&within CH / outside CH\\
\hline
        CH1&480 $\pm$ 130 / 810 $\pm$ 70&180 $\pm$ 40 / 270 $\pm$ 40\\
        CH2&540 $\pm$ 70 / 870 $\pm$ 60&280 $\pm$ 50 / 400 $\pm$ 50\\
        CH3&540 $\pm$ 60 / 900 $\pm$ 50&210 $\pm$ 40 / 370 $\pm$ 50\\
        CH4&630 $\pm$ 50 / 790 $\pm$ 40&300 $\pm$ 50 / 300 $\pm$ 30\\
        CH5&610 $\pm$ 60 / 750 $\pm$ 50&260 $\pm$ 40 / 290 $\pm$ 30\\
\hline   
\end{tabular}
}
\label{area_table}
\end{table}
\begin{figure*}[h!]
\captionsetup[subfloat]{farskip=1pt,captionskip=1pt}
\centering
\subfloat{\includegraphics[width=0.36\textwidth]{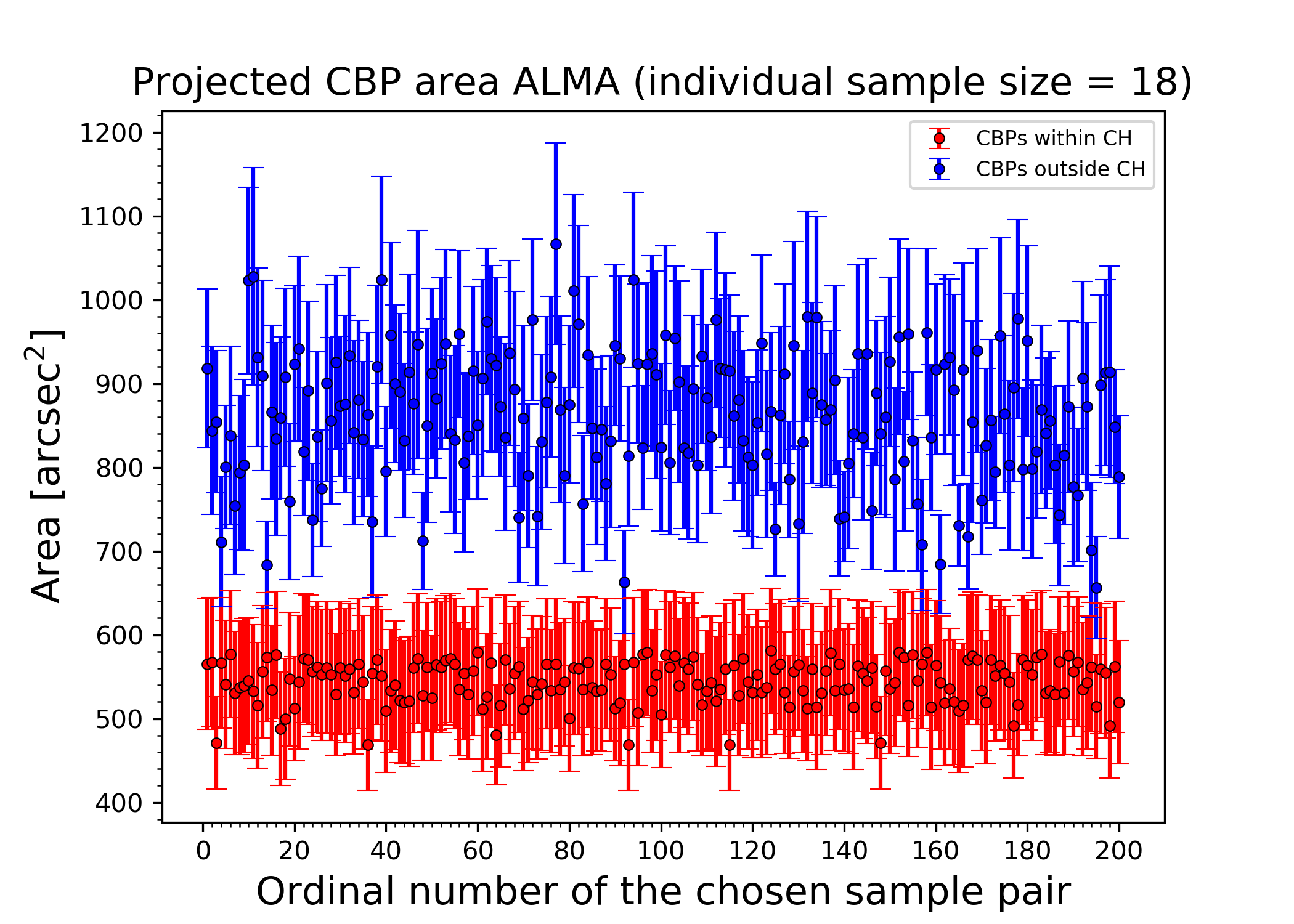}}
\subfloat{\includegraphics[width=0.36\textwidth]{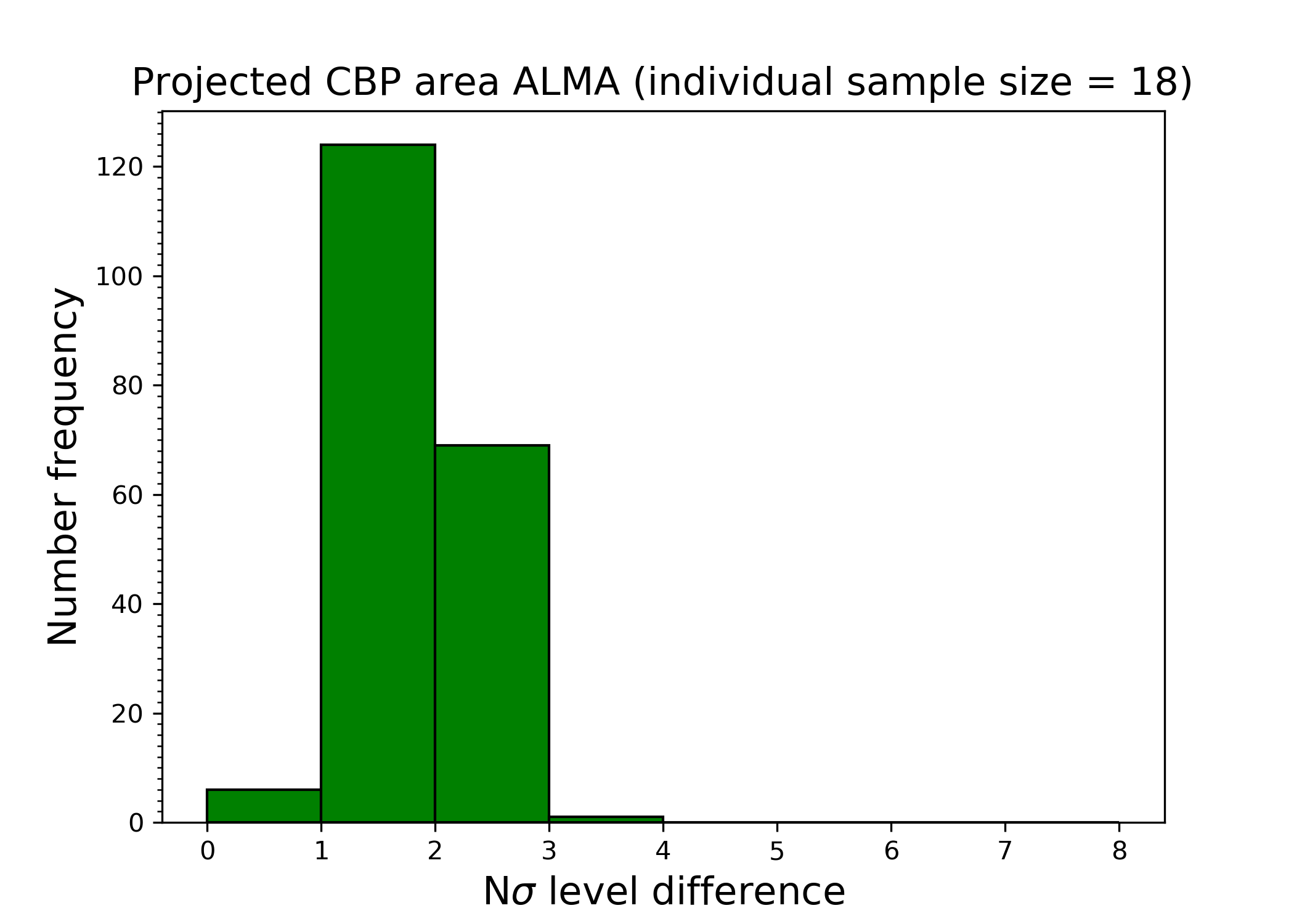}}\\
\subfloat{\includegraphics[width=0.36\textwidth]{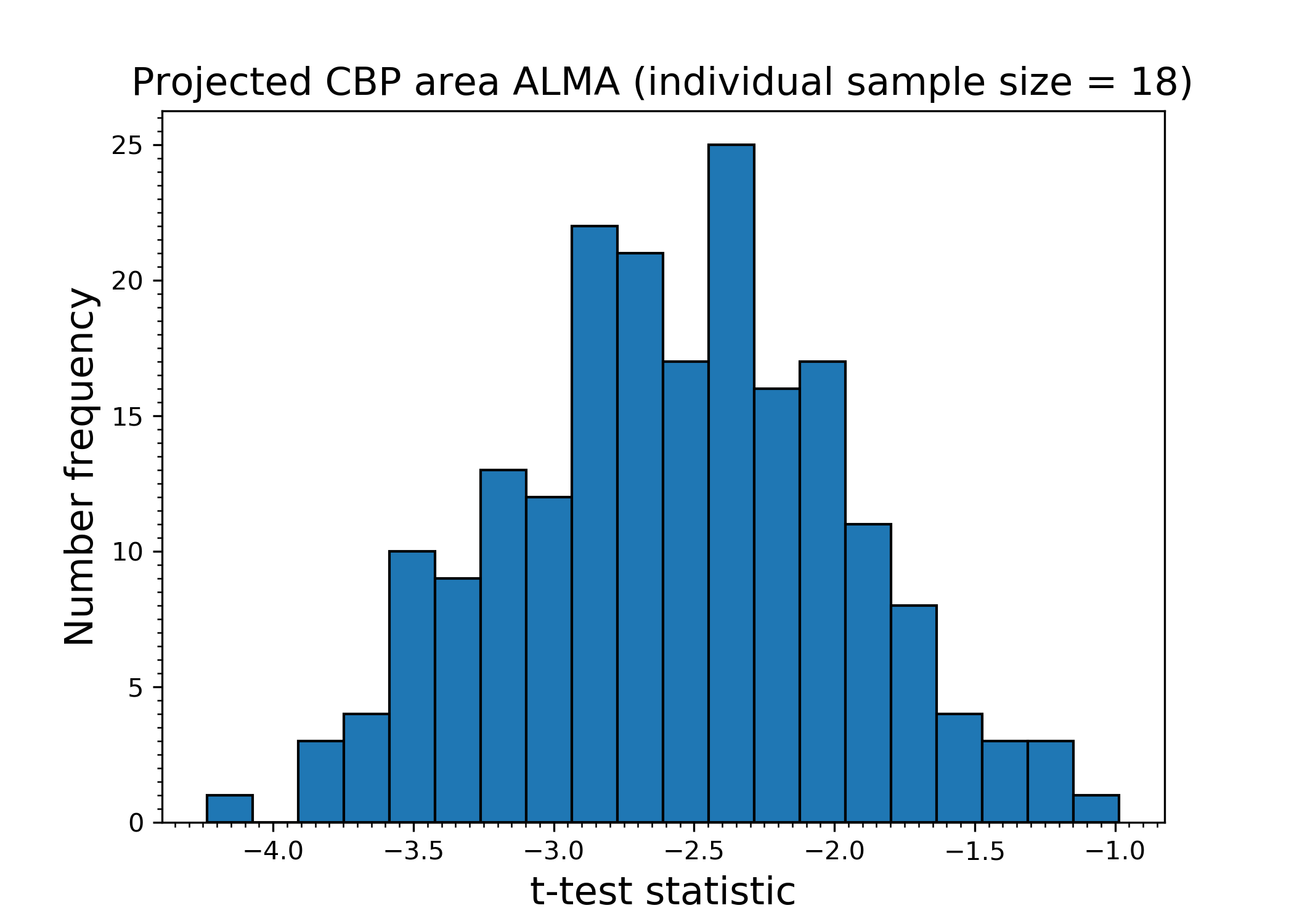}}
\subfloat{\includegraphics[width=0.36\textwidth]{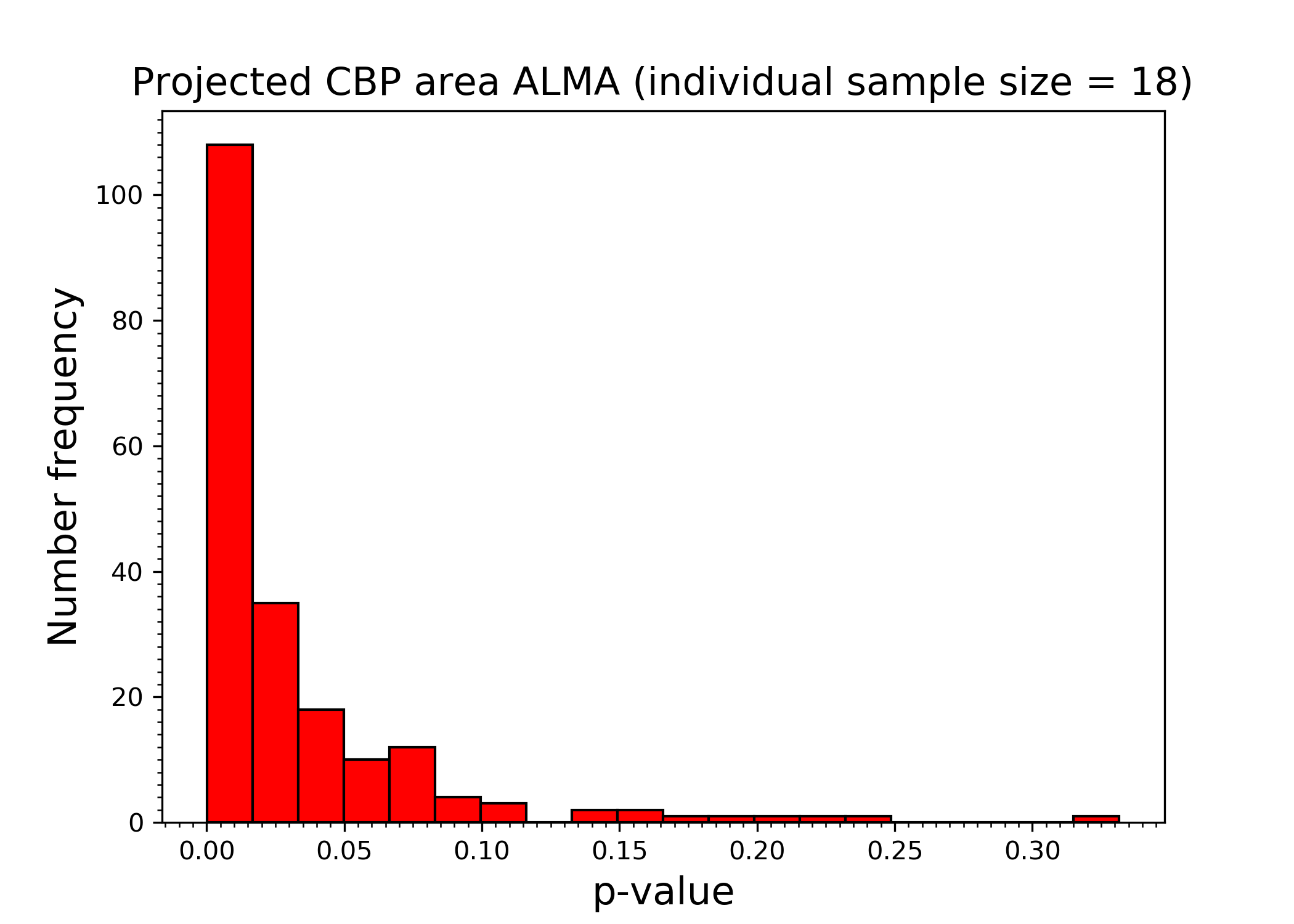}}
\caption{Statistical analysis of projected ALMA Band 6 CBP area. Top row: Left panel shows the mean values of the CBP areas in the ALMA Band 6 image with corresponding standard errors of 200 randomly chosen equal size CBP sample pairs, with one sample containing CBPs within (red) and the other outside (blue) the CH2. The right panel shows histogram of the largest N for which the relation (\ref{Nsigma}) holds true. Bottom row: Left panel shows histograms of the $t$-test statistic values ($t$-values) and the right panel shows the histogram of the $p$-values obtained for the mean values of CBP areas in the ALMA Band 6 image. Individual CBP sample contains 18 randomly chosen CBPs out of the many selected CBPs either within or outside the CH2.}
\label{area_fig}
\end{figure*}
Results from Table \ref{area_table} show a clear difference in the ALMA area between CBPs inside and outside the CHs. Here the CBPs inside all five CHs have on average smaller areas, with the highest total mean value of about 630 arcsec$^2$ found for CBPs within CH4. The smallest value for CBPs within a CH was found for CH1 of about 480 arcsec$^2$, but also with a much higher standard error due to the small number of CBPs within CH1. The maximum total mean area for CBPs outside a CH was found for CH3 with a value of about 900 arcsec$^2$.

Statistical analysis done for the CBP areas using the ALMA Band 6 data is presented in Fig. \ref{area_fig}. The top left panel of Fig. \ref{area_fig} shows a pronounced separation between CBPs within and outside of CH2. The area dispersion for CBPs inside the CH2 is approximately two times smaller than for the ones outside the CH2. Both different CBP group distributions are dispersed around the general mean values mentioned in Table \ref{area_table}, where the mean  area value for CBPs in the quiet Sun outside the CH2 is more than 300 arcsec$^2$ higher than for the ones inside CH2. By changing the individual CBP sample sizes, we found similar behaviour as was previously the case for the mean CBP brightness. We also obtained similar results for the rest of the CHs (Fig. \ref{area_fig_ch1}, \ref{area_fig_ch3}, \ref{area_fig_ch4}, and \ref{area_fig_ch5}), with a larger overlap between the mean values obtained for CH1 and CH5, but depending on the sample size, with a small or almost no area difference visible between CBPs within and outside of CH5.

Next, the top right panel of Fig. \ref{area_fig} points to a very high number of CBP sample pairs having the area difference between 1$\sigma$ and 3$\sigma$. A similar result can be seen in the bottom row of Fig. \ref{area_fig}, where most of the CBP sample pairs have their areas differing between 2$\sigma$ and 3$\sigma$. Moreover, a $p$-value histogram (bottom right panel of Fig. \ref{area_fig}) indicates that there is a maximum number of sample pairs with a $p$-value under 0.05, with a small number of them also having $p$-values above the chosen threshold. This result clearly indicates a rather significant difference in CBP area between CBPs inside and outside CH2. Analysing different sample sizes showed that the area difference was more significant for larger samples. Similar results were found for the rest of the CHs, where the area difference for CH1 (Fig. \ref{area_fig_ch1}) was less significant than the one presented here for CH2, with a maximum closer to 1$\sigma$. Similarly to CH1, CH5 (Fig. \ref{area_fig_ch5}) also showed a not very significant difference around 1.5$\sigma$. However, for CH3 (Fig. \ref{area_fig_ch3}) and CH4 (Fig. \ref{area_fig_ch4}), we found more significant CBP area difference around 2$\sigma$ for CH4 to even above 3$\sigma$ for CH3. 

\subsubsection{SDO/AIA data}
\label{area_sdo_data}
For the SDO/AIA 193 \AA\space data, the maximum CBP area and the overall means of the measured CBP SDO/AIA areas for all selected CHs are presented in the second column of Table \ref{area_table}.
\begin{figure*}[h!]
\captionsetup[subfloat]{farskip=1pt,captionskip=1pt}
\centering
\subfloat{\includegraphics[width=0.36\textwidth]{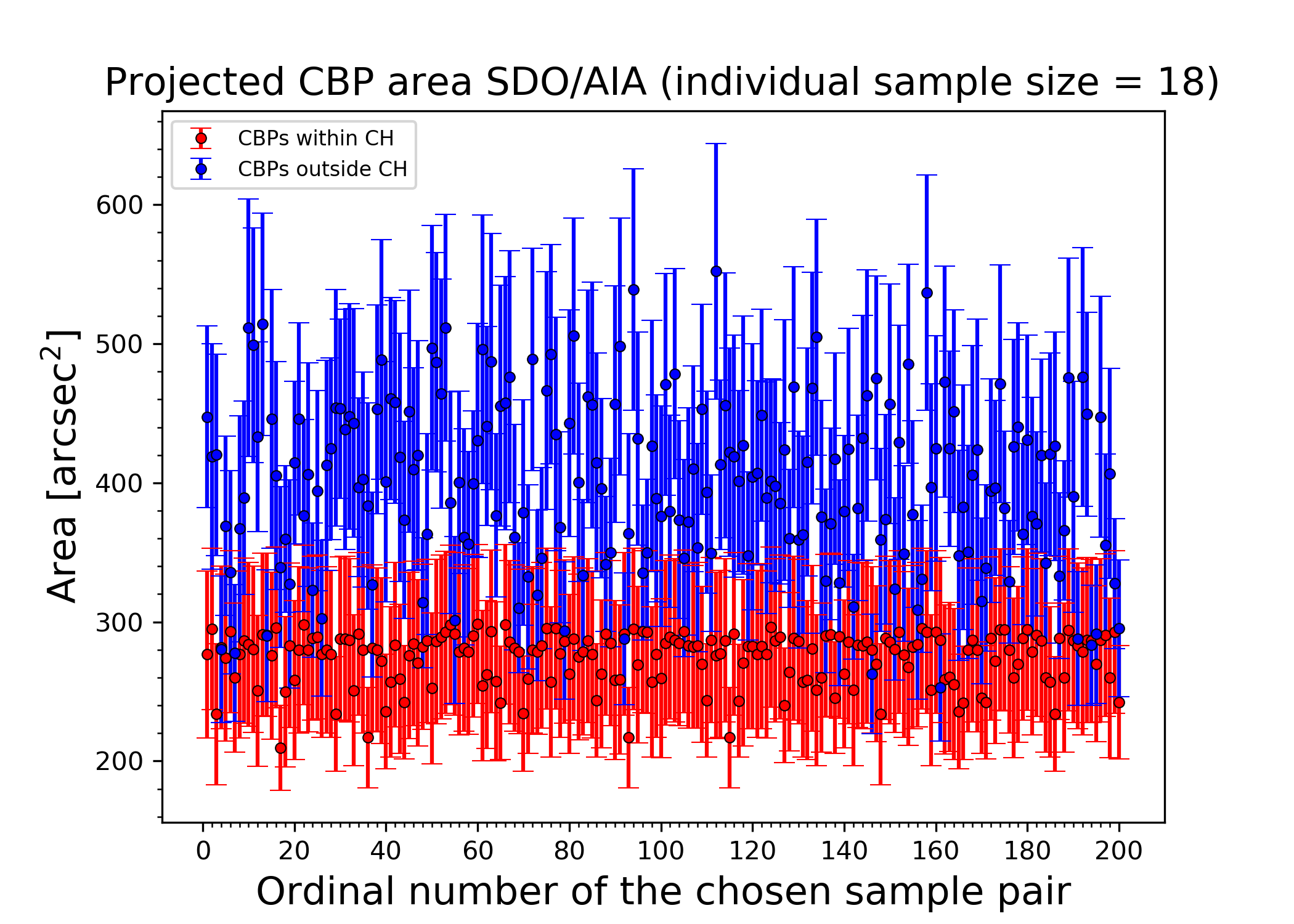}}
\subfloat{\includegraphics[width=0.36\textwidth]{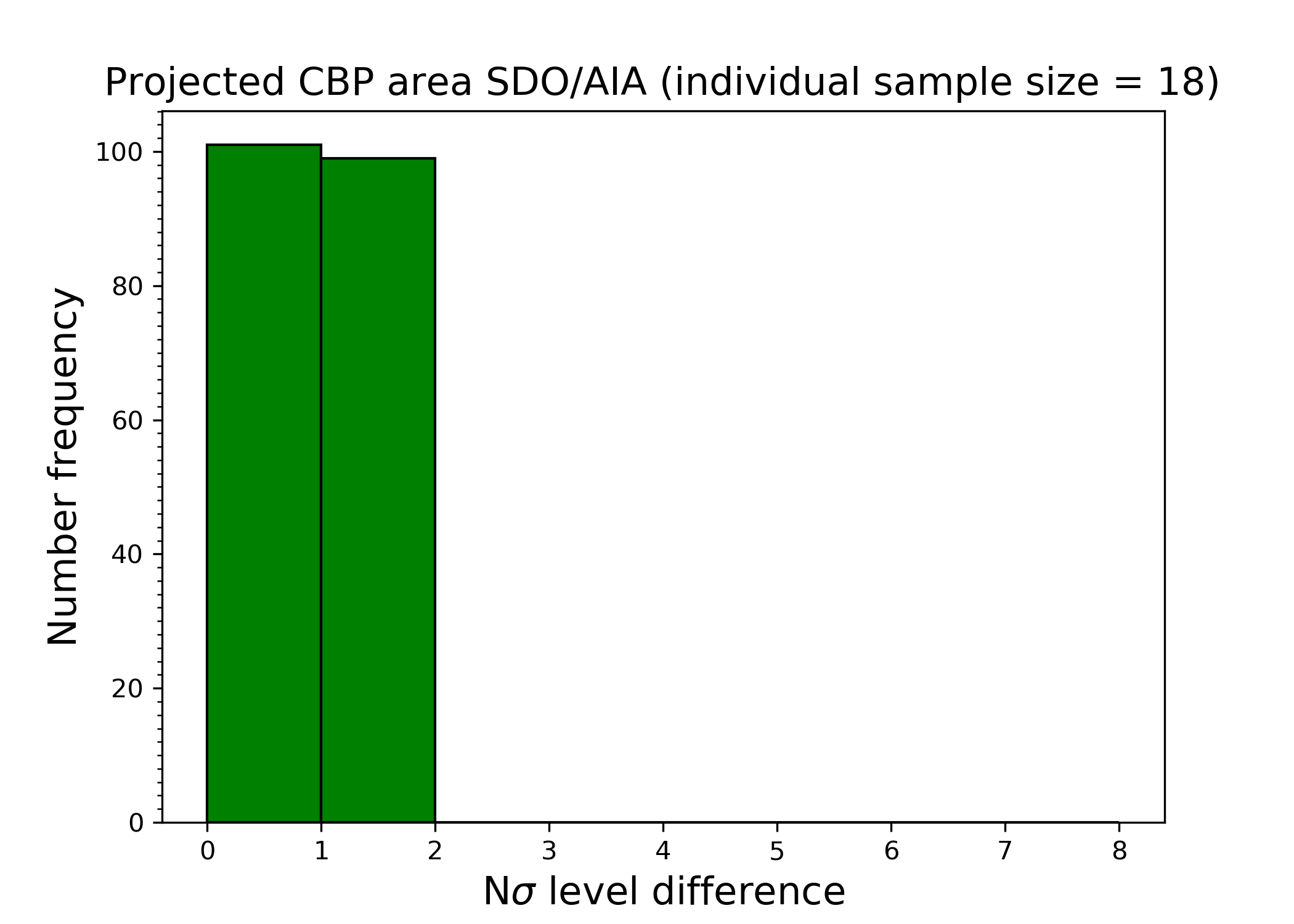}}\\
\subfloat{\includegraphics[width=0.36\textwidth]{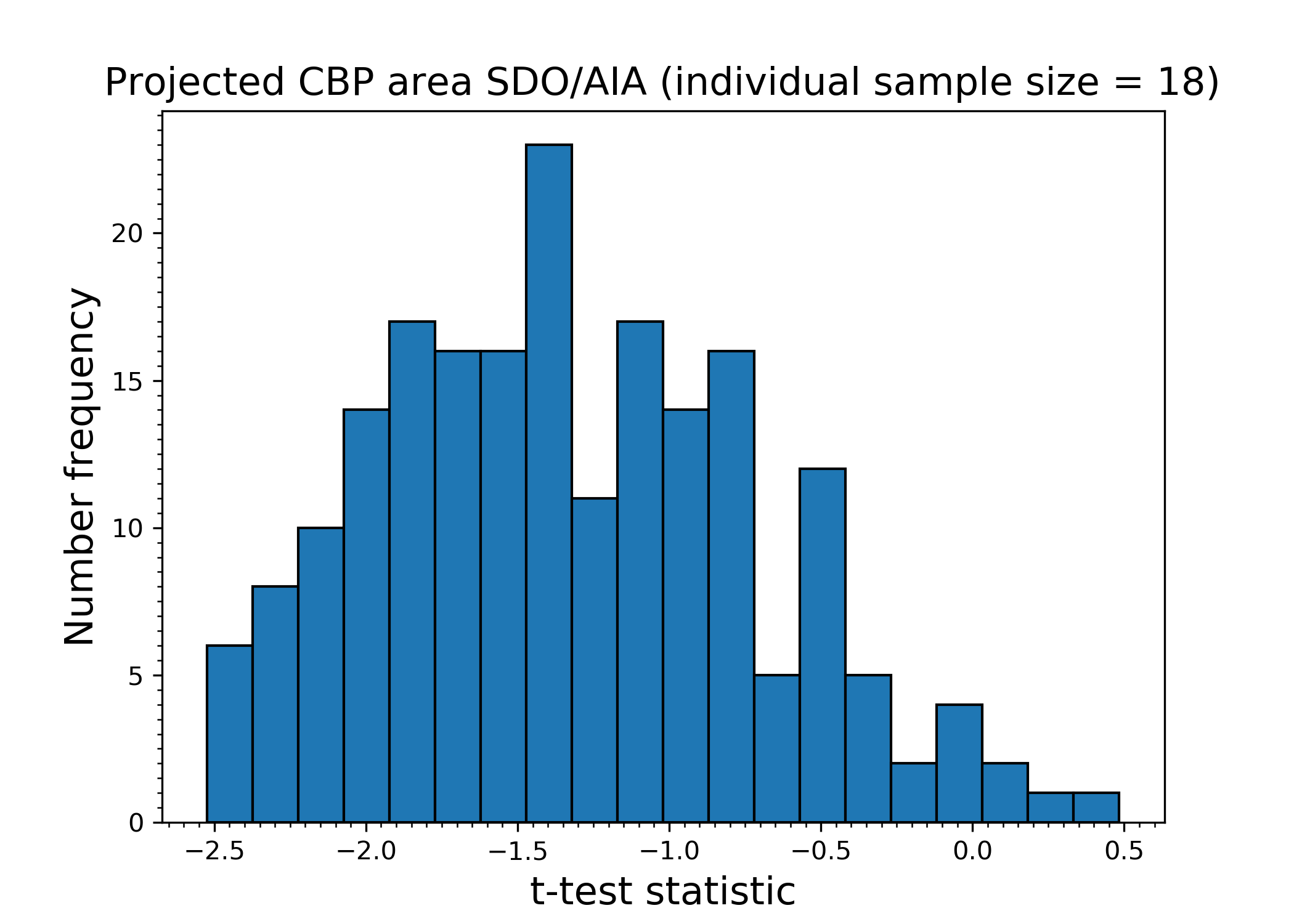}}
\subfloat{\includegraphics[width=0.36\textwidth]{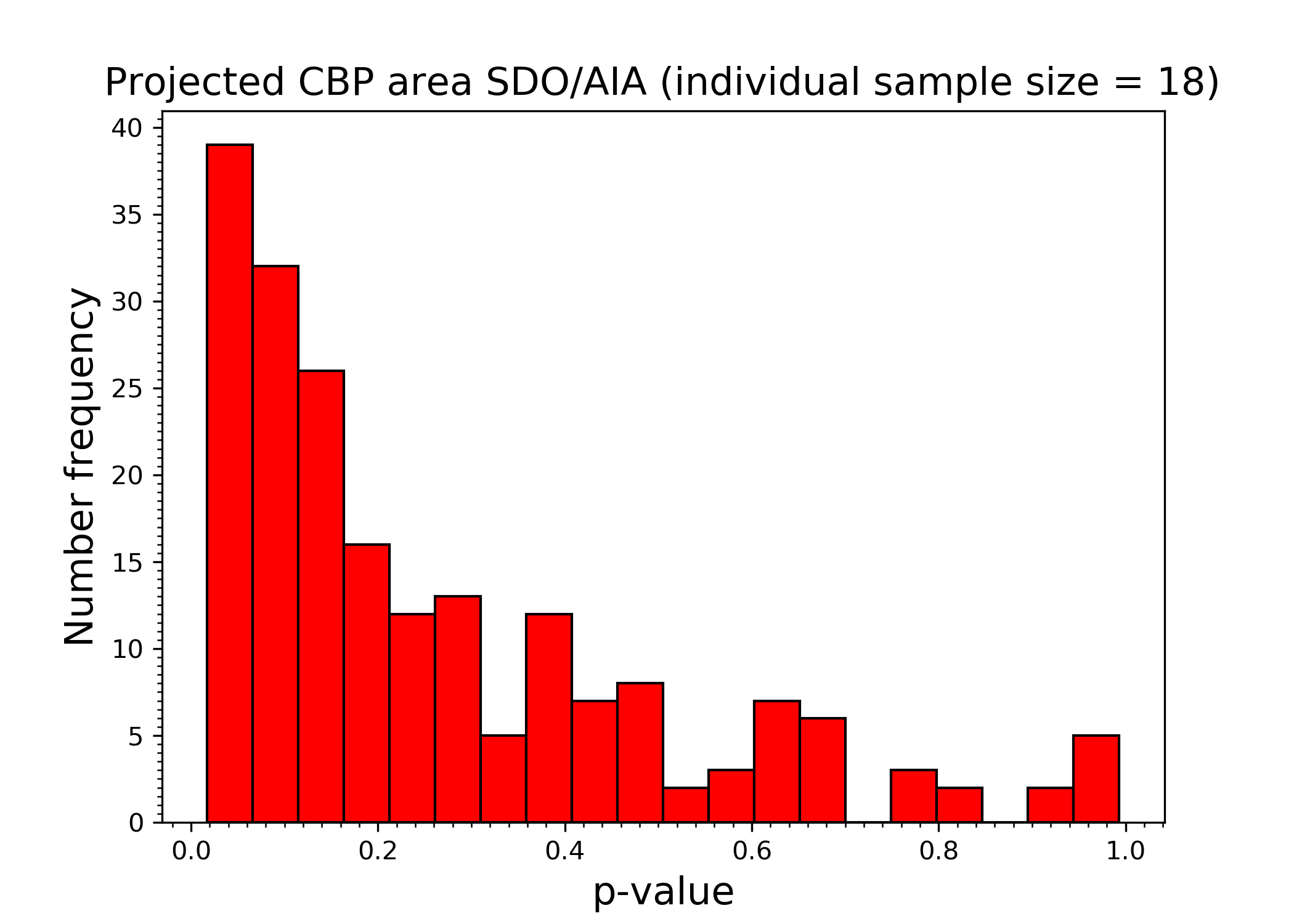}}
\caption{Same as Fig. \ref{area_fig}, but for projected SDO/AIA 193 \AA\space CBP area.}
\label{area_fig1}
\end{figure*}
Based on the measured SDO/AIA CBP areas in Table \ref{area_table}, we found that for the mean value of all the measured SDO/AIA areas for CBPs inside almost all of the CHs have on average smaller values than those outside of CHs. The exception here is CH4, for which the measured CBP areas have similar values, meaning that there is no difference for the SDO/AIA area between CBPs within and outside of CH4 since the general mean area values in Table \ref{area_table} have a perfect overlap. A smaller overlap is found for CH5, for which we see rather similar values for the overall mean areas both inside and outside of CH5, which differ by only 30 arcsec$^2$.

Statistical analysis of the measured SDO/AIA CBP areas is presented in Fig. \ref{area_fig1}. The top left panel of Fig. \ref{area_fig1} indicates a clear separation between CBPs inside and outside of CH2. The mean area dispersion is much smaller for CBPs inside CH2, with a width just above 100 arcsec$^2$ around the mean value of 280 arcsec$^2$, while CBPs outside CH2 have three times higher dispersion around the mean of about 400 arcsec$^2$. Even though we see a separation, there are still some small overlaps visible between the CBPs within and outside of CH2. Results found for the rest three out of four CHs (Fig. \ref{area_fig_ch1_1}, \ref{area_fig_ch3_1}, and \ref{area_fig_ch5_1}) indicate a similar behaviour of the measured SDO/AIA CBP area, with more overlapping seen for CH1 (Fig. \ref{area_fig_ch1_1}) and CH5 (Fig. \ref{area_fig_ch5_1}), especially for smaller CBP sample sizes. The one exception here is CH4 (Fig. \ref{area_fig_ch4_1}), for which we obtained the largest deviation from the results of CH2. Here for CH4 we had no visible separation between CBPs within and outside of it even for the large individual CBP sample sizes. This was already seen in Table \ref{area_table}, where we obtained very similar CBP area values both inside and outside of CH4.

In the top right panel of Fig. \ref{area_fig1}, for CH2 we see that the highest number of the CBP sample pairs have their SDO/AIA areas differing by less than 2$\sigma$. Similarly, the histogram of the obtained $t$-value in the bottom row of Fig. \ref{area_fig1} indicates that the highest number of the CBP sample pairs have CBP areas differing between 1$\sigma$ and 2$\sigma$, with a maximum of the distribution around 1.5$\sigma$. In the $p$-value histogram on the bottom right, we see that only a part of the CBP sample pairs have a $p$-value under 0.05, with a non negligible number of them having $p$-values above this threshold. These results clearly show that the CBP area difference is not too significant, but at the same time it is not too small, which indicates that there is after all a small difference in CBP area still visible. Similar results were found for CH1 (Fig. \ref{area_fig_ch1_1}) and CH3 (Fig. \ref{area_fig_ch3_1}) as well, where the SDO/AIA area difference for CH3 was about 2$\sigma$. On the other hand, for CH4 (Fig. \ref{area_fig_ch4_1}) and CH5 (Fig. \ref{area_fig_ch5_1}), the area difference was less than 1$\sigma$, sometimes even less than 0.5$\sigma$, and the obtained $p$-value distributions were, depending on the sample sizes, more or less equally scattered across all the possible values. For these two CHs specifically, the results indicate that there is extremely small, or even no significant difference in the SDO/AIA CBP area.

In comparison, the ALMA CBP area difference is more significant than the one obtained for the SDO/AIA. This could be the result of a different spatial resolution of the ALMA and SDO/AIA data, where a poorer spatial resolution of the ALMA single-dish images makes it difficult to distinguish detailed morphological CBP features. Therefore, the definition of the ALMA CBP areas is more difficult, and this could have resulted in a higher area difference for ALMA.

\subsubsection{Correlation between projected CBP area and mean CH properties}
\label{correlation1}
Following the analysis done in Subsection \ref{correlation} for the mean CBP brightness, we also find certain correlations between CBP areas and CH properties. Considering first only the CBP areas, we find a relation between relative area ratios ($A_\mathrm{in}/A_\mathrm{out}-1$) for the ALMA Band 6 and SDO/AIA 193 \AA\space data in a form:
\begin{equation}
\frac{A_\mathrm{SDO,in}}{A_\mathrm{SDO,out}}-1\sim 0.59\times\Bigg(\frac{A_\mathrm{ALMA,in}}{A_\mathrm{ALMA,out}}-1\Bigg);\;\; r_\mathrm{S}=0.94
.\end{equation}
Here we see a very strong correlation, similar to one obtained for mean CBP brightness before, where now the result shows that the change in the CBP area at coronal part of a CBP loop is also visible at chromospheric part, and vice versa.

Next, by comparing the CBP areas and the CH areas ($A_\mathrm{CH}$), we find a linear relation for the ALMA Band 6 and SDO/AIA 193 \AA\space data of:
\begin{equation}
\frac{A_\mathrm{ALMA,in}}{A_\mathrm{ALMA,out}}-1 \sim 7.50\times\Bigg(\frac{A_\mathrm{CH}}{10^{10} \mathrm{km}^2}\Bigg);\;\; r_\mathrm{S}=0.33
\end{equation}
\begin{equation}
\frac{A_\mathrm{SDO,in}}{A_\mathrm{SDO,out}}-1\sim 4.26\times\Bigg(\frac{A_\mathrm{CH}}{10^{10} \mathrm{km}^2}\Bigg);\;\; r_\mathrm{S}=0.30
.\end{equation}
Both ALMA and SDO/AIA data show a rather weak correlation between CBP and CH areas. However, by excluding CH3 from consideration, we find a much stronger correlation with $r_\mathrm{S}=0.54$ for the ALMA data and $r_\mathrm{S}=0.67$ for the SDO/AIA data. Similarly as in the Subsection \ref{correlation}, here we can conclude that larger CHs allow larger CBPs to form within the CH boundaries.

Moreover, if take the absolute value of the signed CH magnetic field ($|B_\mathrm{sign}|$), for the ALMA Band 6 and SDO/AIA 193 \AA\space data we find:
\begin{equation}
\frac{A_\mathrm{ALMA,in}}{A_\mathrm{ALMA,out}}-1\sim -1.00\times\Bigg(\frac{|B_\mathrm{sign}|}{\mathrm{Gauss}}\Bigg);\;\; r_\mathrm{S}=-0.20
\end{equation}
\begin{equation}
\frac{A_\mathrm{SDO,in}}{A_\mathrm{SDO,out}}-1\sim 0.38\times\Bigg(\frac{|B_\mathrm{sign}|}{\mathrm{Gauss}}\Bigg);\;\; r_\mathrm{S}=0.12
.\end{equation}
Here we again see that there is almost no correlation visible as it was the case for the mean CBP brightness before. But, if we exclude CH3, we find $r_\mathrm{S}=-0.71$ for the ALMA data and $r_\mathrm{S}=-0.49$ for the SDO/AIA data, where we see a modest to strong anticorrelation. This result shows that the larger $|B_\mathrm{sign}|$ suppresses the CBP areas to smaller values within the CHs.

\section{Discussion}
\label{Discussion}
In this paper, we report measurements of the mean brightness and areas of CBPs within the boundaries of five selected CHs at specific times, as well as outside them in the quiet Sun region. A goal of these measurements was to find if the physical properties differ depending on the region where CBPs reside.

\subsection{Mean CBP brightness}
\label{mean_int_discussion}
In the SDO/AIA images, CBPs appear as bright small-scale coronal loops, while in the ALMA images, those same CBPs correspond to bright point features overlaying the bipolar structures seen in the SDO/HMI magnetograms, as is expected by \citeauthor{Brajsa_2018} (\citeyear{Brajsa_2018}, \citeyear{Brajsa}) and \cite{Madjarska}. The ALMA Band 6, SDO/AIA 193 \AA\space, and SDO/HMI images in Fig. \ref{CH_ALMA_SDO1}, \ref{CH_ALMA_SDO2}, \ref{CH_ALMA_SDO3}, \ref{CH_ALMA_SDO4}, and \ref{CH_ALMA_SDO5} show selected CBPs at the times the images were taken, where we see these exact features with a variety of brightness values.

Figures \ref{mean_int_fig}, \ref{mean_int_fig1}, \ref{mean_int_fig_ch1}, \ref{mean_int_fig_ch1_1}, \ref{mean_int_fig_ch3}, \ref{mean_int_fig_ch3_1}, \ref{mean_int_fig_ch4}, \ref{mean_int_fig_ch4_1}, \ref{mean_int_fig_ch5}, and \ref{mean_int_fig_ch5_1} show a visible separation in the mean intensity between the CBPs within CH boundaries and those outside them in the quiet Sun. Here the CBPs inside all CHs appear to be fainter than the CBPs outside the CHs in both ALMA and SDO data. A better picture of the CBP brightness difference was obtained by the statistical analysis done using the relation (\ref{Nsigma}) and $t$-test method. The analysis resulted in a high significance in the mean brightness difference between the CBPs within and outside of CHs, which was about 2$\sigma$ or more. This indicates that the mean CBP brightness for both ALMA and SDO/AIA data at any time is significantly lower for the CBPs inside a CH in comparison to the ones outside of it. From the analysis, we also obtained a wider dispersion of mean CBP brightness values for CBPs outside of the CHs in the quiet Sun region, which shows that CBPs outside of CHs can have a wider range of possible brightness values, both high and low. This result, with a very significant CBP brightness difference discussed previously, indicates that certain physical conditions, for example the magnetic field strength, of the areas where CBPs reside might affect the brightness properties of the CBPs that we observe at any wavelength/height, especially inside CHs where those conditions might prevent CBPs from reaching high brightness values, or in other words temperatures.

The abundance and strength of open magnetic field within CHs (\citeauthor{Hofmeister_2017} \citeyear{Hofmeister_2017}; \citeauthor{Hofmeister_2019} \citeyear{Hofmeister_2019}; \citeauthor{Tian} \citeyear{Tian}) can in fact have effects on the surrounding closed field, where dipole regions with CBPs are formed. Model results confirm that low-lying loops are mostly present within CHs and that they are on average flatter compared to the outside quiet Sun regions (\citeauthor{Wiegelmann_2004} \citeyear{Wiegelmann_2004}; \citeauthor{Wiegelmann_2005} \citeyear{Wiegelmann_2005}). The radially outgoing external field in CHs may constrain CBPs to smaller heights better than the more random fields in the quiet Sun (see Fig. 3 in \citeauthor{Wiegelmann_2005} \citeyear{Wiegelmann_2005} and coronal magnetic-field topology of CBPs in \citeauthor{Galsgaard} \citeyear{Galsgaard}). A similar result was found in a recent study by \cite{Heinemann_2}, where the CH magnetic field is derived by using bright bipolar structures in the CH in the SDO/AIA 304 \AA\space filtergrams, which were also visible in the SDO/AIA 193 \AA\space filtergrams. The authors found that the strength of the CH magnetic field constrains how large and how far up the bright structures can be, indicating that the appearance of CBPs and the CH magnetic field are linked. This is evident in the anticorrelation between the relative mean CBP brightness ratios and the absolute value of the signed CH magnetic field, previously discussed in Subsection \ref{correlation}. There the higher absolute signed magnetic field results in lower brightness values for CBPs within a CH, meaning that the larger abundance of open unipolar magnetic fields, and their strength, might limit the CBPs to lower temperatures. Even though there is a stronger magnetic field inside CHs, there is less magnetic activity than in the quiet Sun regions outside CHs, for example flux emergence. \cite{Nobrega} found that the flux emergence can indeed enhance the CBP activity, and therefore its intensity and temperature. Hence, less magnetic activity inside CHs might result in lower input in magnetic energy within CBP loop structure at lower heights, producing lower heating of a CBP. One could use differential emission measure to see what the temperature is truly like for CBPs inside and outside CHs and then model the responsible mechanisms accordingly (out of the scope of this work).

Alternatively, the magnetic connectivity environment in a CH is more stable, since the external field is always directed the same way. This might help smaller CBPs last longer against the effects of convective erosion against external fields. Such low-lying CBP loops inside CHs might be related to low plasma flows (\citeauthor{Wiegelmann_2005} \citeyear{Wiegelmann_2005}), revealing a lower temperature/density. In addition, we observe centrally on-disk located features as line-of-sight integrated intensities, and CBPs in the vicinity of low intensity open flux tubes might appear less bright compared to those located outside CHs. Since CBPs seem to appear in layered structures with different temperatures, where hotter loops are overlaying the cooler loops (\citeauthor{Madjarska} \citeyear{Madjarska}), within CHs only the small-scale cooler loops might be present. An emission measure analysis of the CBPs using multiple SDO filters may help decide what combination of factors is driving the intensity differences (beyond the scope of this study).

We also must not exclude the possibility of the CH morphology, for example CH area, also causing the observed CBP brightness difference. This is evident in the correlation between the relative mean CBP brightness ratio and CH area, where in Subsection \ref{correlation} we obtained a rather strong correlation when CH3 was excluded. Therefore, our result indicates that larger CHs might allow CBPs to obtain higher temperatures and altitudes.

Moreover, we report a slightly smaller difference in the measured mean CBP brightness seen for the ALMA data in comparison to the SDO/AIA data, which was independent of the CBP sample size. Our assumption is that the spatial resolution of the ALMA Band 6 single-dish images might be the main cause of such a result, resulting in the more uncertain boundary estimation, and therefore the estimation of the CBP brightness. This effect will be discussed in more detail in Subsection \ref{area_discussion}.

Furthermore, we report that the individual CBP sample size influences the mean brightness difference, as well as the area difference, seen between the CBPs inside and outside of CHs. Taking larger sample sizes closer to the maximal number of CBPs found inside a CH resulted in a more pronounced difference seen between the CBPs inside and outside CHs, where the mean value range barely, or did not even overlapped at all. Therefore, we recommend using larger samples containing the number of CBPs closer to a number of them inside the CH of interest to maximise the observed difference in physical CBP properties between these two different groups of CBPs, like the results presented throughout this paper.

\subsection{Projected CBP area}
\label{area_discussion}
The measured CBP areas presented in Fig. \ref{area_fig}, \ref{area_fig1}, \ref{area_fig_ch1}, \ref{area_fig_ch3}, \ref{area_fig_ch3_1}, \ref{area_fig_ch4}, and \ref{area_fig_ch5} also showed a visible separation between the CBPs inside and outside all five CHs. We find that CBPs inside the boundaries of CHs have on average smaller areas than those outside in the quiet Sun, that have a wider range of possible areas, similarly to the previously discussed mean CBP brightness. The area difference between the CBPs in different regions was not as significant as it was the case with the mean CBP brightness. This could be due to the fact that we observe two different physical properties, brightness temperature (ALMA) and EUV emission (SDO/AIA), for which we cannot a priori assume that their change between different solar regions behaves equally. For the ALMA data, we indeed obtained a rather significant CBP area difference, with the exception of CH1, but for the SDO/AIA data, this difference was for most of the CHs not too significant, with the exception of CH3 having a very significant CBP area difference. Nevertheless, the obtained area difference was still large enough to indicate that the CBP areas might be influenced by certain properties of the surrounding region, for example the magnetic field.

We discussed previously in Subsection \ref{mean_int_discussion} that CBPs within CHs are linked with the CH magnetic field in a way that the CBP properties are confined by the open magnetic field (\citeauthor{Heinemann_2} \citeyear{Heinemann_2}; \citeauthor{Wiegelmann_2004} \citeyear{Wiegelmann_2004}; \citeauthor{Wiegelmann_2005} \citeyear{Wiegelmann_2005}). Not only is the CBP brightness influenced by this confinement, which was discussed previously, but the size of CBPs as well, where the CH magnetic field limits the maximum possible size of the CBPs within the CH. This behaviour is evident from the anticorrelation between the relative CBP area ratio and the absolute signed CH magnetic field from Subsection \ref{correlation1}. Here the result points to a possibility that the larger abundance of open magnetic field, as well as its overall strength, limit the CBPs to smaller sizes.

Moreover, we found that the CH area might influence the observed CBP area. This was obtained through the correlation of the CBP area with the CH area in Subsection \ref{correlation1}, where we found a rather strong correlation when CH3 was excluded. Similarly as in the Subsection \ref{correlation}, here the obtained correlation indicates that larger CHs allow larger CBPs to form. We have to mention here that all of the obtained correlations between the CBP and CH properties in Subsection \ref{correlation} and \ref{correlation1} in fact do suffer from the small number of data points taken into consideration. Therefore, we cannot make strong conclusions for the found correlations until we have a much larger set of different CHs, which will give us a better picture of the possible influence of CH on the CBP properties (future work).

Considering the significance of the CBP area difference analysed previously, we found that CH4 and CH5 have a non significant difference in the CBP area even for large individual CBP sample sizes. This was found specifically for the SDO/AIA data, where the area difference between the CBPs within and outside of these CHs was less than 1$\sigma$, while the ALMA data still showed some significance. In the case of CH4, no difference is seen in the CBP area (Fig. \ref{area_fig_ch4_1}), which could be the result of CH4 having a very large area, and this thus allows large CBPs to exist such as those found outside the CH4 in the quiet Sun regions. In the case of CH5, on the other hand (Fig. \ref{area_fig_ch5_1}), there is no clear evidence as to why we have a non-significant CBP area difference. The area of the CH5 is the second smallest out of the five CHs, which based on the previous discussion should limit the CBP size to smaller values. However, the absolute signed magnetic field of CH5 is relatively small in comparison to most of the other CHs, which then should allow larger CBPs to form. The future analysis will confirm what was the true nature of the obtained results for CH4 and CH5, and which CH property could have the dominating effect on the CBPs within these two CHs.

Furthermore, the CBP area is generally influenced by the viewing angle, but since we limited ourselves only on the central regions of the solar disk, this should reduce the influence of the viewing angle on the observed area difference because the CBP loops are less likely to be curved or tilted at high angles. However, even here some of the central CBPs with peculiar morphologies, due to non-linear magnetic field, can have parts of their loop structure curved and tilted. Moreover, the area difference might be more influenced by the spatial resolution and the characteristic temperature of the used imager channels. The large difference in the spatial resolution between ALMA and SDO/AIA could have produced a significantly higher CBP area difference for ALMA. This resolution issue might be supported by the fact that we used similar methods of CBP extraction for both ALMA and SDO/AIA. Moreover, since the CBPs do not look the same when they are observed in different parts of the spectrum using different ALMA mm and SDO/AIA EUV imager channels, this could also be true for the observed changes of the CBP features. Future studies will look more into these possible issues using observations in more than one imager channel and the interferometric ALMA observations.

So far, we have only considered CH properties, for example the magnetic field and area, having the main influence on the CBP properties (mean brightness and area). However, we note that CH1 and CH3, which were both outliers in some of the discussions, are also the only two CHs with active regions near their boundaries. Perhaps the presence of the nearby strong fields of active region can further affect CBP properties within CHs. To explore this hypothesis, we would need to use a larger sample of CHs and find a way to analyse the field of active regions and its influence on CBPs within CHs (beyond the scope of this study).

\section{Conclusion and prospects}
\label{conclude}
In this study we find significant differences in the area and mean brightness between CPBs located within and outside five CHs. We describe the analysis applied on the SDO/AIA 193 \AA\space and the millimetre ALMA Band 6 data in detail. Based on the obtained results, we conclude that CBPs inside the CHs have lower brightness on average than the CBPs outside of them in the quiet Sun region, but they are also smaller in area on average. We find this to be true for CBPs visible both at chromospheric heights in the ALMA data and at coronal heights in the SDO/AIA data. 

Also, we find that there could be an influence of CH properties, for example the magnetic field inside the CH boundaries and its area, on the observed properties of the CBPs. Based on the observed CBP differences, as well as the found (anti-)correlation between the CBP and CH properties, we may conclude that the strength and the abundance of the open CH magnetic field and the CH area could influence the CBP properties within CHs. In addition, the different ratio between the open and closed magnetic flux within and outside CHs might hold important information on the formation of CHs themselves. Here CBPs within CHs can represent the missing link to better understand the general appearance of CHs. Indeed there exists a link between CBPs and the appearance of CHs, where the change in CBP morphology changes the CH boundaries on small scales (\citeauthor{Madjarska2} \citeyear{Madjarska2}; \citeauthor{Madjarska3} \citeyear{Madjarska3}; \citeauthor{Subramanian} \citeyear{Subramanian}). Still, we do not fully understand how CBPs may affect the large-scale appearance of CHs and the generation of high-speed streams. Although the conclusions have thus far been made on the basis of a small CH sample, they can be further strengthened with additional data of larger CH samples, which will be used in future studies to improve the statistics.

Given the effect of the spatial resolution of the ALMA Band 6 single-dish images on the carried measurements of the CBP physical properties discussed in Sect. \ref{Discussion}, we conclude that, in order to obtain more precise measurements, the interferometric ALMA data should be used instead of the single-dish data that we used in our work. Such data will be of great importance for the future study of the evolution of CBPs inside and outside CHs through time at chromospheric heights.

A detailed analysis of the mean CBP brightness and area, in addition to CH properties, for a longer time period, instead of just one specific time as we observed, will be carried out in future work. Moreover, we plan to follow the evolution of brightness, morphological, and magnetic properties of not only CBPs, but CHs as well. Here we plan to use the interferometric ALMA data mentioned previously, with the available SDO/AIA and SDO/HMI data at the corresponding times of the ALMA observations. Further studies will focus on the magnetic field and morphology of CHs in great detail in order to confirm if the CH magnetic and morphological properties influence the observed CBP physical properties, mainly CBP brightness and sizes. If the influence appears to be true, as our results indicate, we will investigate, in more detail, how the strength and structure of the magnetic field around the CBPs inside CHs, in addition to the shape and size of the CHs, affect the observed CBP properties.

\begin{acknowledgements}
This work has been supported by the Croatian Science Foundation as part of the „Young Researchers’ Career Development Project - Training New Doctorial Students” under the project 7549 “Millimeter and submillimeter observations of the solar chromosphere with ALMA”.  It has also received funding from the Horizon 2020 project SOLARNET (824135, 2019–2022). The support from the Austrian-Croatian Bilateral Scientific Projects ”Comparison of ALMA observations with MHD-simulations of coronal waves interacting with coronal holes” and ”Multi-Wavelength Analysis of Solar Rotation Profile” is also acknowledged. In this work ALMA data ADS/JAO.ALMA\#2016.1.00050.S, ADS/JAO.ALMA\#2016.1.00202.S, ADS/JAO.ALMA\#2017.1.00009.S, ADS/JAO.ALMA\#2017.1.01138.S and ADS/JAO.ALMA\#2018.1.00199.S were used. We also acknowledge the use of the ALMA Solar Ephemeris Generator (\citeauthor{Skokic_ephemeris} \citeyear{Skokic_ephemeris}). ALMA is a partnership of ESO (representing its member states), NSF (USA) and NINS (Japan), together with NRC (Canada), MOST and ASIAA (Taiwan), and KASI (Republic of Korea), in cooperation with the Republic of Chile. The Joint ALMA Observatory is operated by ESO, AUI/NRAO and NAOJ. We are grateful to the ALMA project for making solar observing with ALMA possible. SDO is the first mission launched for NASA’s Living With a Star (LWS) Program. This research used  version 3.1.3 (10.5281/zenodo.3633844) of the Matplotlib open source software package (\citeauthor{Hunter} \citeyear{Hunter}), version 1.7.3 (10.5281/zenodo.5725464) of the SciPy open source software package (\citeauthor{Virtanen} \citeyear{Virtanen}) and version 2.0.5 (10.5281/zenodo.4292495) of the SunPy open source software package (\citeauthor{SunPy} \citeyear{SunPy}).
\end{acknowledgements}

\onecolumn
\begin{appendix}
\section{CBP edge detection and background subtraction}
\begin{figure*}[!htp]
\captionsetup[subfloat]{farskip=1pt,captionskip=1pt}
\centering
\subfloat{\includegraphics[width=0.65\textwidth]{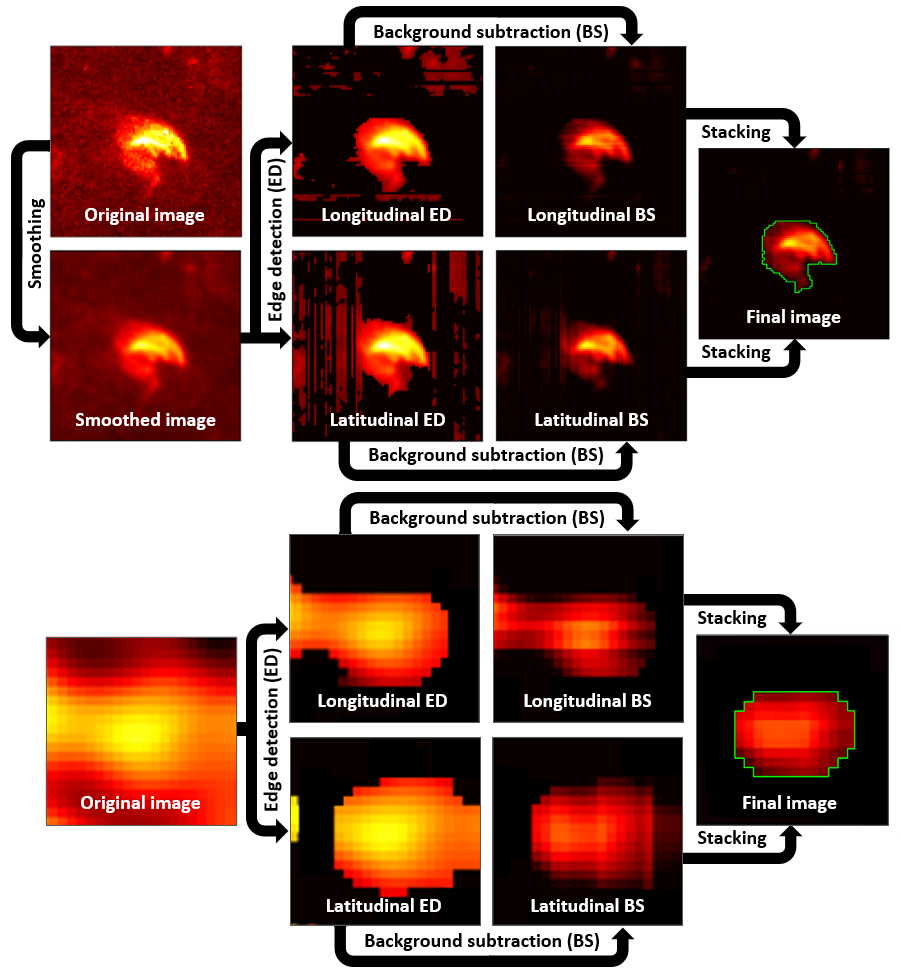}}
\caption{Same as Fig. \ref{ED_ALMA_SDO1}, but for CBP example inside of CH.}
\label{ED_ALMA_SDO2}
\end{figure*}

\clearpage
\section{SDO and ALMA images with detected CHs and CBPs}
\label{intensitymaps}
\begin{figure*}[!htp]
\captionsetup[subfloat]{farskip=1pt,captionskip=1pt}
\centering
\subfloat{\includegraphics[width=0.82\textwidth]{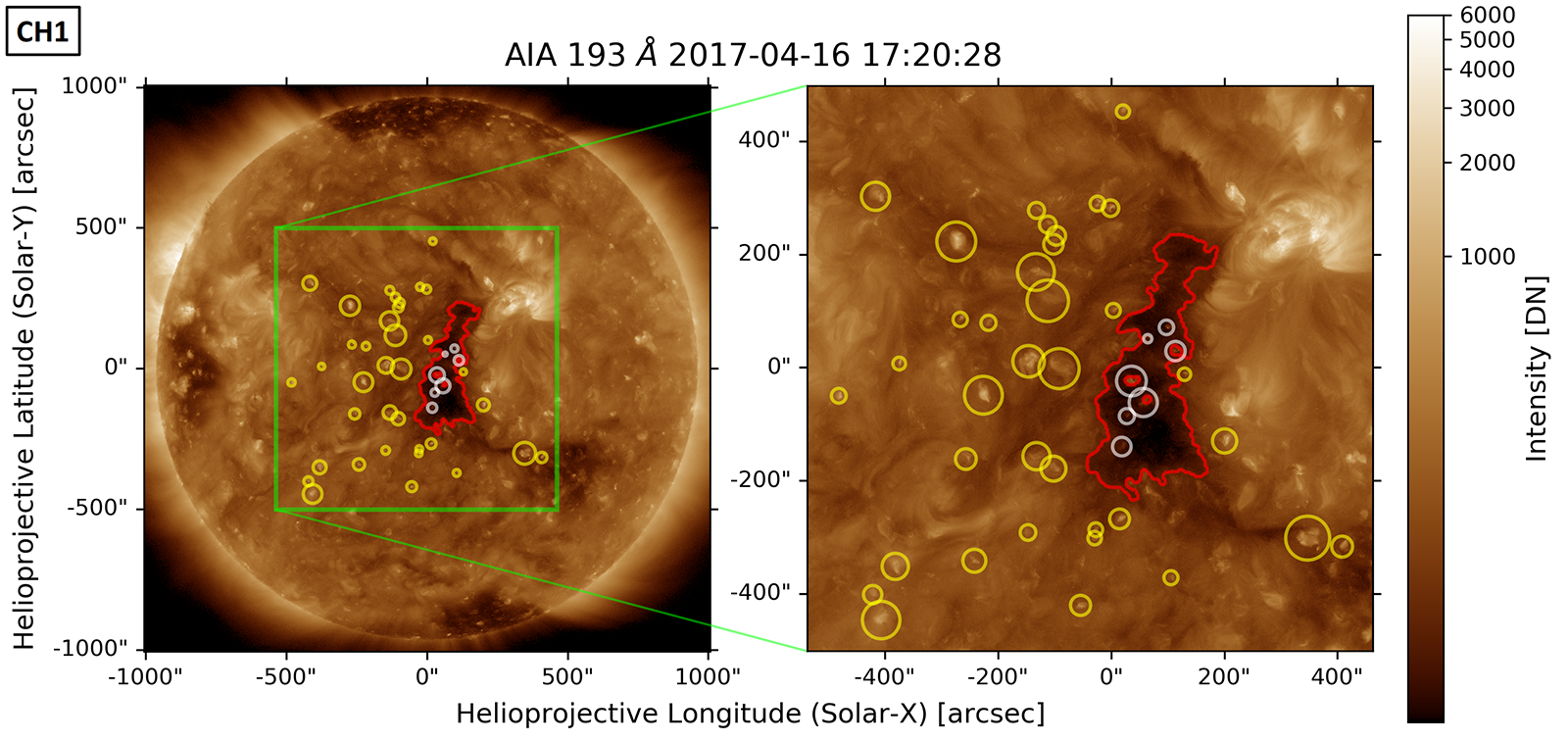}}\\\vspace{0.5em}
\subfloat{\includegraphics[width=0.82\textwidth]{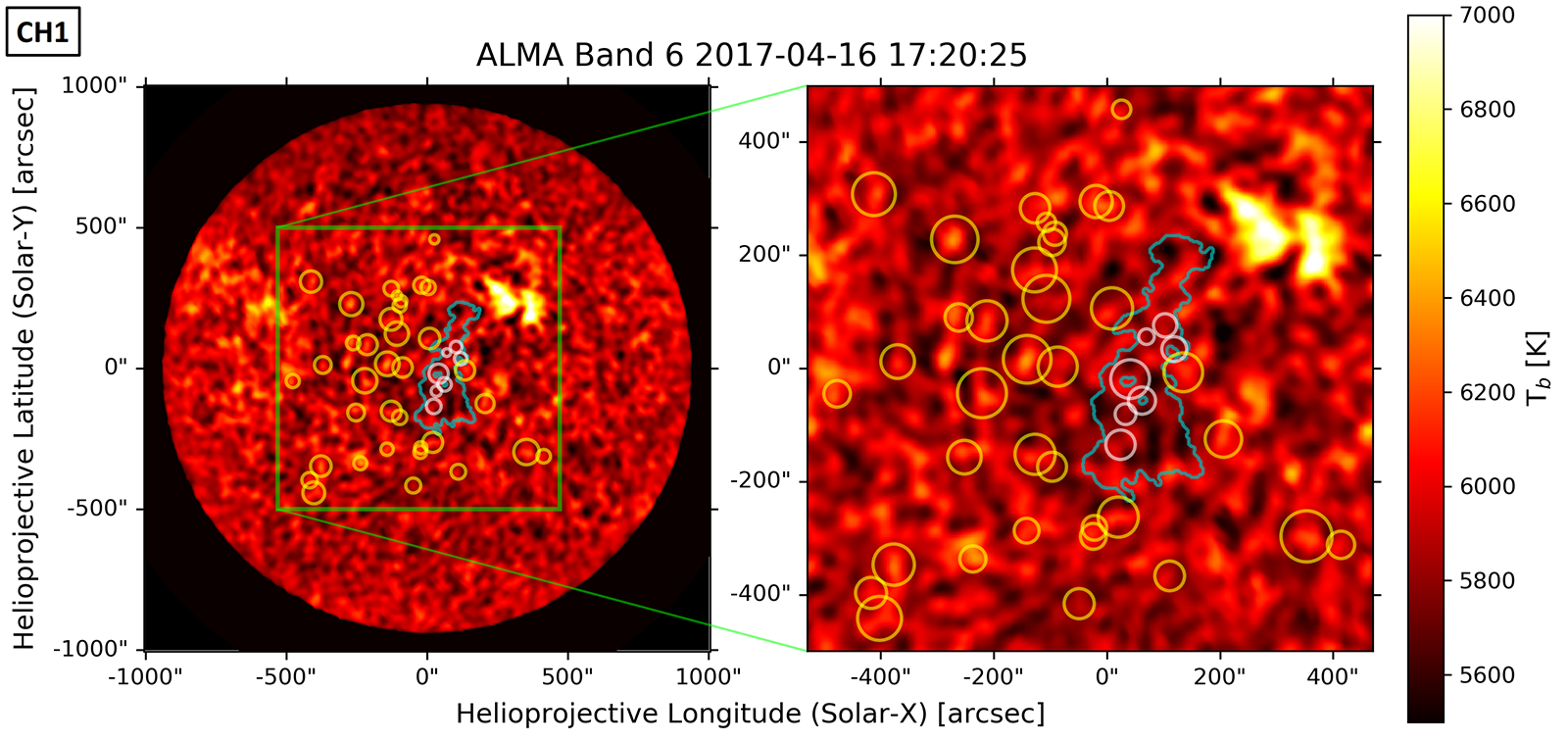}}\\\vspace{0.5em}
\subfloat{\includegraphics[width=0.82\textwidth]{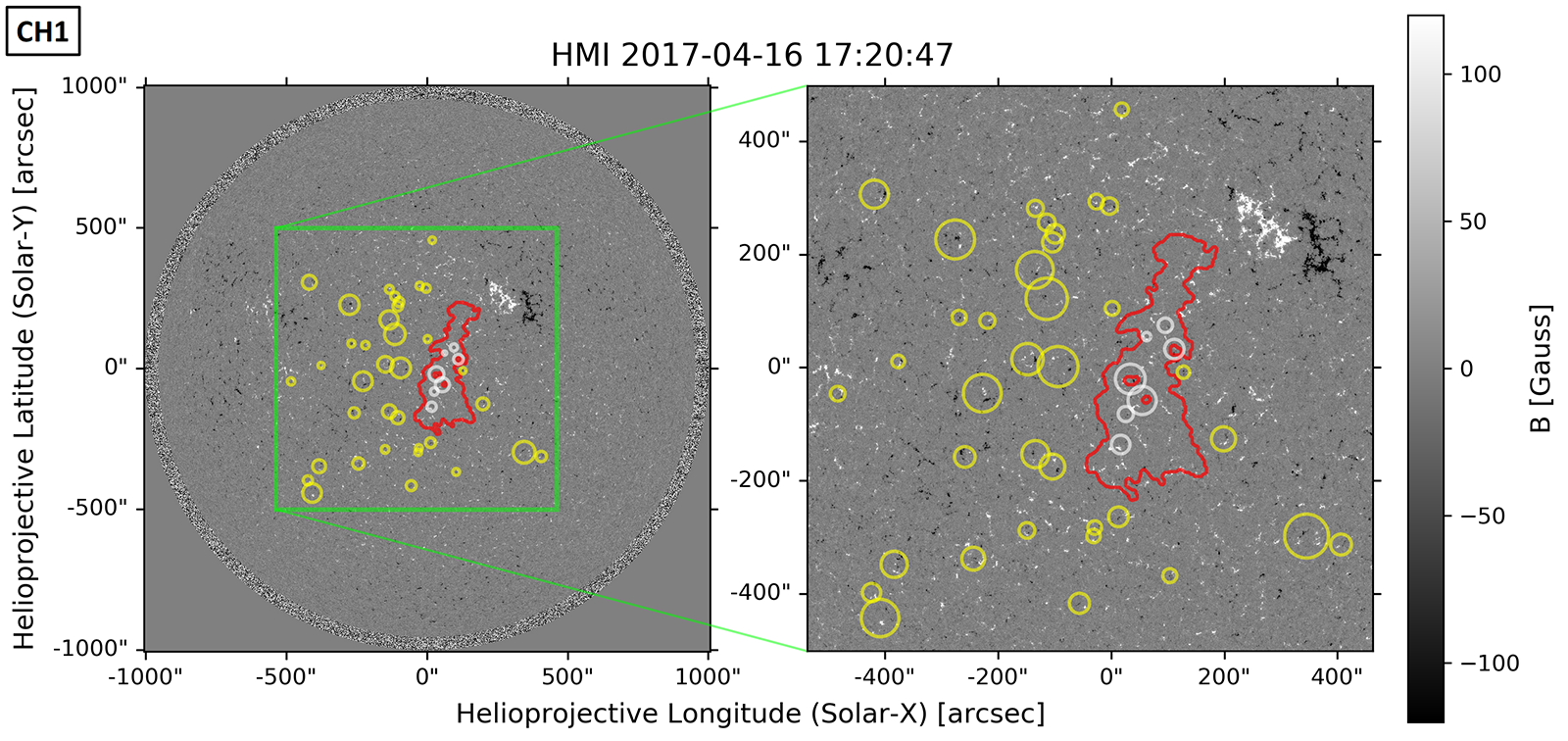}}
\caption{Same as Fig. \ref{CH_ALMA_SDO1}, but for CH1.}
\label{CH_ALMA_SDO2}
\end{figure*}

\begin{figure*}[!htp]
\captionsetup[subfloat]{farskip=1pt,captionskip=1pt}
\centering
\subfloat{\includegraphics[width=0.82\textwidth]{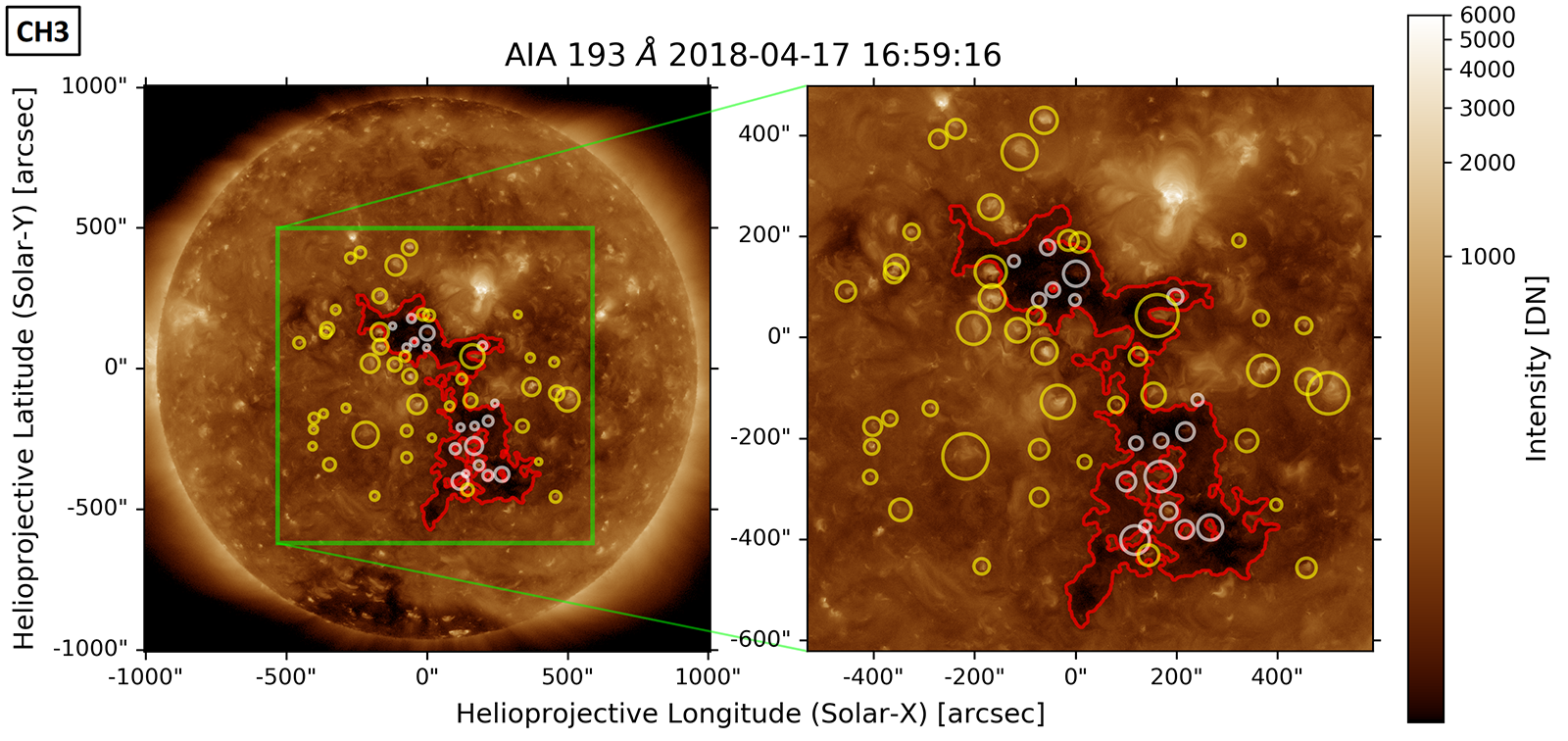}}\\\vspace{0.5em}
\subfloat{\includegraphics[width=0.82\textwidth]{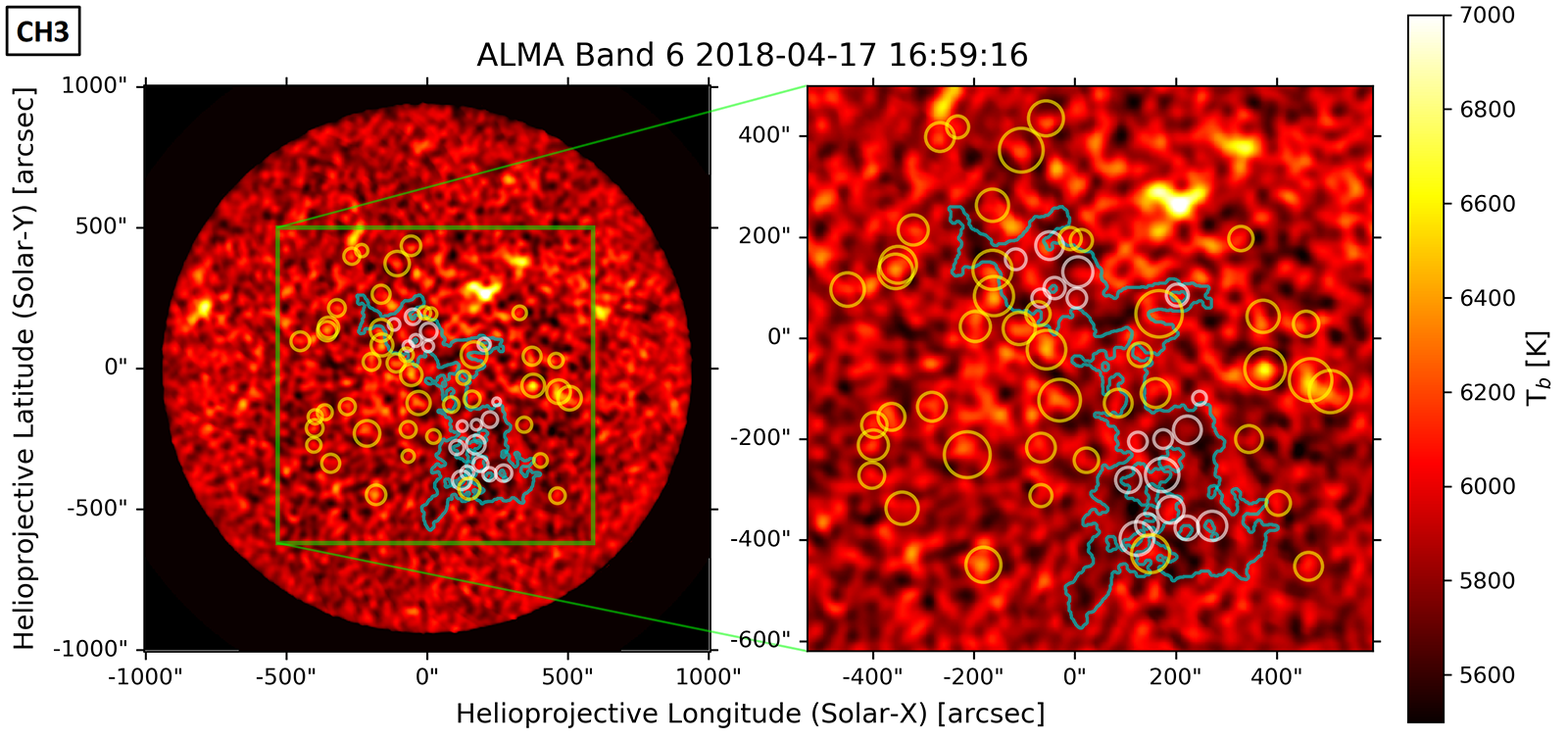}}\\\vspace{0.5em}
\subfloat{\includegraphics[width=0.82\textwidth]{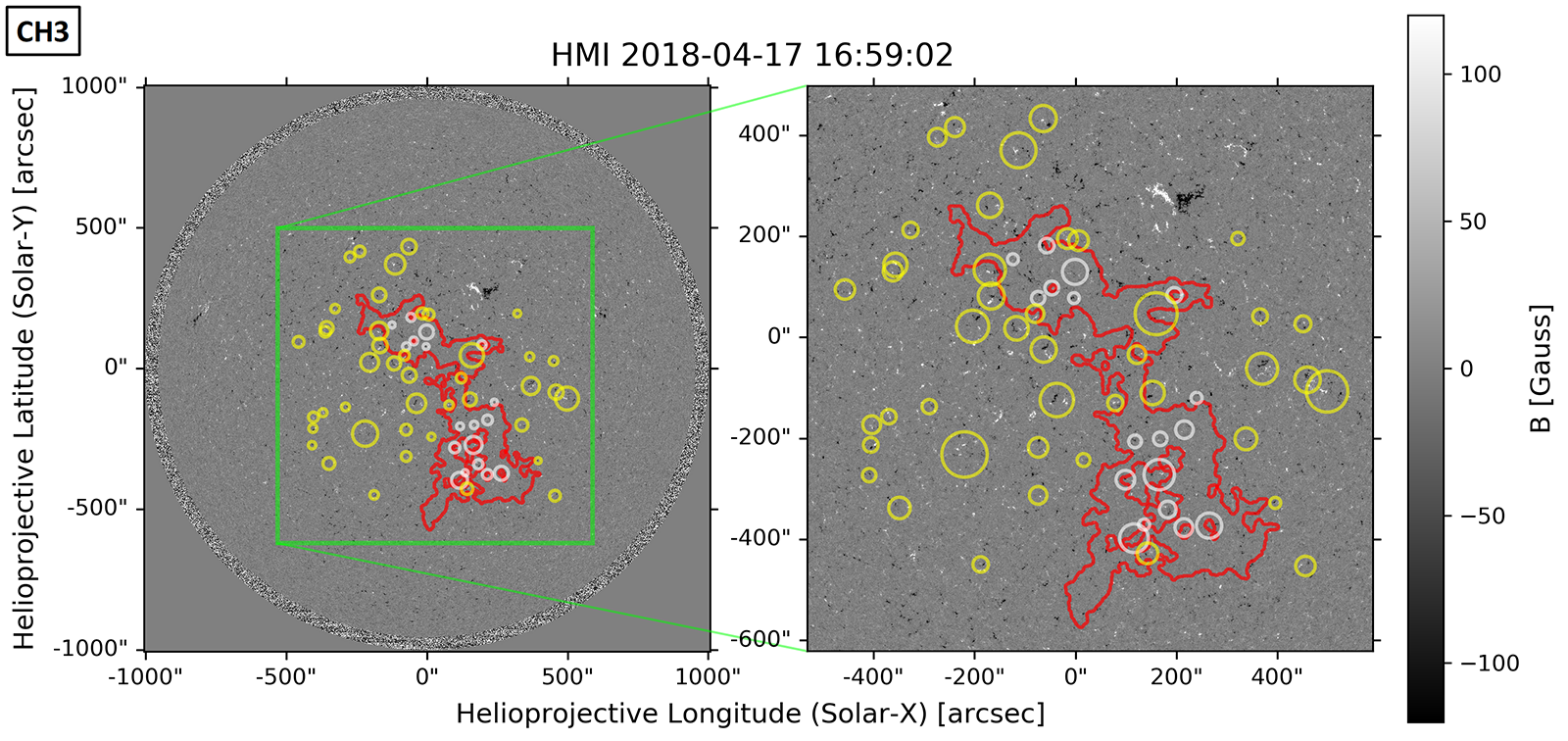}}
\caption{Same as Fig. \ref{CH_ALMA_SDO1}, but for CH3.}
\label{CH_ALMA_SDO3}
\end{figure*}

\begin{figure*}[!htp]
\captionsetup[subfloat]{farskip=1pt,captionskip=1pt}
\centering
\subfloat{\includegraphics[width=0.82\textwidth]{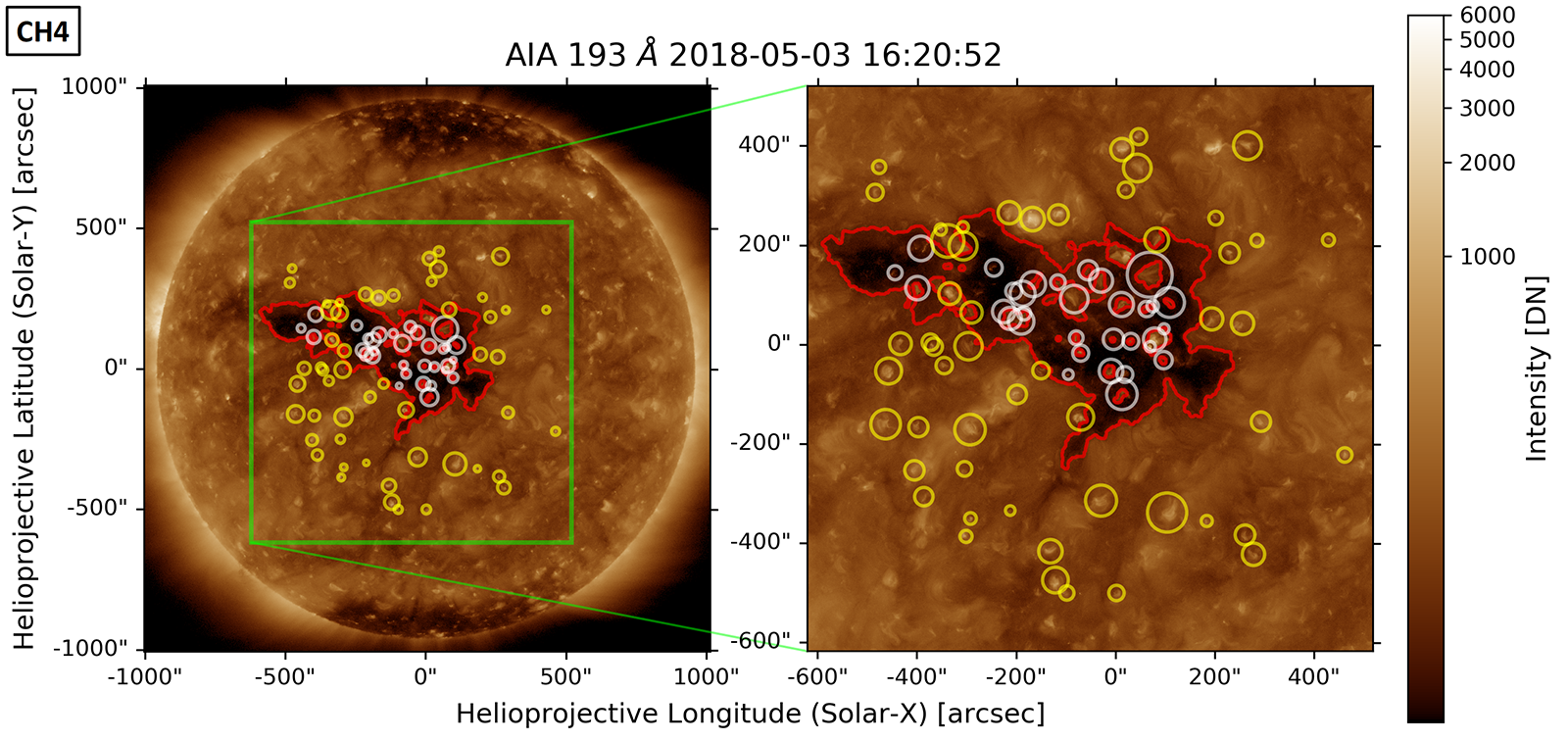}}\\\vspace{0.5em}
\subfloat{\includegraphics[width=0.82\textwidth]{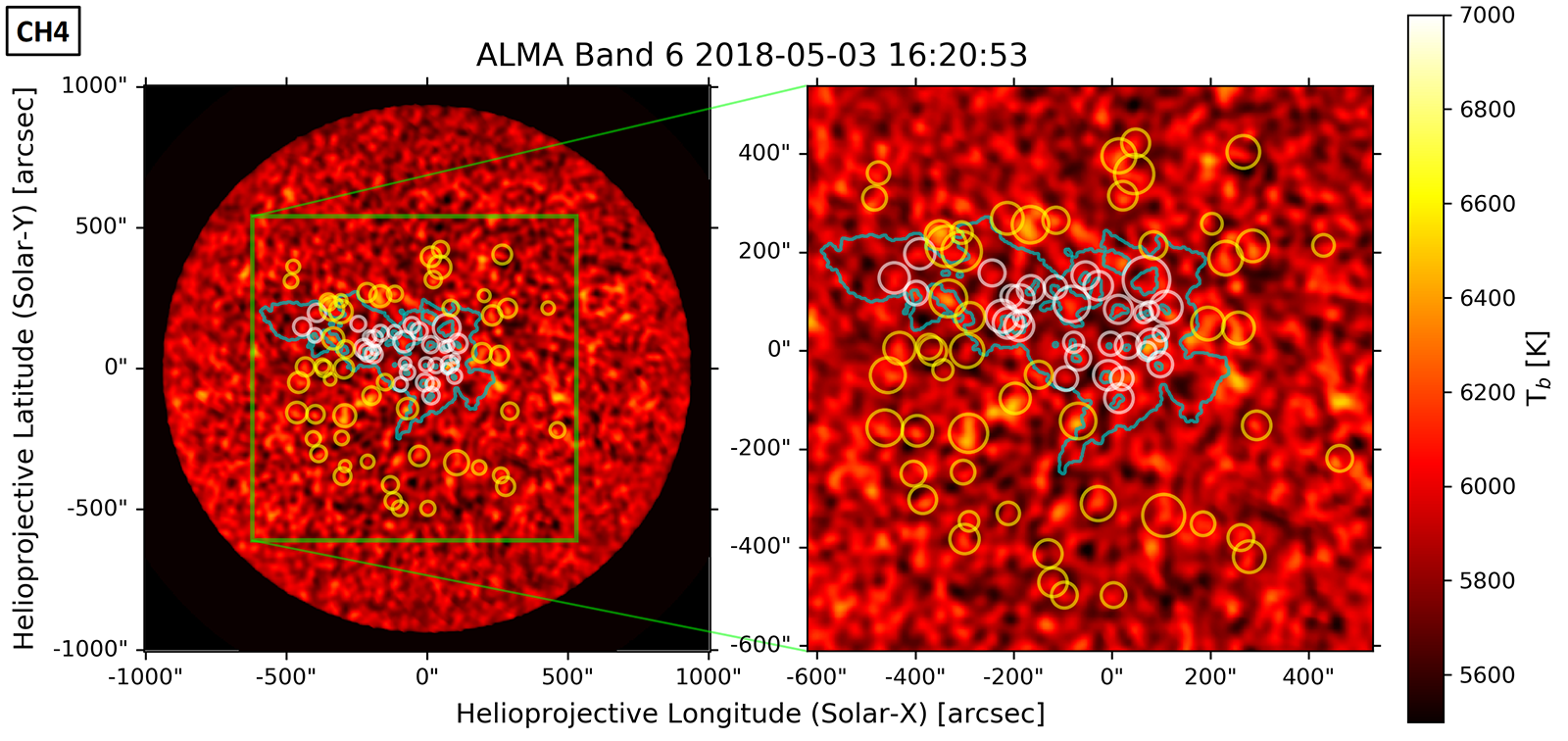}}\\\vspace{0.5em}
\subfloat{\includegraphics[width=0.82\textwidth]{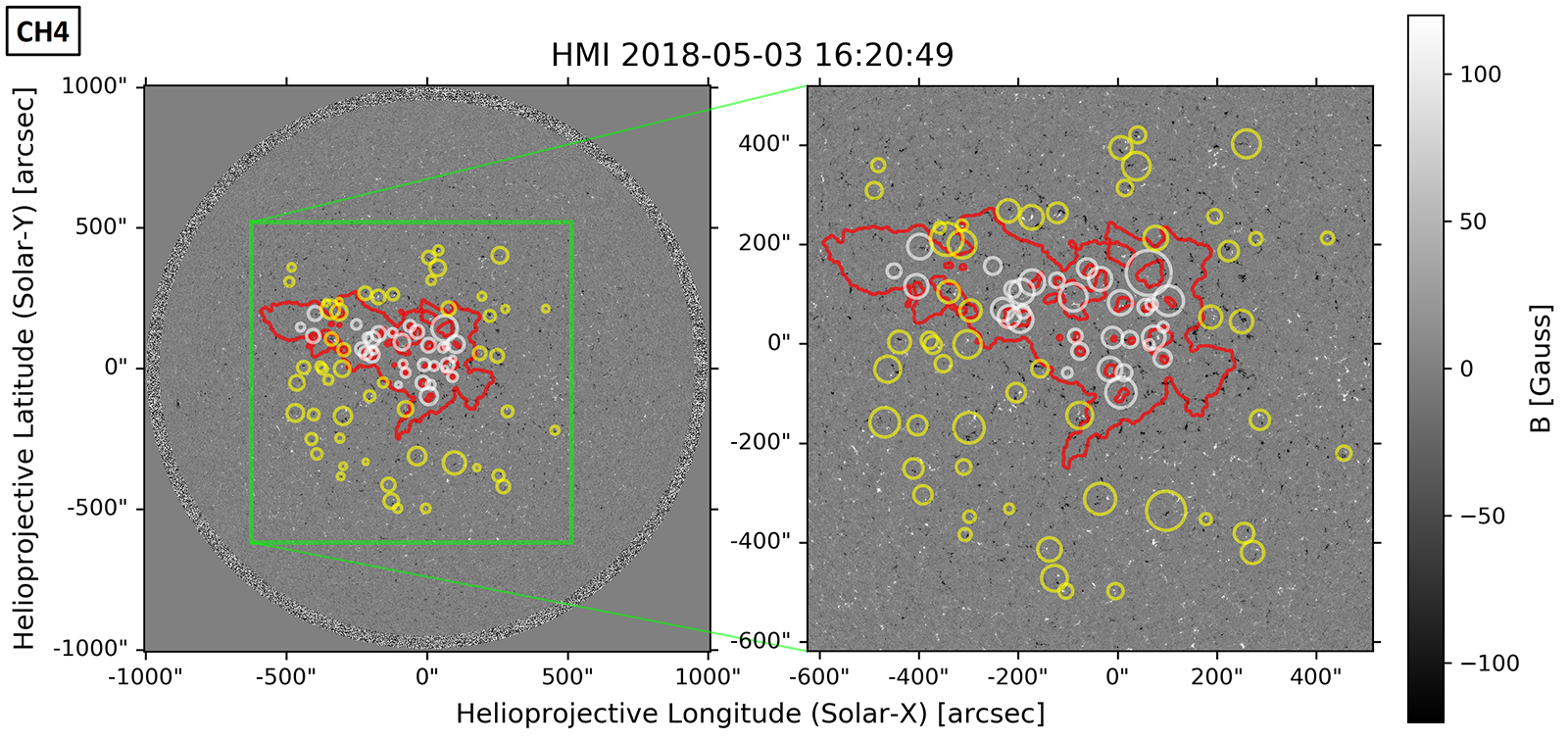}}
\caption{Same as Fig. \ref{CH_ALMA_SDO1}, but for CH4.}
\label{CH_ALMA_SDO4}
\end{figure*}

\begin{figure*}[!htp]
\captionsetup[subfloat]{farskip=1pt,captionskip=1pt}
\centering
\subfloat{\includegraphics[width=0.82\textwidth]{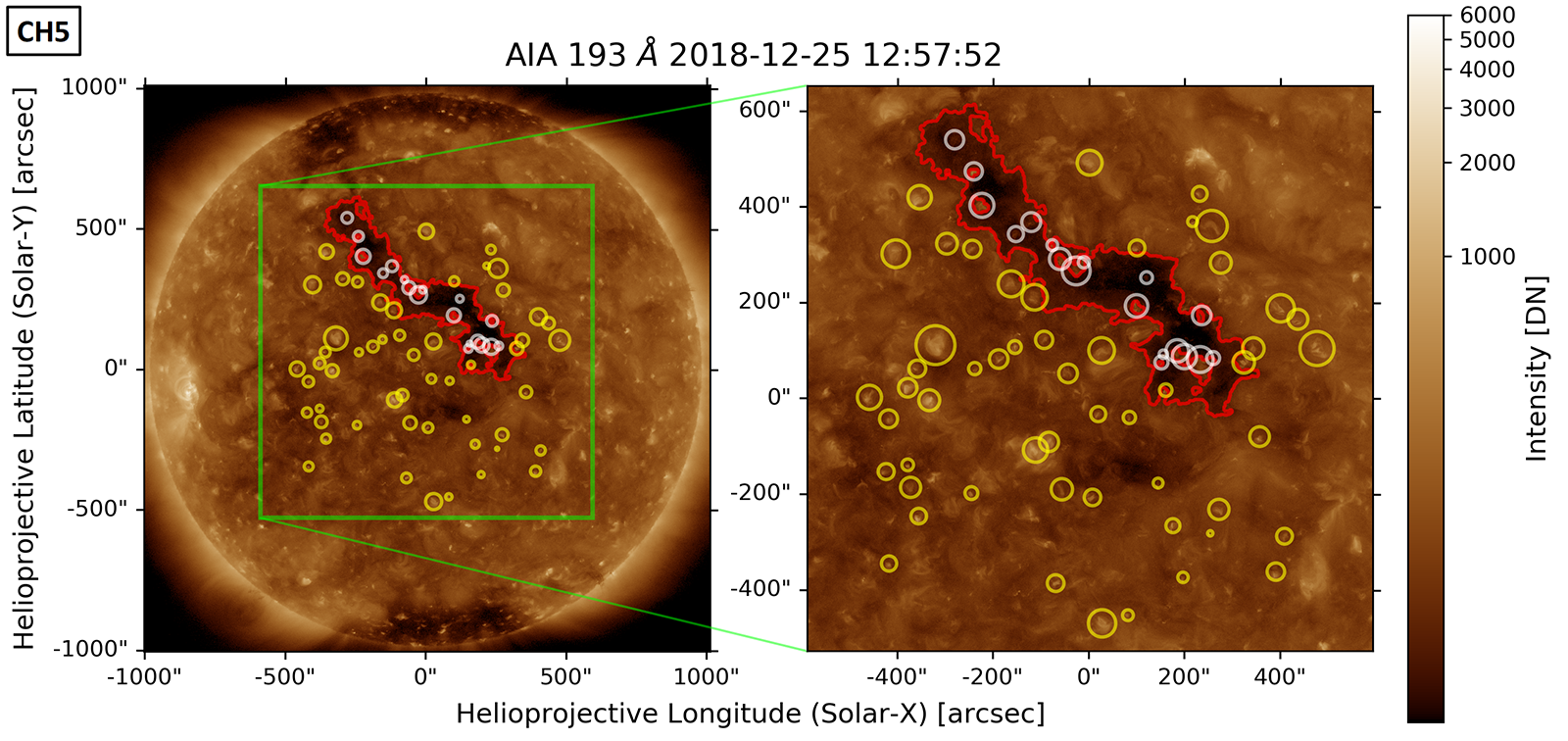}}\\\vspace{0.5em}
\subfloat{\includegraphics[width=0.82\textwidth]{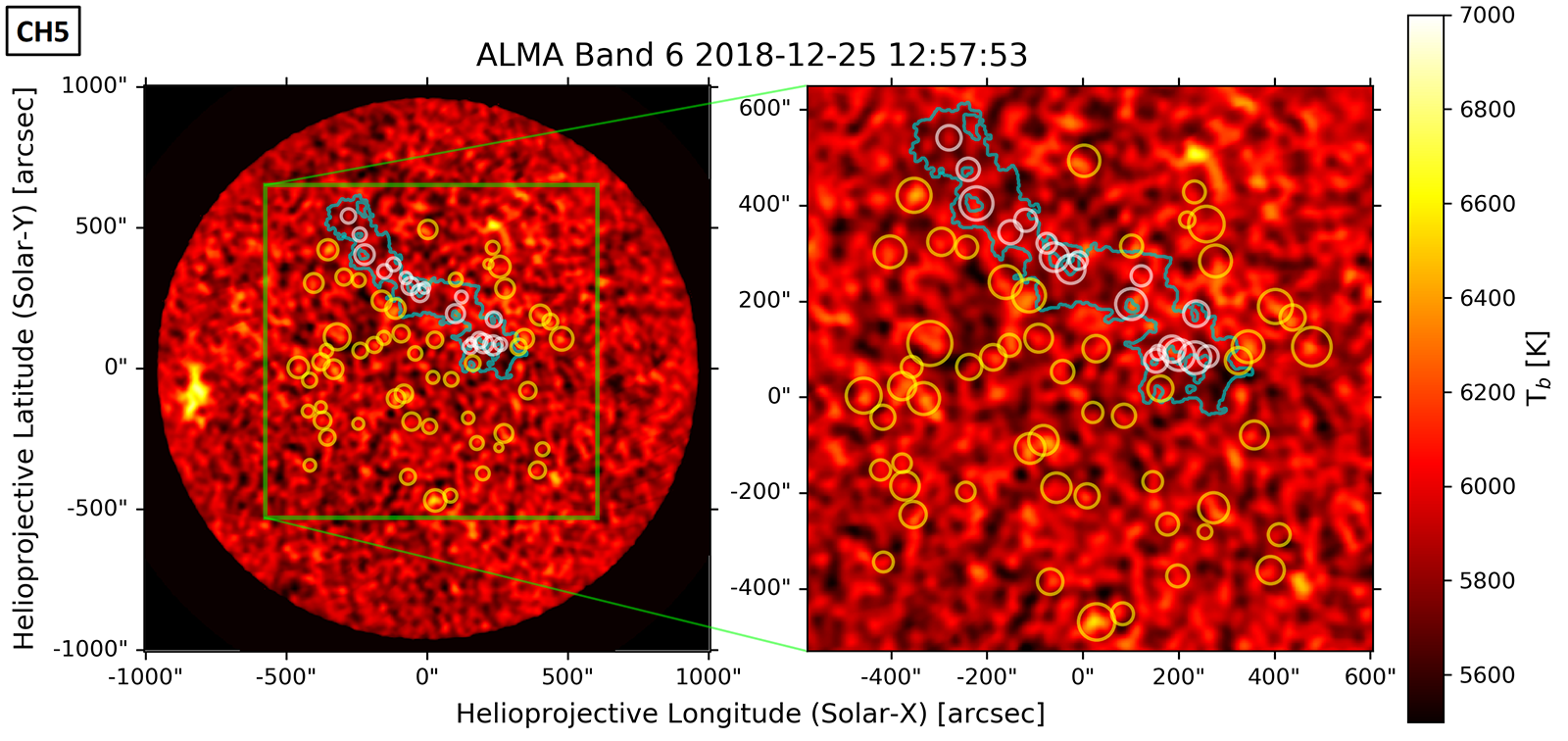}}\\\vspace{0.5em}
\subfloat{\includegraphics[width=0.82\textwidth]{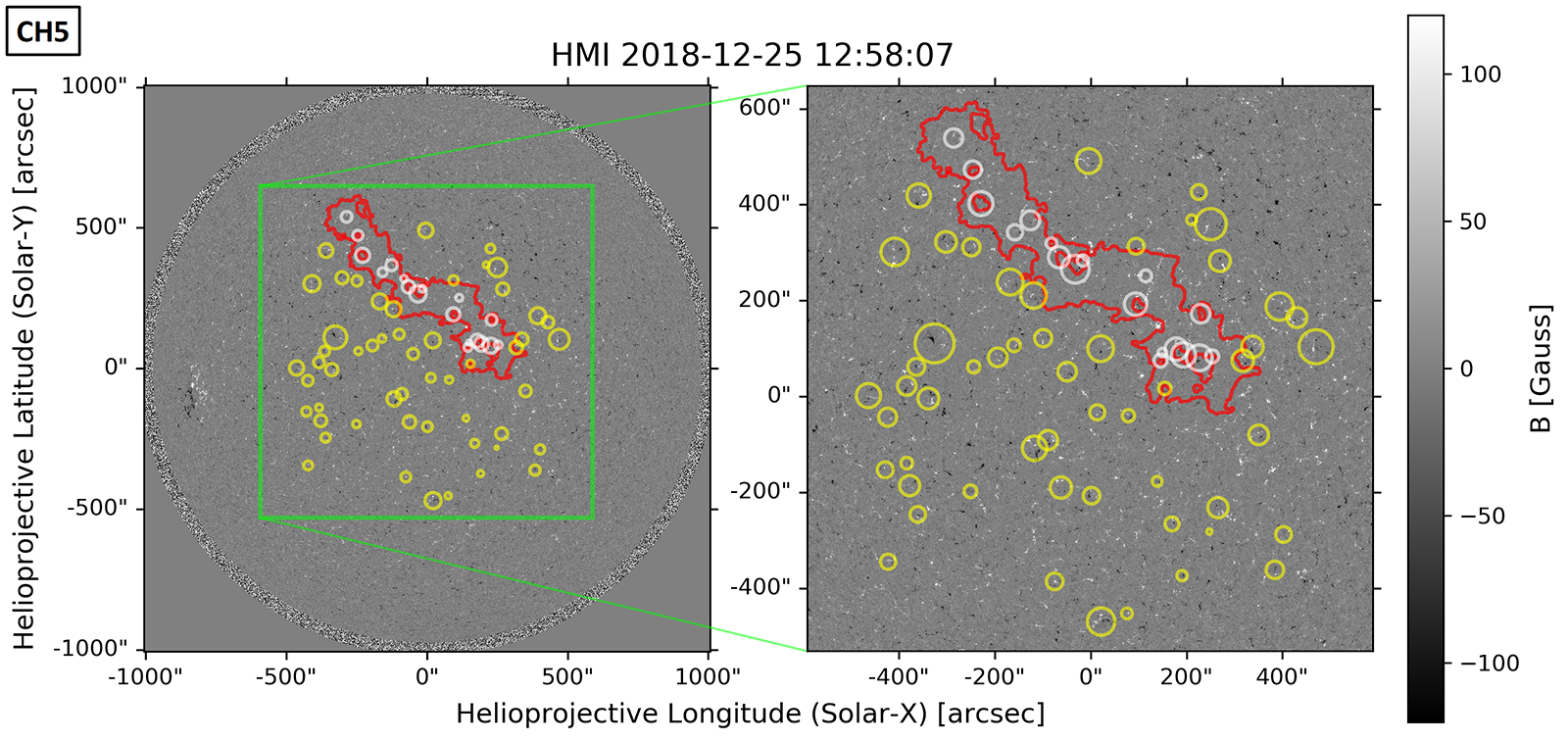}}
\caption{Same as Fig. \ref{CH_ALMA_SDO1}, but for CH5.}
\label{CH_ALMA_SDO5}
\end{figure*}

\clearpage
\section{Statistical analysis of the physical properties for CBPs within and outside CH1}
\label{CH1}
\begin{figure*}[h!]
\captionsetup[subfloat]{farskip=1pt,captionskip=1pt}
\centering
\subfloat{\includegraphics[width=0.36\textwidth]{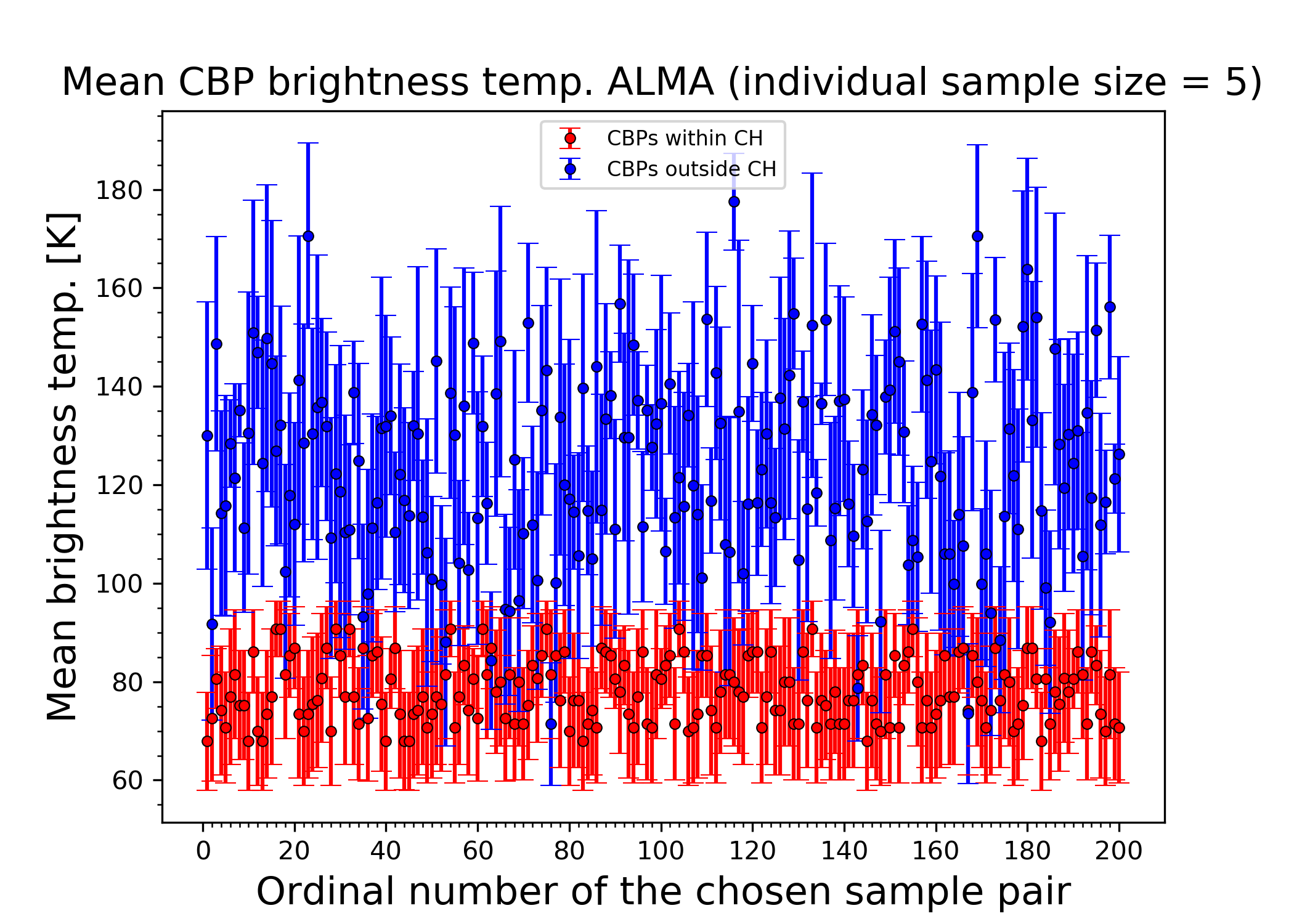}}
\subfloat{\includegraphics[width=0.36\textwidth]{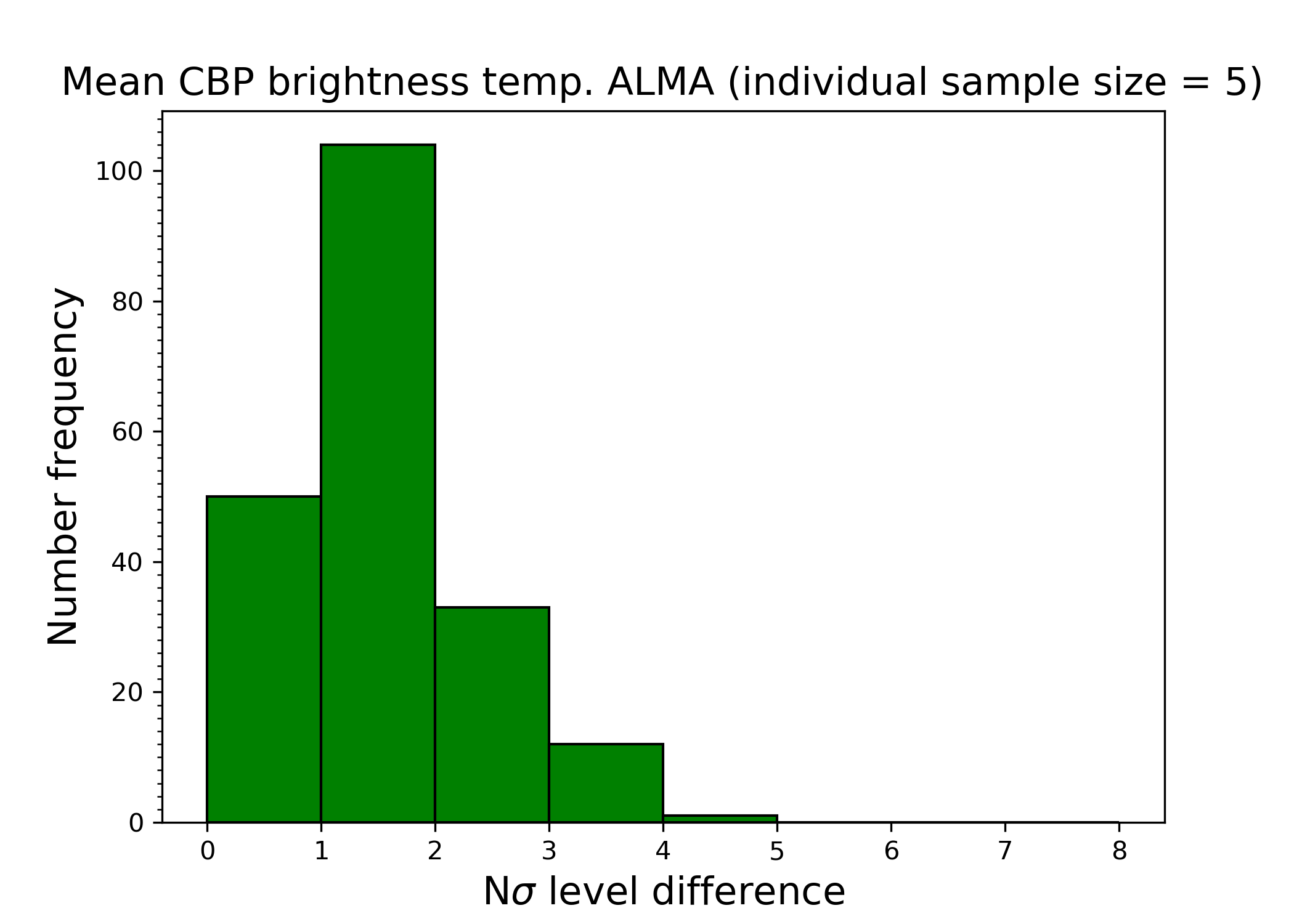}}\\
\subfloat{\includegraphics[width=0.36\textwidth]{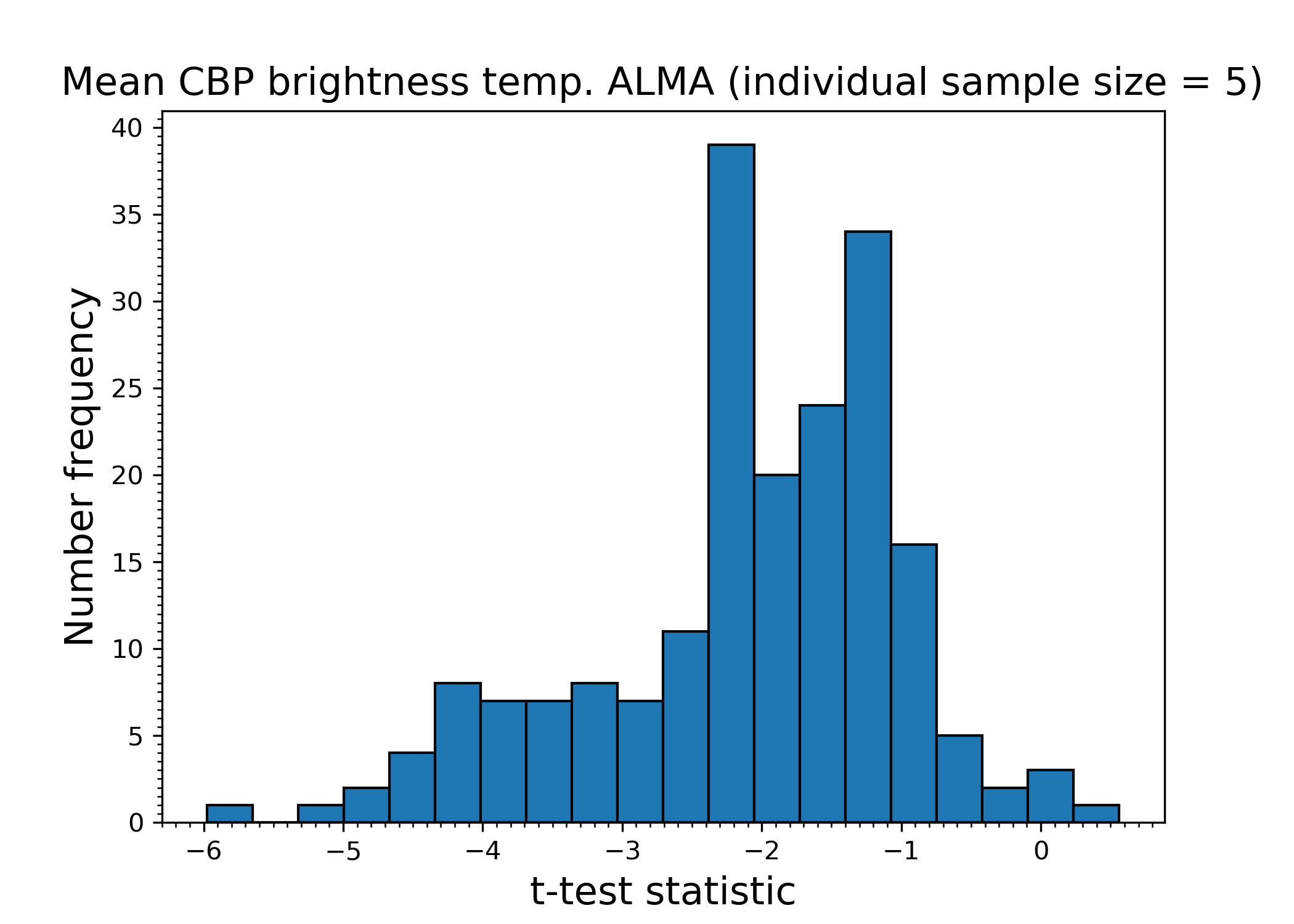}}
\subfloat{\includegraphics[width=0.36\textwidth]{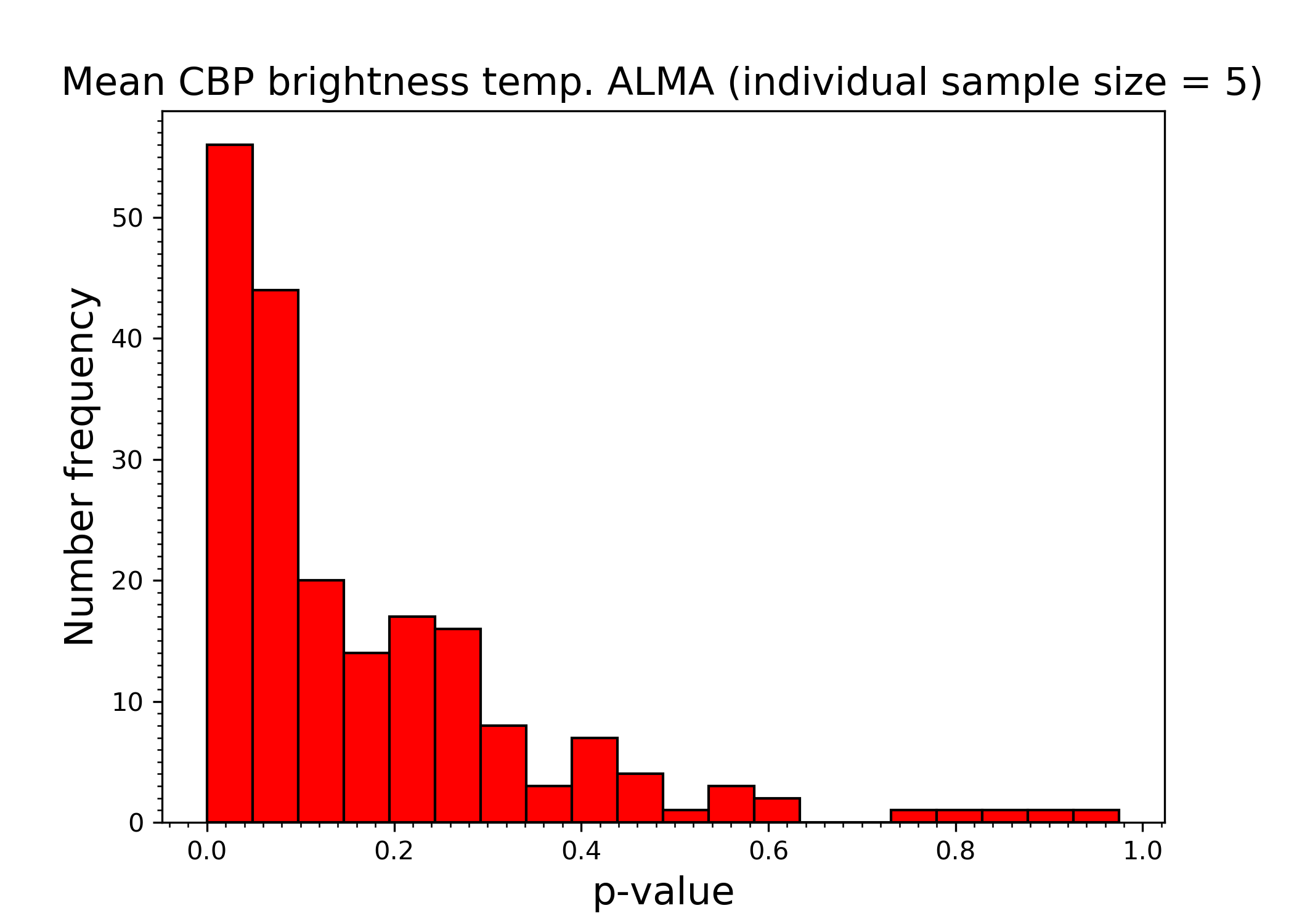}}
\caption{Statistical analysis of mean ALMA Band 6 CBP brightness temperature. Top row: Left panel shows the mean values of the mean CBP intensities in the ALMA Band 6 image with corresponding standard errors of 200 randomly chosen equal size CBP sample pairs, with one sample containing CBPs within (red) and the other outside (blue) the CH1. The right panel shows histogram of the largest N for which the relation (\ref{Nsigma}) holds true. Bottom row: Left panel shows histogram of the $t$-test statistic values ($t$-values) and the right panel shows the histogram of the $p$-values obtained for the mean values of CBP mean intensities in the ALMA Band 6 image. Individual CBP sample contains 5 randomly chosen CBPs out of the many selected CBPs either within or outside the CH1.}
\label{mean_int_fig_ch1}
\end{figure*}

\begin{figure*}[h!]
\captionsetup[subfloat]{farskip=1pt,captionskip=1pt}
\centering
\subfloat{\includegraphics[width=0.36\textwidth]{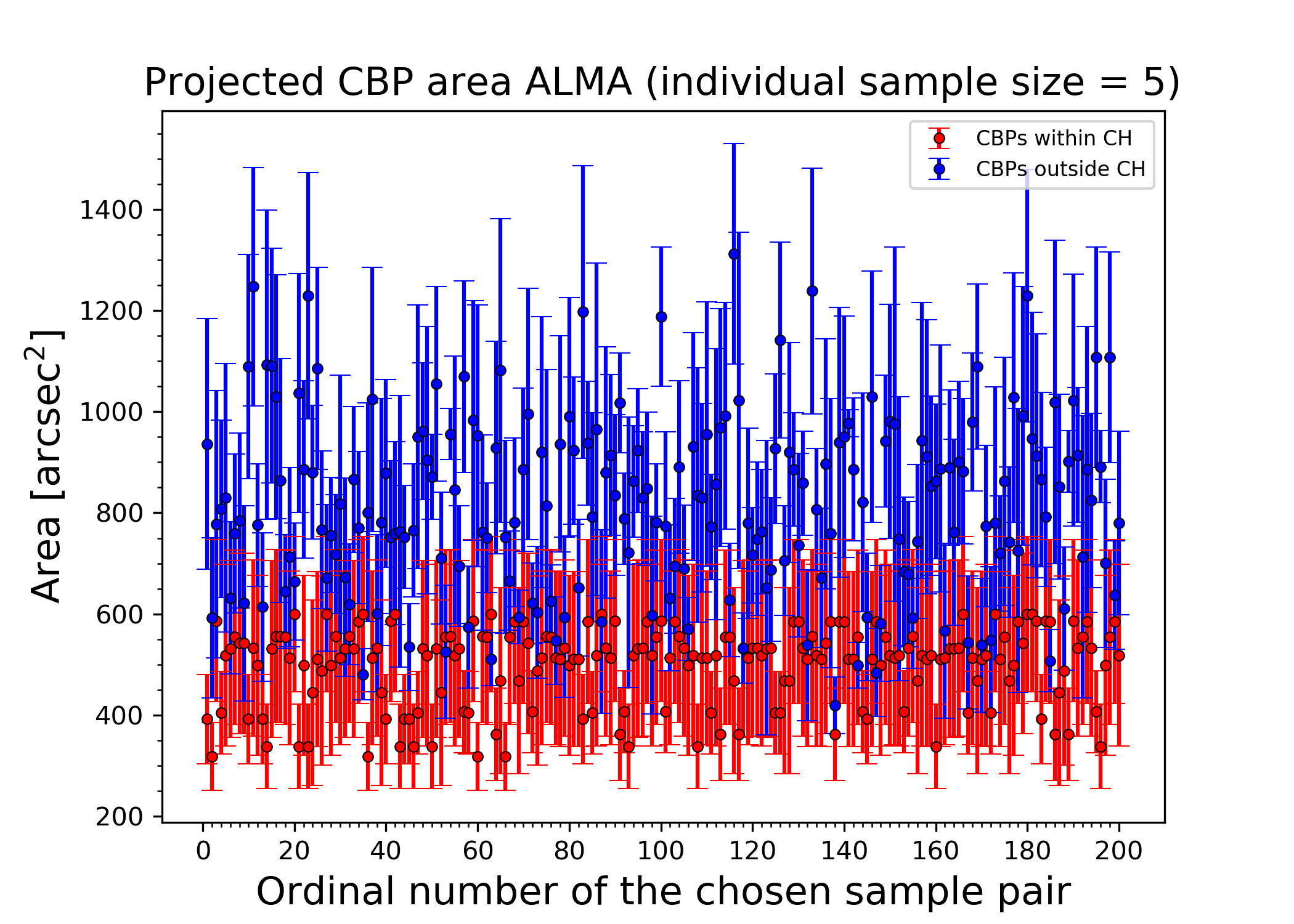}}
\subfloat{\includegraphics[width=0.36\textwidth]{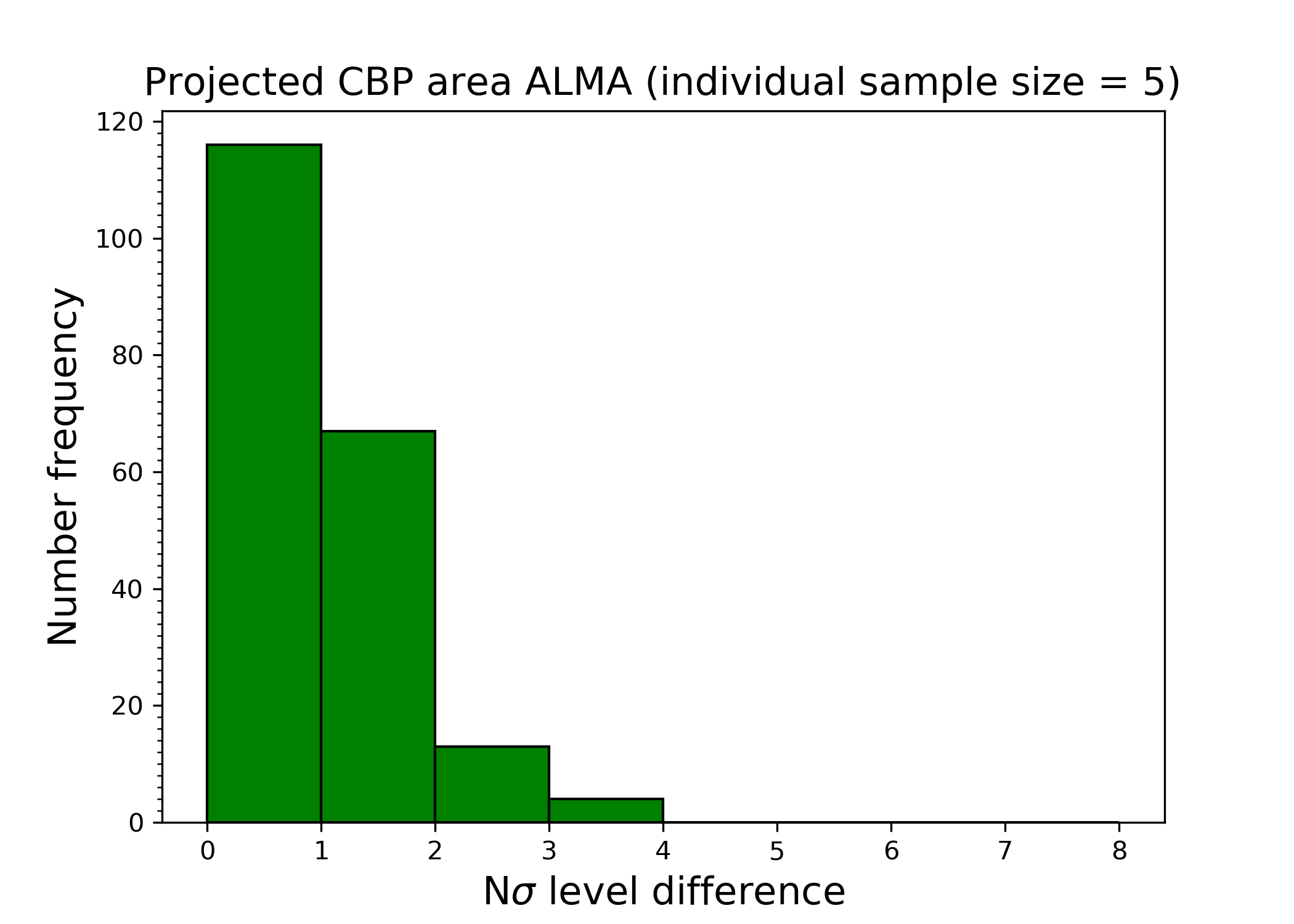}}\\
\subfloat{\includegraphics[width=0.36\textwidth]{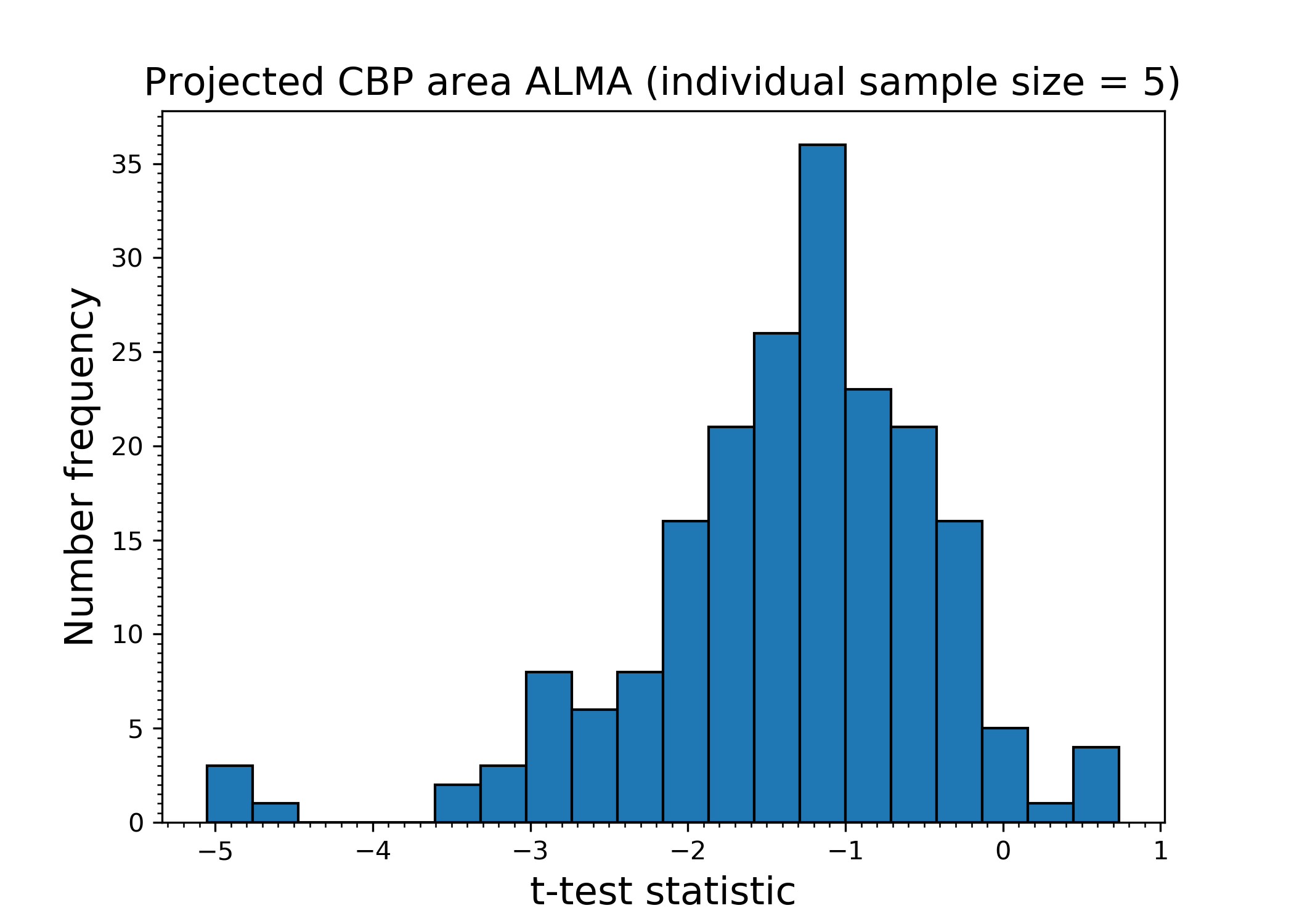}}
\subfloat{\includegraphics[width=0.36\textwidth]{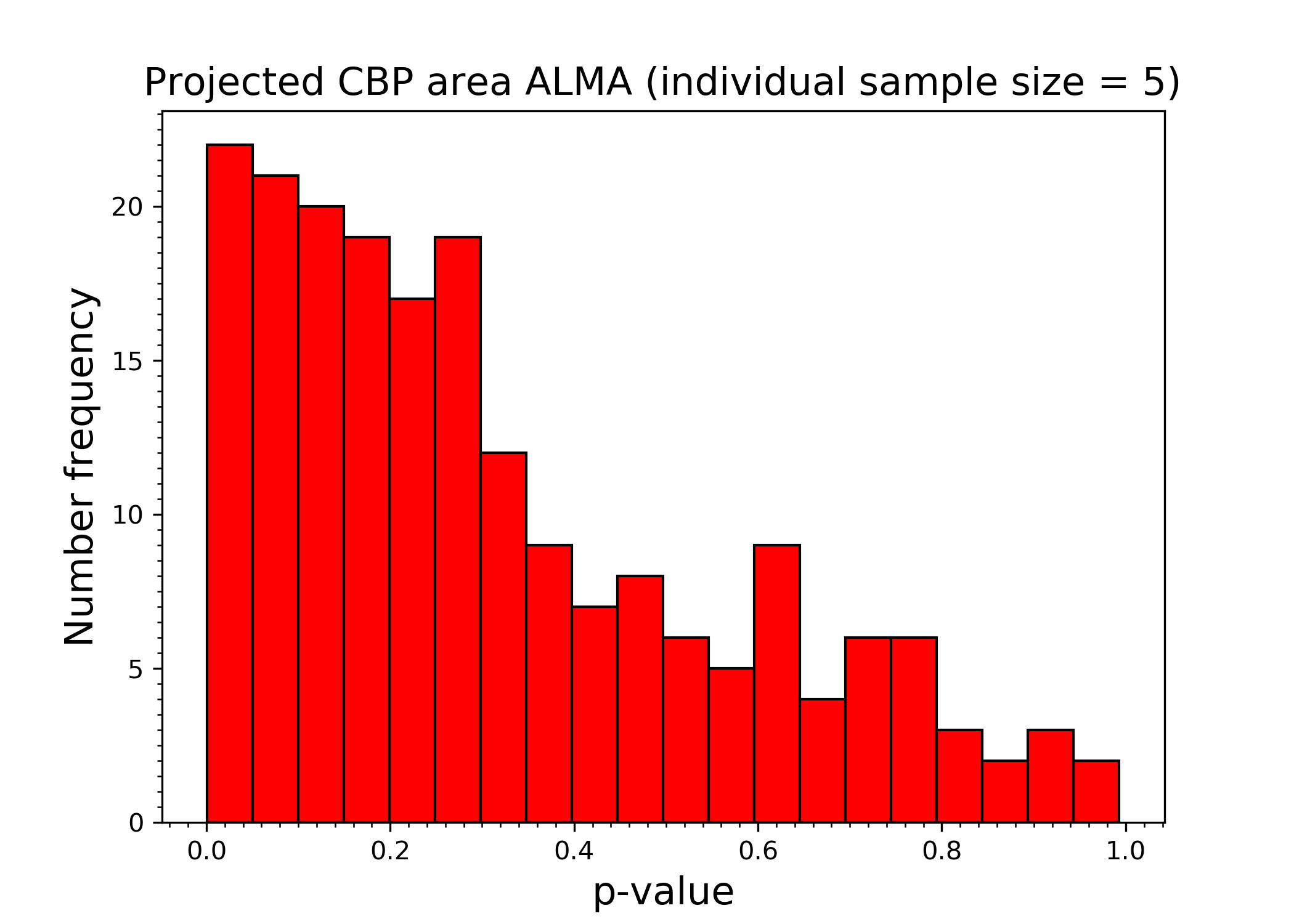}}
\caption{Same as Fig. \ref{mean_int_fig_ch1}, but for projected ALMA Band 6 CBP area.}
\label{area_fig_ch1}
\end{figure*}

\begin{figure*}[h!]
\captionsetup[subfloat]{farskip=1pt,captionskip=1pt}
\centering
\subfloat{\includegraphics[width=0.36\textwidth]{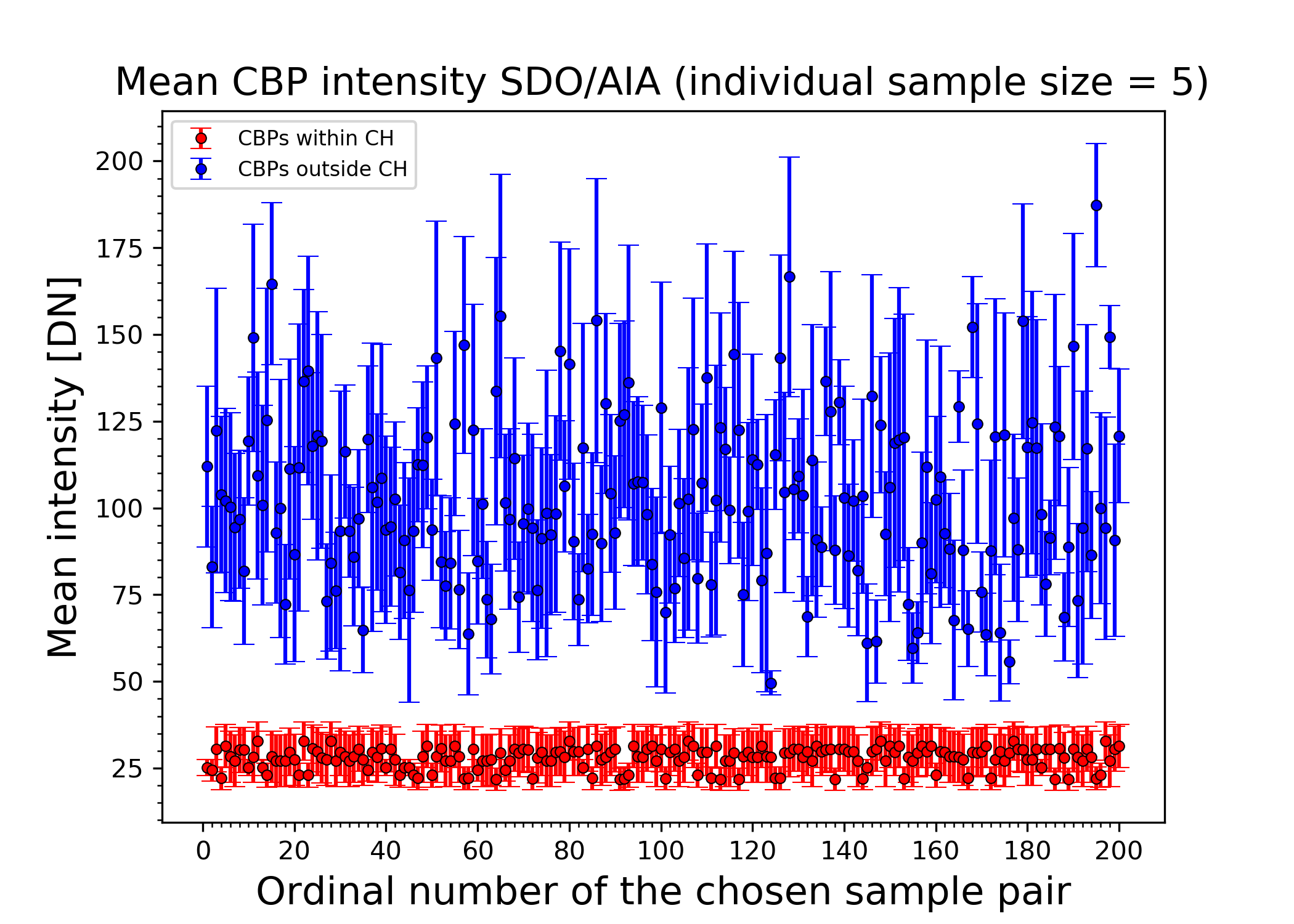}}
\subfloat{\includegraphics[width=0.36\textwidth]{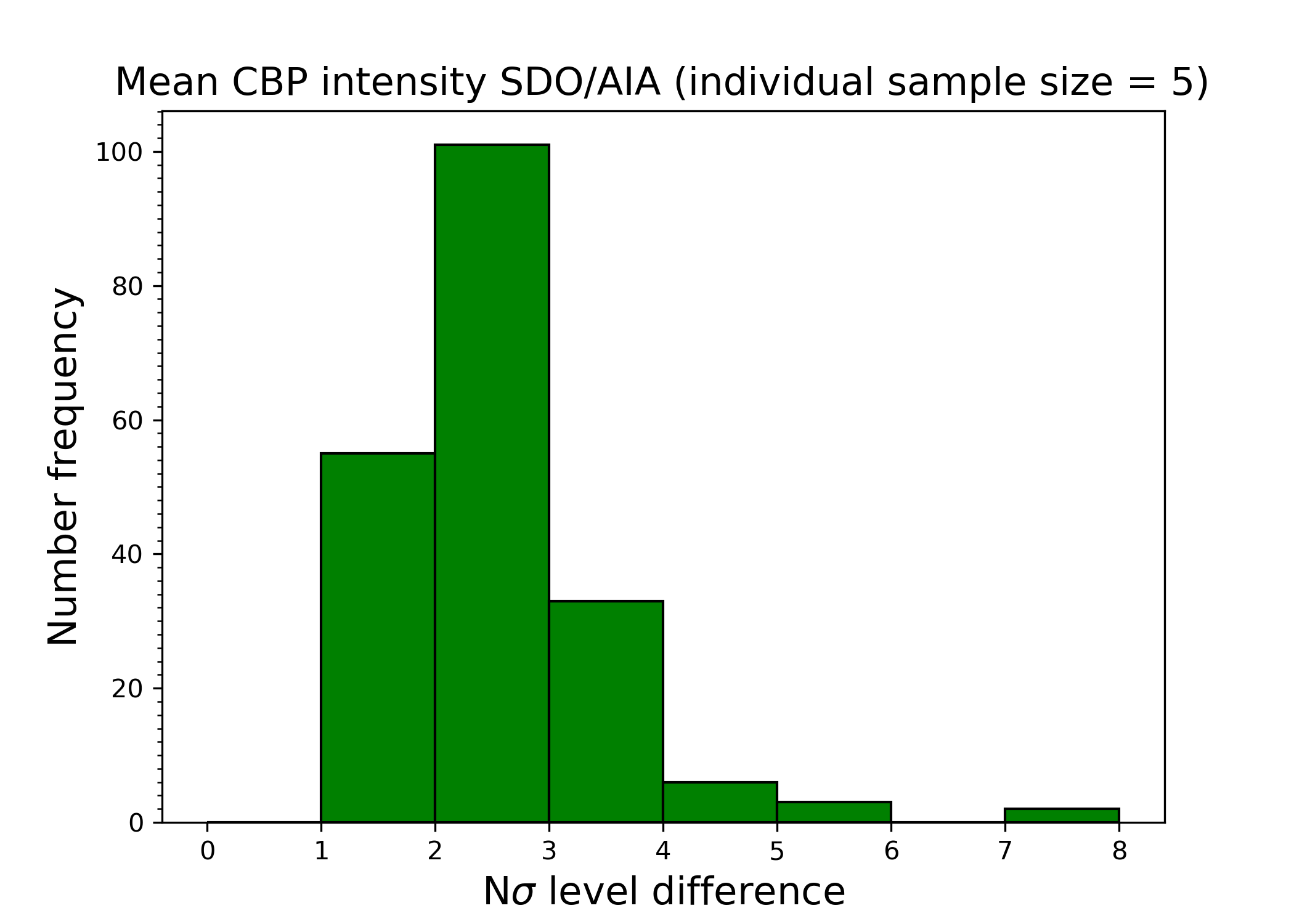}}\\
\subfloat{\includegraphics[width=0.36\textwidth]{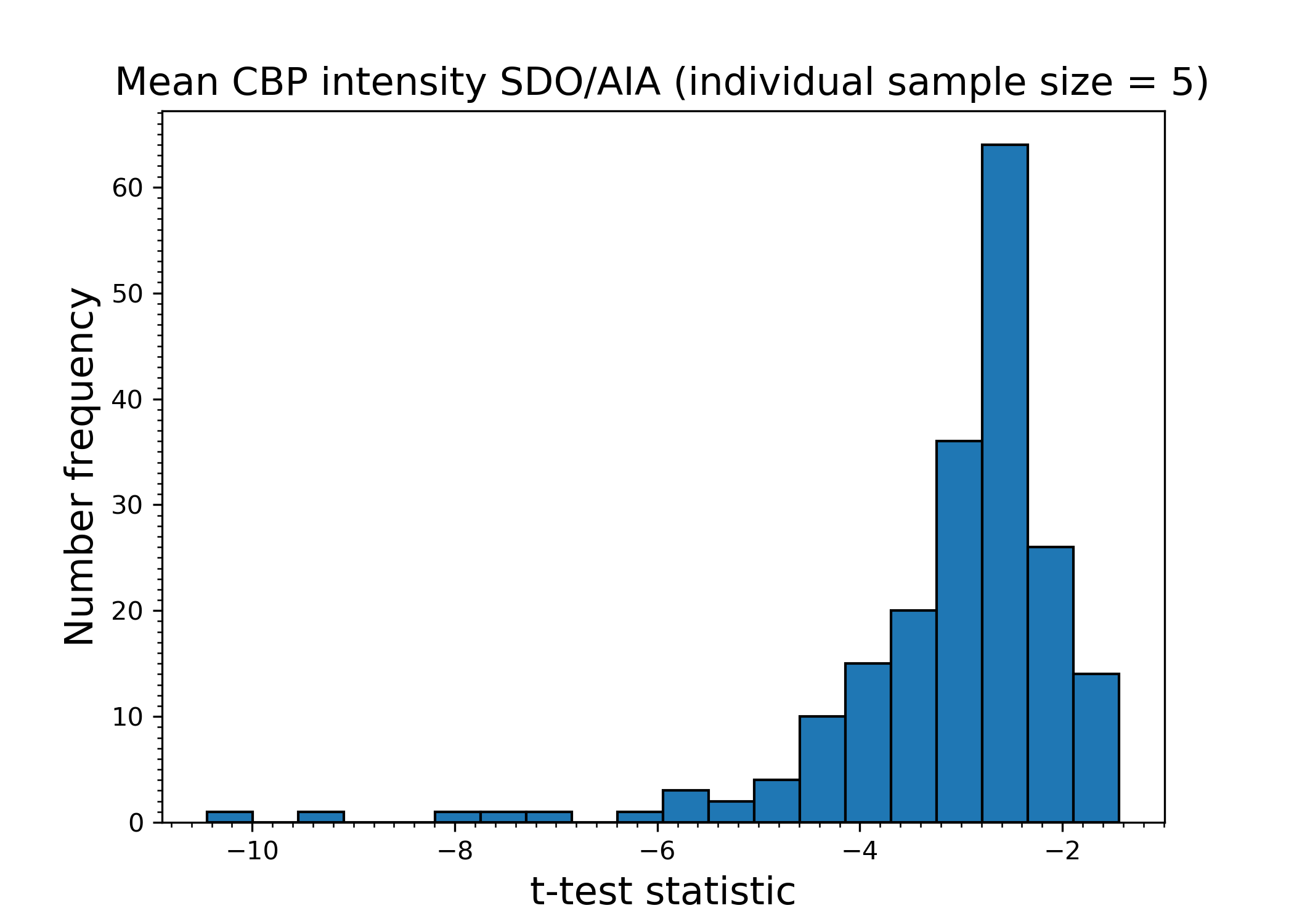}}
\subfloat{\includegraphics[width=0.36\textwidth]{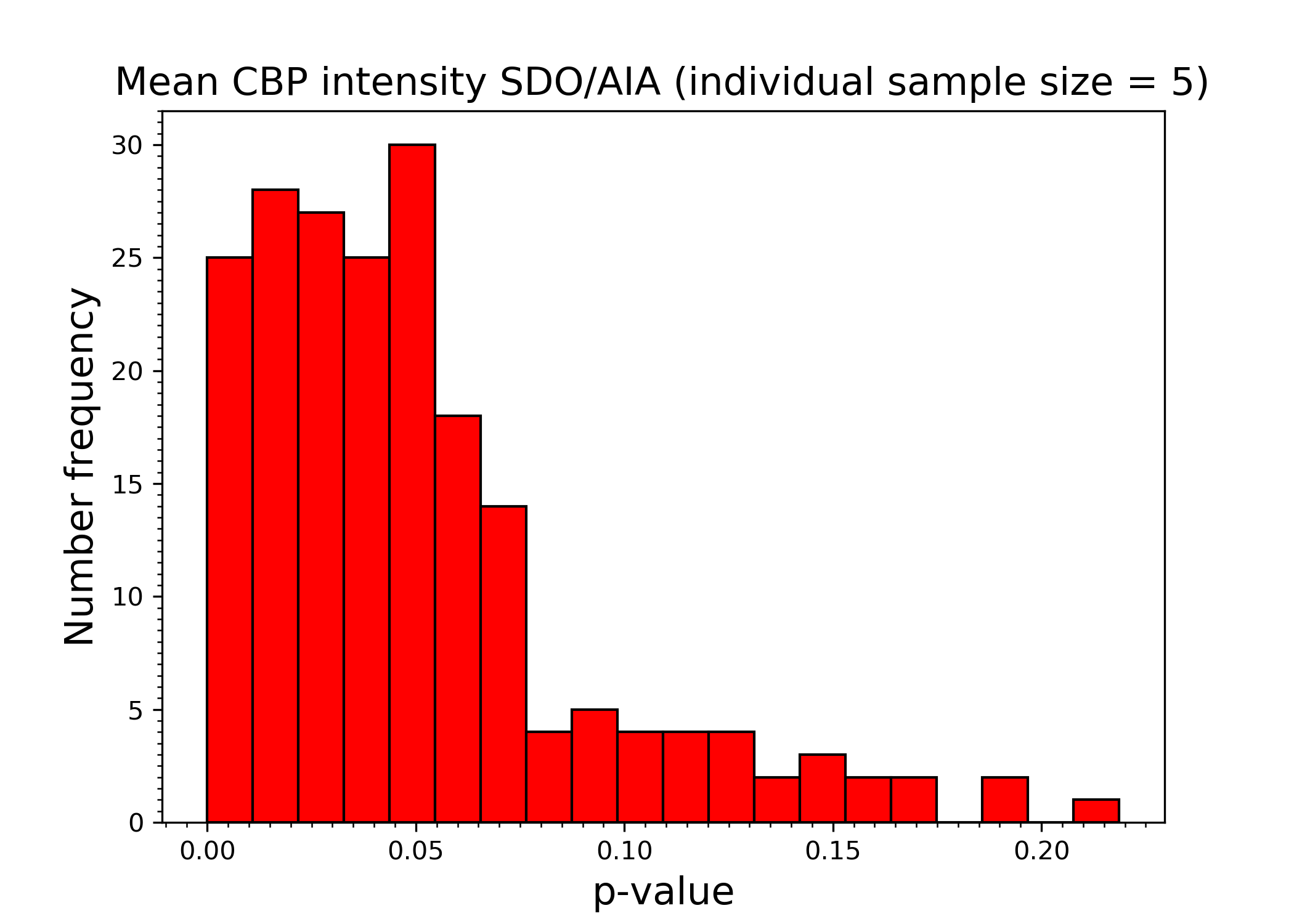}}
\caption{Same as Fig. \ref{mean_int_fig_ch1}, but for mean SDO/AIA 193 \AA\space CBP intensity.}
\label{mean_int_fig_ch1_1}
\end{figure*}

\begin{figure*}[h!]
\captionsetup[subfloat]{farskip=1pt,captionskip=1pt}
\centering
\subfloat{\includegraphics[width=0.36\textwidth]{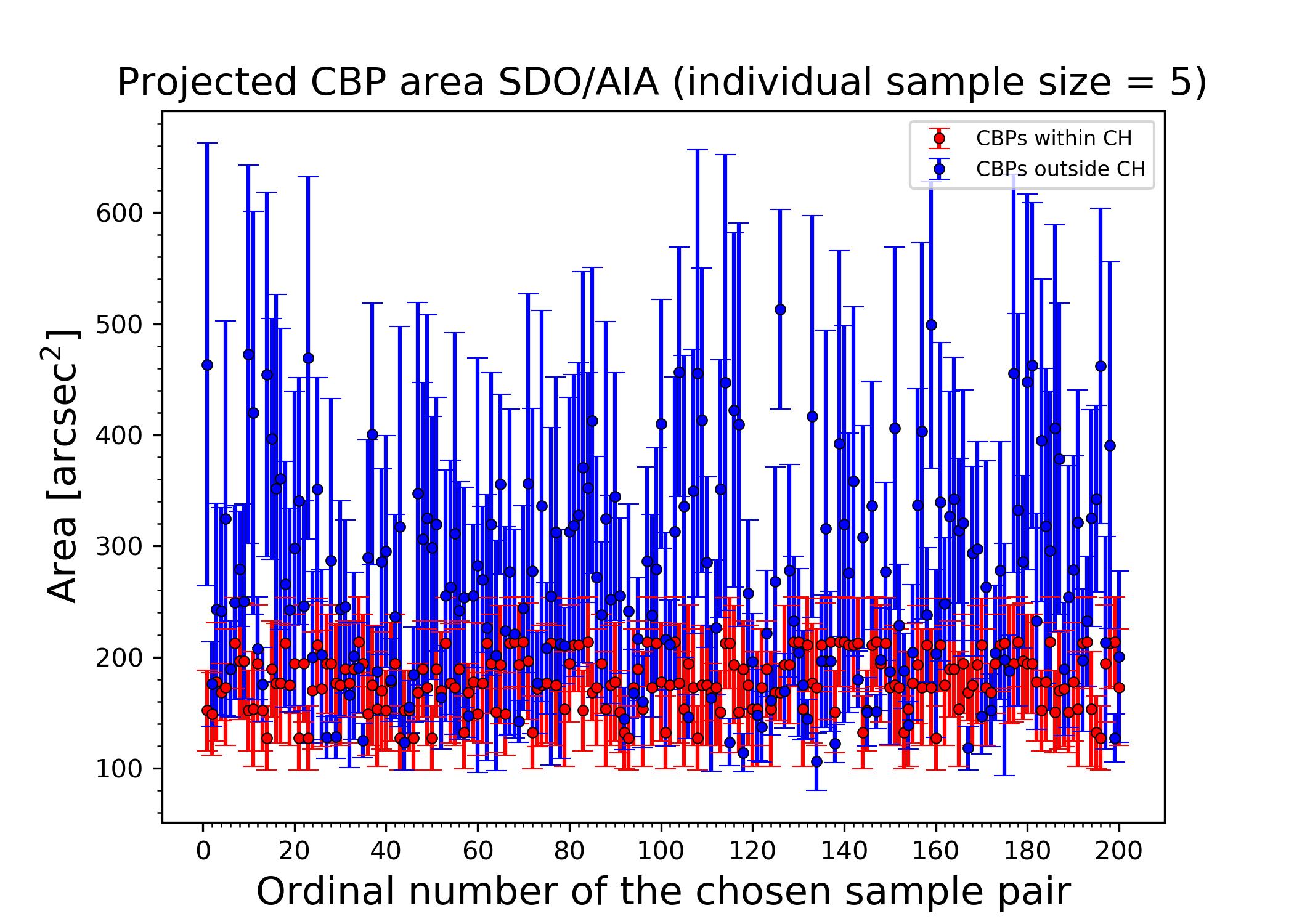}}
\subfloat{\includegraphics[width=0.36\textwidth]{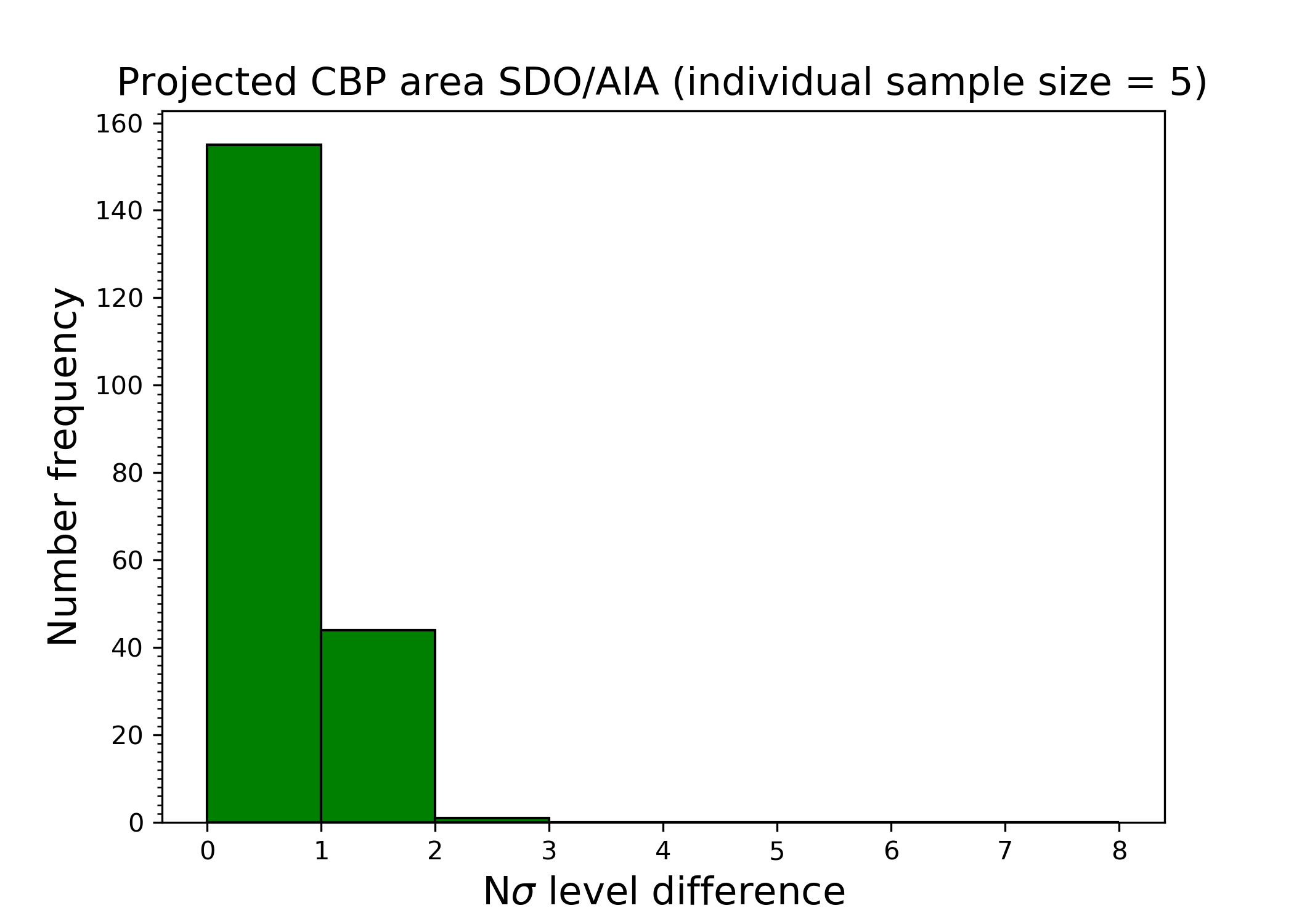}}\\
\subfloat{\includegraphics[width=0.36\textwidth]{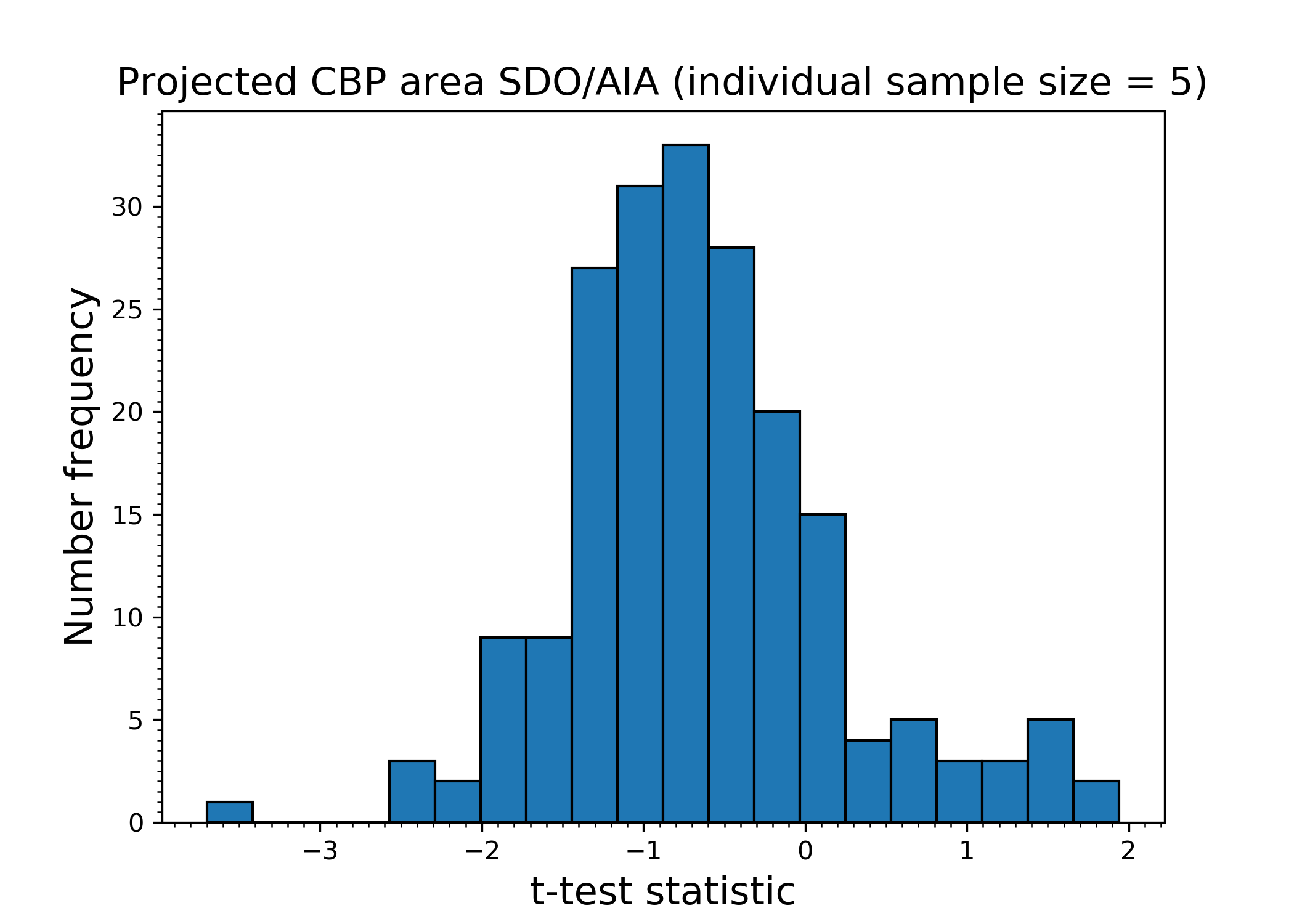}}
\subfloat{\includegraphics[width=0.36\textwidth]{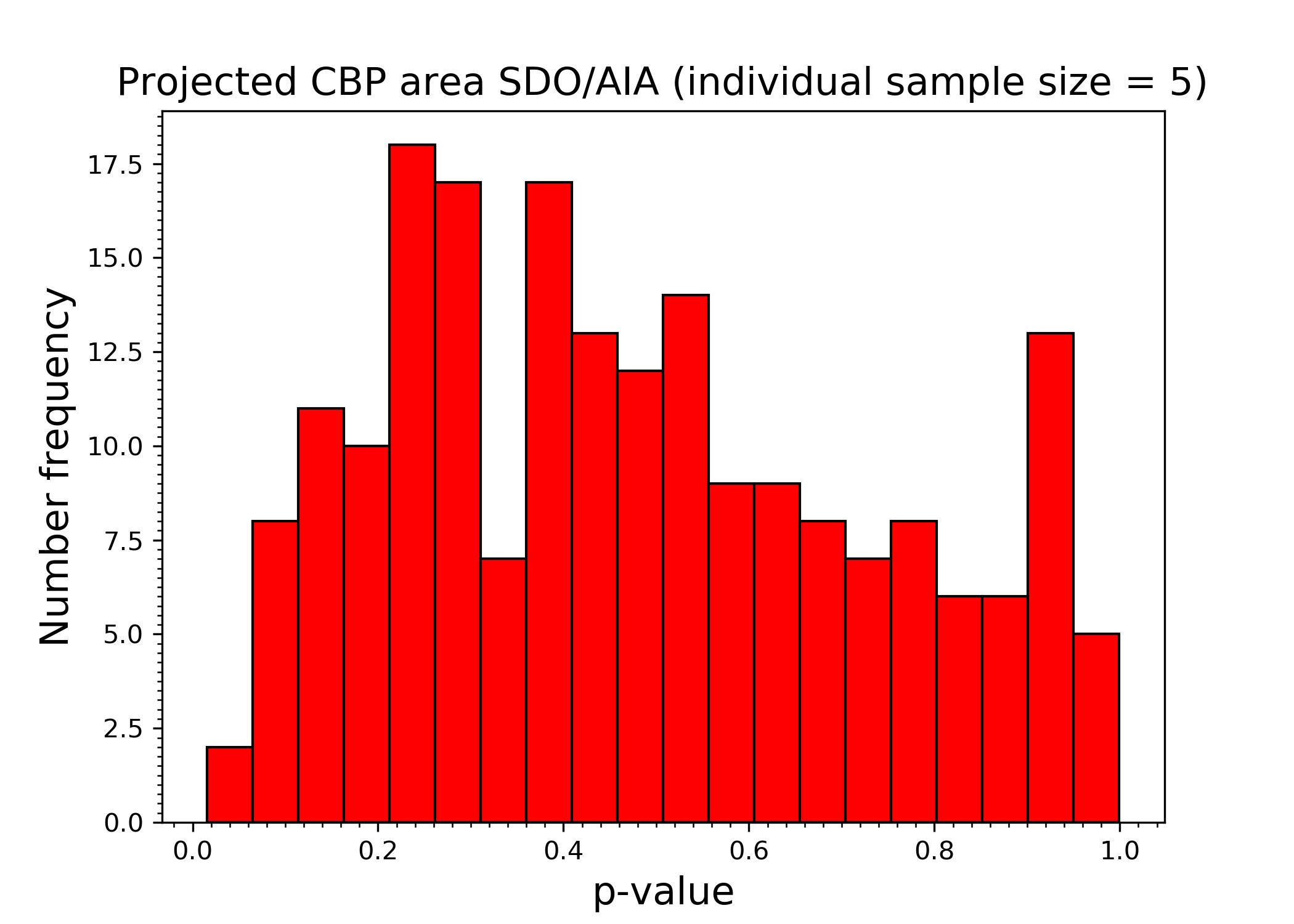}}
\caption{Same as Fig. \ref{mean_int_fig_ch1}, but for projected SDO/AIA 193 \AA\space CBP area.}
\label{area_fig_ch1_1}
\end{figure*}

\clearpage
\section{Statistical analysis of the physical properties for CBPs within and outside CH3}
\label{CH3}
\begin{figure*}[h!]
\captionsetup[subfloat]{farskip=1pt,captionskip=1pt}
\centering
\subfloat{\includegraphics[width=0.36\textwidth]{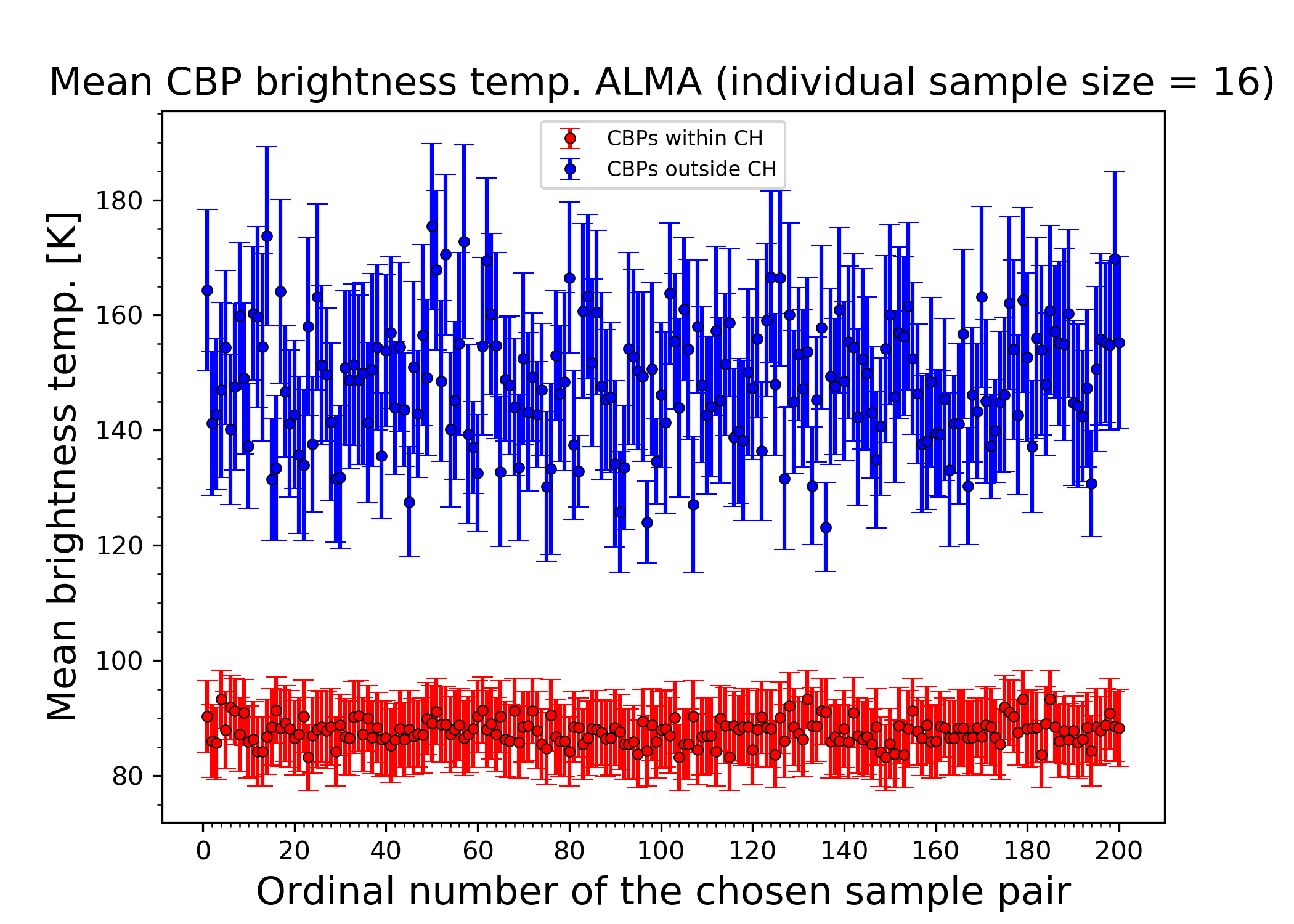}}
\subfloat{\includegraphics[width=0.36\textwidth]{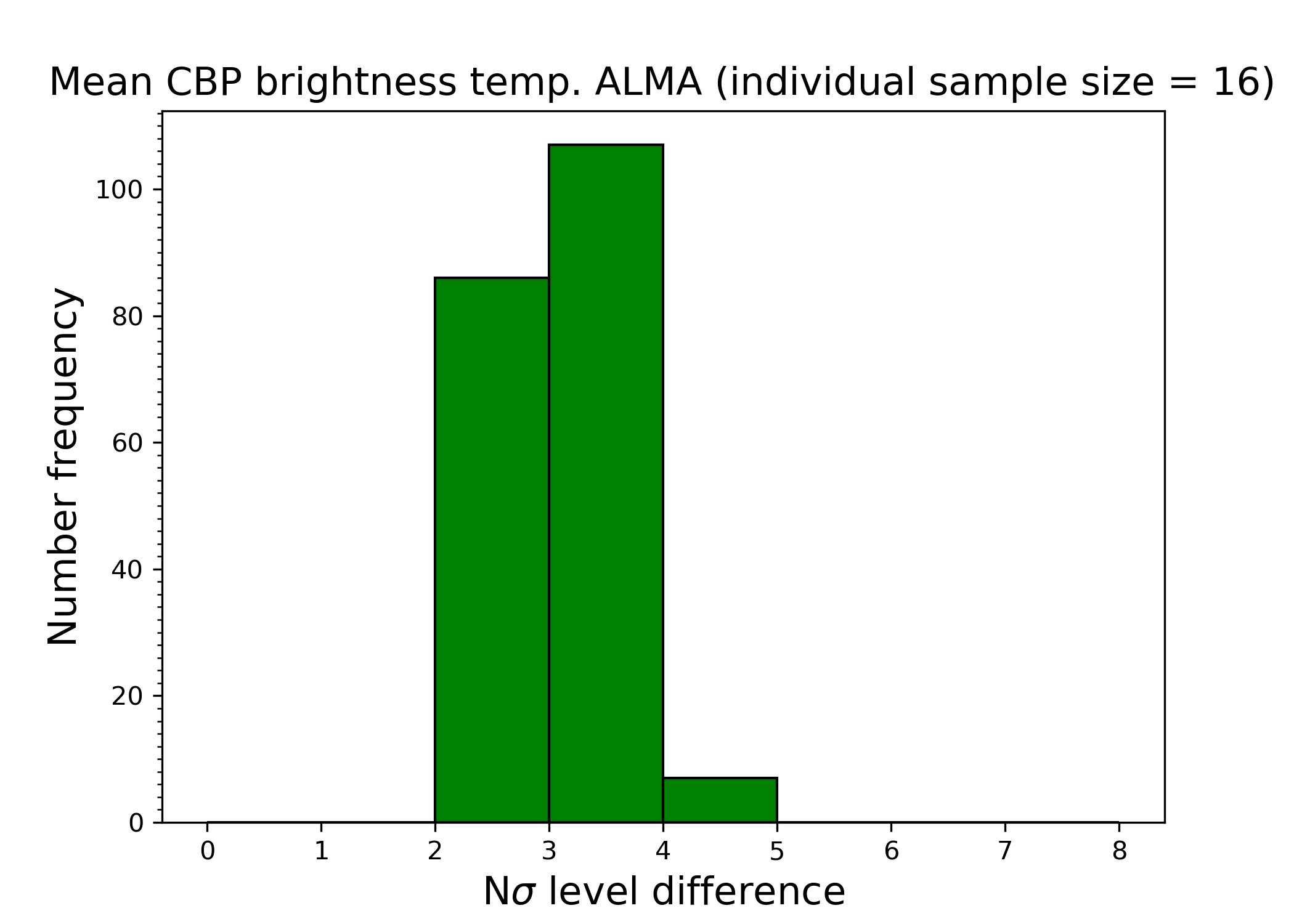}}\\
\subfloat{\includegraphics[width=0.36\textwidth]{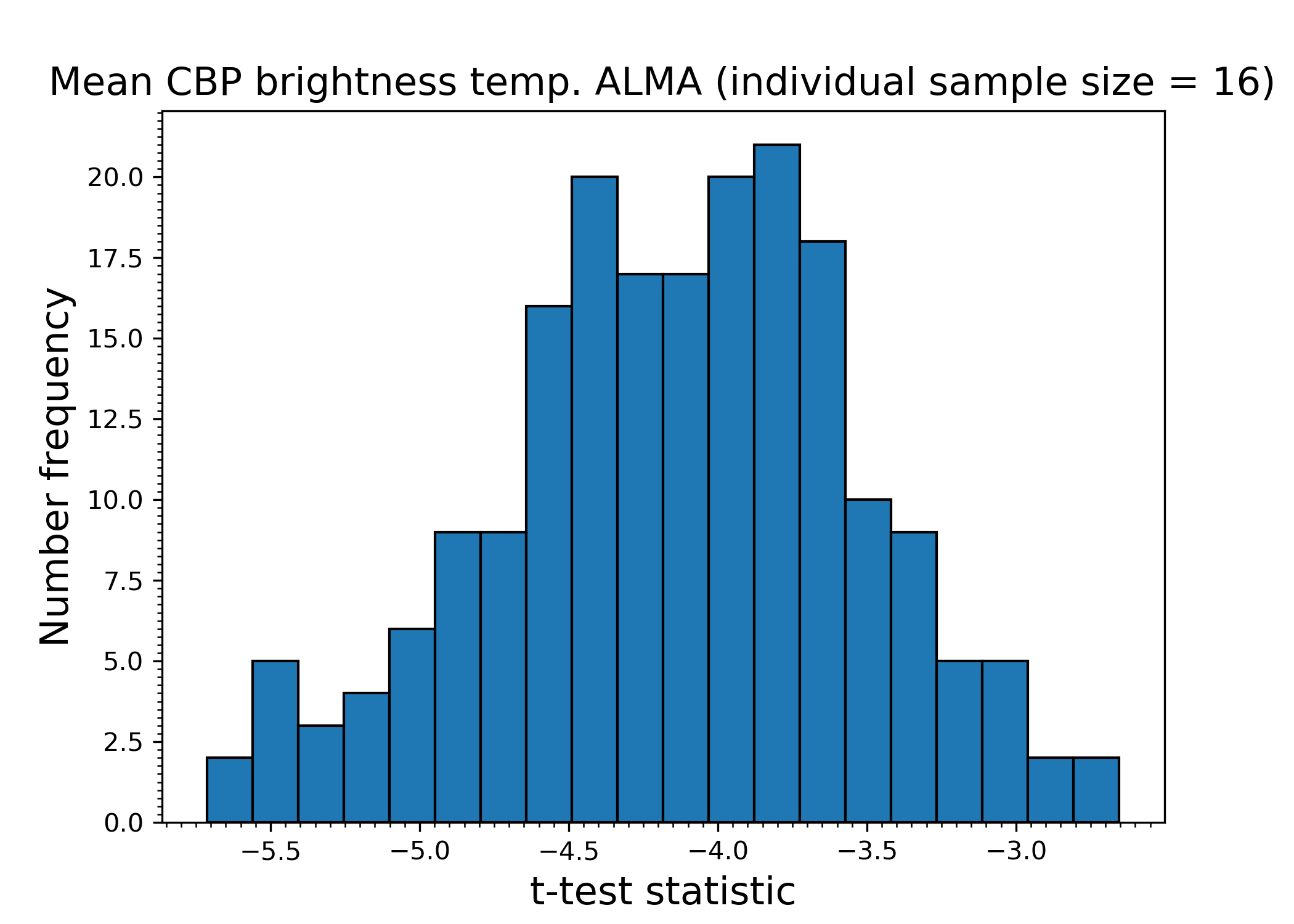}}
\subfloat{\includegraphics[width=0.36\textwidth]{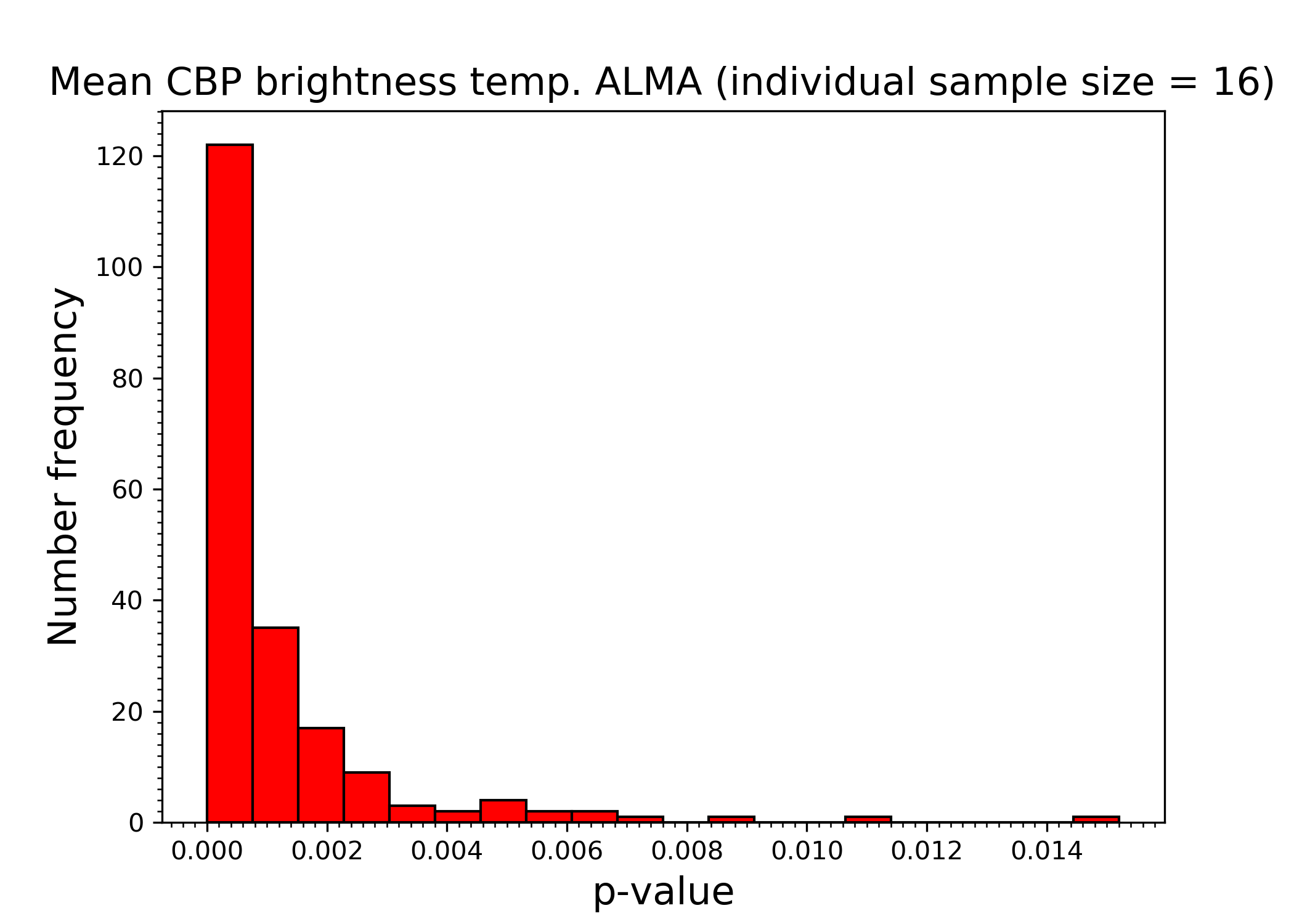}}
\caption{Same as Fig. \ref{mean_int_fig_ch1}, but for CH3, with individual CBP sample containing 16 CBPs.}
\label{mean_int_fig_ch3}
\end{figure*}

\begin{figure*}[h!]
\captionsetup[subfloat]{farskip=1pt,captionskip=1pt}
\centering
\subfloat{\includegraphics[width=0.36\textwidth]{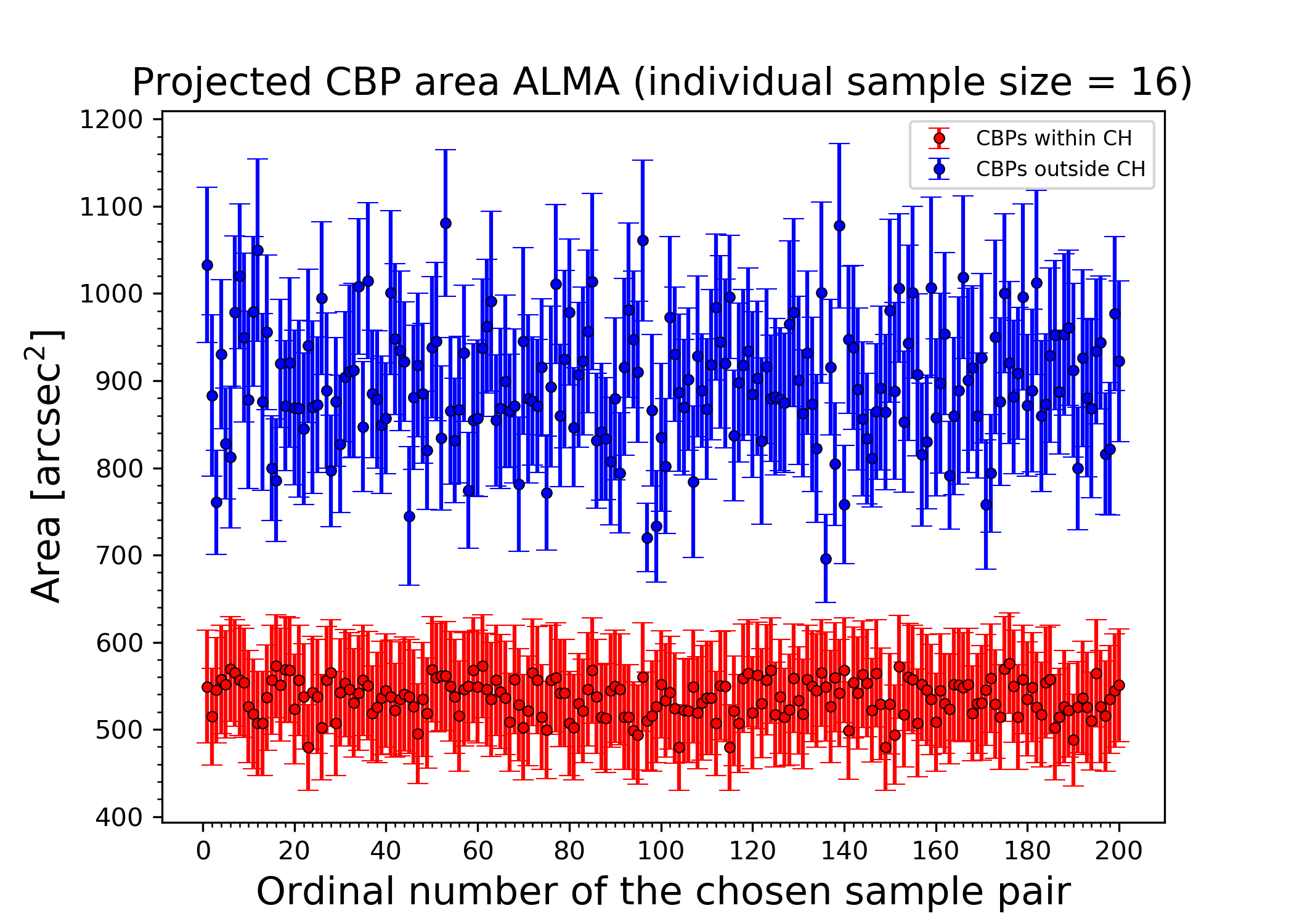}}
\subfloat{\includegraphics[width=0.36\textwidth]{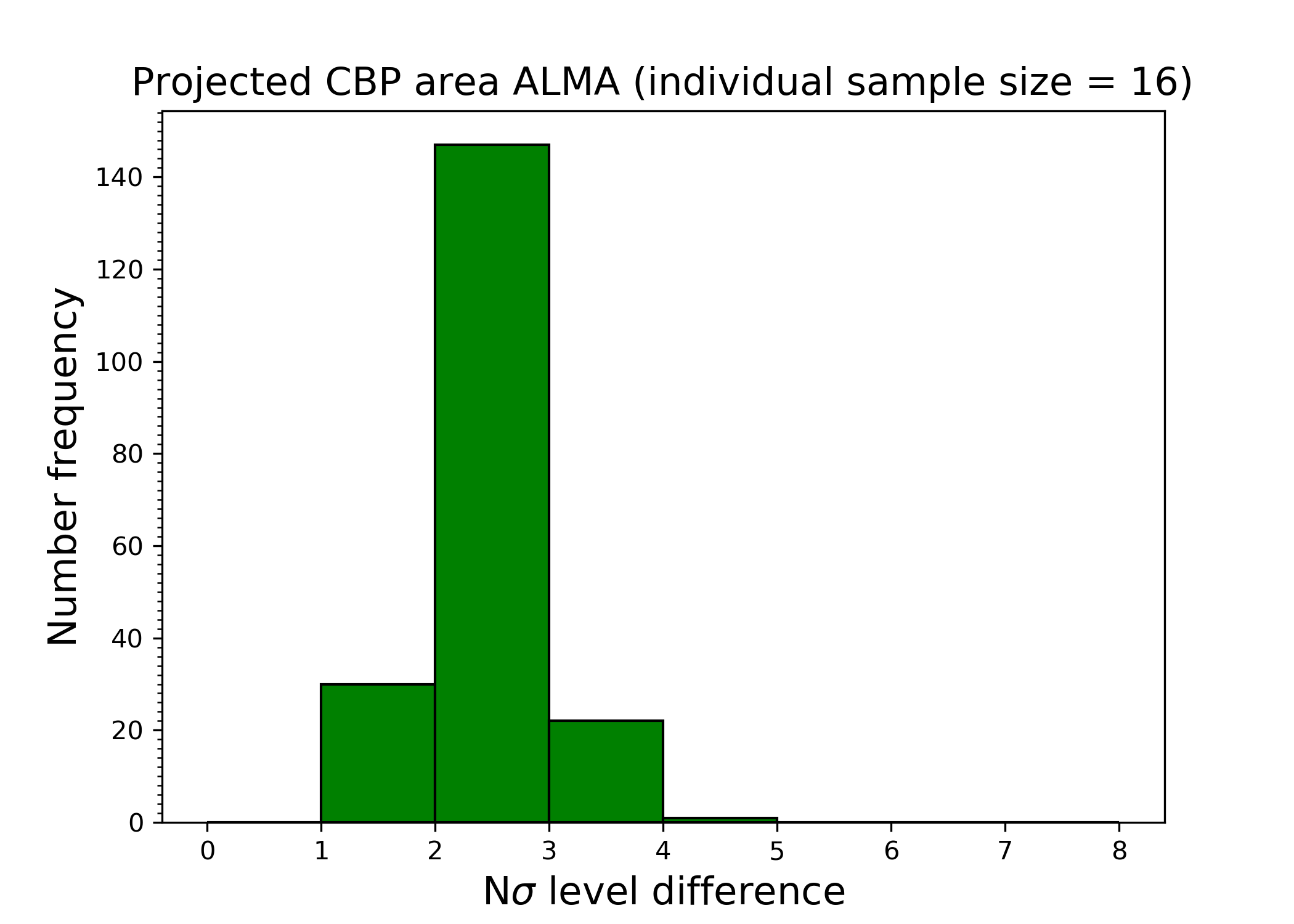}}\\
\subfloat{\includegraphics[width=0.36\textwidth]{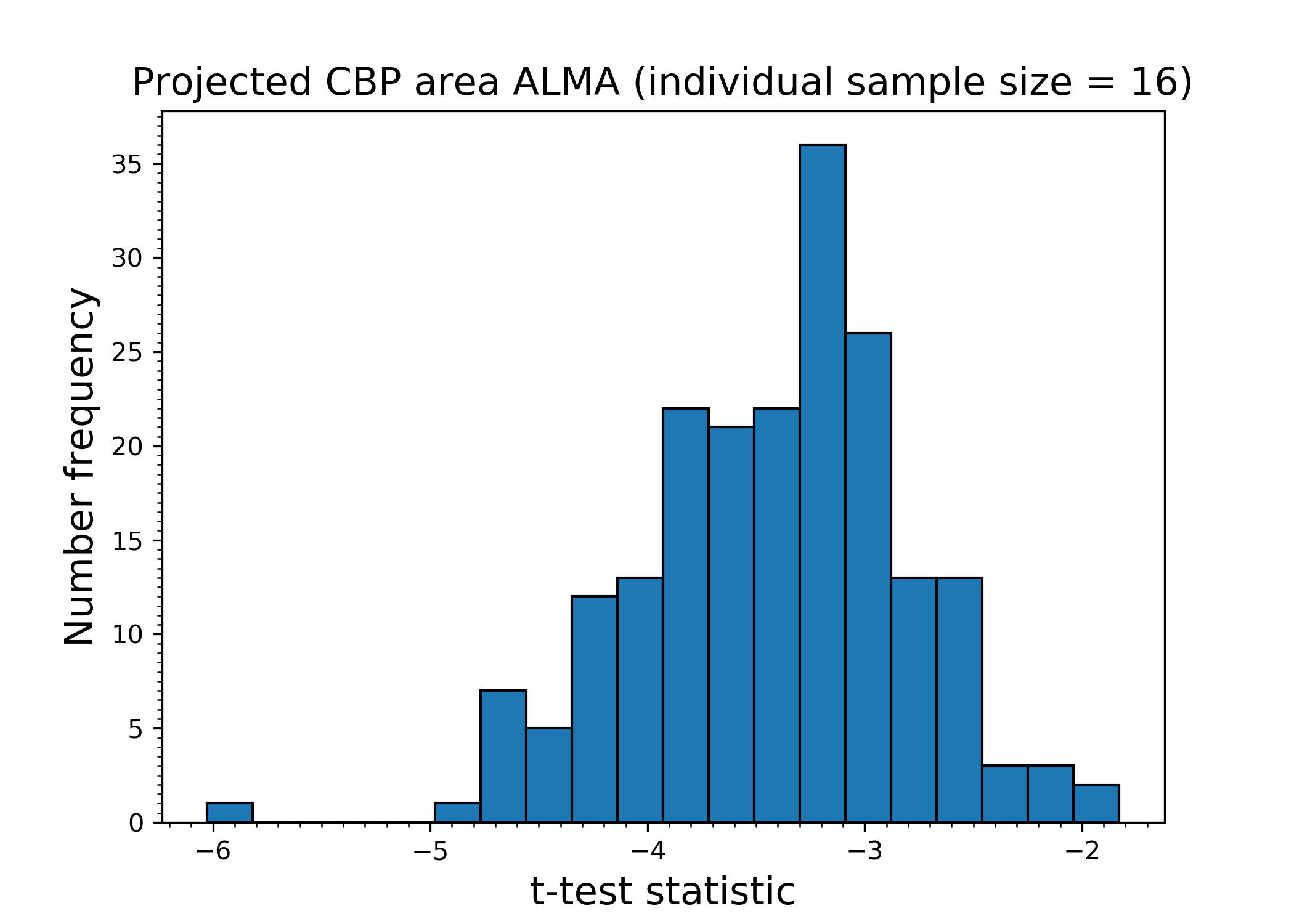}}
\subfloat{\includegraphics[width=0.36\textwidth]{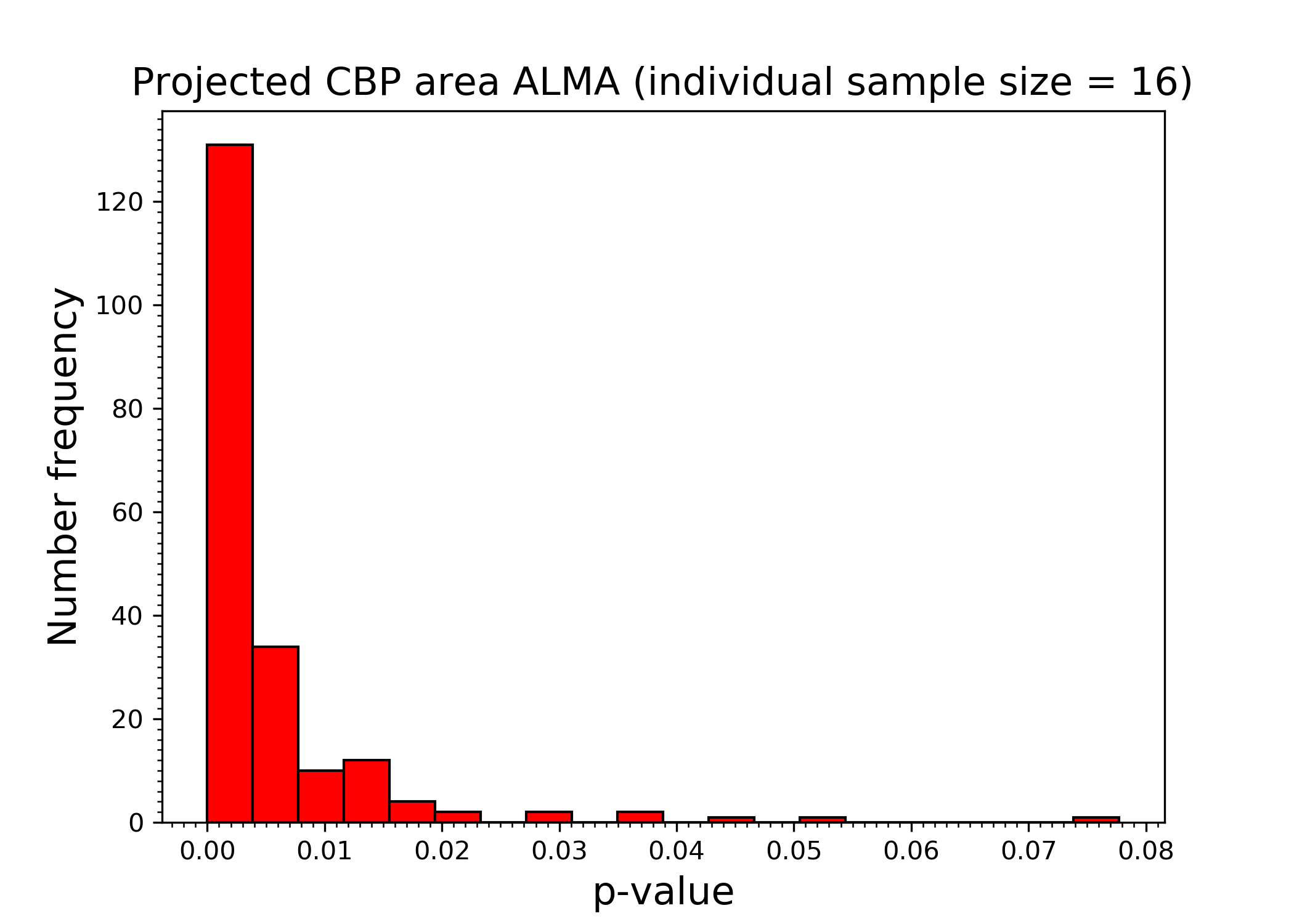}}
\caption{Same as Fig. \ref{mean_int_fig_ch1}, but for projected ALMA Band 6 CBP area and CH3, with individual CBP sample containing 16 CBPs.}
\label{area_fig_ch3}
\end{figure*}

\begin{figure*}[h!]
\captionsetup[subfloat]{farskip=1pt,captionskip=1pt}
\centering
\subfloat{\includegraphics[width=0.36\textwidth]{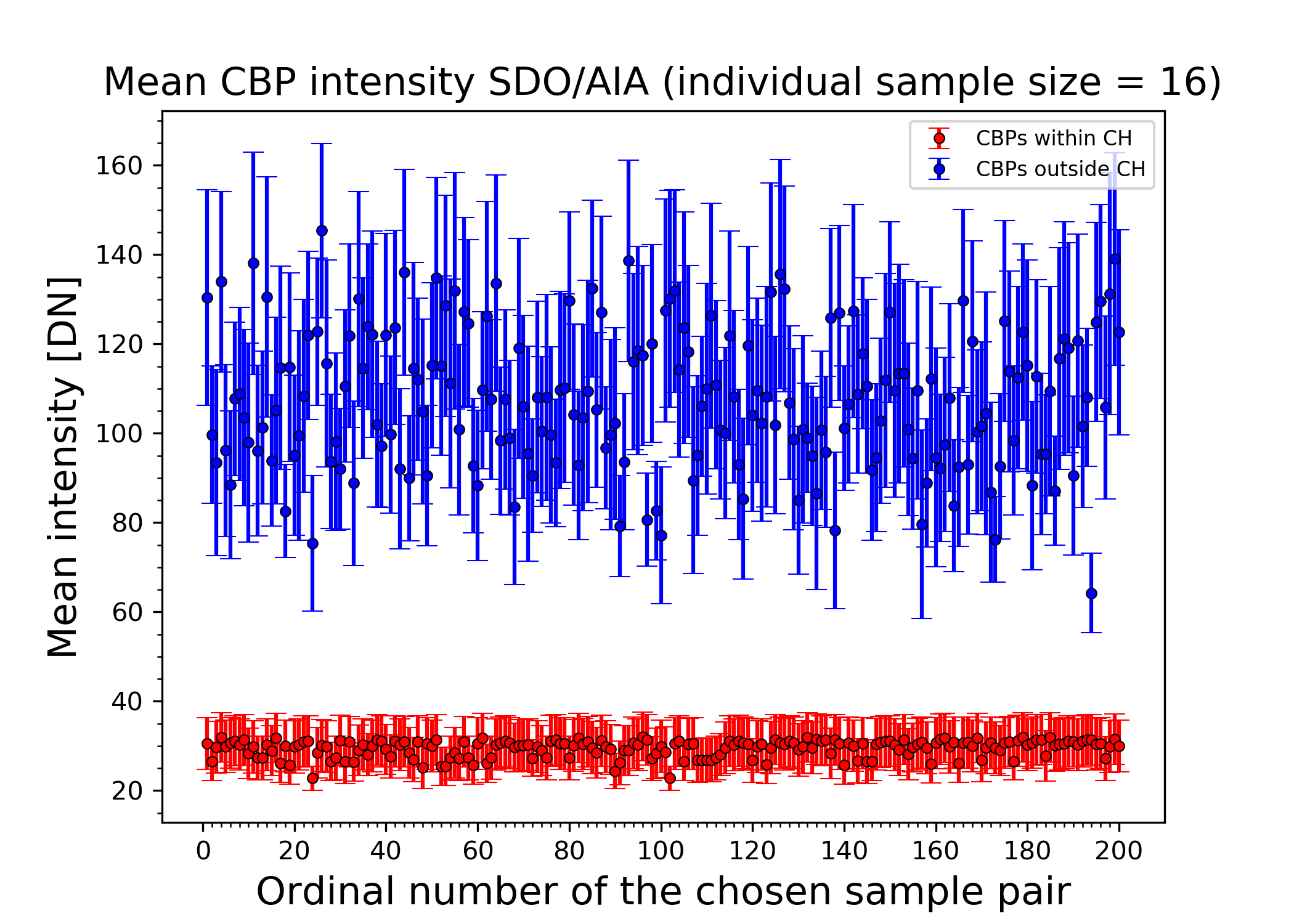}}
\subfloat{\includegraphics[width=0.36\textwidth]{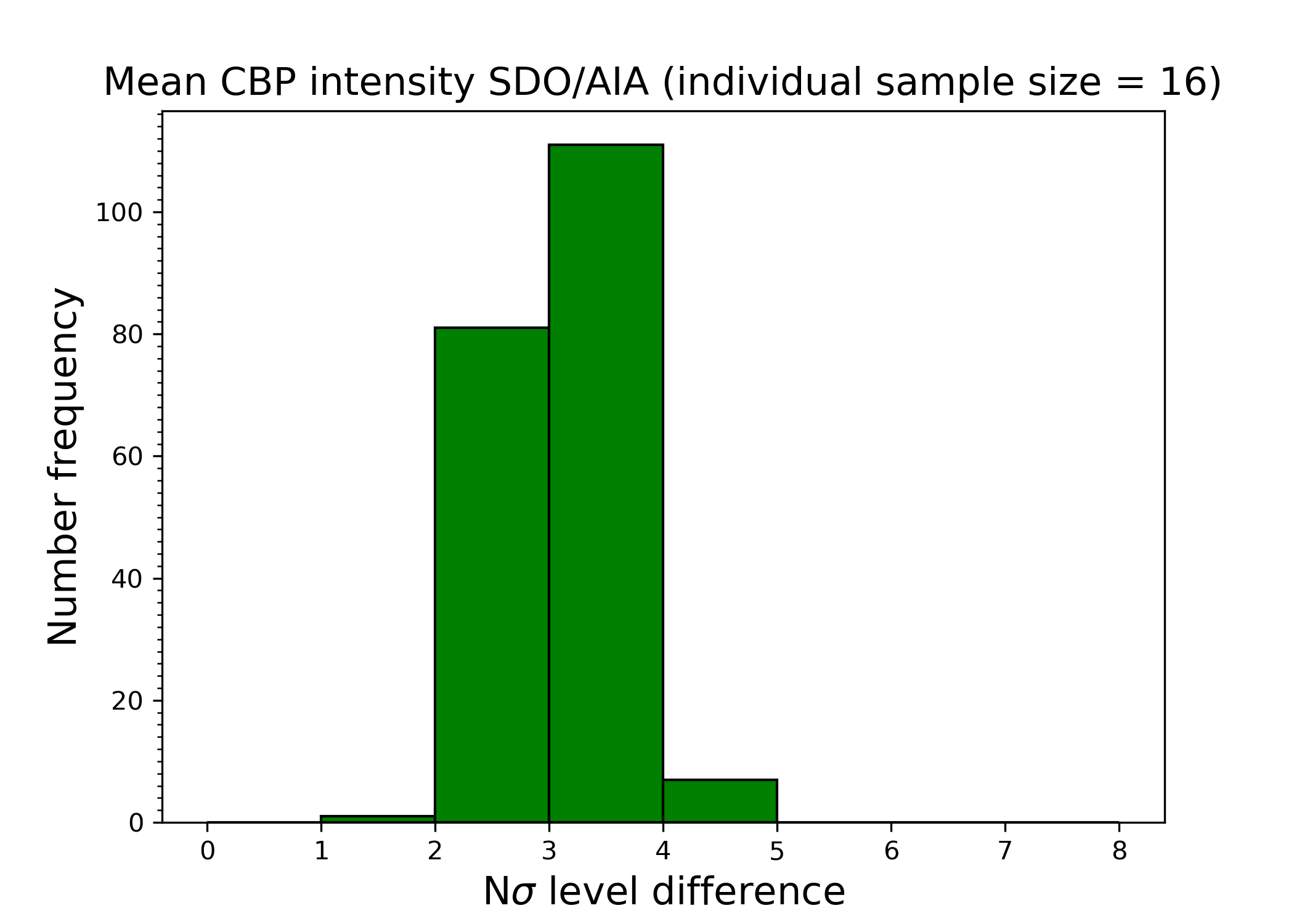}}\\
\subfloat{\includegraphics[width=0.36\textwidth]{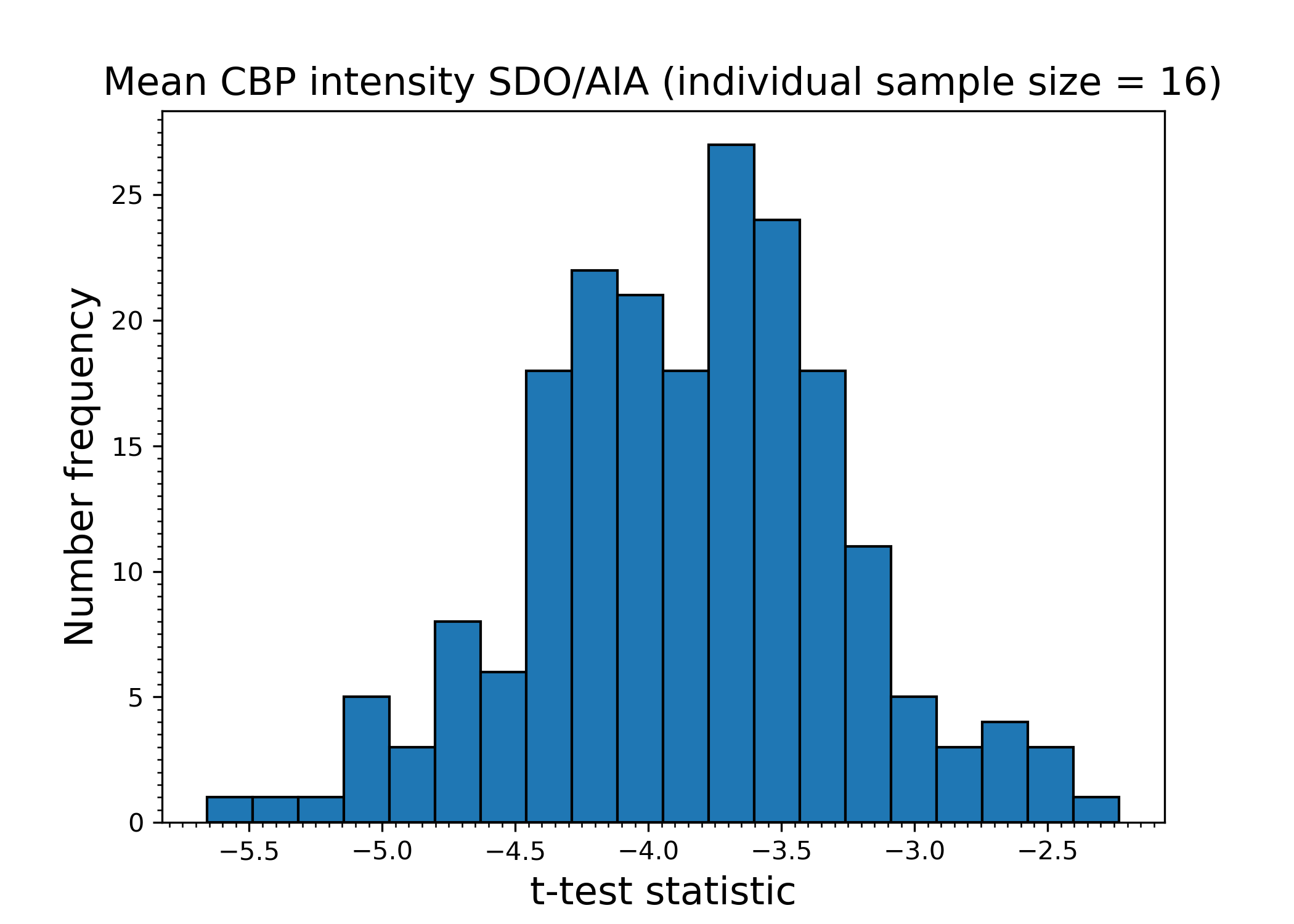}}
\subfloat{\includegraphics[width=0.36\textwidth]{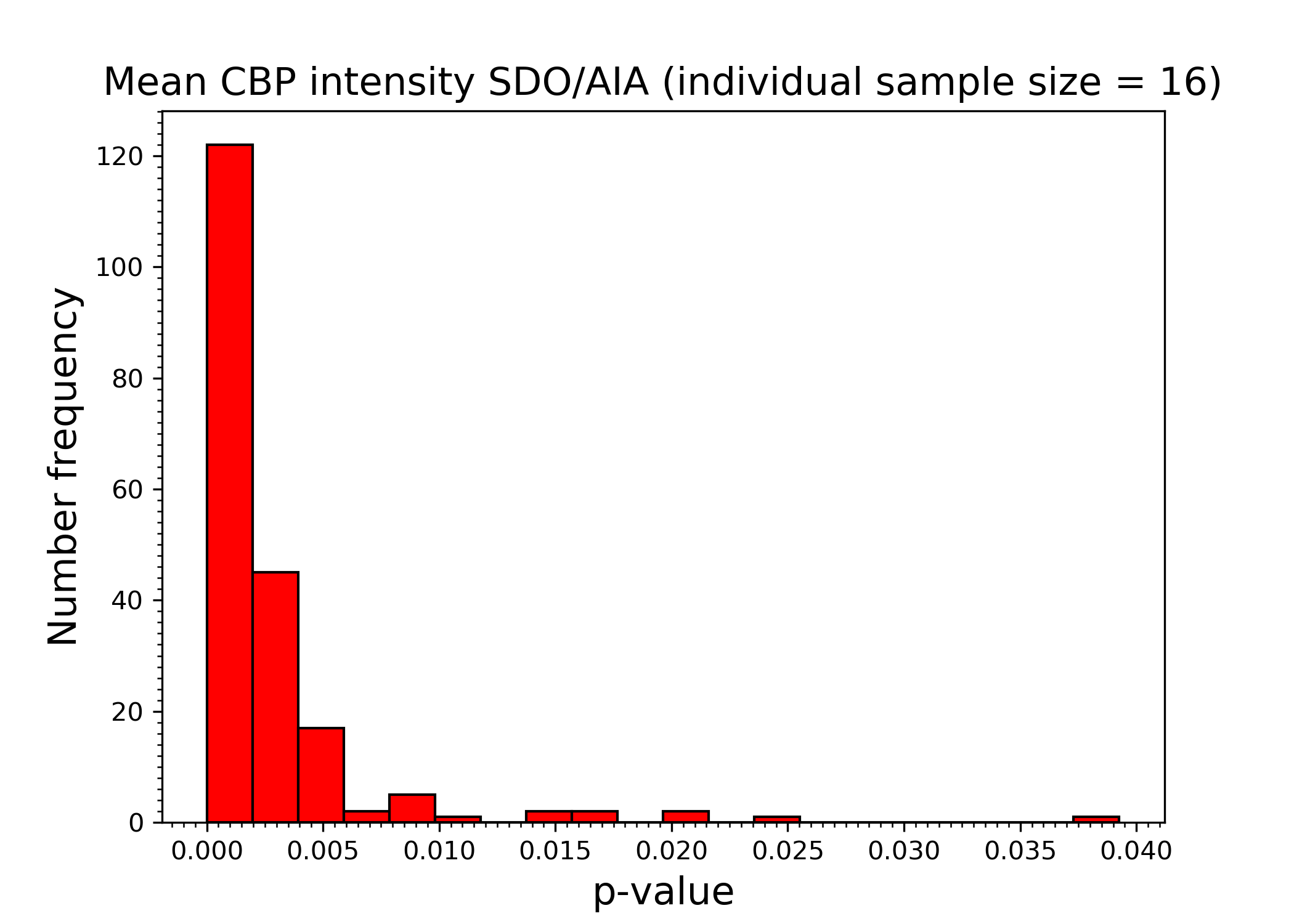}}
\caption{Same as Fig. \ref{mean_int_fig_ch1}, but for mean SDO/AIA 193 \AA\space CBP intensity and CH3, with individual CBP sample containing 16 CBPs.}
\label{mean_int_fig_ch3_1}
\end{figure*}

\begin{figure*}[h!]
\captionsetup[subfloat]{farskip=1pt,captionskip=1pt}
\centering
\subfloat{\includegraphics[width=0.36\textwidth]{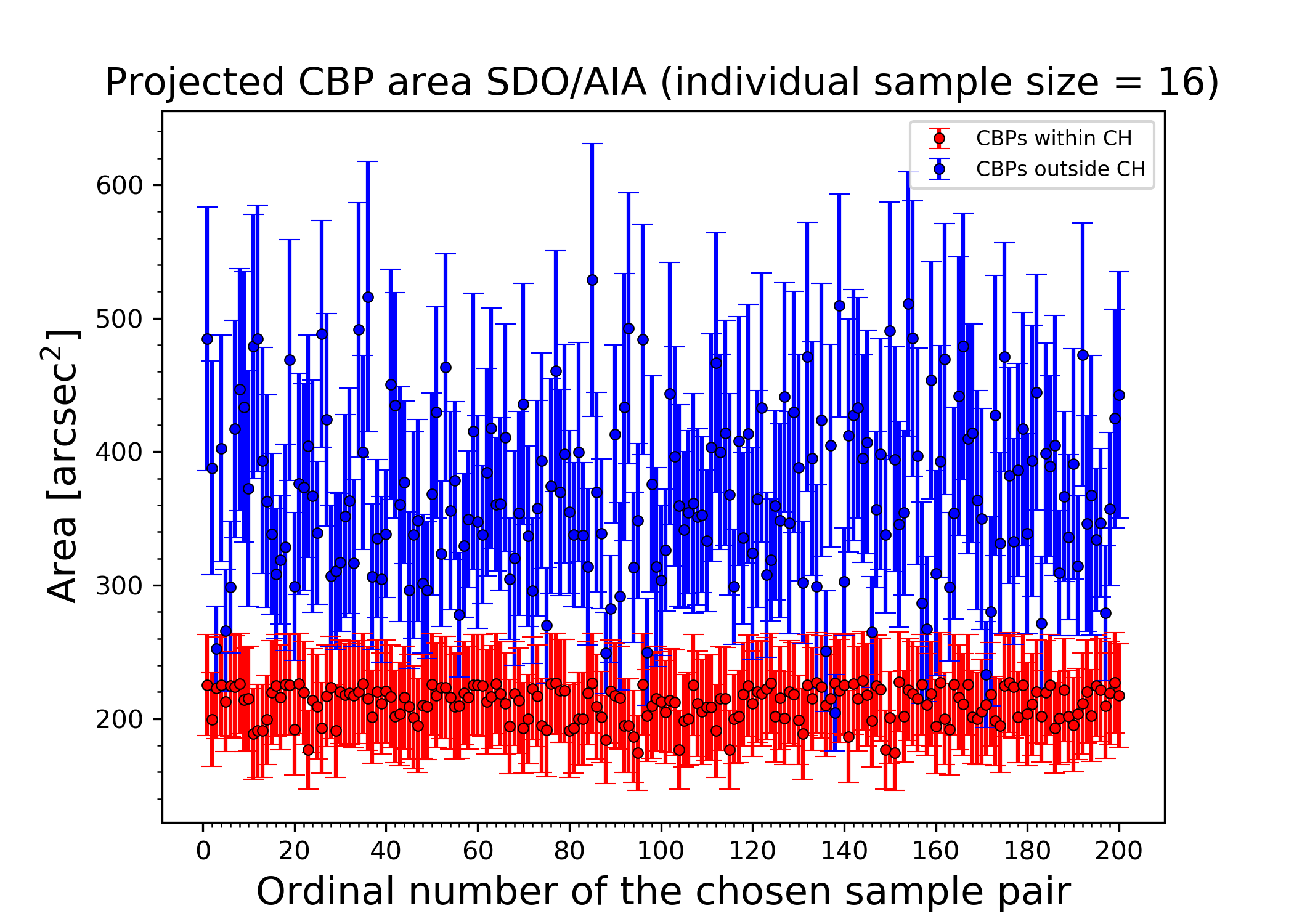}}
\subfloat{\includegraphics[width=0.36\textwidth]{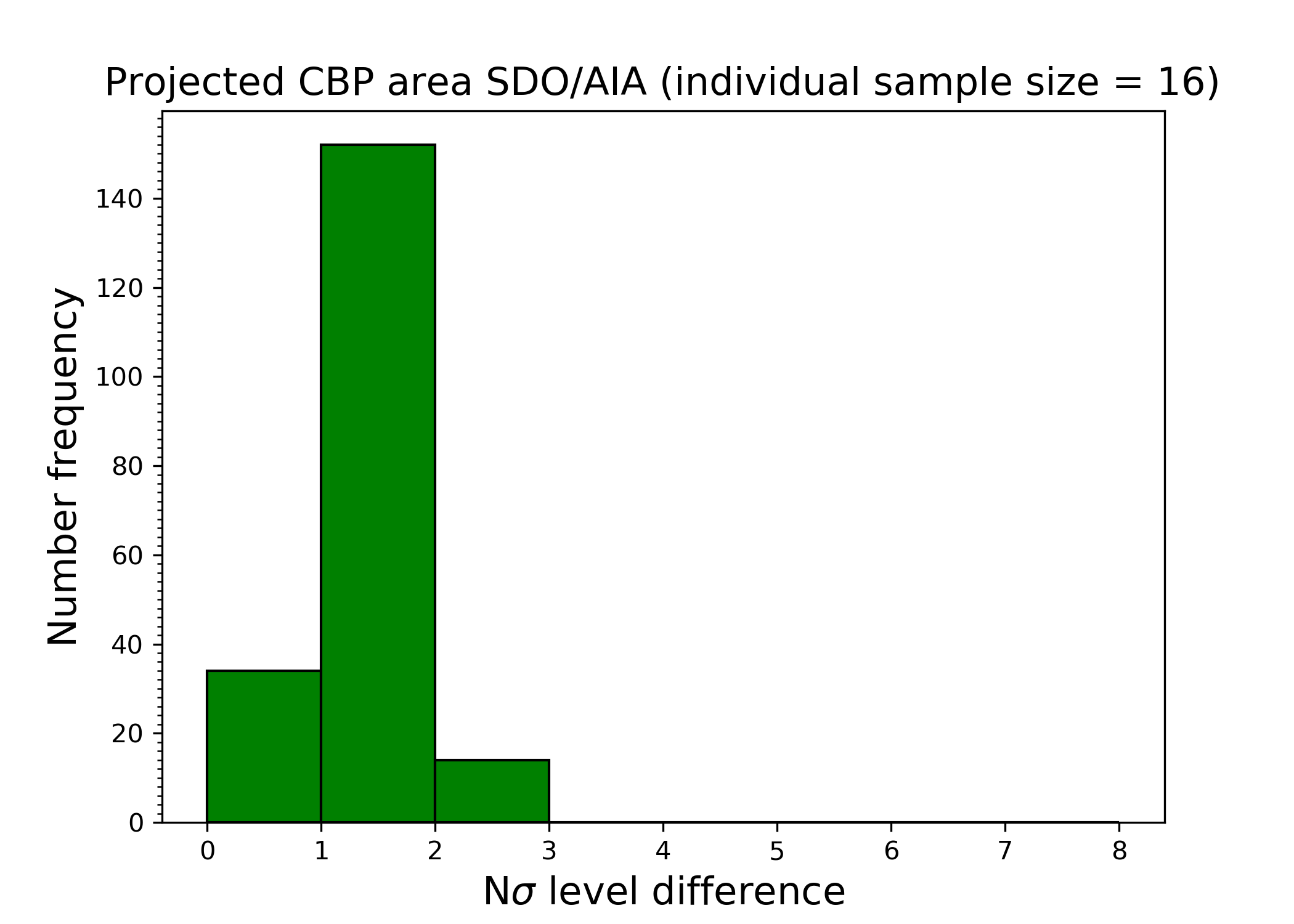}}\\
\subfloat{\includegraphics[width=0.36\textwidth]{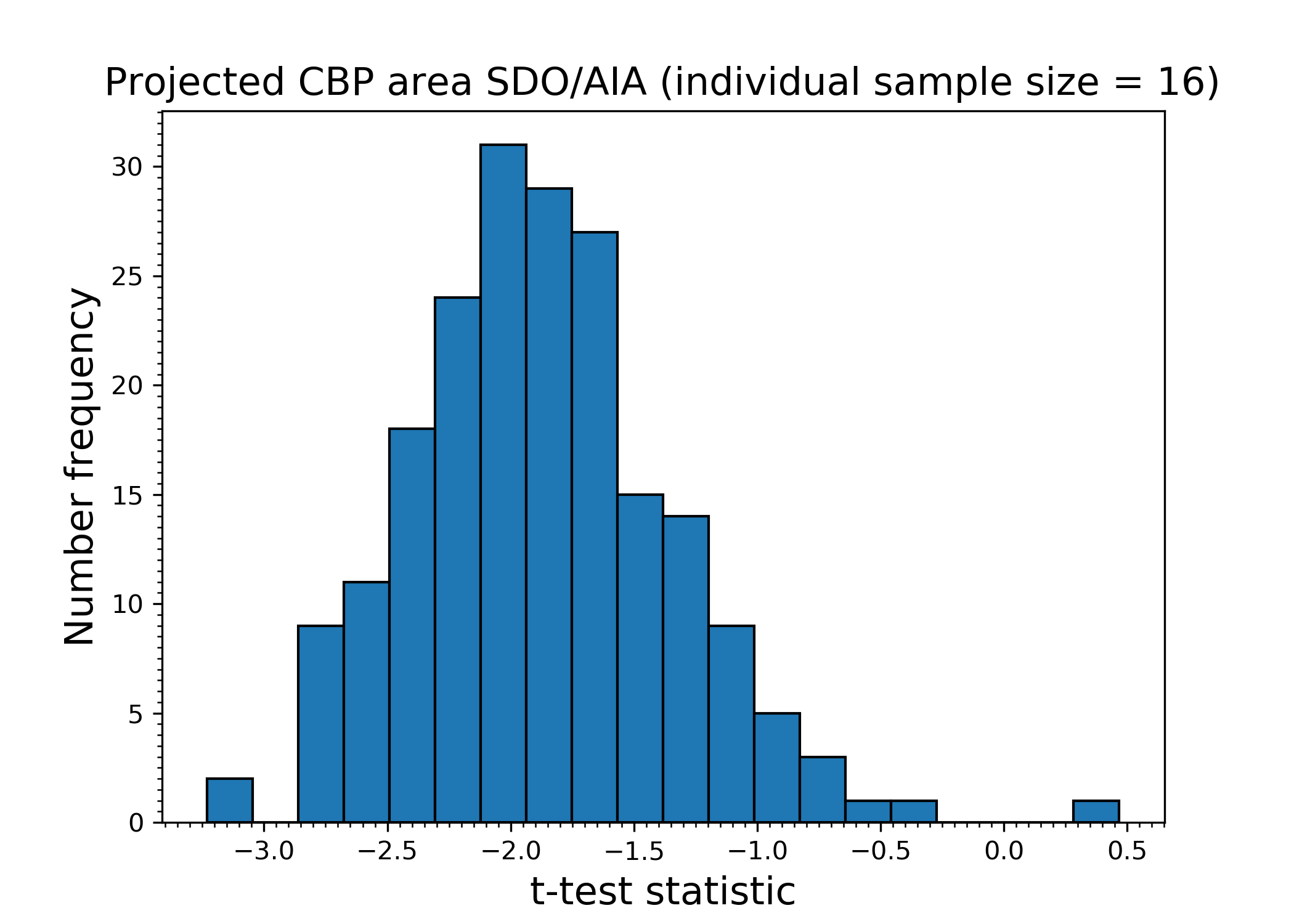}}
\subfloat{\includegraphics[width=0.36\textwidth]{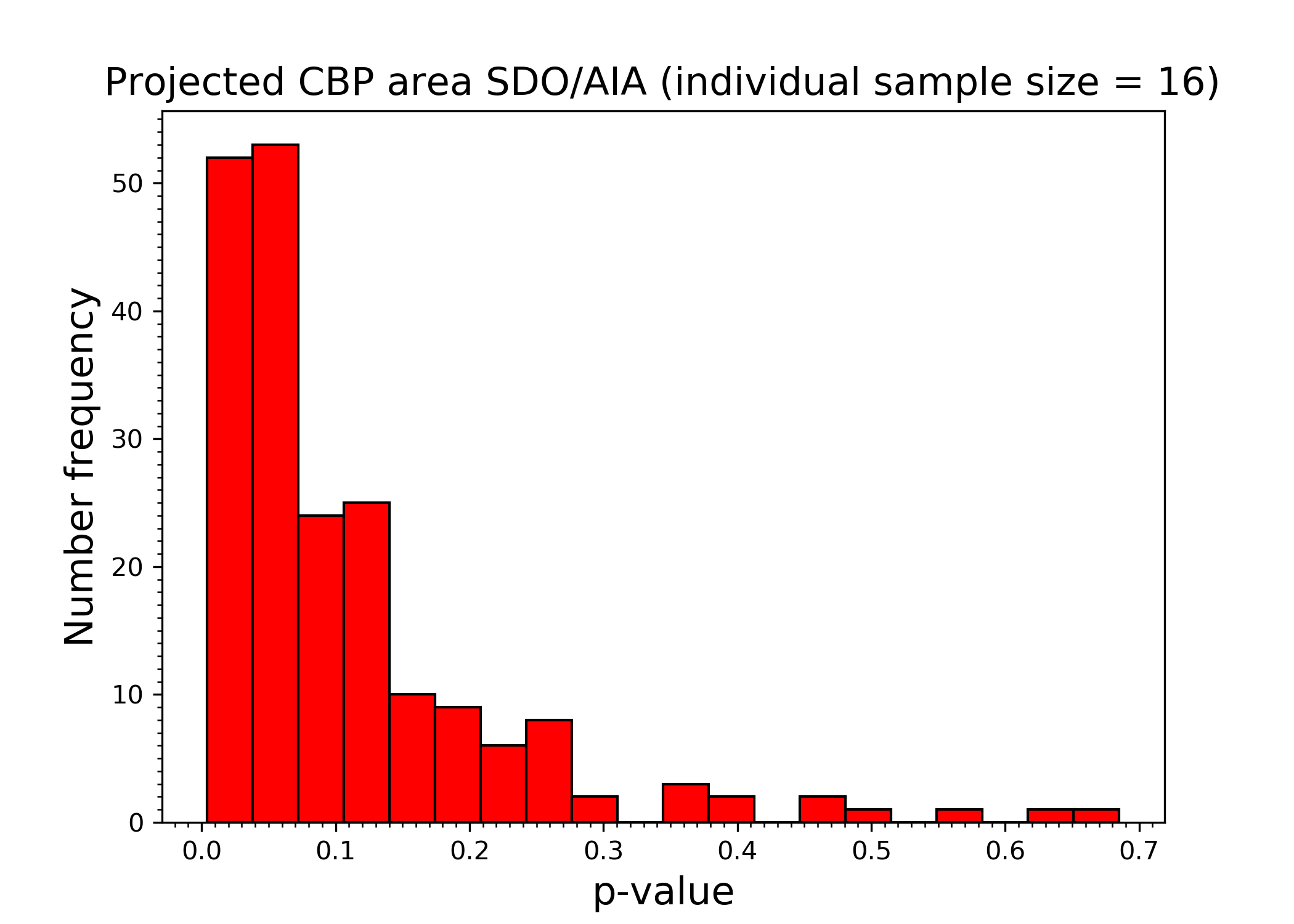}}
\caption{Same as Fig. \ref{mean_int_fig_ch1}, but for projected SDO/AIA 193 \AA\space CBP area and CH3, with individual CBP sample containing 16 CBPs.}
\label{area_fig_ch3_1}
\end{figure*}

\clearpage
\section{Statistical analysis of the physical properties for CBPs within and outside CH4}
\label{CH4}
\begin{figure*}[h!]
\captionsetup[subfloat]{farskip=1pt,captionskip=1pt}
\centering
\subfloat{\includegraphics[width=0.36\textwidth]{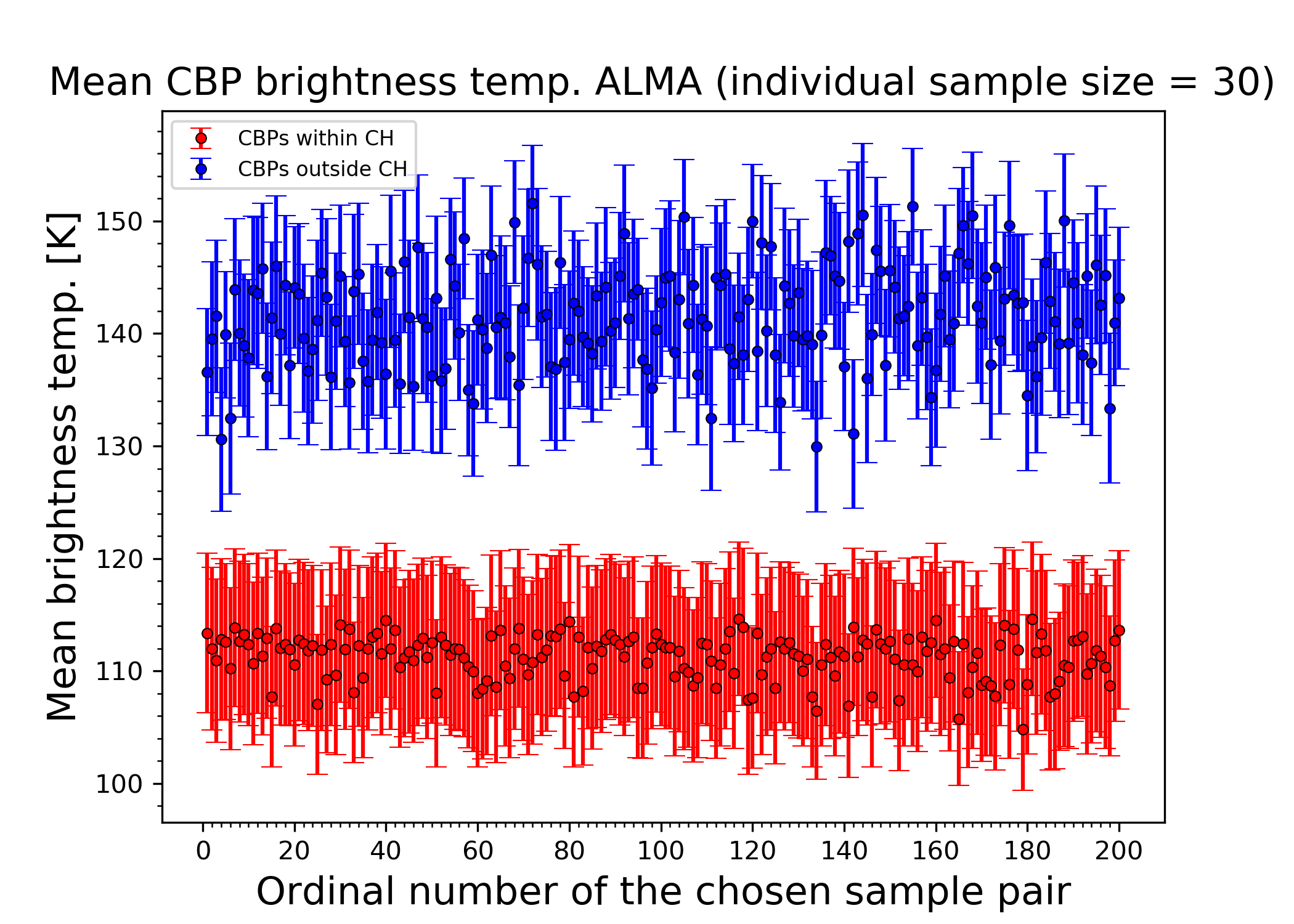}}
\subfloat{\includegraphics[width=0.36\textwidth]{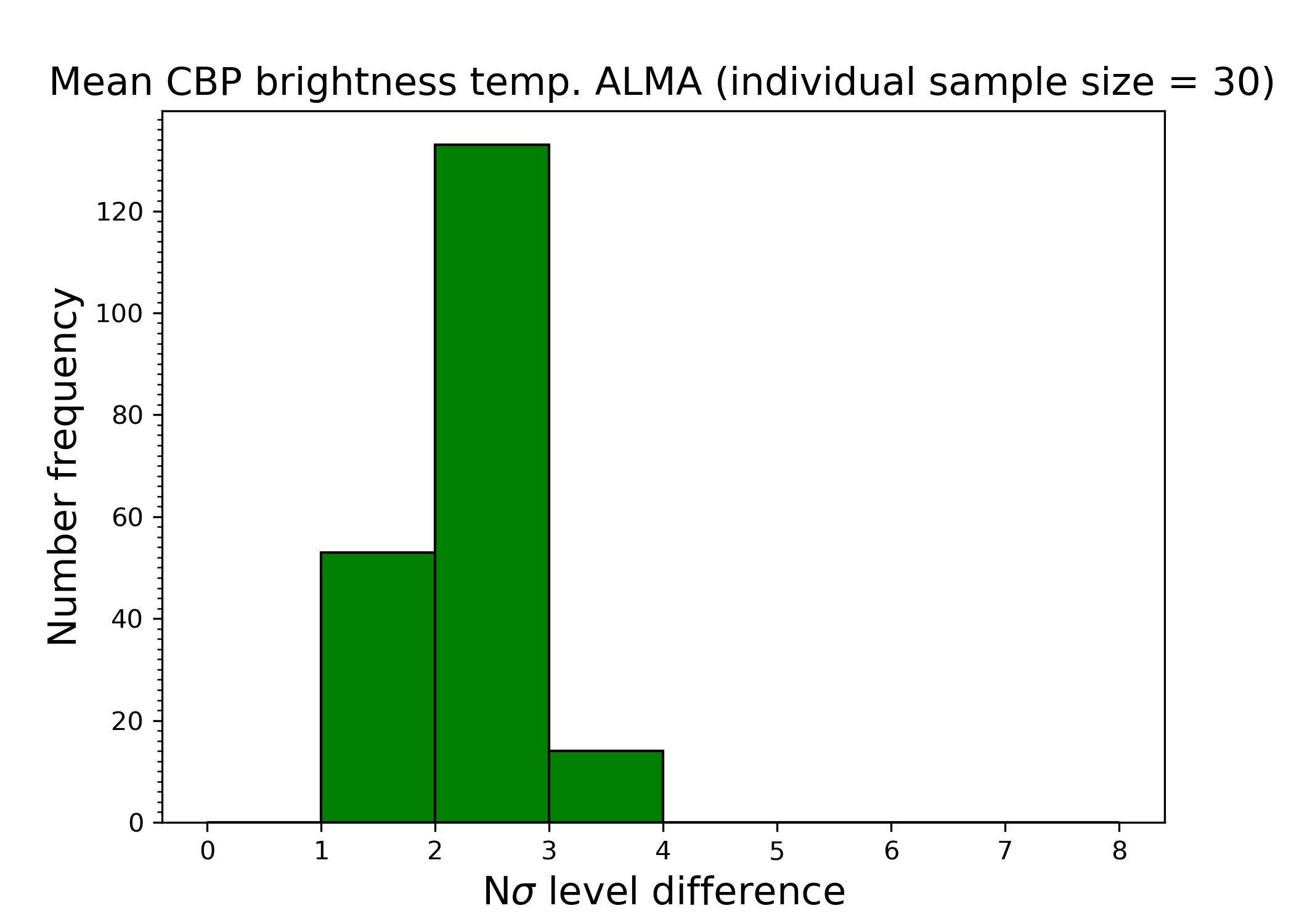}}\\
\subfloat{\includegraphics[width=0.36\textwidth]{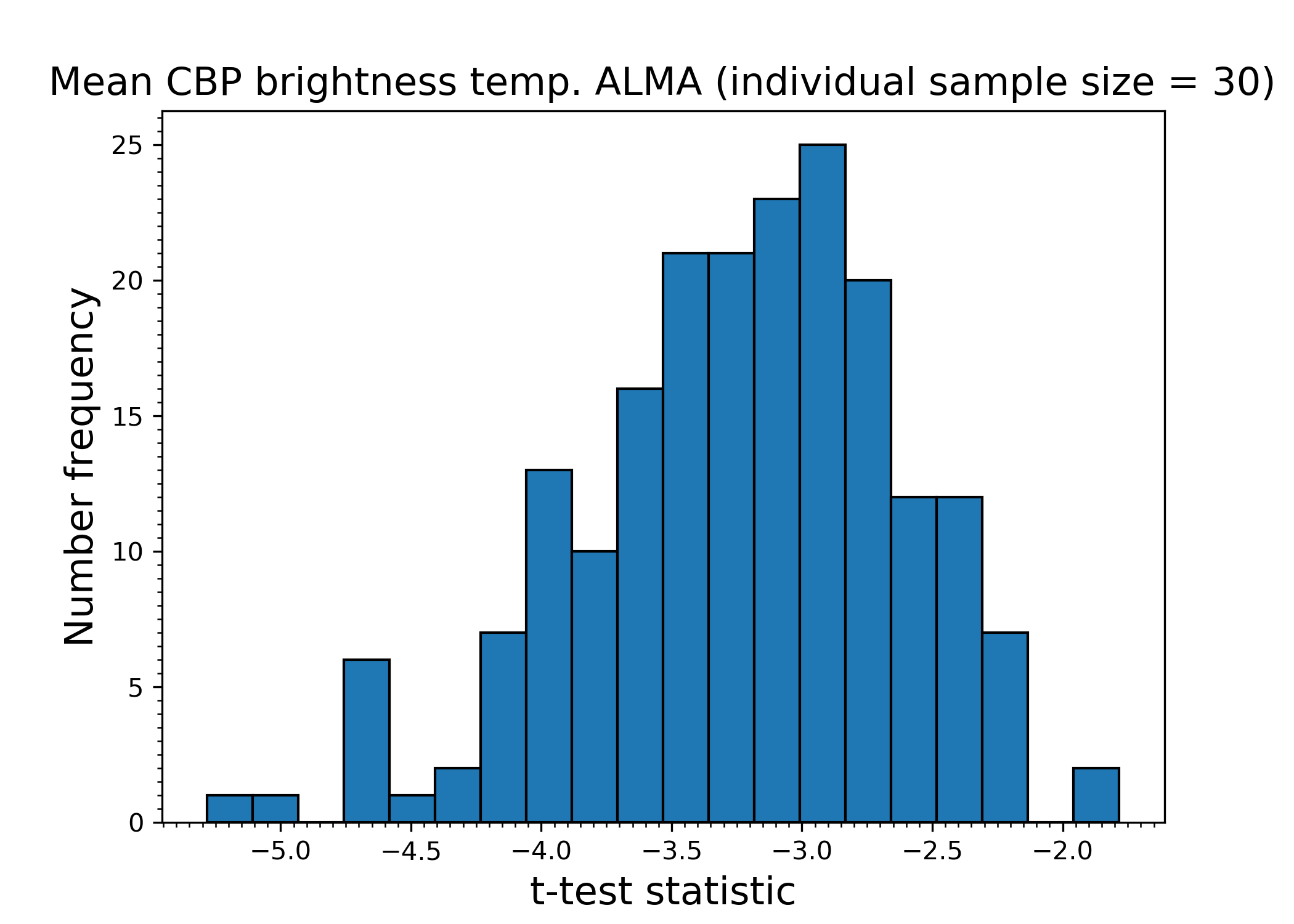}}
\subfloat{\includegraphics[width=0.36\textwidth]{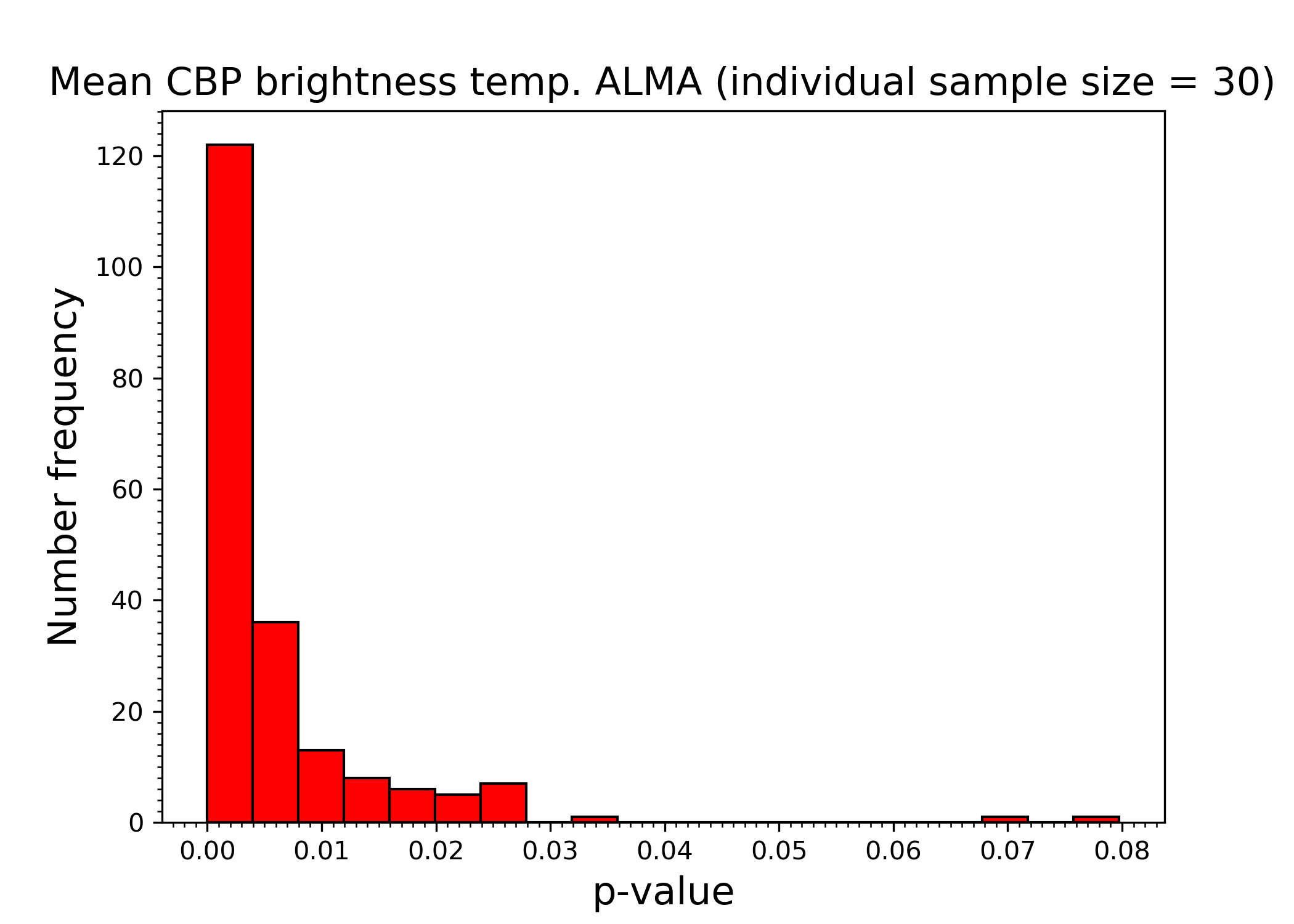}}
\caption{Same as Fig. \ref{mean_int_fig_ch1}, but for CH4, with individual CBP sample containing 30 CBPs.}
\label{mean_int_fig_ch4}
\end{figure*}

\begin{figure*}[h!]
\captionsetup[subfloat]{farskip=1pt,captionskip=1pt}
\centering
\subfloat{\includegraphics[width=0.36\textwidth]{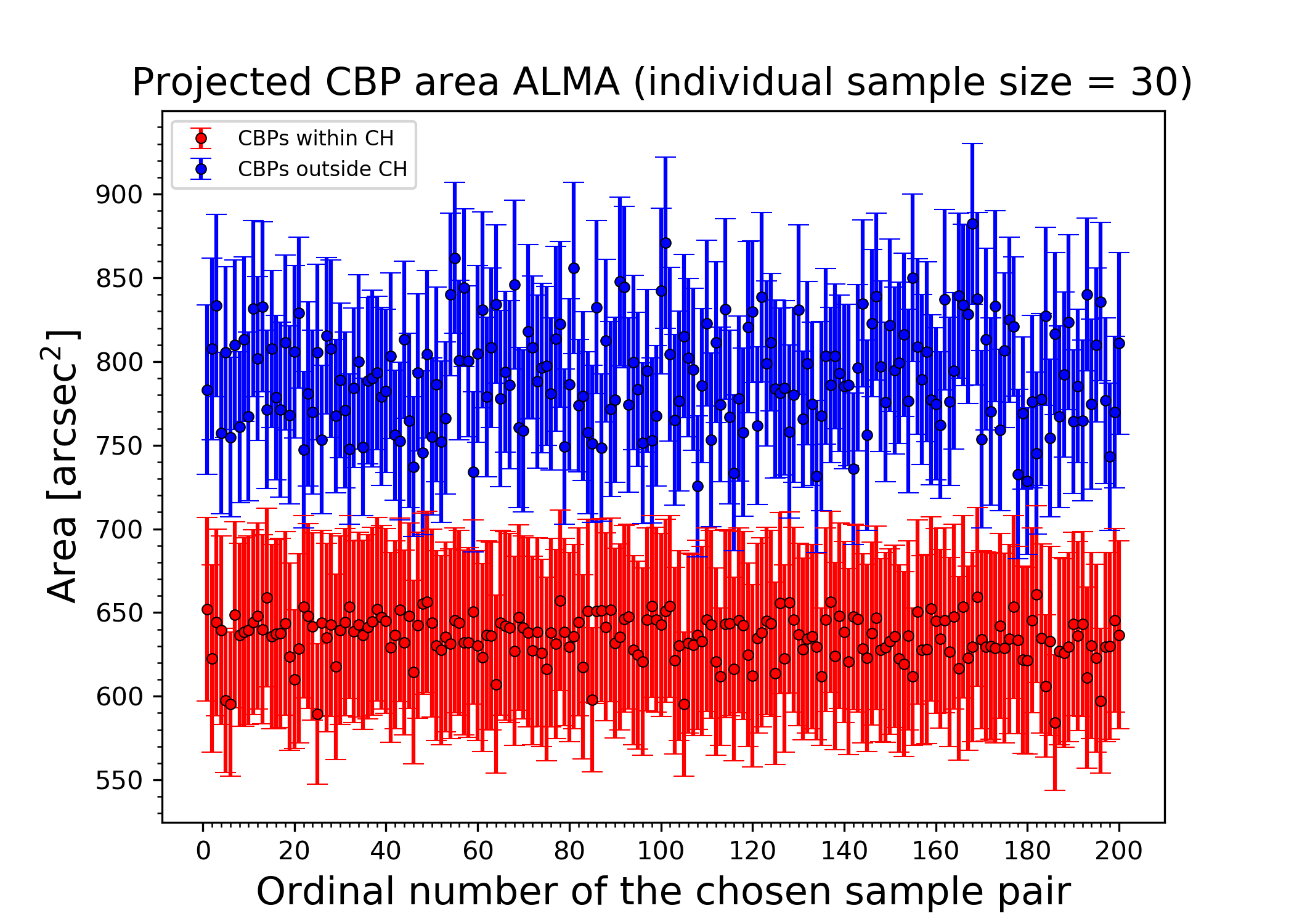}}
\subfloat{\includegraphics[width=0.36\textwidth]{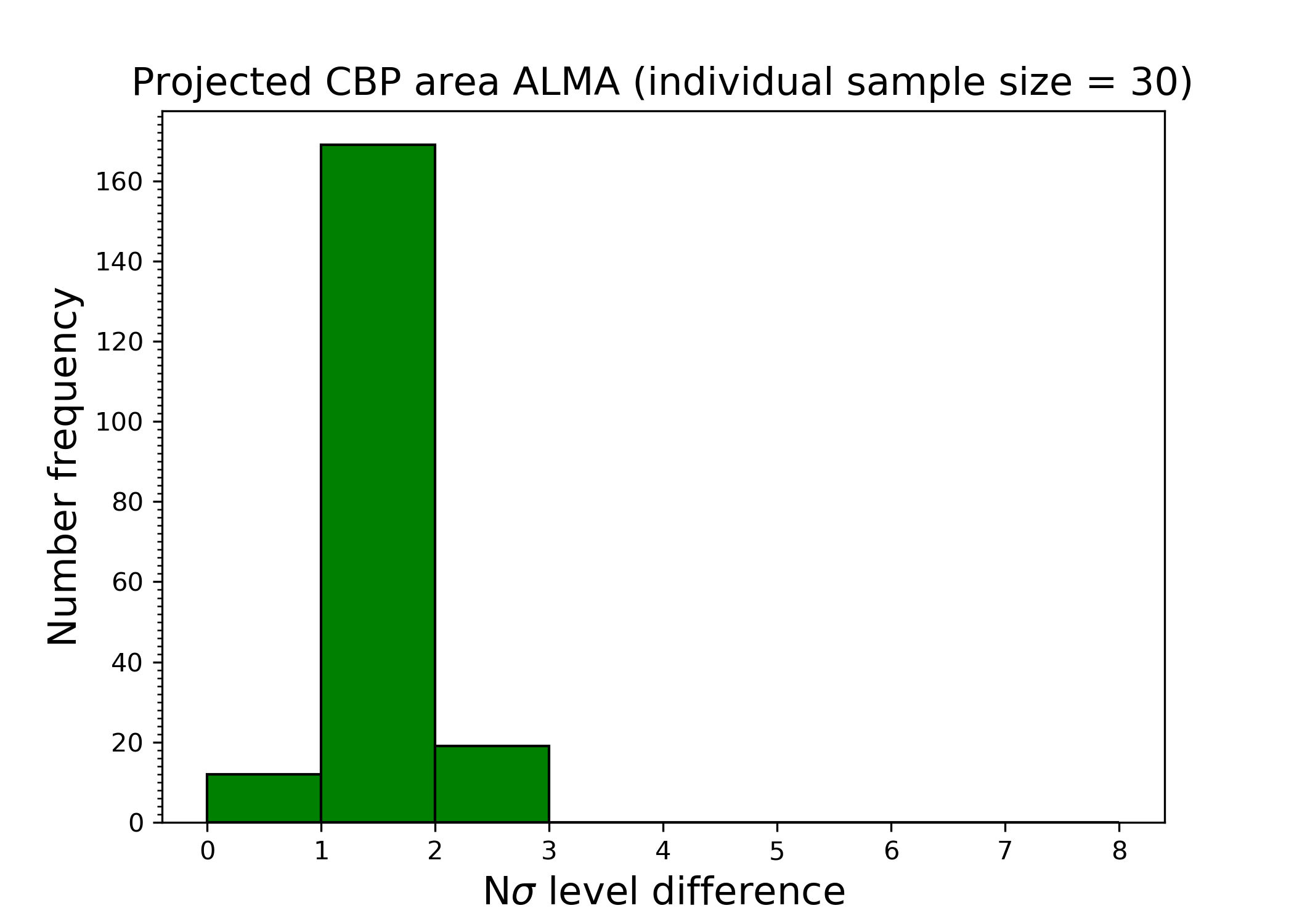}}\\
\subfloat{\includegraphics[width=0.36\textwidth]{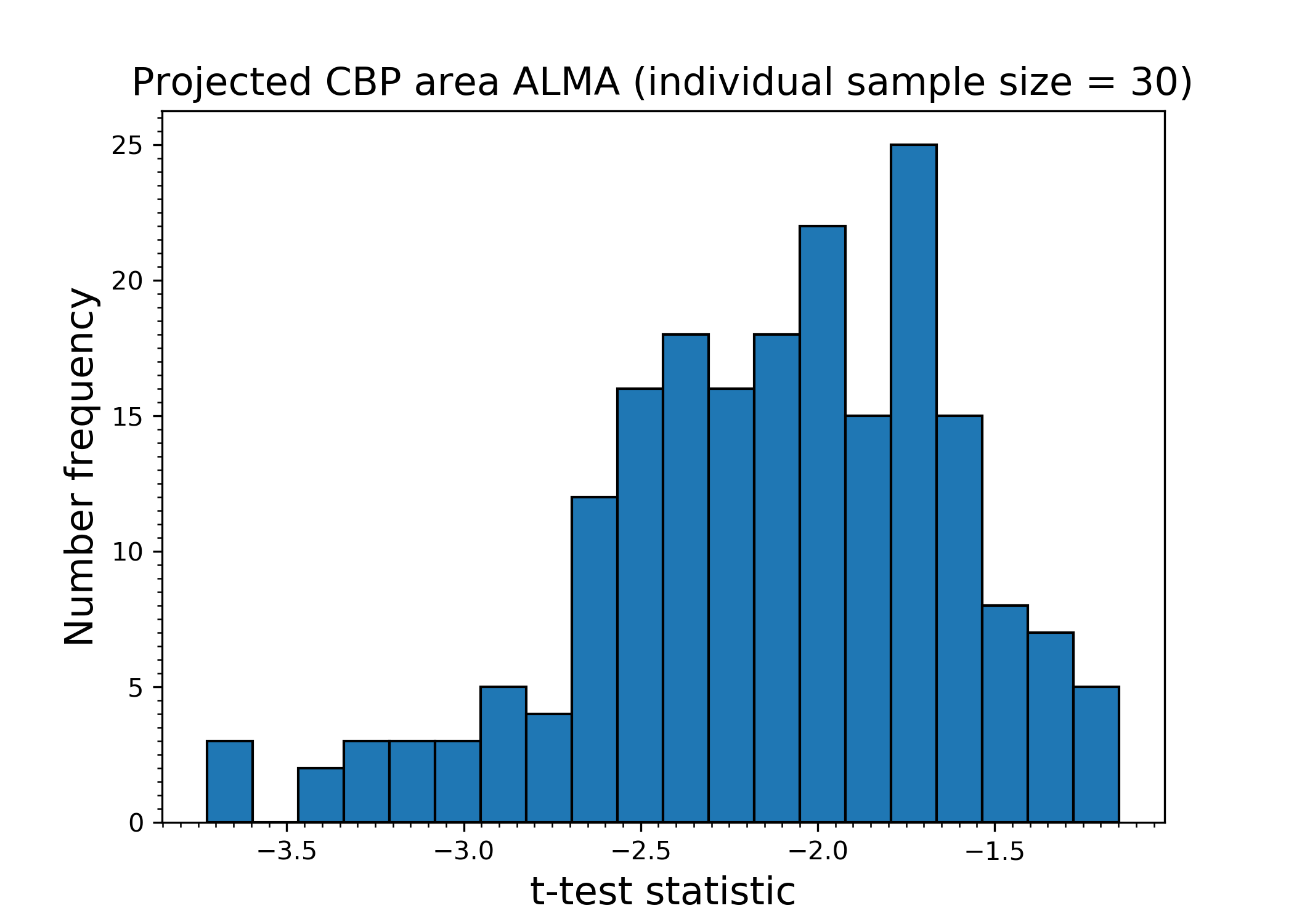}}
\subfloat{\includegraphics[width=0.36\textwidth]{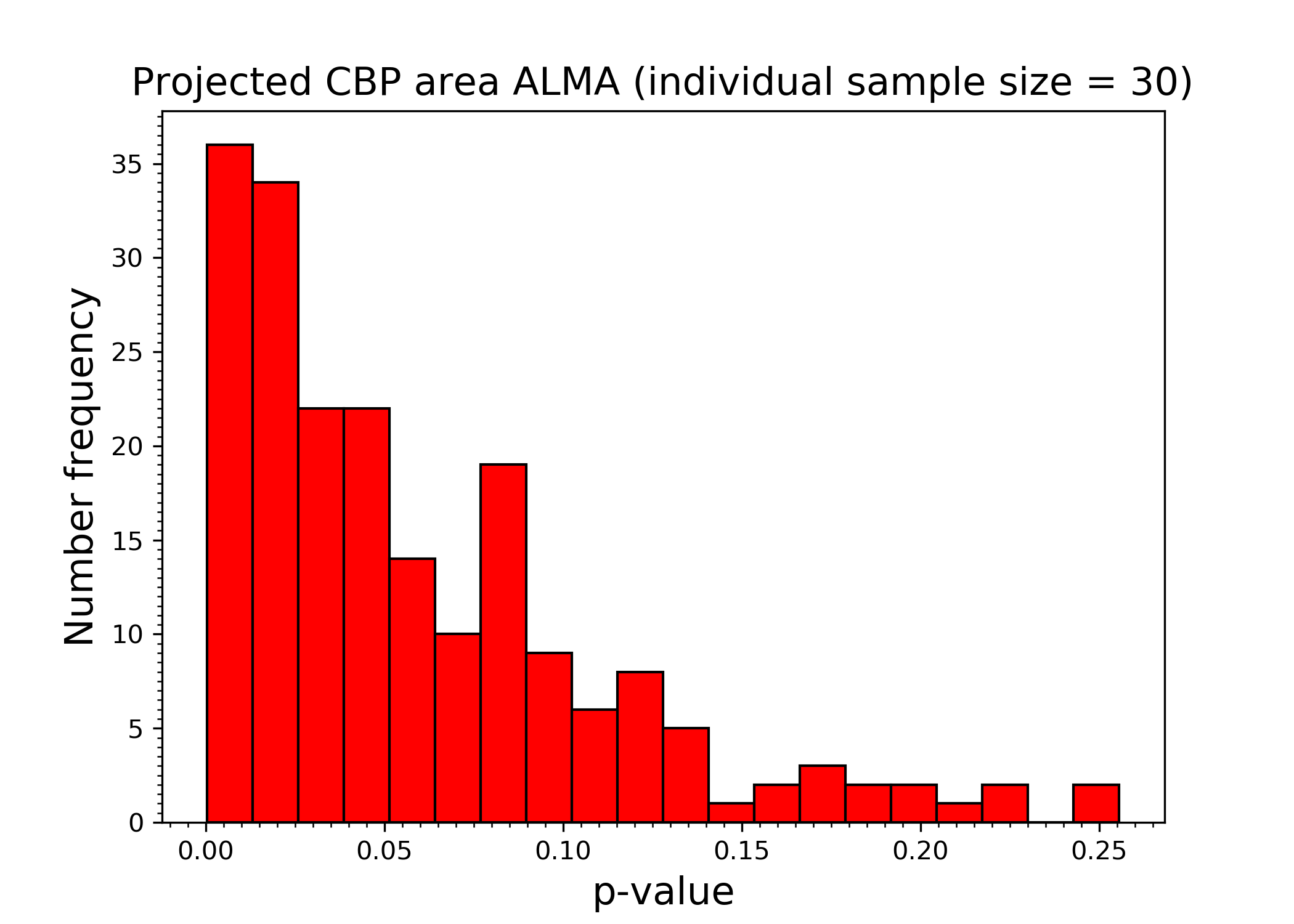}}
\caption{Same as Fig. \ref{mean_int_fig_ch1}, but for projected ALMA Band 6 CBP area and CH4, with individual CBP sample containing 30 CBPs.}
\label{area_fig_ch4}
\end{figure*}

\begin{figure*}[h!]
\captionsetup[subfloat]{farskip=1pt,captionskip=1pt}
\centering
\subfloat{\includegraphics[width=0.36\textwidth]{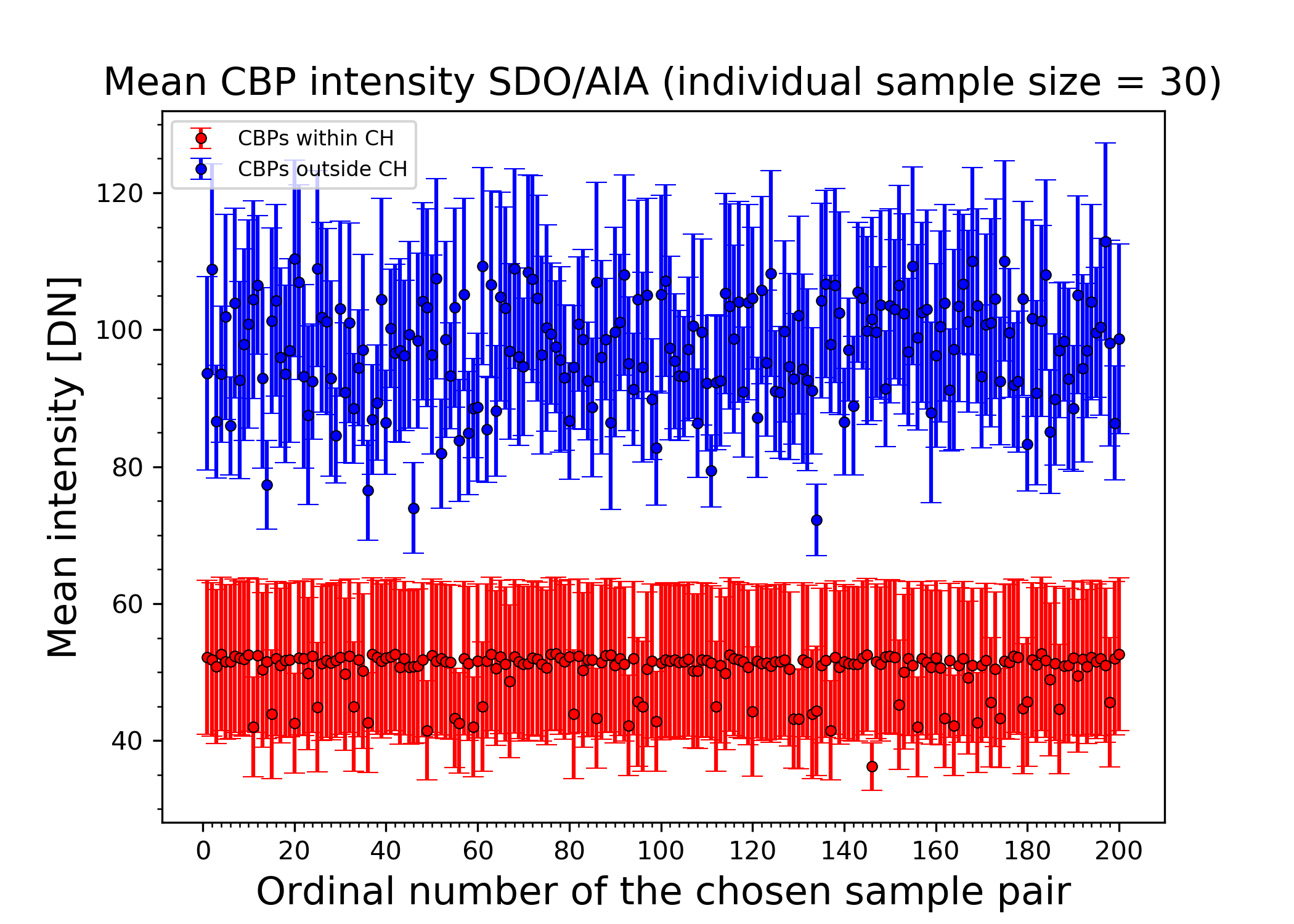}}
\subfloat{\includegraphics[width=0.36\textwidth]{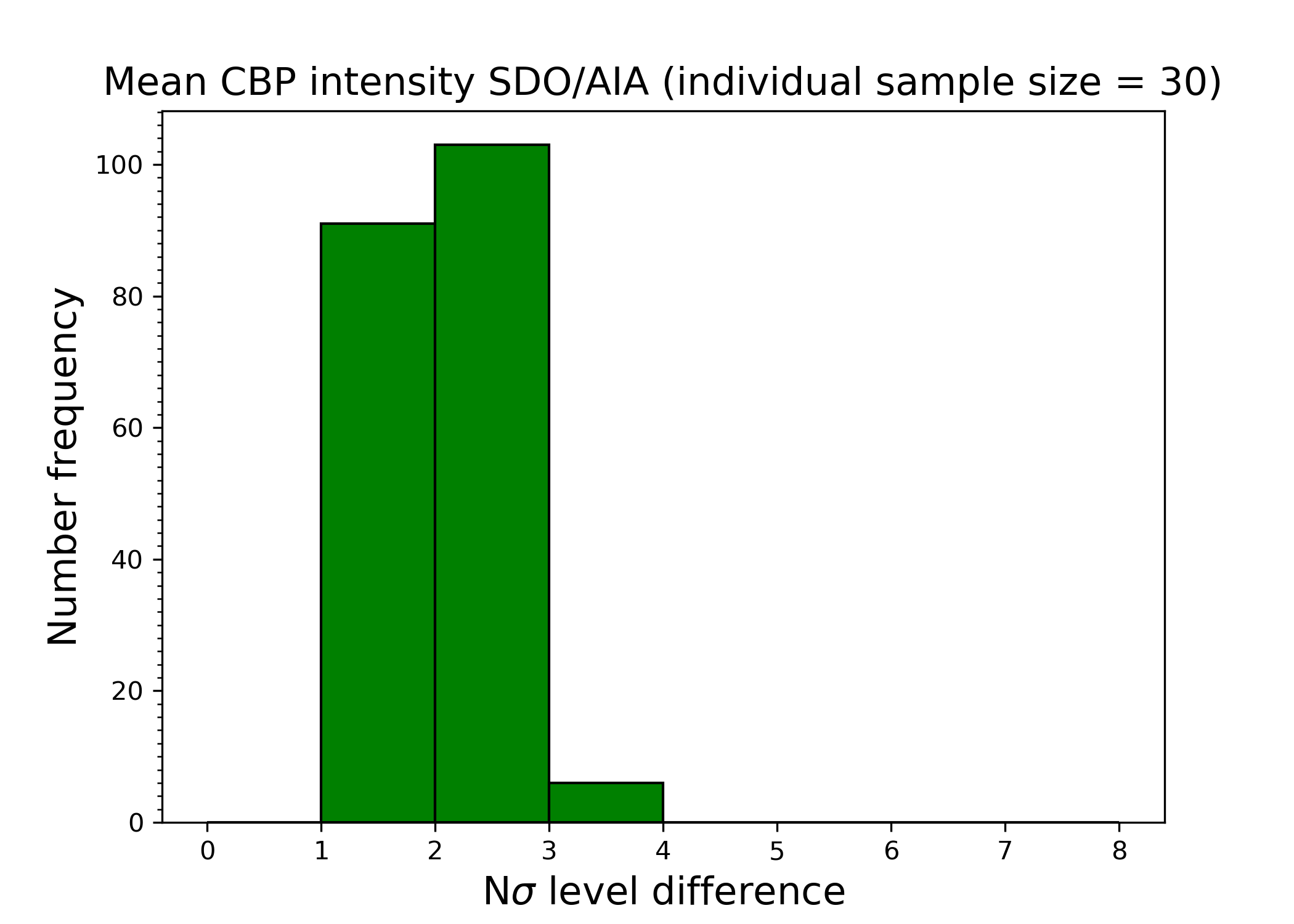}}\\
\subfloat{\includegraphics[width=0.36\textwidth]{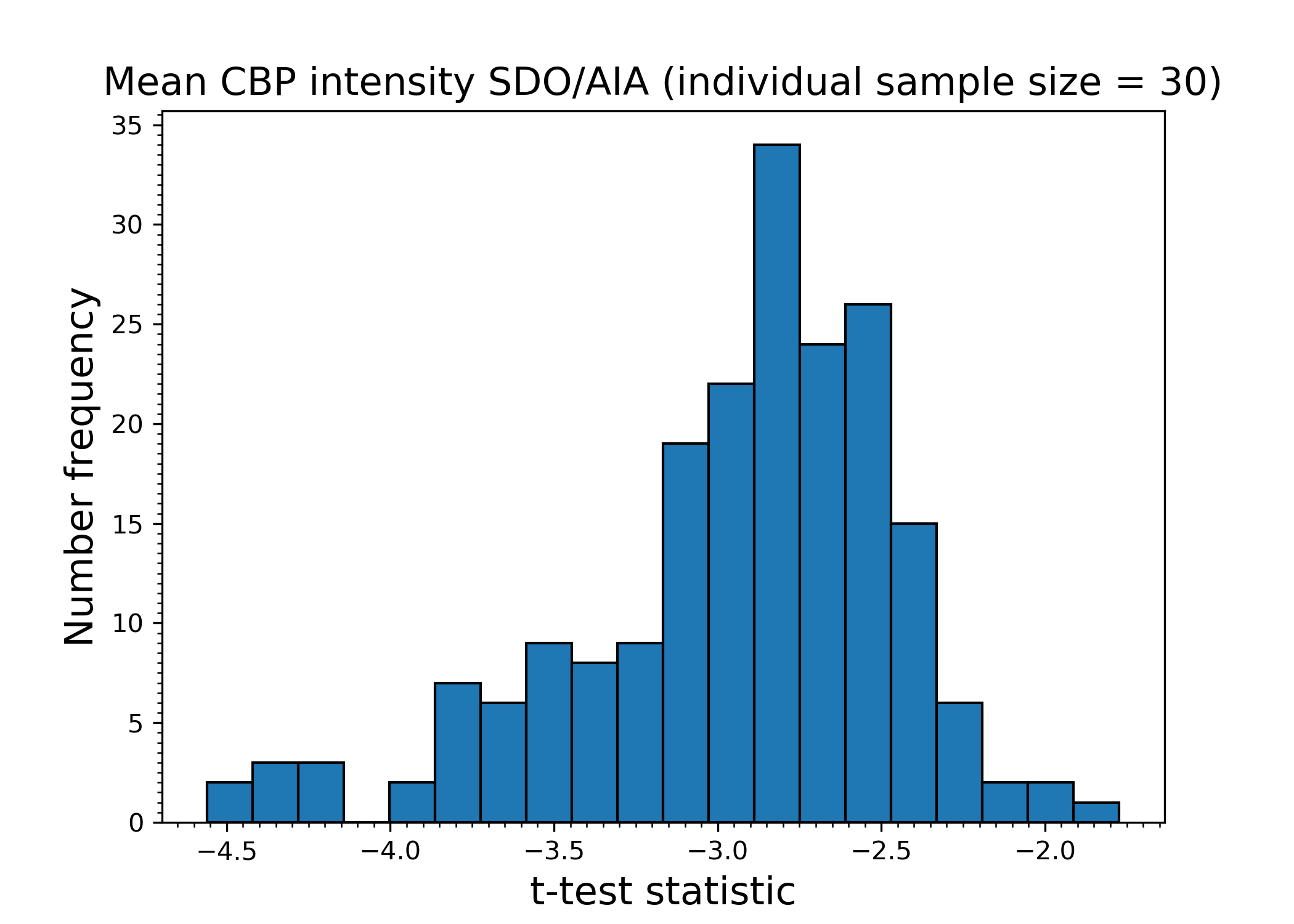}}
\subfloat{\includegraphics[width=0.36\textwidth]{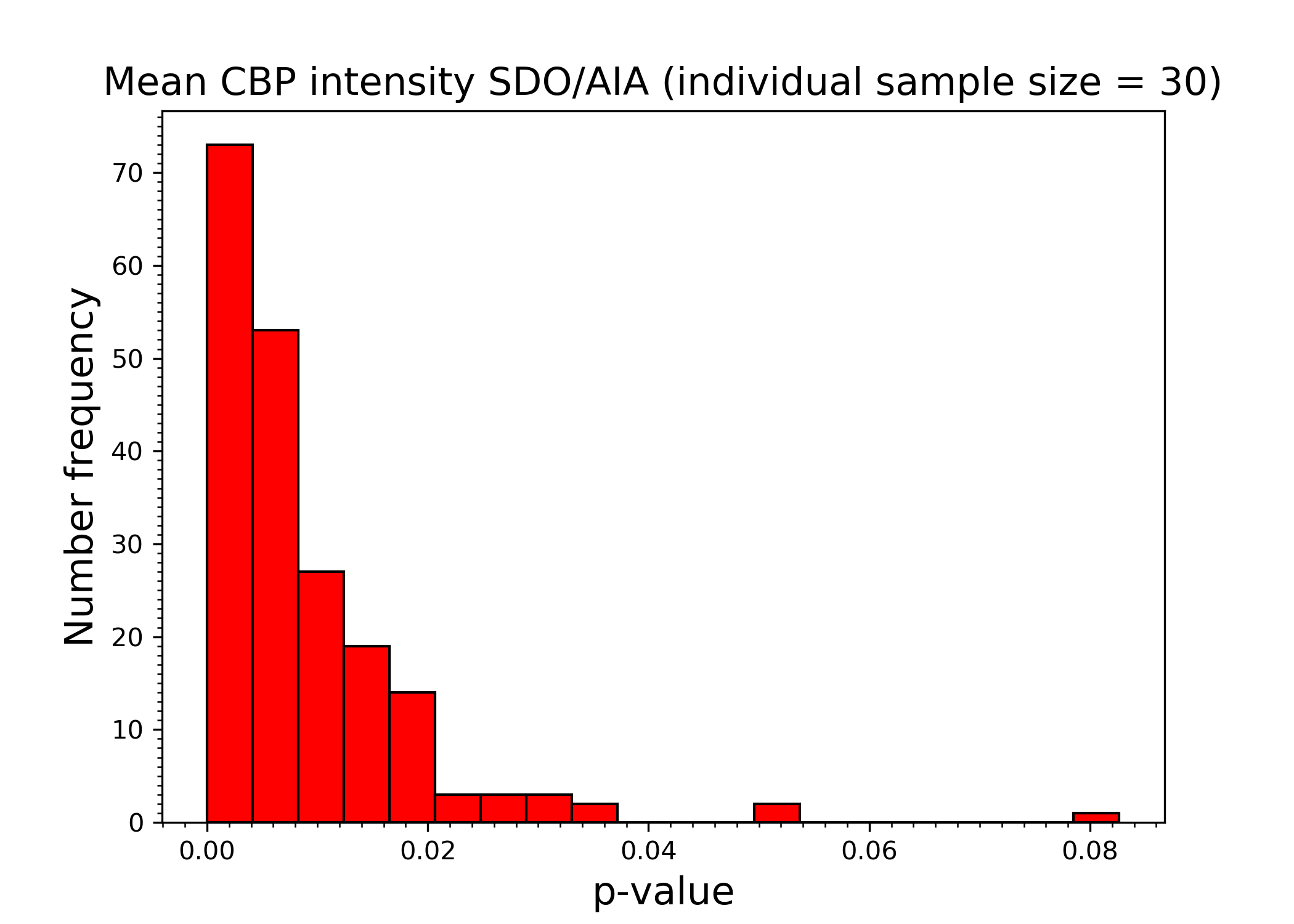}}
\caption{Same as Fig. \ref{mean_int_fig_ch1}, but for mean SDO/AIA 193 \AA\space CBP intensity and CH4, with individual CBP sample containing 30 CBPs.}
\label{mean_int_fig_ch4_1}
\end{figure*}

\begin{figure*}[h!]
\captionsetup[subfloat]{farskip=1pt,captionskip=1pt}
\centering
\subfloat{\includegraphics[width=0.36\textwidth]{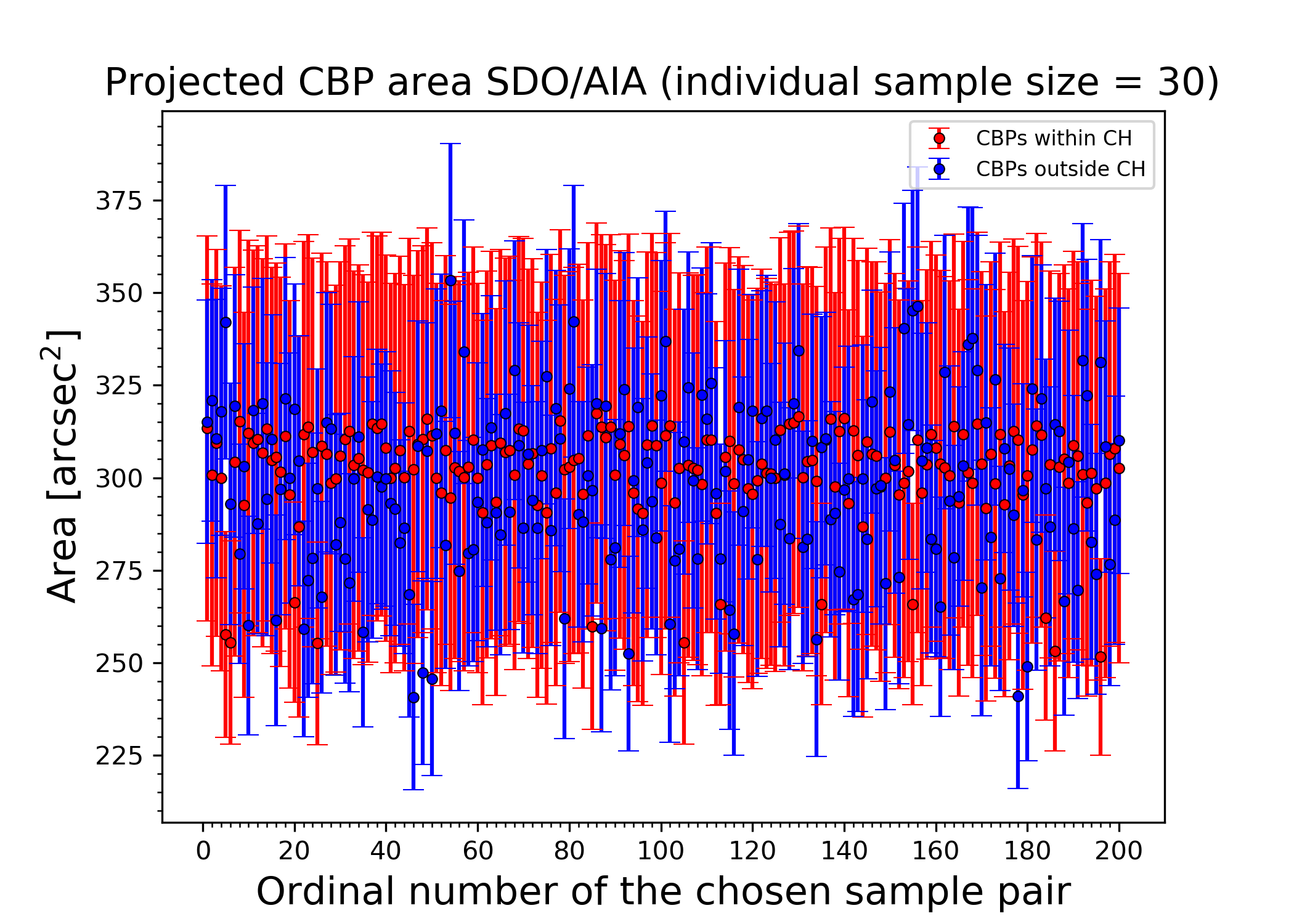}}
\subfloat{\includegraphics[width=0.36\textwidth]{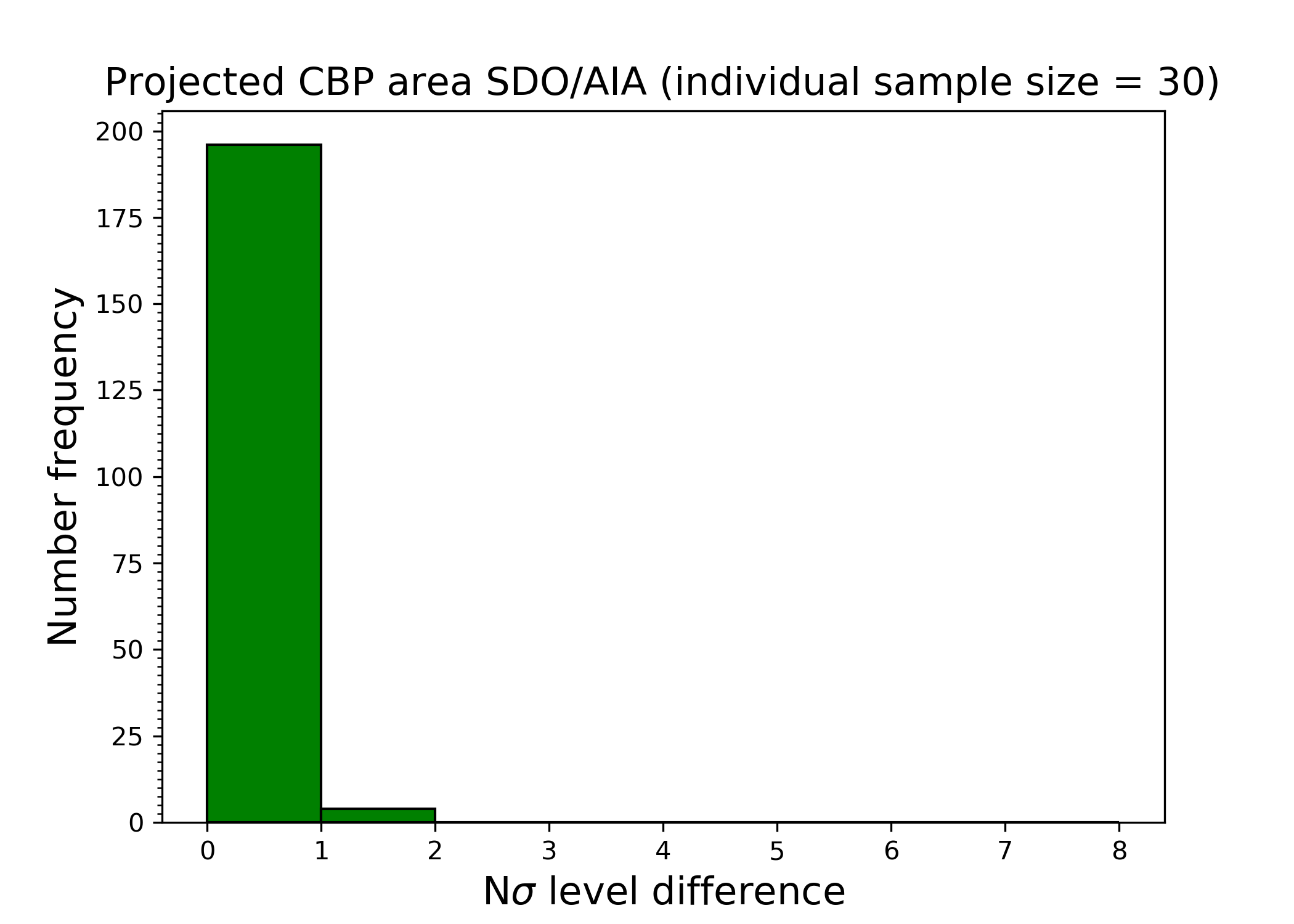}}\\
\subfloat{\includegraphics[width=0.36\textwidth]{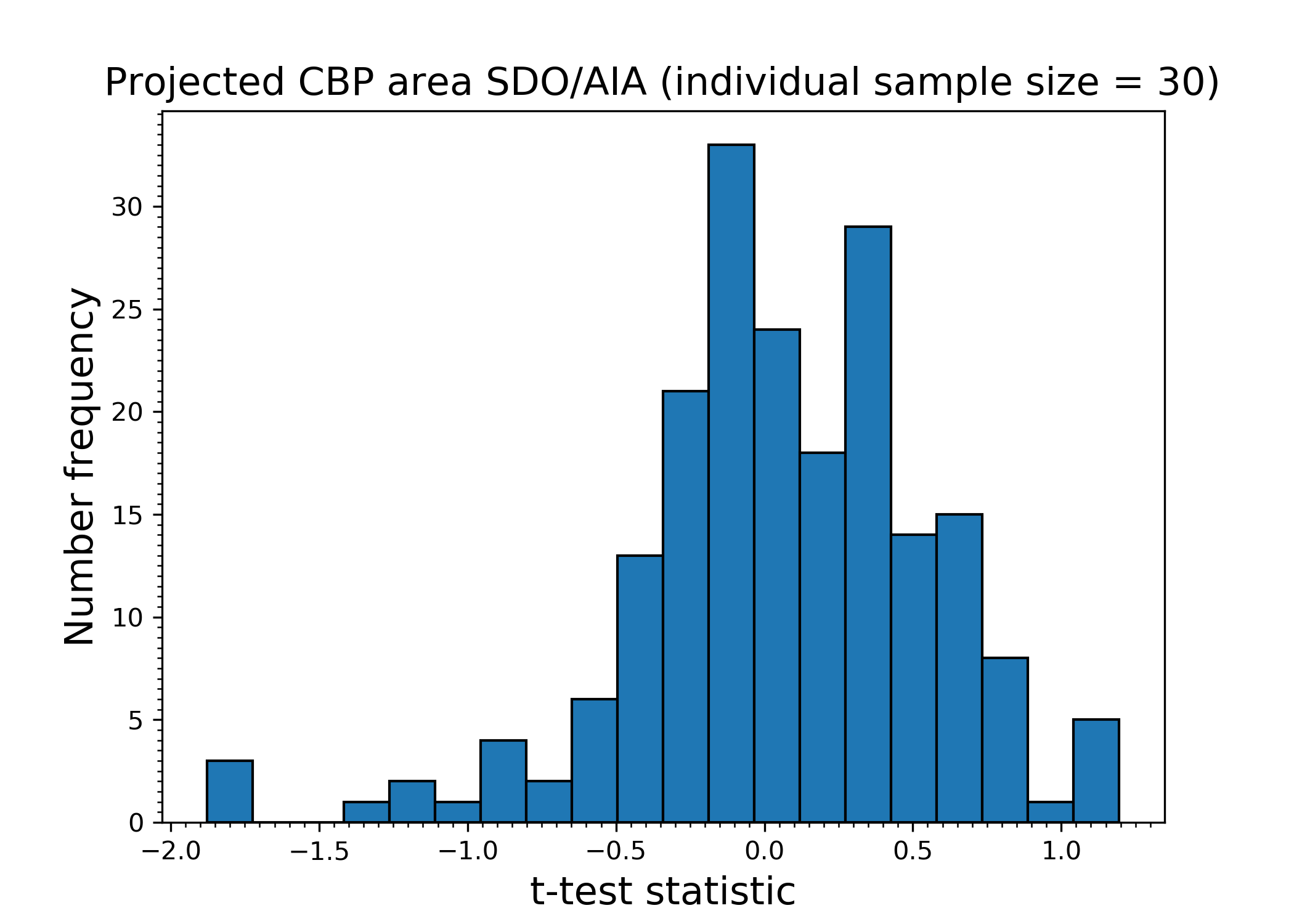}}
\subfloat{\includegraphics[width=0.36\textwidth]{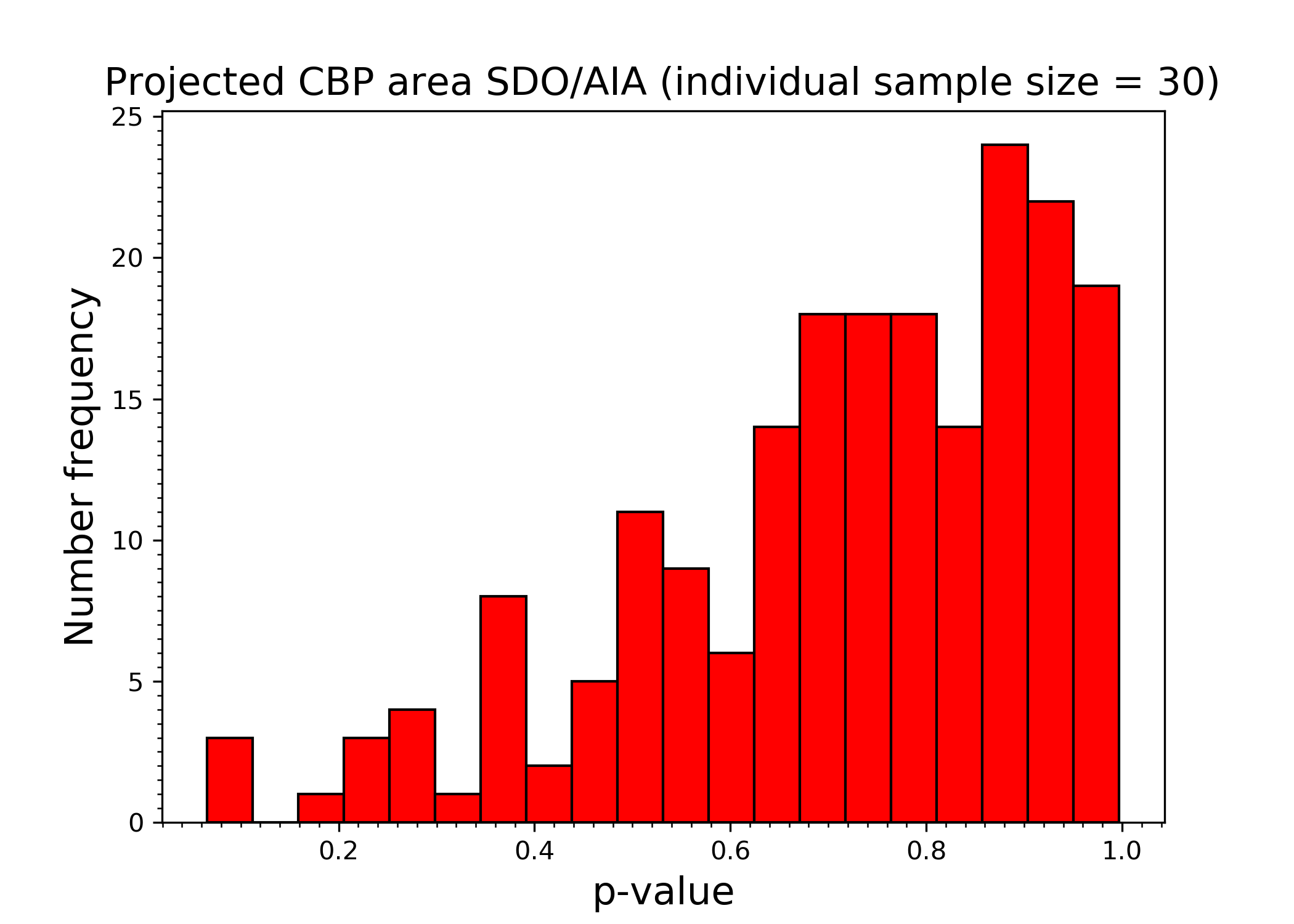}}
\caption{Same as Fig. \ref{mean_int_fig_ch1}, but for projected SDO/AIA 193 \AA\space CBP area and CH4, with individual CBP sample containing 30 CBPs.}
\label{area_fig_ch4_1}
\end{figure*}

\clearpage
\section{Statistical analysis of the physical properties for CBPs within and outside CH5}
\label{CH5}
\begin{figure*}[h!]
\captionsetup[subfloat]{farskip=1pt,captionskip=1pt}
\centering
\subfloat{\includegraphics[width=0.36\textwidth]{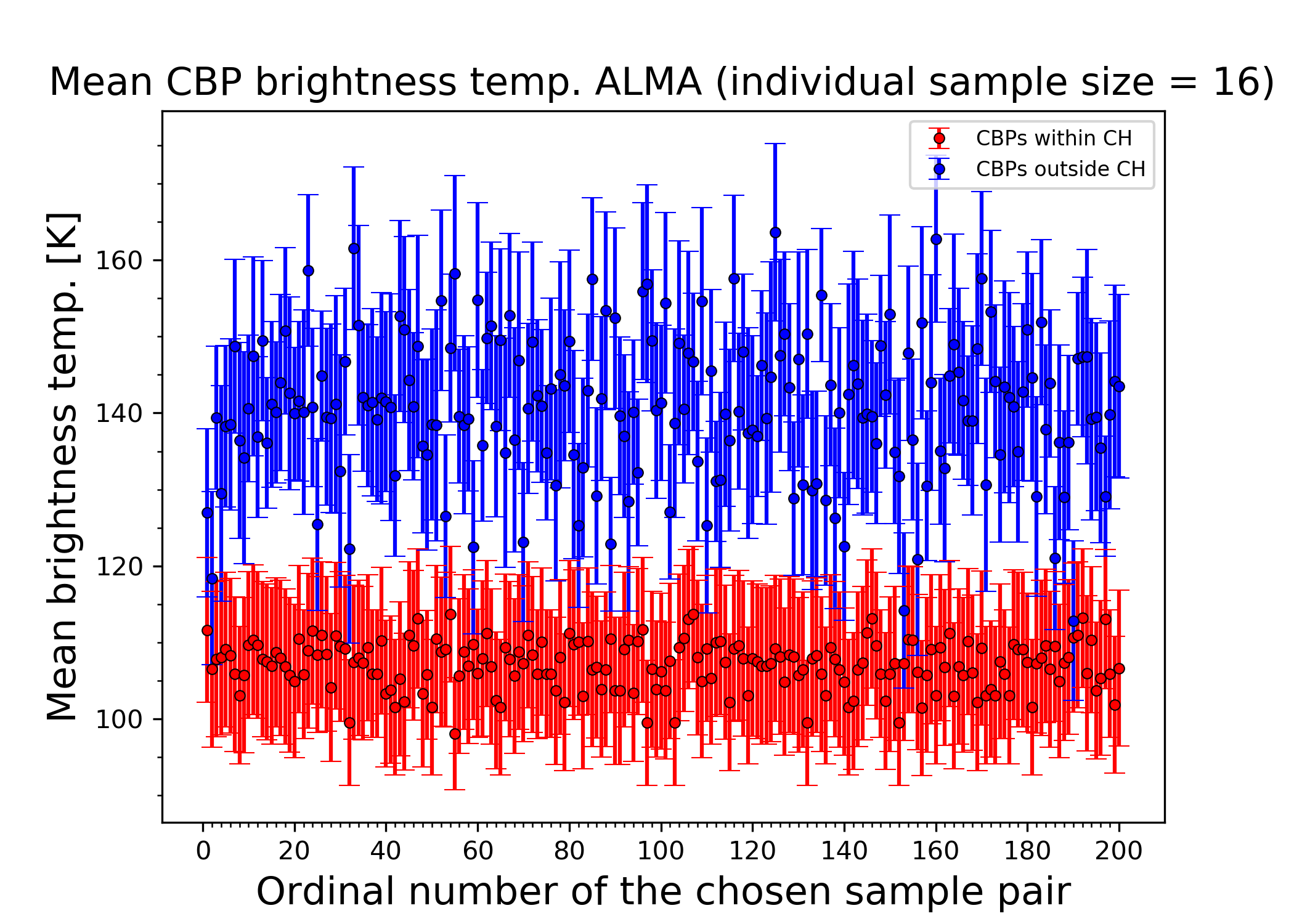}}
\subfloat{\includegraphics[width=0.36\textwidth]{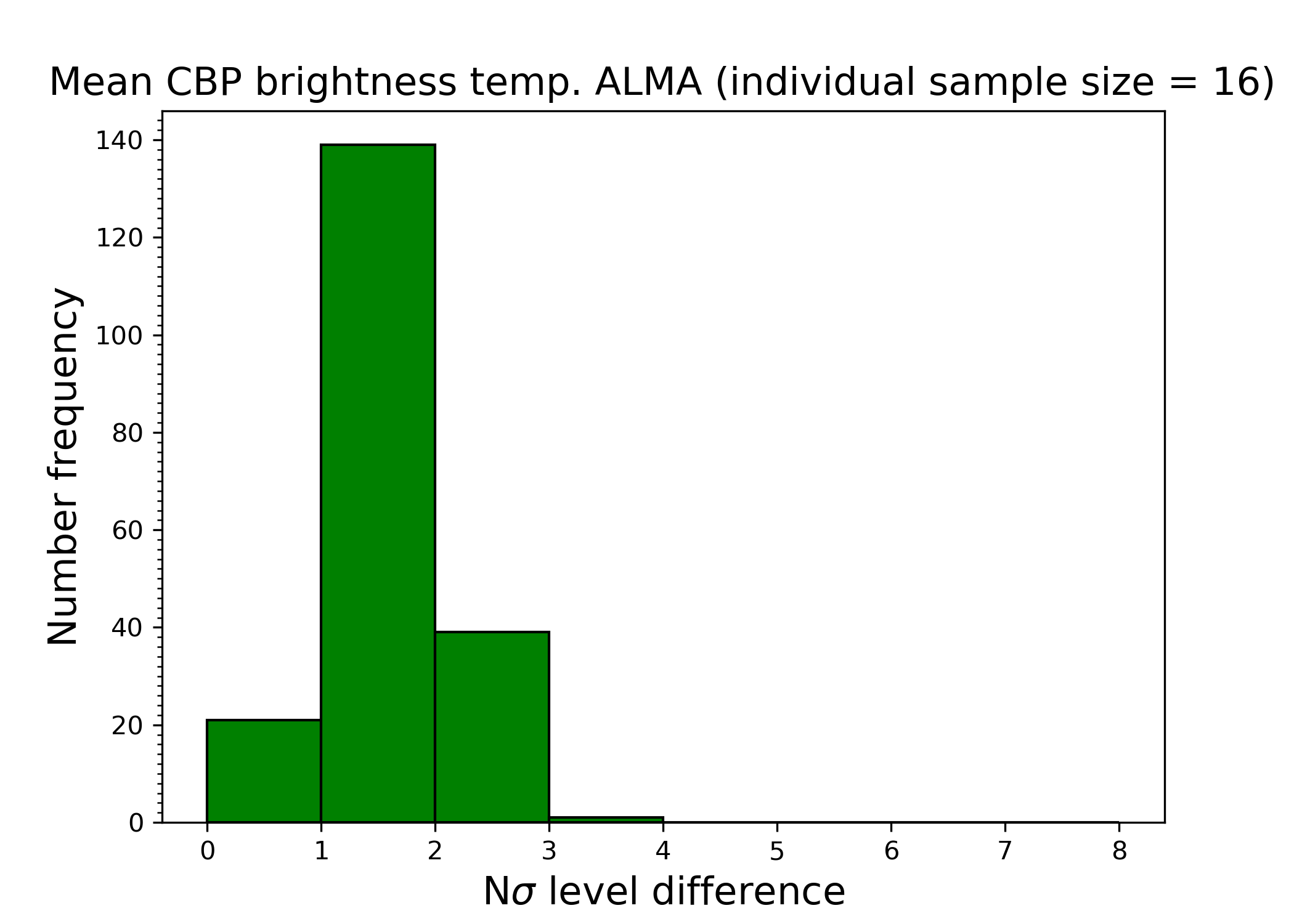}}\\
\subfloat{\includegraphics[width=0.36\textwidth]{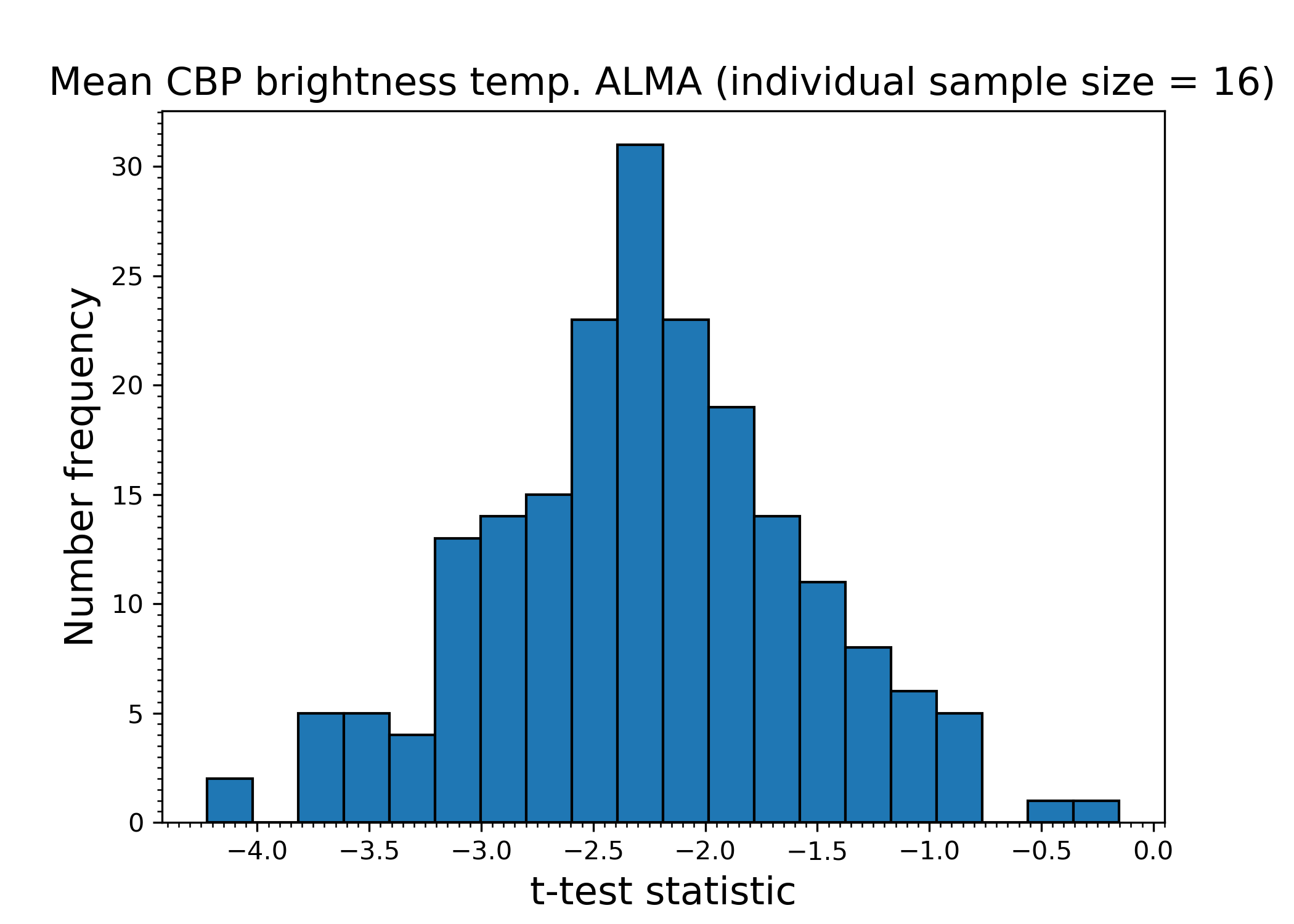}}
\subfloat{\includegraphics[width=0.36\textwidth]{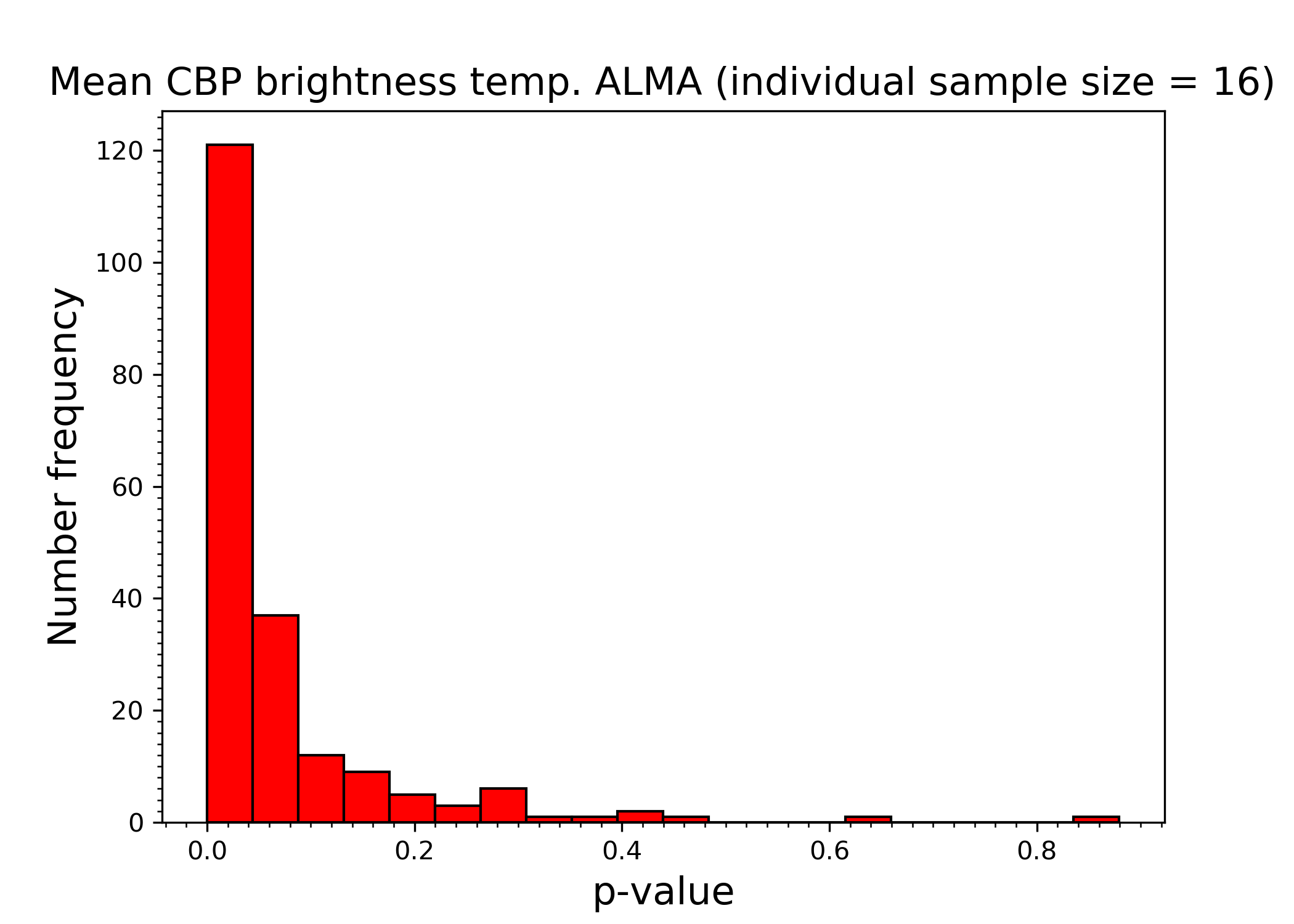}}
\caption{Same as Fig. \ref{mean_int_fig_ch1}, but for CH5, with individual CBP sample containing 16 CBPs.}
\label{mean_int_fig_ch5}
\end{figure*}

\begin{figure*}[h!]
\captionsetup[subfloat]{farskip=1pt,captionskip=1pt}
\centering
\subfloat{\includegraphics[width=0.36\textwidth]{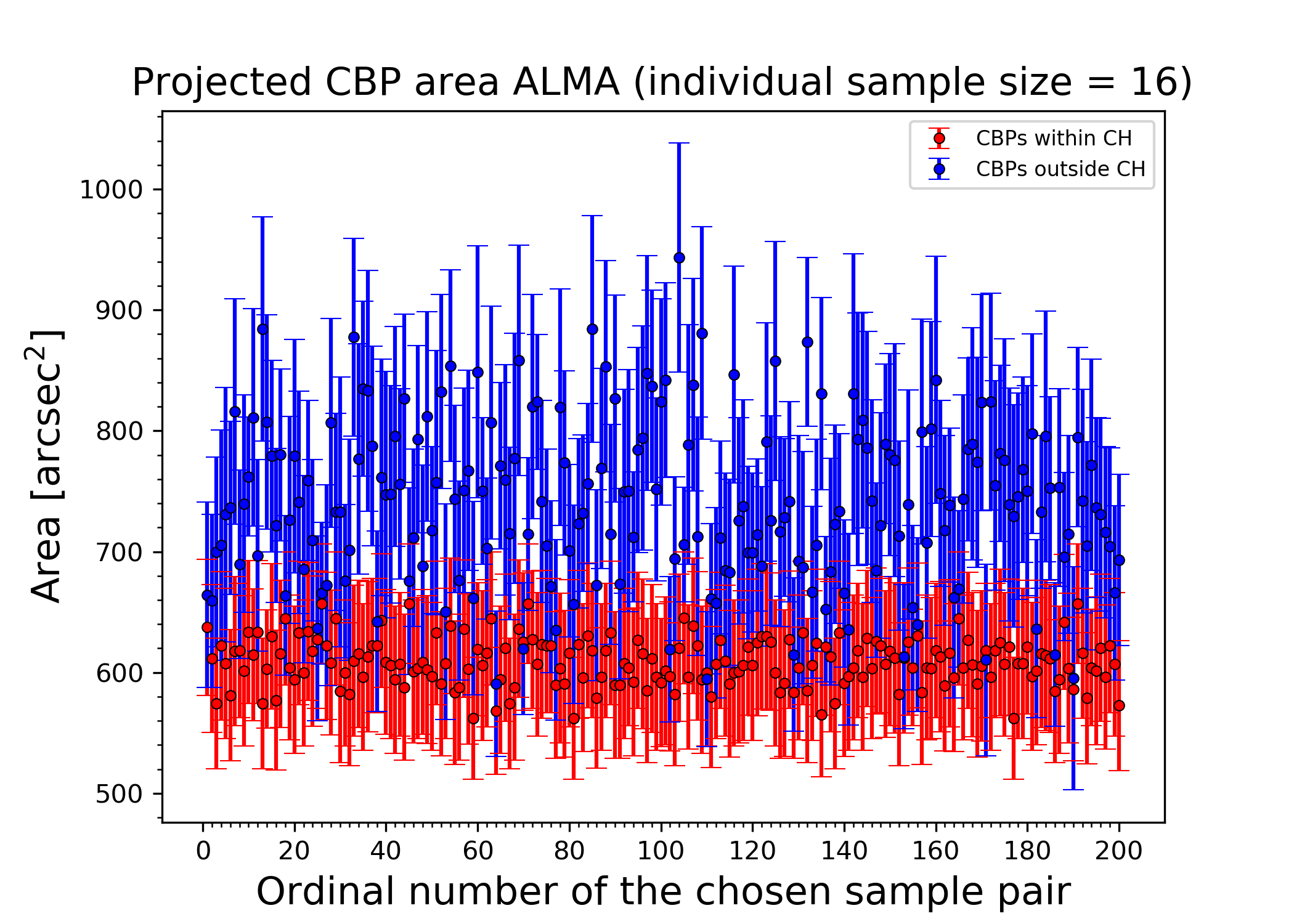}}
\subfloat{\includegraphics[width=0.36\textwidth]{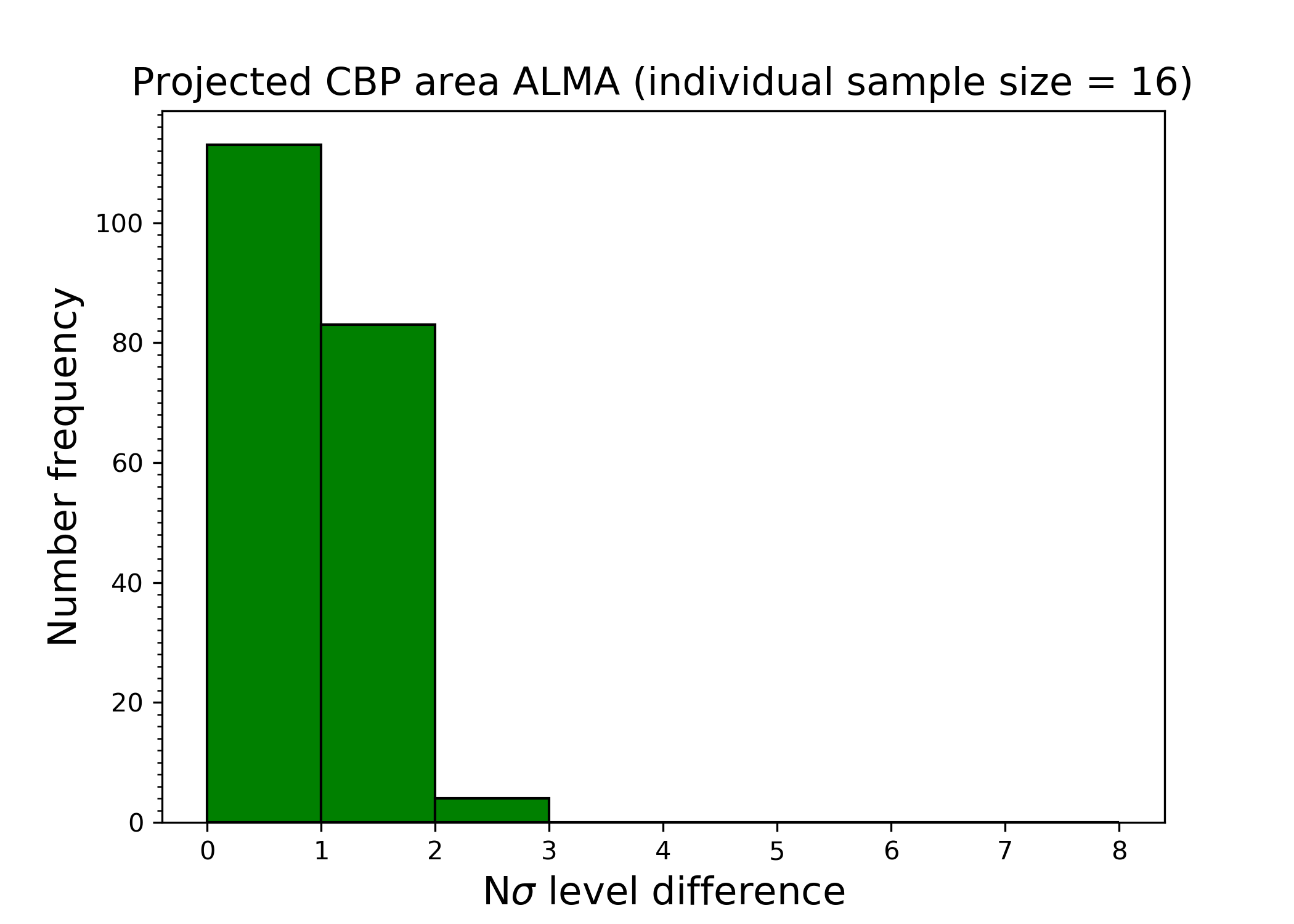}}\\
\subfloat{\includegraphics[width=0.36\textwidth]{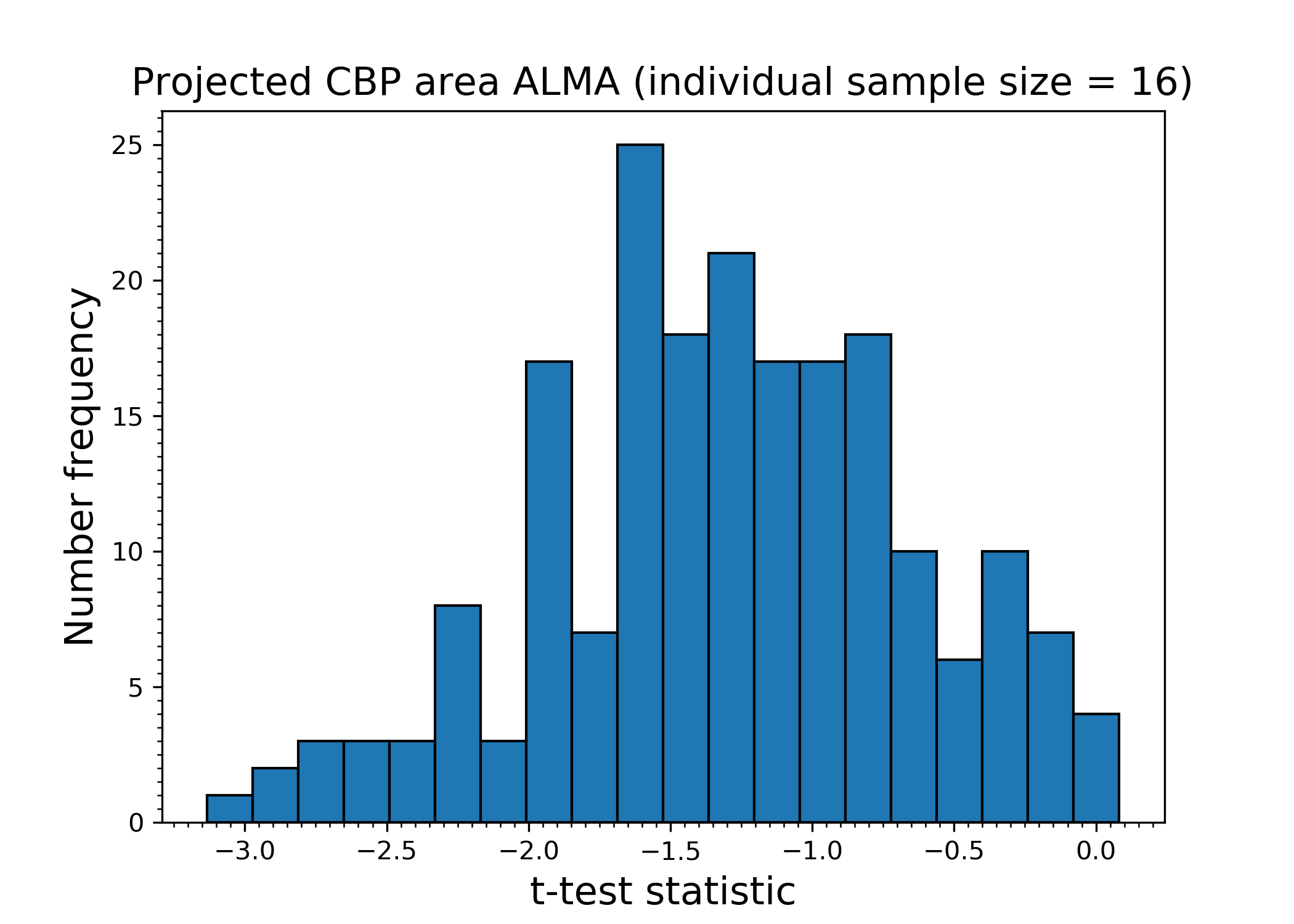}}
\subfloat{\includegraphics[width=0.36\textwidth]{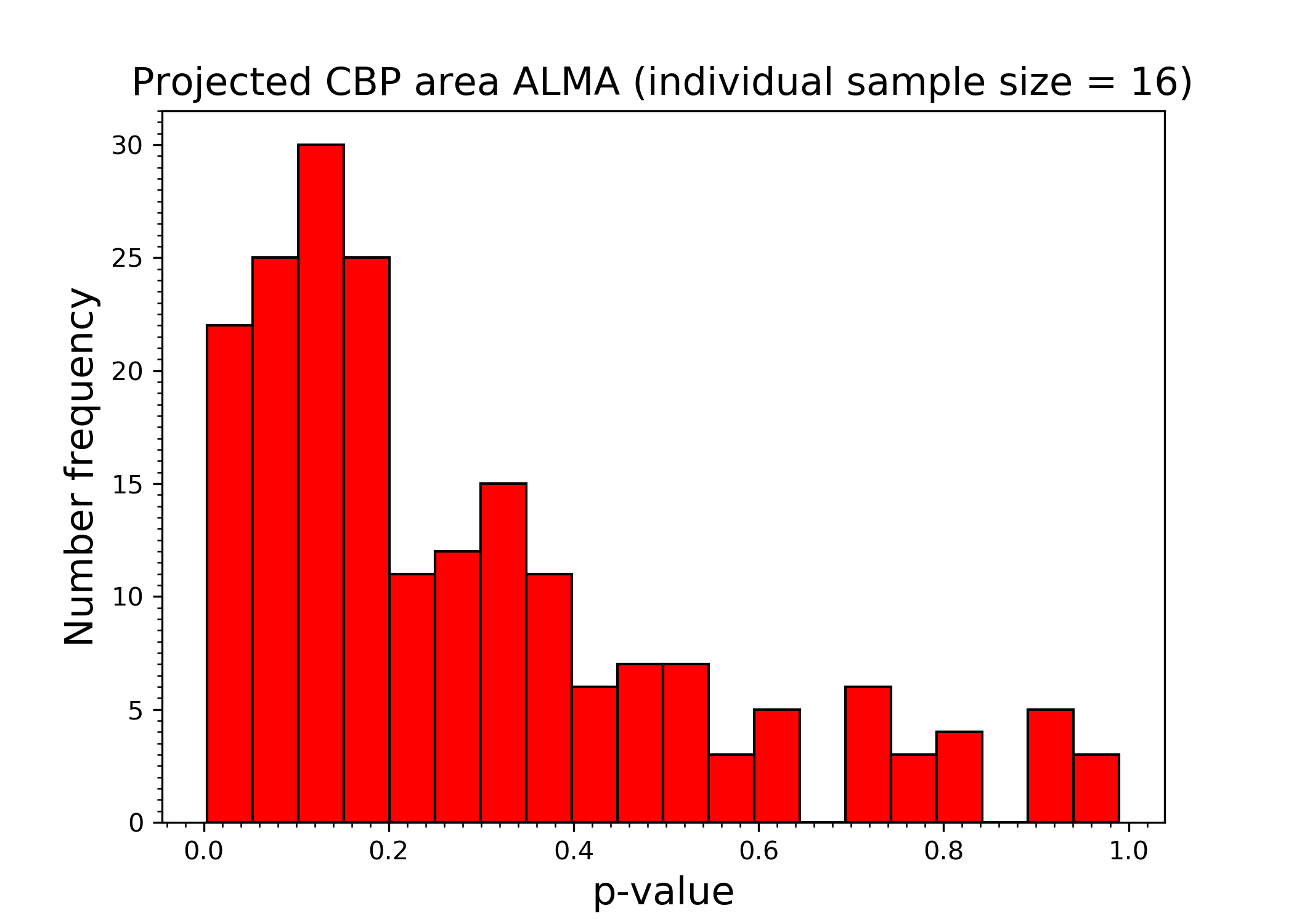}}
\caption{Same as Fig. \ref{mean_int_fig_ch1}, but for projected ALMA Band 6 CBP area and CH5, with individual CBP sample containing 16 CBPs.}
\label{area_fig_ch5}
\end{figure*}

\begin{figure*}[h!]
\captionsetup[subfloat]{farskip=1pt,captionskip=1pt}
\centering
\subfloat{\includegraphics[width=0.36\textwidth]{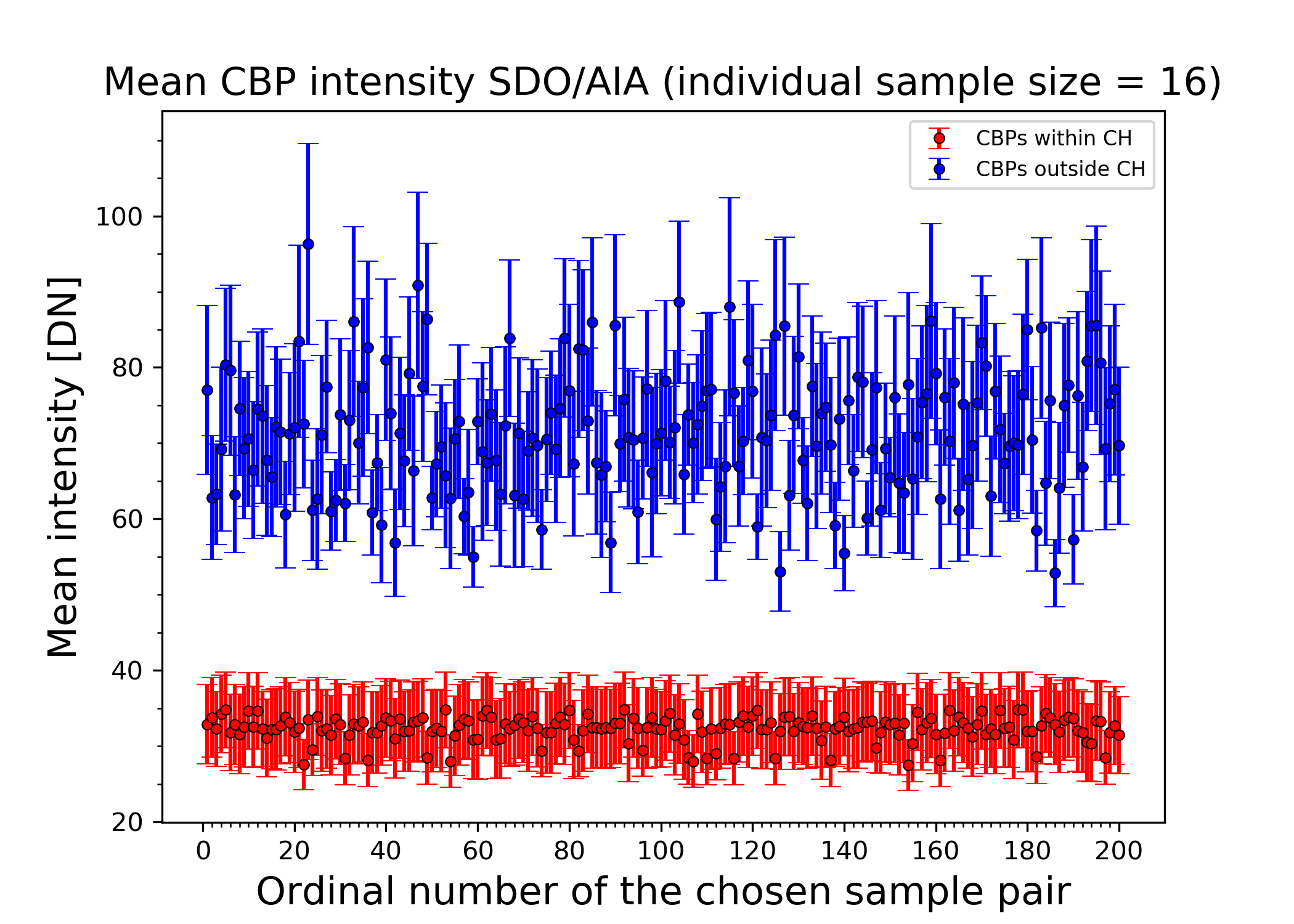}}
\subfloat{\includegraphics[width=0.36\textwidth]{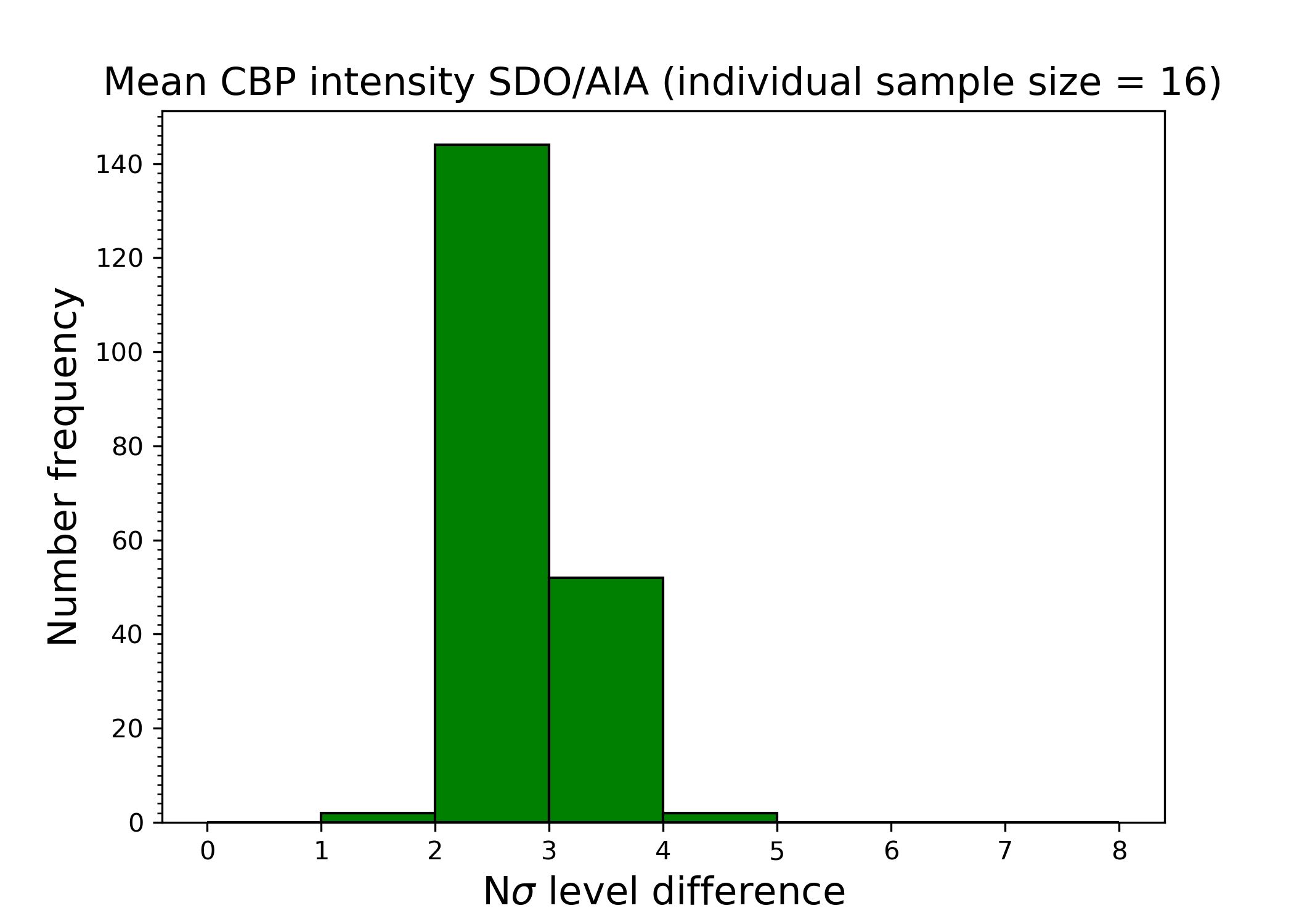}}\\
\subfloat{\includegraphics[width=0.36\textwidth]{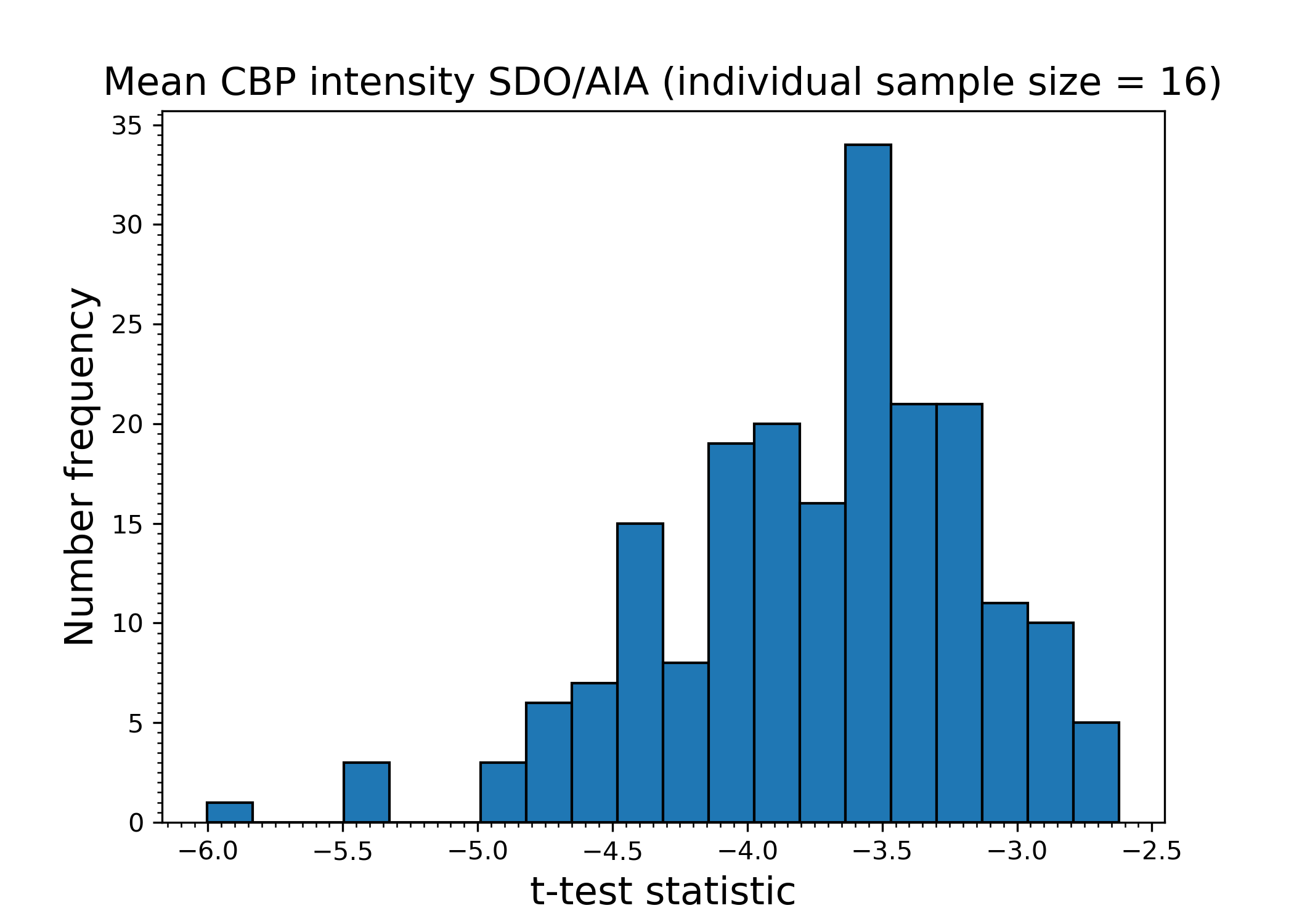}}
\subfloat{\includegraphics[width=0.36\textwidth]{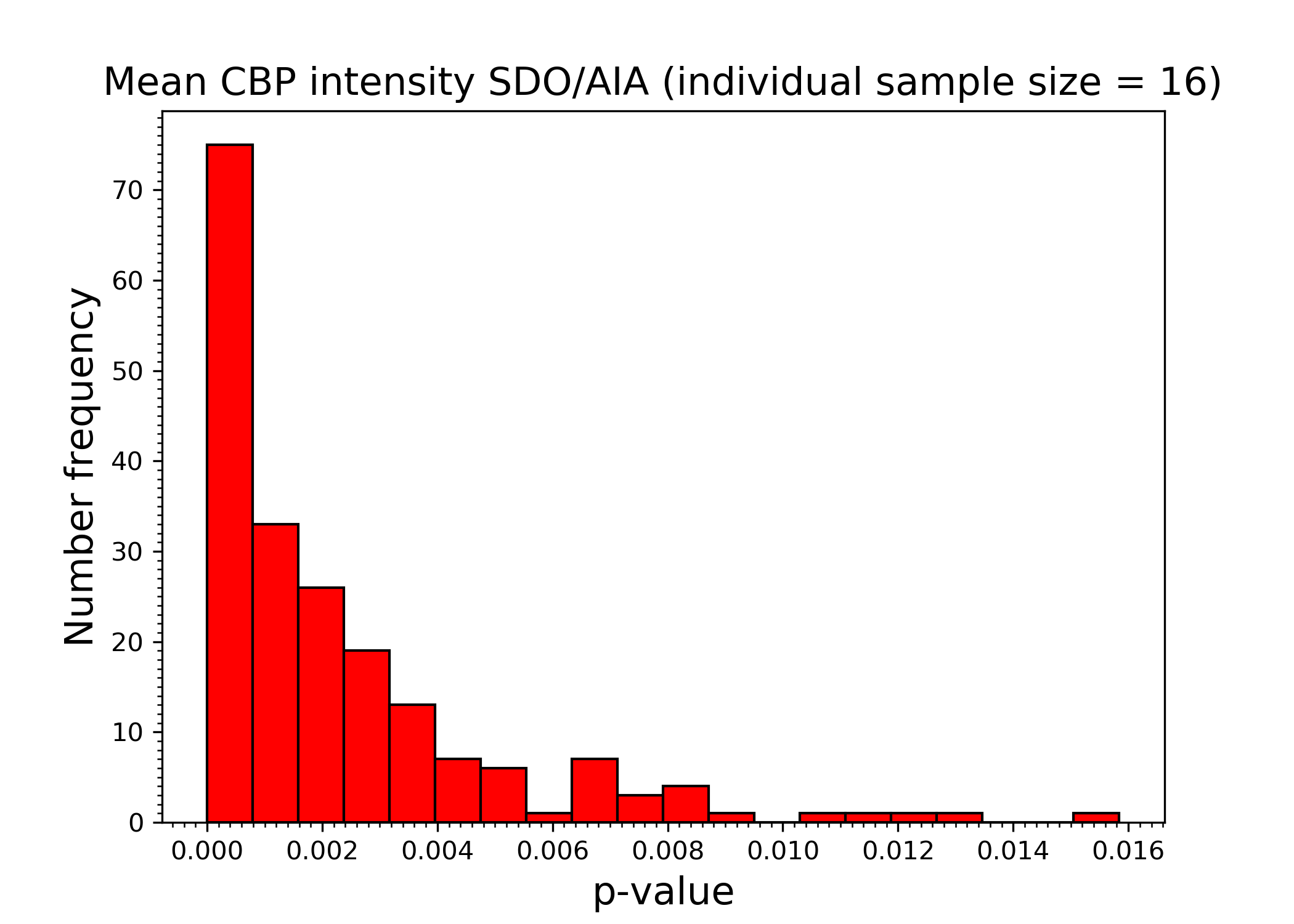}}
\caption{Same as Fig. \ref{mean_int_fig_ch1}, but for mean SDO/AIA 193 \AA\space CBP intensity and CH5, with individual CBP sample containing 16 CBPs.}
\label{mean_int_fig_ch5_1}
\end{figure*}

\begin{figure*}[h!]
\captionsetup[subfloat]{farskip=1pt,captionskip=1pt}
\centering
\subfloat{\includegraphics[width=0.36\textwidth]{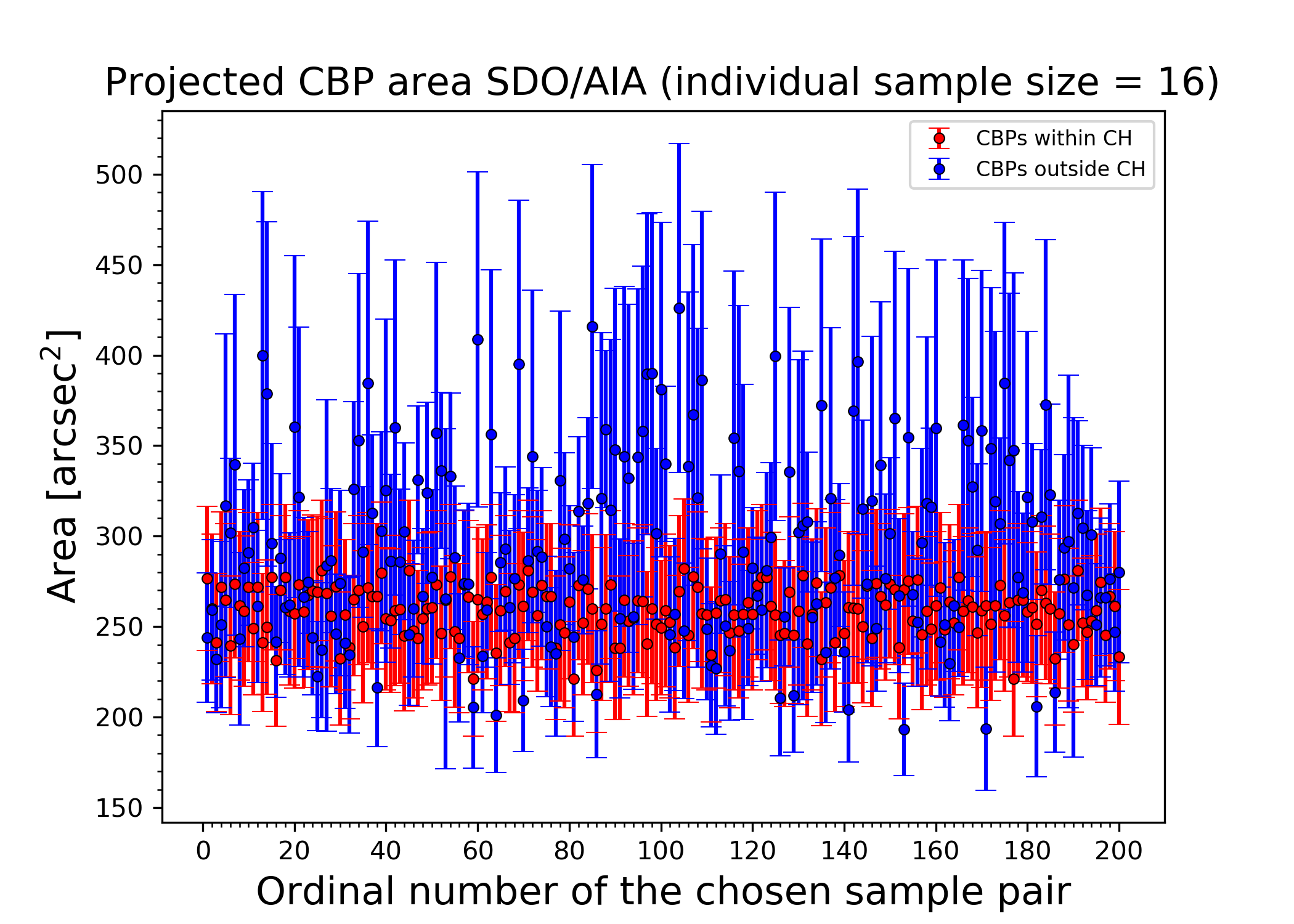}}
\subfloat{\includegraphics[width=0.36\textwidth]{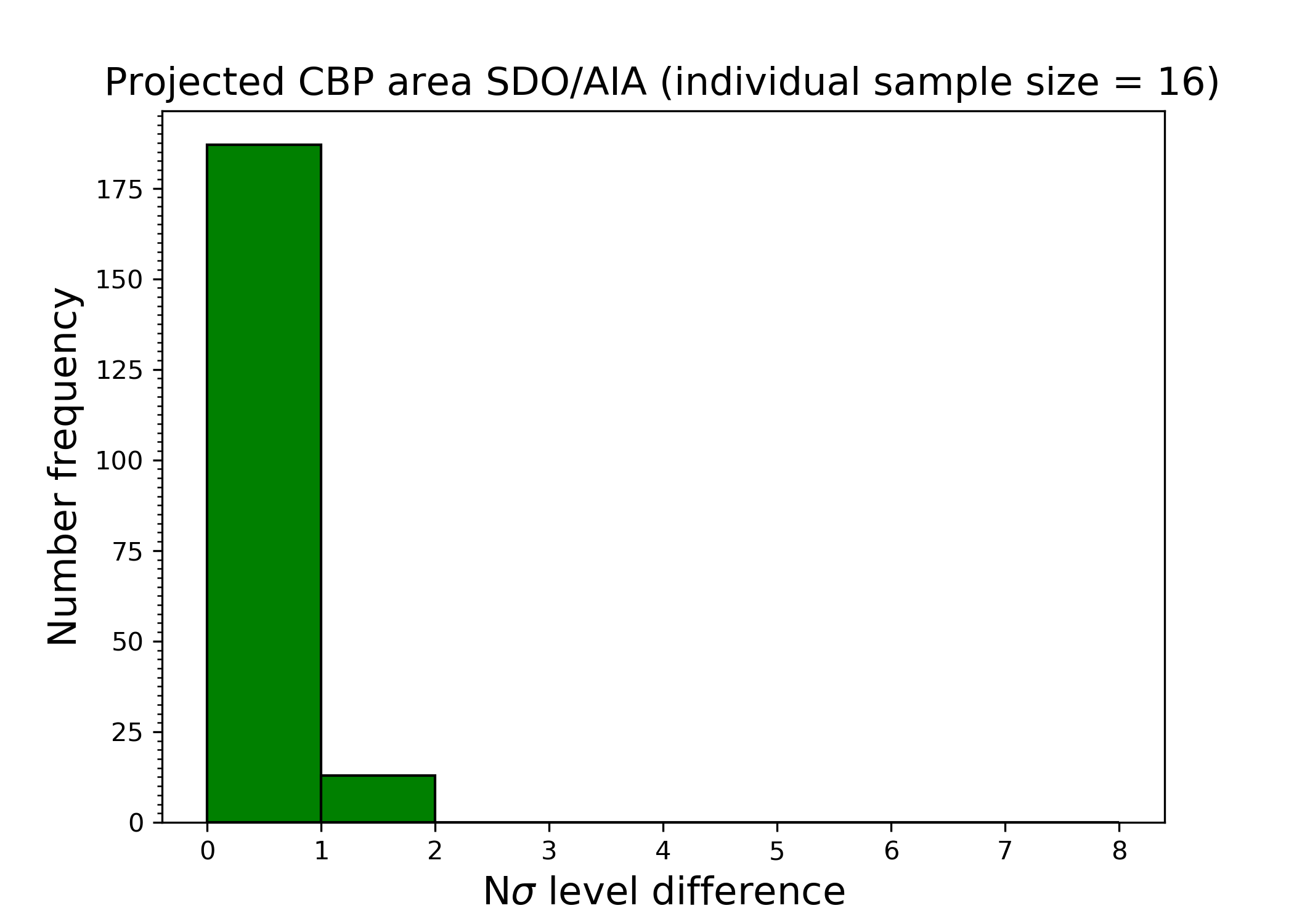}}\\
\subfloat{\includegraphics[width=0.36\textwidth]{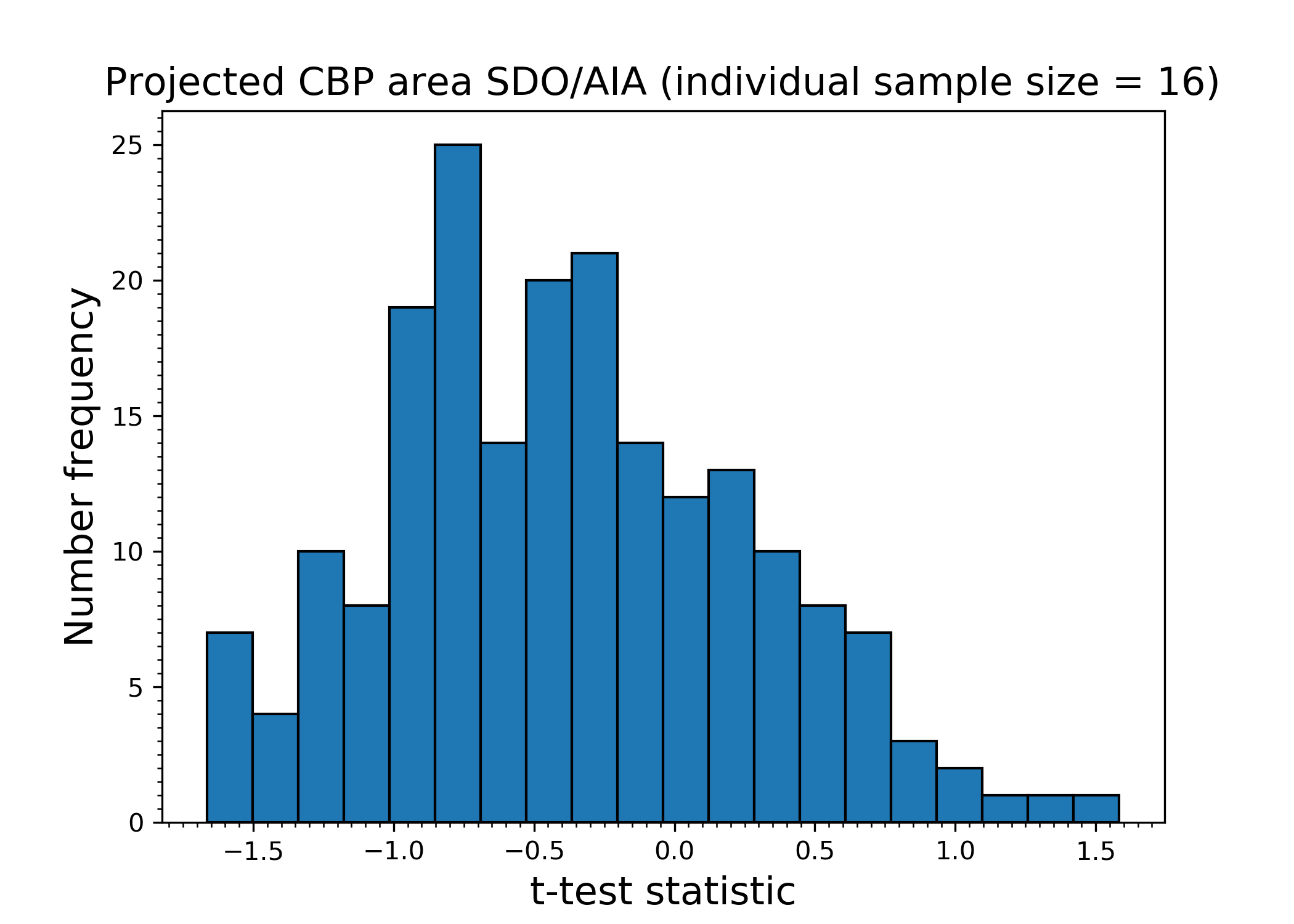}}
\subfloat{\includegraphics[width=0.36\textwidth]{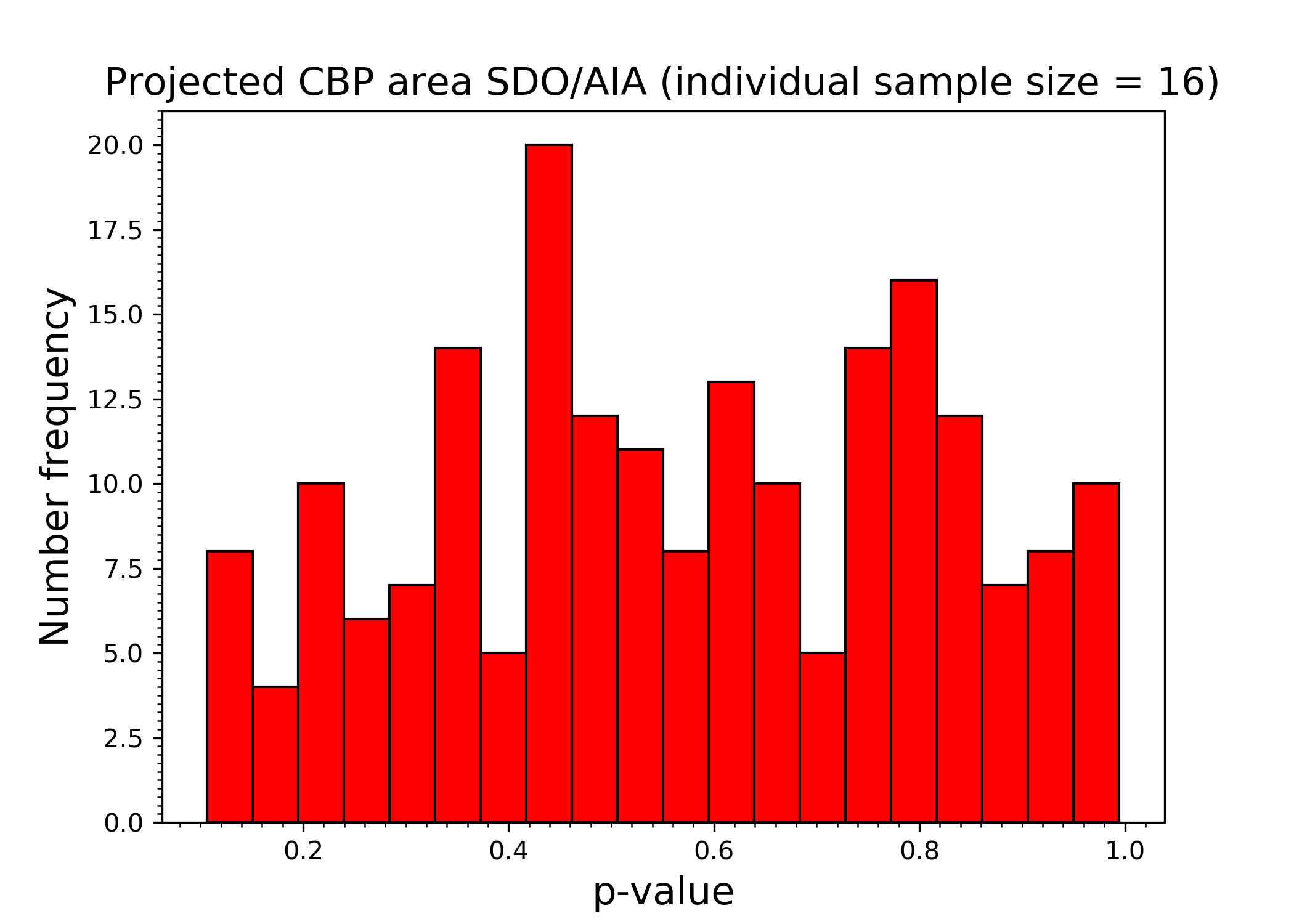}}
\caption{Same as Fig. \ref{mean_int_fig_ch1}, but for projected SDO/AIA 193 \AA\space CBP area and CH5, with individual CBP sample containing 16 CBPs.}
\label{area_fig_ch5_1}
\end{figure*}

\end{appendix}

\end{document}